\documentclass[3p,times,11pt]{elsarticle}

\usepackage[colorlinks,citecolor=blue,linktoc=all,linkcolor=cyan]{hyperref}
\usepackage{graphicx}
\usepackage{hyperref}

\usepackage[T1]{fontenc}
\usepackage{dsfont}               %
\usepackage{mathrsfs}             %
\usepackage{slashed}              %
\usepackage{amsmath}
\usepackage{amssymb}
\usepackage{amsbsy}
\usepackage{amsfonts}

\numberwithin{equation}{section}
\numberwithin{table}{section}
\numberwithin{figure}{section}

\journal{Progress in Particle and Nuclear Physics}

\usepackage{titlesec}
\usepackage{sectsty}
\titleformat{\section}{\normalfont\Large\bfseries}{\thesection}{1em}{}
\titleformat{\subsection}{\normalfont\large\bfseries}{\thesubsection}{1em}{}
\titleformat{\subsubsection}{\normalfont\normalsize\bfseries}{\thesubsubsection}{1em}{}
\usepackage[english]{babel}
\usepackage{graphics}
\usepackage{tabularx}
\usepackage{fancyhdr}
\usepackage{pifont}
\usepackage{float}
\usepackage{delarray}
\usepackage{rotating}
\usepackage{makeidx}
\usepackage[normalem]{ulem}
\usepackage{color}
\usepackage[multiple]{footmisc}
\usepackage{xspace}

\usepackage{wrapfig}
\usepackage{tikz}
\usetikzlibrary{patterns}
\usepackage{pgfplots}
\pgfplotsset{compat=1.12}

\usepackage{array}
\usepackage{longtable}
\usepackage{calc}
\usepackage{multirow}
\usepackage{hhline}
\usepackage{ifthen}
\usepackage{sectsty}

\usepackage{amsmath}
\usepackage{amsfonts}
\usepackage{amssymb}
\usepackage{mathrsfs}
\usepackage{dcolumn}%
\usepackage{bm}%
\usepackage{amsfonts}
\usepackage[caption=false]{subfig} %

\usepackage{grffile} %
\usepackage{lineno}
\usepackage{xcolor}

\newcommand*\Bell{\ensuremath{\boldsymbol\ell}}

\newcommand{\sqrtsnn}{\ensuremath{\sqrt{s_{_{NN}}}}}
\newcommand{\sqrtseN}{\ensuremath{\sqrt{s_{_{eN}}}}}
\newcommand{\sqrtsep}{\ensuremath{\sqrt{s_{_{ep}}}}}

\newcommand{\pT}{\ensuremath{P_T}\xspace}
\newcommand{\kT}{\ensuremath{k_T}\xspace}

\newcommand{\ccbar}{\ensuremath{c\overline{c}}\xspace}

\newcommand{\jpsi}{\ensuremath{J/\psi}\xspace}

\newcommand{\ups}{\ensuremath{\Upsilon}\xspace}
\newcommand{\upsg}{\ensuremath{\Upsilon(1S)}\xspace}
\newcommand{\upsp}{\ensuremath{\Upsilon(2S)}\xspace}
\newcommand{\upspp}{\ensuremath{\Upsilon(3S)}\xspace}

\newcommand{\pp}{\ensuremath{pp}\xspace}
\newcommand{\epem}{{\ensuremath{e^+e^-}}\xspace}

\newcommand{\ppbar}{\ensuremath{p\bar{p}}\xspace}
\newcommand{\pA}{{\ensuremath{pA}}\xspace}

\newcommand{\pPb}{\ensuremath{p\rm{Pb}}\xspace}

\newcommand{\PbPb}{\rm{PbPb}\xspace}

\newcommand{\Lint}{\ensuremath{\int \mathcal L}\xspace}

\newcommand{\invfb}{\ensuremath{\textrm{fb}^{-1}}\xspace}

\newcommand*{\eg}{{e.g.}\xspace}
\newcommand*{\ie}{{i.e.}\xspace}

\usepackage[normalem]{ulem}
\newcommand{\old}[1]{\!\xspace}
\newcommand{\new}[1]{{\color{red}#1}}
\newcommand{\newb}[1]{{\color{black}#1}}

\DeclareMathAlphabet{\pazocal}{OMS}{zplm}{m}{n}
\newcommand{\Q}{\pazocal{Q}}

\DeclareFontFamily{OT1}{pzc}{}
\DeclareFontShape{OT1}{pzc}{m}{it}{<-> s * [1.10] pzcmi7t}{}
\DeclareMathAlphabet{\mathpzc}{OT1}{pzc}{m}{it}

\newcommand{\ce}[1]{Eq.~(\ref{#1})}
\newcommand{\cf}[1]{{Fig.~\ref{#1}}}
\newcommand{\ct}[1]{{Table~\ref{#1}}}

\newcommand{\xB}{x_{\scriptscriptstyle B}}
\newcommand{\sT}{{\scriptscriptstyle T}}
\renewcommand{\d}{\mathrm{d}}

\def\ReA     {\ensuremath{R_{eA}}\xspace}

\def\ReAu    {\ensuremath{R_{e\rm Au}}\xspace}

\def\pPb  {$p\mathrm{Pb}$}

\def\lapproxeq{\lower .7ex\hbox{$\;\stackrel{\textstyle                                                    <}{\sim}\;$}}                        \def\gapproxeq{\lower .7ex\hbox{$\;\stackrel{\textstyle                                                    >}{\sim}\;$}}

\usepackage[shortcuts]{extdash} %

\newcommand{\cmark}{\textcolor{green}{\ding{51}}}%
\newcommand{\xmark}{\textcolor{red}{\ding{55}}}%

\begin{document}
\vspace*{-2cm}
\begin{frontmatter}
\title{Physics case for quarkonium studies at the Electron Ion Collider}

\date{\today}

\author[1]{Dani\"el~Boer\fnref{Editor}}
\author[2]{Chris A. Flett\fnref{Editor}}
\author[2,3,4]{Carlo~Flore\fnref{Editor}}
\author[5]{Daniel~Kiko\l a\fnref{Editor}}
\author[2]{Jean-Philippe~Lansberg\fnref{Editor}}
\author[2]{Maxim Nefedov\fnref{Editor}}
\author[2,6]{Charlotte~Van~Hulse\fnref{Editor}}

\fntext[Editor]{Editor}
\author[7]{Shohini Bhattacharya}%
\author[1,2]{Jelle Bor} %
\author[7b]{Mathias Butenschoen}%
\author[2,8]{Federico Ceccopieri}%
\author[9,10]{Longjie Chen}
\author[11]{Vincent Cheung}%
\author[4]{Umberto D'Alesio}%
\author[12]{Miguel Echevarria}%
\author[7,13]{Yoshitaka Hatta}%
\author[14]{Charles E.~Hyde}%
\author[12,15]{Raj Kishore}%
\author[16]{Leszek Kosarzewski}%
\author[17]{C\'{e}dric Lorc\'{e}}
\author[15,18]{Wenliang Li} %
\author[19]{Xuan Li}%
\author[1,4]{Luca Maxia}%
\author[20]{Andreas Metz}%
\author[21]{Asmita Mukherjee}%
\author[2]{Carlos Mu\~noz Camacho}%
\author[4]{Francesco Murgia}%
\author[17,22]{Pawel Nadel-Turonski} %
\author[4]{Cristian Pisano}%
\author[23]{Jian-Wei Qiu}%
\author[24]{Sangem Rajesh}%
\author[25]{Matteo Rinaldi}%
\author[26,27]{Jennifer Rittenhouse West}%
\author[28]{Vladimir Saleev}%
\author[29]{Nathaly Santiesteban}
\author[1,30]{Chalis Setyadi}
\author[31]{Pieter Taels}%
\author[7]{Zhoudunmin Tu}%
\author[19]{Ivan Vitev}%
\author[11,32]{Ramona Vogt}%
\author[33,34]{Kazuhiro Watanabe}%
\author[35,36]{Xiaojun Yao}%
\author[2,37]{Yelyzaveta Yedelkina}%
\author[9,10]{Shinsuke Yoshida}

\address[1]{Van Swinderen Institute for Particle Physics and Gravity, University of Groningen, 9747 AG Groningen, The Netherlands}
\address[2]{Universit\'e Paris-Saclay, CNRS, IJCLab, 91405 Orsay, France}
\address[3]{Dipartimento di Fisica, Universit\`a di Torino, and INFN Sezione di Torino, Via P.~Giuria 1, I-10125 Torino, Italy}
\address[4]{Dipartimento di Fisica, Universit\`a di Cagliari, and INFN Sezione di Cagliari, Cittadella Univ., I-09042 Monserrato (CA), Italy}
\address[5]{Faculty of Physics, Warsaw University of Technology, plac Politechniki 1,00-661, Warszawa, Poland}
\address[6]{University of Alcal\'{a}, Alcal\'{a} de Henares (Madrid), Spain}
\address[7]{Physics Department, Brookhaven National Laboratory, Bldg. 510A, Upton, NY 11973, USA}
\address[7b]{II. Institut f\"ur Theoretische Physik, Universit\"at Hamburg, Luruper Chaussee 149, 22761 Hamburg, Germany}

\address[8]{Universit\'e de Li\`ege, B4000, Li\`ege, Belgium}
\address[9]{Key Laboratory of Atomic and Subatomic Structure and Quantum Control (MOE), Guangdong Basic Research Center of Excellence for Structure and Fundamental Interactions of Matter, Institute of Quantum Matter, South China Normal University, Guangzhou 510006, China}
\address[10]{Guangdong-Hong Kong Joint Laboratory of Quantum Matter, Guangdong Provincial Key Laboratory of Nuclear Science, Southern Nuclear Science Computing Center, South China Normal University, Guangzhou 510006, China}
\address[11]{Nuclear and Chemical Sciences Division, Lawrence Livermore National Laboratory, Livermore, CA 94551 USA}
\address[12]{Department of Physics \& EHU Quantum Center, University of the Basque Country UPV/EHU, Apartado 644, 48080 Bilbao, Spain}
\address[13]{RIKEN BNL Research Center, Brookhaven National Laboratory, Upton, NY 11973, USA}
\address[14]{Department of Physics, Old Dominion University, Norfolk, VA 23529, USA}
\address[15]{Center for Frontiers in Nuclear Science, Stony Brook University, Stony Brook, NY 11794, USA}
\address[16]{Department of Physics, The Ohio State University, Columbus, Ohio 43210, USA}
\address[17]{CPHT, CNRS, Ecole polytechnique, Institut Polytechnique de Paris, 91120 Palaiseau, France}
\address[18]{Stony Brook University, Stony Brook, NY 11794, USA}
\address[19]{Los Alamos National Laboratory, Los Alamos, NM 87545, USA}
\address[20]{Department of Physics,  Temple University,  Philadelphia,  PA 19122, USA}
\address[21]{Department of Physics, Indian Institute of Technology Bombay, Powai, Mumbai 400076, India}
\address[22]{University of South Carolina, Columbia, SC 29208, USA}
\address[23]{Theory Center, Jefferson Laboratory, Newport News, VA 23606, USA}
\address[24]{Vellore Institute of Technology, Vellore, Tamil Nadu 632014, India}
\address[25]{Dipartimento di Fisica. Universit\`a degli studi di Perugia, and INFN Sezione di Perugia. Via A.~Pascoli, Perugia, 06123, Italy}
\address[26]{Lawrence Berkeley National Laboratory, Berkeley, CA 94720, USA}
\address[27]{University of California at Berkeley, Berkeley, CA 94709, USA}
\address[28]{Joint Institute for Nuclear Research, Dubna, Russia}
\address[29]{University of New Hampshire, Durham, NH 03824, USA}
\address[30]{Department of Physics, Universitas Gadjah Mada, BLS 21 Yogyakarta, Indonesia}
\address[31]{Department of Physics, University of Antwerp, Groenenborgerlaan 171, 2020 Antwerpen, Belgium}
\address[32]{Department of Physics and Astronomy, University of California at Davis, Davis, CA 95616, USA}
\address[33]{SUBATECH UMR 6457 (IMT Atlantique, Universit\'e de Nantes, IN2P3/CNRS), 4 rue Alfred Kastler, 44307 Nantes, France}
\address[34]{Faculty of Science and Technology, Seikei University, Musashino, Tokyo 180-8633, Japan}
\address[35]{Center for Theoretical Physics, Massachusetts Institute of Technology, Cambridge, MA 02139 USA}
\address[36]{InQubator for Quantum Simulation, Department of Physics, University of Washington, Seattle, Washington 98195, USA}
\address[37]{School of Physics, University College Dublin, Dublin 4, Ireland}

\vspace*{2cm}
\begin{abstract}
The physics case for quarkonium-production studies
accessible at the  US Electron Ion Collider is described.

\end{abstract}
\end{frontmatter}

\tableofcontents

\section{Introduction}
The Electron-Ion Collider (EIC) accelerator and detector systems are currently designed following the elaboration of an outstanding physics case aimed at further exploring the nucleon and nucleus partonic structure. The interested reader will find it useful to consult %
reviews {of the EIC}~\cite{Accardi:2012qut,AbdulKhalek:2021gbh}. %

Bound states of heavy quark-antiquark pairs, $Q\bar Q$, \textit{i.e.}~quarkonia, allow for a detailed study of  basic properties of quantum chromodynamics (QCD), the theory of the strong interaction. 
Indeed, charmonia and bottomonia have played a crucial role in the establishment of QCD as the theory of the strong interaction, given the clean signature they provide in different observables.
On the theory side, the main origin of the simplifications is the hierarchy $m_{Q}\gg \Lambda_{\rm QCD}$, with $m_{Q}$ the mass of a heavy quark, meaning that for processes {occuring} at this scale (or higher), a perturbative expansion {in $\alpha_s$} of QCD %
is allowed.
{In parallel,} the non-perturbative effects associated with the formation of the bound state can be factorised. %

Heavy quarkonia are multiscale systems. Besides $m_{Q}$ and $\Lambda_{QCD}$, one needs to consider, in addition, the scale of the typical momentum transfer between heavy constituent quarks ($m_Q v$), $v$ being the velocity of the heavy quarks in the rest frame of the bound state, and the scale of their binding energy ($m_Q v^2$), all of which become widely separated in the limit $m_{Q}\to \infty$.
At this point, the non-relativistic nature of the system comes into play, allowing for the development of different effective theories of QCD that attempt to more adequately describe the production of the bound state in the presence of different relevant scales as well as models such as the Colour-Singlet (CS) Model~\cite{Chang:1979nn,Berger:1980ni,Baier:1981uk,Baier:1983va} or Colour-Evaporation Model (CEM)~\cite{Halzen:1977rs,Fritzsch:1977ay}. 

The multitude of existing theoretical approaches to describe quarkonium production reflects the fact that unfortunately, up to now, there is no universal physics picture of this process accepted by the community that would provide a satisfactory description of all available experimental data~\cite{Lansberg:2019adr}. This complicates the 
{use} of quarkonia as tools for precision studies. Heavy quarkonia nevertheless remain useful to uncover new facets of the structure of nucleons and nuclei which we review in this document. 

{In this context,}  measurements of various {quarkonium-production} observables in {electron-proton~}($ep$) and {electron-nucleus~}($eA$) collisions at the EIC could provide crucial experimental clues to finally settle  the quarkonium-production-mechanism debate.

{Important targets for the EIC experimental programme are {vector-}quarkonium-polarisation observables and cross-section measurements of $C=+1$ states, like the $\eta_c$ and $\chi_{c,b}$. These play a central role in the current debate about the heavy-quarkonium-production mechanism and yet corresponding precise data from $ep$ collisions at HERA are simply lacking.
Such measurements would hopefully clear up the quarkonium-production debate and allow one to fully employ quarkonium data at the EIC as tools.}
{Before discussing how quarkonia can be used as tools to study nucleons, let us recall that the} multi-dimensional structure of nucleons is parameterised by different hadronic functions, which encode the dynamics of partons at different levels of complexity. These span from the one-dimensional (1D) parton distribution functions (PDFs), to the five-dimensional (5D) Wigner distributions --or generalised transverse-momentum-dependent distributions (GTMDs)--, to mention a few. {These also} incorporate a variety of spin and momentum correlations between the parton (or partons) participating in the hard subprocess and its (their) parent hadron. 
Depending on the considered scattering process and the measured kinematics, different hadronic functions enter the relevant cross sections. 
{Among them, let us cite the transverse-momentum-dependent PDFs (TMD PDFs or TMDs), arising from TMD factorisation~\cite{Collins:2011zzd} and which provide information on the distribution of partons inside the nucleon as a function of both their longitudinal 
and transverse momentum.}
{In} the case of quarkonium production at {transverse momenta, $\pT$, small compared to their mass} and for {specific} other kinematical end{-}point regions, new TMD functions {related to the produced quarkonium and referred to as shape functions}%
,  {are expected to} enter the cross-section formula besides the TMD PDFs of the initial-state hadron(s). This reflects the interplay between radiation of soft gluons and effects of the formation of the ${Q}\bar{Q}$ bound state. %
{Their impact on the phenomenology remains at present unknown}.

Much progress has been made in the determination of the above-mentioned PDFs, achieving different levels of success. Currently, the gluon distributions in general remain %
much less explored than their quark analogues. In this context, quarkonia arise as a powerful handle to remedy this situation since{, in the vast majority of 
situations, } the ${Q}\bar{Q}$ pair {at the origin of the quarkonium comes from photon-gluon (gluon-gluon) fusion in $ep$ (resp. $pp$) collisions, whereas deep inelastic scattering {(DIS)} or Drell-Yan-pair production are sensitive to gluon only through radiative QCD corrections.}  %

However, it has been shown~\cite{Collins:2007nk, Rogers:2010dm} that factorisation of observables {--cross sections, angular modulations, spin asymmetries,\dots--} in terms of TMD PDFs is less universal than that in terms of 
{standard} ({\it collinear})
PDFs and that {consequently}  {such a factorisation} could be violated {in back-to-back-hadron production} in proton-proton ($pp$) collisions. In $ep$ and $eA$ collisions, there is no anticipated violation of TMD factorisation, at least for 
{inclusive single-}hadron production, so {quarkonium} measurements  will likely be easier to interpret {in terms of gluon TMDs at the EIC} 
{rather than} at hadron colliders.    

{Quarkonia are also key players in exclusive reactions. This is not surprising as e}xclusive meson production involving a hard scale 
is one of the main processes to access Generalised Parton Distributions (GPDs). 
Gluon GPDs are in particular accessible via exclusive 
heavy-quarkonium production~\cite{Diehl:2003ny,Ivanov:2004vd}.
These GPDs provide information on the distribution of gluons inside the nucleon simultaneously as a function of their longitudinal momentum and their transverse position. They also provide information on the  angular momentum of the gluons inside the nucleon, about which very little is known to date. Furthermore, exclusive heavy-quarkonium production near the production threshold was suggested~\cite{Kharzeev:1995ij,Kharzeev:1998bz} as a tool for constraining the gluon condensate in the nucleon, itself linked to the nucleon mass, albeit {with} some unavoidable model dependence. 

At small momentum fraction, $x$, the differential exclusive electro- and photoproduction cross sections of quarkonia 
can be expressed in terms of particular products of integrals of GTMDs. 
{In single-quarkonium production, when a collinear expansion is applied,} the cross section reduces to expressions in terms of GPDs, {see for example \cite{Hatta:2017cte,Boer:2023mip}}. However, in general, {especially} beyond single-particle production, it %
provides {additional} information on GTMDs and offers an opportunity to learn more about the combined three-momentum and spatial distributions of gluons inside a nucleon. 
Moreover, while there is a direct relation between 
{exclusive} photoproduction case {in $ep$ collisions} and %
in ultra-peripheral %
$pp$ and $p$Pb collisions~(UPCs), %
{studies} at \newb{the} EIC would allow one to probe in more details the transverse-momentum dependence of the GTMDs. 

To date, {the detector simulations for} the EIC physics case connected to quarkonium physics has been limited to \jpsi\ { and $\Upsilon$} exclusive production %
as reported in the EIC Yellow report~\cite{AbdulKhalek:2021gbh}. Whereas, as we discussed above, quarkonium production is still the object of intense debates within the community\footnote{We guide interested readers to the following reviews~\cite{Kramer:2001hh,QuarkoniumWorkingGroup:2004kpm,Lansberg:2006dh,Brambilla:2010cs,ConesadelValle:2011fw} which address HERA and Tevatron results, to more recent ones~\cite{Andronic:2015wma,Lansberg:2019adr} addressing progress made thanks to the RHIC and LHC data and to to the HEPData database (\href{https://www.hepdata.net/}{\tt https://www.hepdata.net/}), a dedicated repository of quarkonium measurements up to 2012 (\href{http://hepdata.cedar.ac.uk/review/quarkonii/}{\tt http://hepdata.cedar.ac.uk/review/quarkonii/} to~\cite{Andronic:2013zim} and to~\cite{Tang:2020ame} for experimental quarkonium data.}, there is no doubt that it can play a crucial role in the scientific success of the EIC. As %
{was} recently done for the High Luminosity LHC phase~\cite{Chapon:2020heu}, we gather in this review what we believe to be the most complete list of quarkonium studies that can be carried out at the EIC along with their motivation.

The document is organised as follows. {In Section~\ref{sec:eh_general}, the EIC accelerator system and the first EIC detector, ePIC, as currently envisioned are presented. After a description of the kinematics of lepton-hadron collisions, the importance and a theoretical treatment of QED radiation are discussed, to then end with a note on feed-down from $b$-quark production in the study of charmonium. In Section~\ref{sec:why_quarkonia}, 
the different theoretical descriptions and measurements related to the production mechanism of quarkonia are presented. First, the various existing theoretical formalisms in collinear factorisation are discussed. Then, the legacy of existing measurements 
and the potential of future measurements at the EIC in constraining these formalisms are presented. Subsequently, it is shown that TMD observables can also teach us about quarkonium formation and polarisation and predictions for the EIC are made. Finally, the effect of final-state interactions on the production of quarkonium in lepton-nucleus colissions is touched upon.} Section~\ref{sec:ep} focuses on the studies accessible in electron-proton collisions in order to advance our knowledge of the nucleon partonic structure %
{and then} moves on to studies with nuclear beams, which is a unique feature of the EIC, and which will allow us to make a giant leap {forward} into a new precision era of the {partonic structure of nuclei.} %
{We underscore throughout this comprehensive review the diverse ways in which the EIC will utilise quarkonia to probe hadronic and nuclear physics and, conversely, will itself be a powerful tool for probing quarkonia.} 

\section{Generalities about quarkonium studies at the EIC}
\label{sec:eh_general}
\subsection{The proposed EIC accelerator system}
The {EIC}
{is an upcoming particle accelerator that will deliver} intense beams of longitudinally polarised electrons and polarised light nuclei (p, d, $^3$He) as well as unpolarised heavier ions, ranging up to uranium. It will produce electron-ion collisions at the highest energy and at the highest rate %
{ever achieved}. 

The EIC will be constructed in Brookhaven National Laboratory 
using a few key elements of the currently operating Relativistic Heavy Ion Collider (RHIC)~\cite{RHIC,Jacobs:2004qv}, such as the hadron ring
and the RHIC Electron-Beam-Ion-Source (EBIS)~\cite{Beebe:2015iga}.
The collider will be supplemented with a %
{new} electron ring, which will contain continuously injected,  polarised electrons with an energy from 5~GeV up to 18~GeV. 
The coverage in centre-of-mass energy will range from 
28~GeV to 141~GeV for 
 lepton-proton collisions, while for lepton-ion collisions an upper energy of 89~GeV/nucleon will be reached.
The %
{expected} instantaneous luminosity depends on the centre-of-mass energy and will range from $10^{33}$ to $10^{34}$~cm$^{-2}$s$^{-1}$ for electron-proton collisions,  with a maximum value expected for $\sqrtsep = 105$~GeV. For electron-ion reactions, it will be on the order of $10^{34}$~cm$^{-2}$s$^{-1}$. Such figures will correspond to integrated luminosities %
{of} the order of $10$ to $100$ fb$^{-1}$ per year. The designed average polarisation of electron, proton and $^3\text{He}$ beams is %
{of} the order of $70\%$.  

\begin{figure}[hbt!] %
\centering
\includegraphics[width=0.6\linewidth]{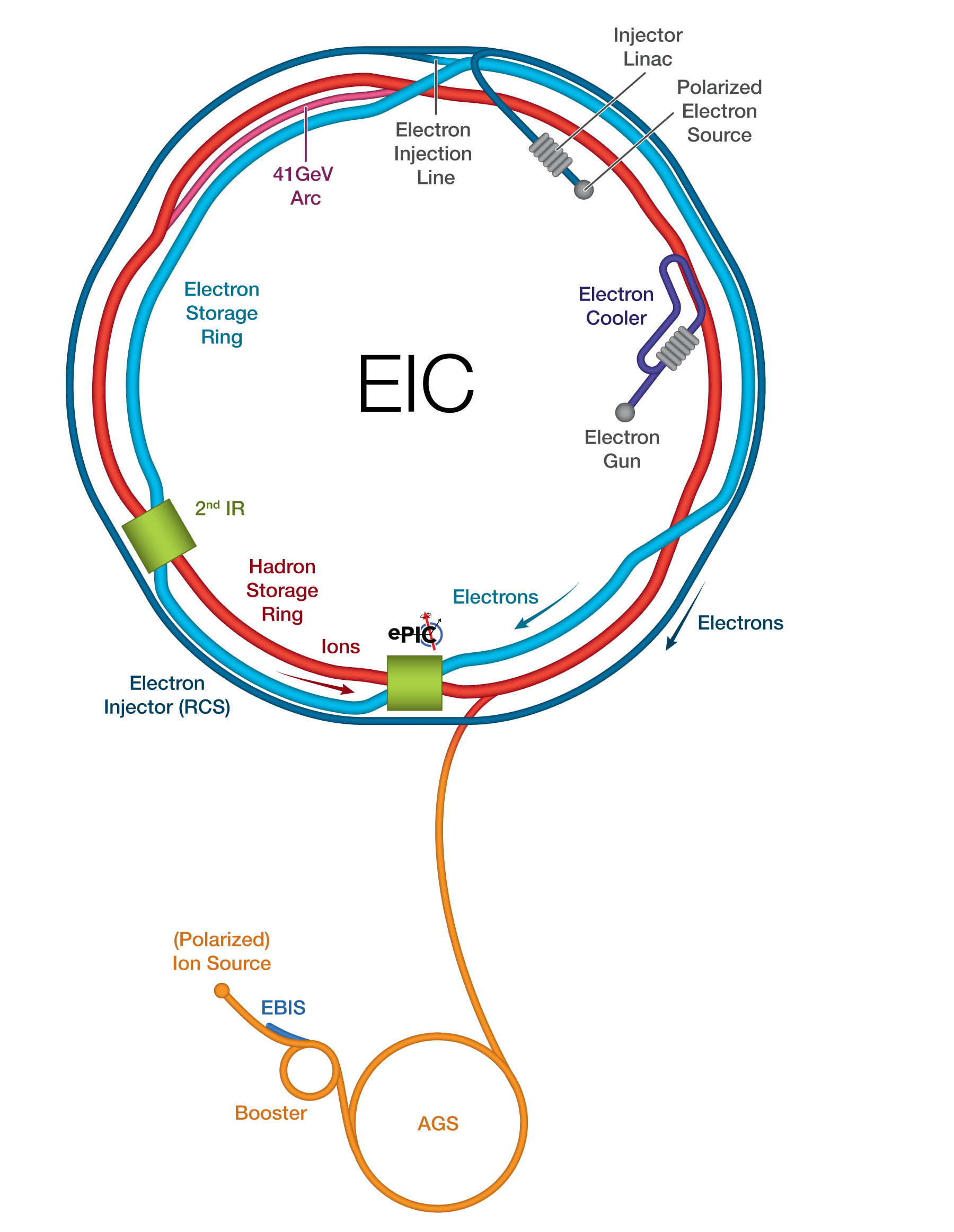}
\caption{A schematic drawing of the planned EIC~\cite{ePIC:web:page}. %
}
\label{fig:eic}
\end{figure}

At present, the installation of a first EIC detector is foreseen 
at interaction point 6 (IP6).
A second interaction point (IP8) can, at %
{any} stage, host a second and complementary detector. 
The second interaction point could accommodate a design with a secondary focus, which in combination with forward spectrometry would
allow for an extension of the acceptance towards the detection of particles at very small polar angles.
The interaction points will re-use the existing large detector halls, currently occupied by the STAR and sPHENIX experiments. 
The first collisions at the EIC are expected in {the} early 2030s.

\subsection{The proposed EIC detector}

\subsubsection{Requirements for an EIC detector in the context of quarkonium studies}

The specification of an EIC detector is determined by the kinematics of the electron-ion scattering (see \cf{fig:eic:generic:det}) and the observables and processes of interest. It should address the full range of physics outlined in the 
EIC White Paper~\cite{Accardi:2012qut}, the NAS report~\cite{NAP25171} and the EIC Yellow Report~\cite{AbdulKhalek:2021gbh}.
The basic requirements include~\cite{Willeke:2021ymc, AbdulKhalek:2021gbh} 4$\pi$ hermeticity with large 
acceptance in pseudorapidity, $\eta$, of about -4 < $\eta$ < 4, very good momentum resolution both in the central, forward and backward regions, %
{very good} energy resolution in electromagnetic calorimeters and particle identification capabilities up to $50$~GeV %
in momentum. Such a setup allows %
{one to study processes} over a wider range of four-momentum transfer $Q$. In addition, measurements of heavy-flavour hadron production demand a microvertexing detector that provides good impact-parameter resolution. 

The detector technologies and configuration implementation will be known once the detector design is finalised.
However, existing high-energy experiments (for example ALICE at the LHC and STAR at RHIC) indicate that an EIC detector that fulfils the aforementioned requirements will have capabilities for \jpsi\ and $\Upsilon(nS)$ measurements via their $e^+e^-$ decay channel~~\cite{STAR:2019vkt,STAR:2018smh,STAR:2020igu,ALICE:2021dtt,ALICE:2012vpz,Kosarzewski:2021knw}. 
The precision of quarkonium reconstruction will strongly depend on the hardware configuration. For example, an internal silicon tracker could generate additional combinatorial background arising from conversions $\gamma \to e^+e^-$, limiting precision for low-mass quarkonia at low \pT. Moreover, the energy loss of electrons due to Bremsstrahlung in the detector material deteriorates the mass resolution. It may complicate, if not make impossible, separation of {the} \upsg, %
\upsp and \upspp %
state{s}. Measurements of other quarkonium states (for instance $\chi_c$ or $\chi_b$) add constraints for the experimental apparatus. Studies of decays involving photon radiation (e.g., $\chi_{c}(1P) \to \jpsi + \gamma$) would require an electromagnetic calorimeter able to isolate a soft photon and measure its energy with appropriate resolution. In addition, a muon detector would significantly extend capabilities for quarkonium studies. This is %
{briefly} discussed in Sec.~\ref{sec:muon-detector}.

Three different designs, ATHENA~\cite{athena}, CORE~\cite{core} and ECCE~\cite{ecce}, were proposed.
The main difference between the ATHENA and ECCE design consists of the magnet, providing respectively a 3.0 T and 1.4 T magnetic field.
The distinguishing characteristic of the CORE detector is the compactness of the detector, obtained through exploitation of technological advances.
From the proposed designs, the ECCE proposal was selected as baseline for the first EIC detector, %
{with} improvements to the proposal %
at present under development.
This first EIC detector received the name electron-proton/ion collider (ePIC) detector.
A description of the ePIC detector in its current design state is given below.

\begin{figure}[htp] %
\centering
\includegraphics[width=0.6\linewidth]{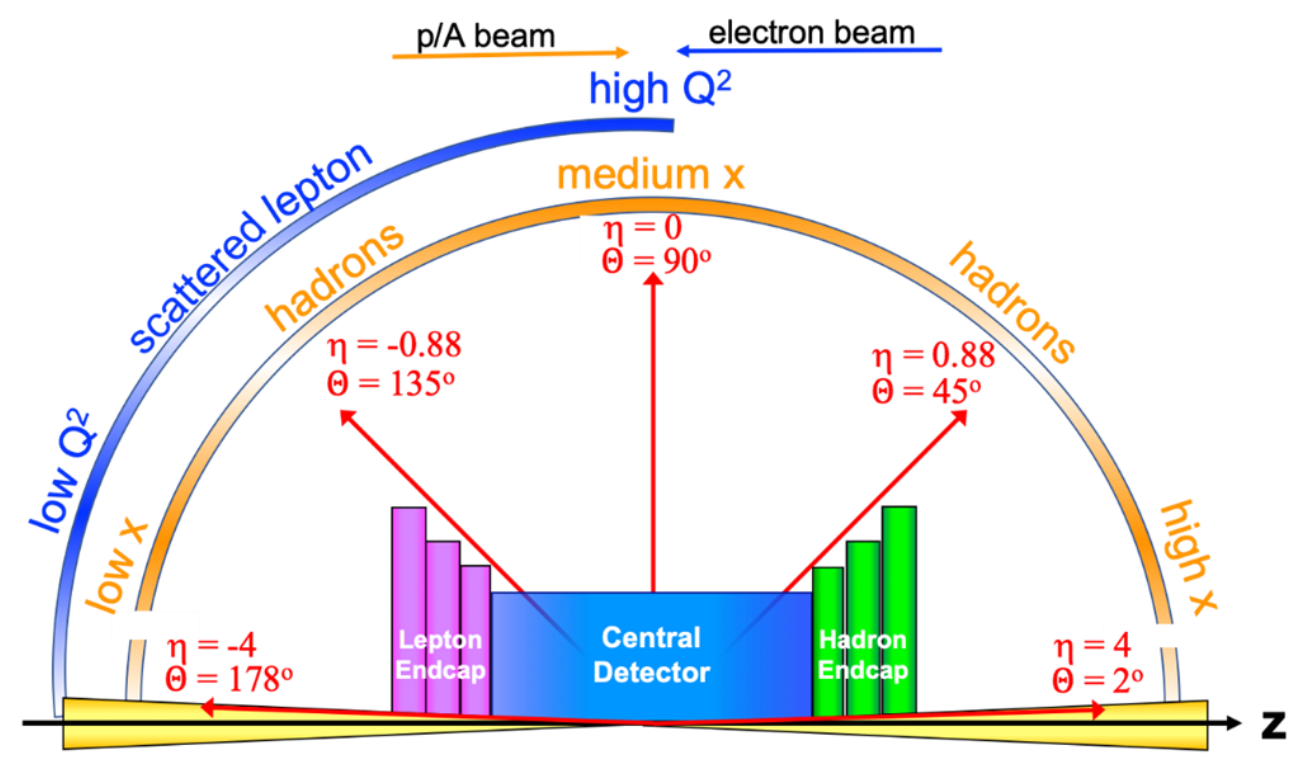} 
\caption{A schematic of the expected lepton and hadron kinematic distributions in EIC reactions and related detector requirements. Figure taken from~\cite{Willeke:2021ymc}.}
\label{fig:eic:generic:det}
\end{figure}

\label{sec:detector}

\subsubsection{The ePIC detector}

\begin{figure}[htbp]
\begin{center}
{\includegraphics[width=0.70\textwidth]{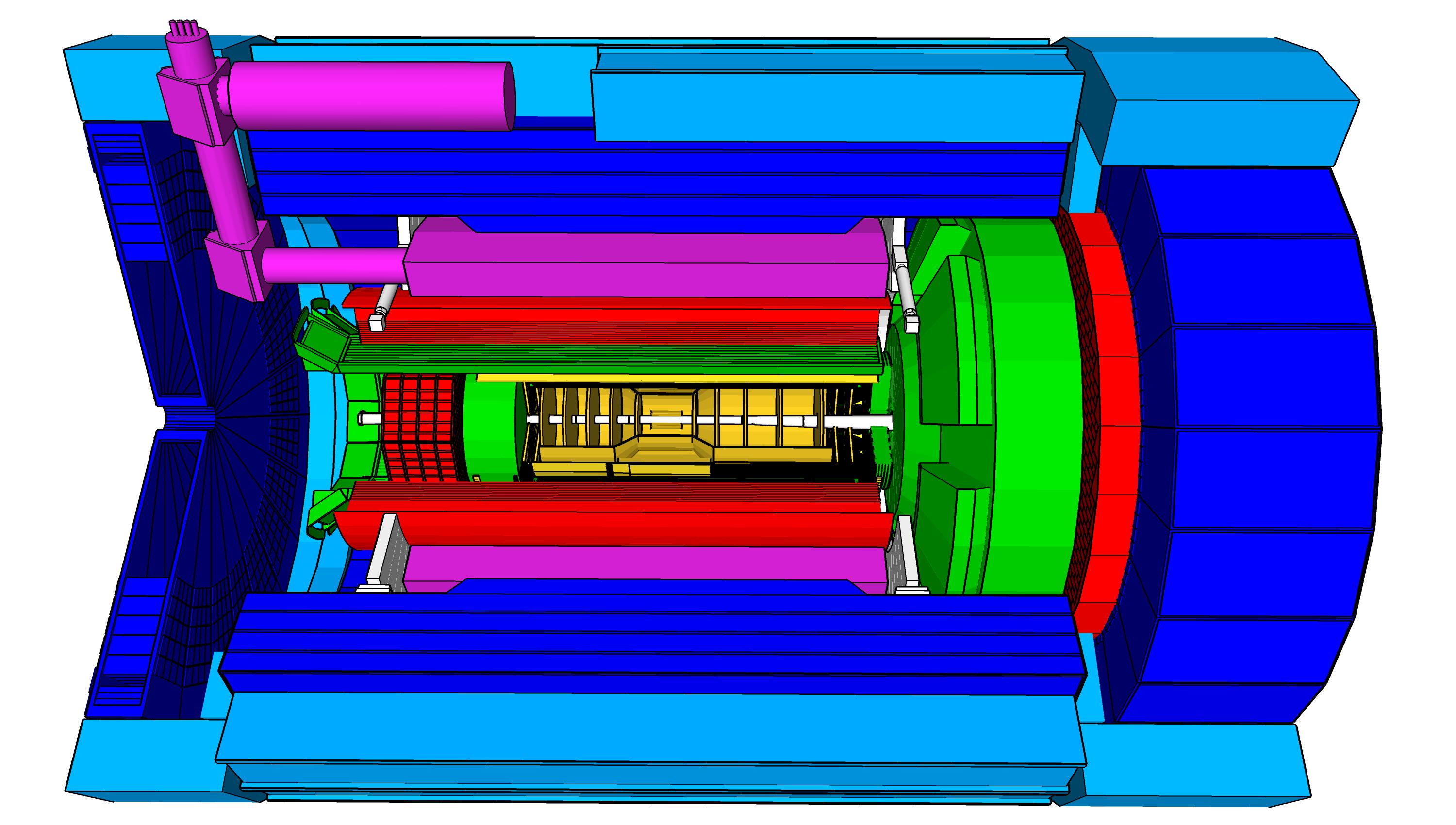}}
\caption{Drawing of the ePIC detector as %
{envisioned} at present~\cite{ePIC:web:page}.  %
{The hadron beam comes in from the left and defines the forward direction, while the lepton beam comes in from the right.} { The 1.7 T magnet is indicated in magenta; the tracking detectors are shown in yellow; the electromagnetic and hadronic calorimeters are represented, respectively, in red and darker blue, and the particle-identification Cherenkov detectors are drawn in green.} %
}
\label{fig:epic_det}
\end{center}
\end{figure}

The central barrel of the ePIC detector, as currently %
{envisioned}, is depicted in figure~\ref{fig:epic_det}. Here, the hadron
beam comes in from the left and defines the forward-going direction.
The central barrel is around %
{10} m long and %
{5} m in diameter, providing a full coverage
in azimuthal angle and a coverage in polar angle between $0^{\circ}$ and $178^{\circ}$, corresponding to a pseudorapidity coverage
between -4 and 4.
In addition to the barrel detector, detectors in the far-forward and far-backward regions are foreseen.
The far-backward region will contain a luminosity monitor and two detectors to tag
low-$Q^2$ events. The far-forward region will contain a series of detectors aimed at detecting particles produced close to the beam line and as such will be instrumental %
{to} the reconstruction of an extensive set of diffractive processes and tagged measurements, such as proton reconstruction in exclusive processes, tagging of the two spectator protons when investigating %
{the} neutron structure through lepton-$^3$He interactions and tagging {of} respectively the neutron and $\Lambda$\newb{-}baryon \newb{decay particles} when probing the pion and kaon %
structure %
in lepton-proton interactions. 
The far-forward system will consist of a B0 spectrometer, %
{containing} an electromagnetic calorimeter and trackers {for respectively the tagging of photons and reconstruction of charged particles},
Roman Pots %
{and} off-momentum detectors, %
{performing charged-particle reconstruction}, %
{as well as} 
Zero-Degree Calorimeters, %
{capable of detecting} photons and neutrons.

In the central barrel, track and vertex reconstruction will be performed by
silicon monolithic active pixel sensors placed close to the beam line {and interaction point}, while at a further distance
micro-pattern gaseous detectors (%
micro-Resistive Well %
{and} Micro-Mesh Gaseous Structure) and AC-coupled low-gain
avalanche diodes will contribute to track reconstruction.
The tracking detectors will be embedded into a 1.7~T magnetic field.
Such {a} setup will provide the momentum resolution needed to %
{fulfil} the EIC physics programme.

Electromagnetic calorimeters cover the backward, central and forward regions of the central barrel, providing
electron and photon detection as well as hadron suppression.
In the backward region, a high-precision lead-tungstate calorimeter read out by silicon photo-multipliers is foreseen.
The detector will be %
{critical to} the reconstruction of (scattered) electrons,
improving the reconstruction precision over that obtained from tracking detectors only,
and in the identification of these electrons, by suppressing the background contribution strongly.
This contribution originates mostly from charged pions.
In the central region, %
a lead-scintillator imaging calorimeter is foreseen.
For the forward region, an electromagnetic calorimeter will be integrated with the forward hadronic calorimeter.
The system %
{focuses} on the containment of high-energetic particle showers while at the same time providing a
good energy resolution for lower-energetic particles.
Particle identification requires %
{a good position resolution, in particular in the electromagnetic calorimeter}. This will be provided by constructing the electromagnetic calorimeter
out of segments{,} of scintillating fibres embedded in tungsten powder\newb{,} 
smaller than the Moli\`{e}re radius. This will also result in a good shower %
{separation} at high pseudorapidity.

{In the central region, a hadronic calorimeter }%
 will allow for the detection of neutral hadrons and as such {will} improve the resolution of jet reconstruction. %
Given the good momentum resolution of the central trackers, the central hadronic calorimeter system will not have
an impact on the reconstruction of charged particles.
The forward hadronic calorimeter, which forms an integrated system with the electromagnetic calorimeter, will
consist of layers of alternating tiles of scintillating material and steel, while towards the end of the detector
the steel is replaced by tungsten in order to serve as %
tail catcher of the shower and thus
maximise the interaction length within the available space.
{Also in the backward region, an hadronic calorimeter will be installed, with the aim to serve as tail catcher of particle signals.}

Detectors based on the detection of Cherenkov light will be used for the identification of
charged pions, kaons and protons, while also contributing to electron identification.
In addition, the %
{aforementioned} AC-coupled low-gain
avalanche diodes will provide particle identification in the low-momentum region, below $\sim$ 2~GeV, based on the detection of the
time of flight of a particle.
In the backward region, a {proximity-focusing} ring-imaging Cherenkov (RICH) detector with aerogel as radiator %
will be used. %
Because of the tight space constraints, a
DIRC -- detection of internally reflected Cherenkov light -- detector will be incorporated  in the central region.
The forward region will contain a dual RICH detector, with %
{an aerogel radiator}  for the low-momentum particles
and C$_2$F$_6$ for the high-energetic ones, covering the momentum range up to 50~GeV. 

No muon detectors are foreseen for the ePIC detector. While first studies, performed for the ATHENA and ECCE proposals, indicate that the reconstruction of %
{\jpsi} mesons from exclusive processes through their $e^+e^-$ decay should be possible with the ePIC detector, there are neither studies for other quarkonium states nor for inclusive or semi-inclusive processes. 
Here, dedicated	muon detectors might be	needed.	This is	discussed in the following sub-section.

\subsubsection{The case for a muon detector for quarkonium studies at the EIC}
\label{sec:muon-detector}

Measurement of vector-quarkonium production using their di-muon decay provides significant benefits. The energy loss of muons due to interaction{s} with detector material is much smaller than that of electrons. This leads typically to a better momentum resolution 
of the muons than of the electrons{,} and 
therefore the resolution of the quarkonium mass reconstructed in the $\mu^+\mu^-$ channel is better compared to the $e^+e^-$ one. The LHCb and CMS experiments provide a case in point as the performance of their muon detectors facilitated a rich and fruitful quarkonium physics program, which included that of %
\upsg, \upsp, \upspp and other quarkonium states such as the $\chi_c$ and $\chi_b$ via their radiative decays into vector quarkonia. Additional measurement{s} via the $\mu^+\mu^-$ decay channel would also essentially double the available statistics as the branching ratios into $\mu^+ \mu^-$ and $e^+e^-$ are %
{nearly} the same and enable analyses of rare decays (for example, $\chi_c \to \jpsi \mu\mu$). With a proper design,  
 studies via the di-muon channel benefit from a lower combinatorial background, thus improving the statistical precision of the measurement. In addition, they provide a cross check %
 {of the} $e^+e^-$ results, which should in turn reduce  systematic uncertainties.  

In summary, a muon detector would significantly extend capabilities for quarkonium studies at the EIC. The present ePIC design does not consider muon-identification instrumentation, but
possibilities for an enhanced muon identification can be investigated for ePIC. Moreover, 
the incorporation of dedicated muon-identification detectors in
the design phase of the $2^{\rm nd}$ EIC detector can 
{vastly extend} quarkonium measurement capacities {in the manner described above}. %

\subsection{Kinematics and QED radiative corrections}

\subsubsection{Kinematics of electron-hadron reactions}
\label{sec:kinem-SIDIS}
In this section, we collect %
basic kinematical definitions useful for the description of lepton-hadron reactions. The next section is devoted to how QED radiative corrections on the lepton side can affect the resolution on various kinematic variables and {to} possible ways to address this problem.

{Let us consider} the {inclusive} production of an identified hadron ${\cal H}$, which in the context of this review is most likely to be a quarkonium, in electron-{nucleon} ($eN$) scattering:
\begin{equation}
e(\ell) + N({P_N}) \to e(\ell^{\prime}) + {\cal H}(P_{\cal H}) + X,
\label{eq:process:e-pA}
\end{equation}
{For electron-nucleus ($eA$) scattering}, the momentum ${P_N}$ usually denotes the average momentum of a single nucleon. %
{ Depending on the experimental possibilities, one can}  %
tag the outgoing electron with the momentum $\ell^\prime$ or 
consider the reaction inclusive w.r.t. the final-state electron. If the {momentum} $\ell^\prime$ has been measured, %
{one can define} the momentum transfer $q=\ell-\ell^\prime$ with $q^2 = -Q^2$ and the following Lorentz-invariant kinematic variables become experimentally accessible:
\begin{equation}
\xB = \frac{Q^2}{2P_N\cdot q}\,, \qquad y = \frac{P_N\cdot q}{P_N\cdot \ell}\, , \qquad z = \frac{P_N\cdot P_{\cal H}}{P_N\cdot q}\,{,} \label{eq:SIDIS:kin-var}
\end{equation}
{where} {$z$ is referred to as the elasticity and $y$ as the {inelasticity} %
{that} should not be confused with the rapidity}.
\begin{figure}[hbt!]
    \centering
    \includegraphics[width=0.45\textwidth]{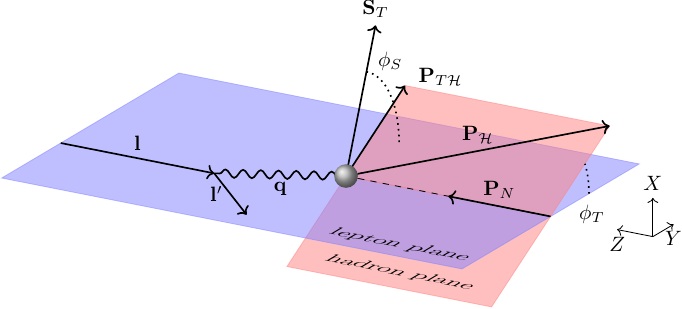} 
    \includegraphics[width=0.45\textwidth]{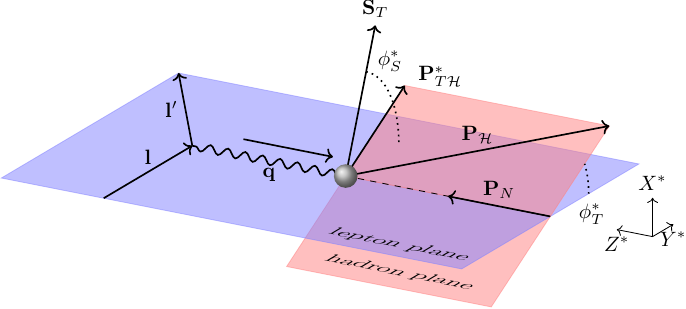}
    \caption{{Momenta and angles} in two reference frames commonly used to describe the (SI)DIS process: {\it laboratory frame} (left) and {\it photon-hadron {frame}} (right). 
    }
    \label{fig:eP_frames}
\end{figure}

Among frame-dependent variables, one usually distinguishes the transverse momentum of the hadron ${\bf P}_{T{\cal H}}$ in the {\it laboratory frame} (see %
\cf{fig:eP_frames} {(left)}{)}, in which the initial electron $e(\ell)$ and %
{nucleon $N(P_N)$} collide head on, defining the ${Z}$ (collision) axis, from the transverse momentum ${\bf P}_{T{\cal H}}^{*}$  of the hadron ${\cal H}$ in the {\it photon-hadron frame} (see %
\cf{fig:eP_frames} {(right)}), where three-momenta ${\bf q}$ and ${\bf P}_\newb{N}$ are aligned with the $Z$ axis of this frame\footnote{Different photon-hadron frames are related by %
{a} boost along {the} $Z$ axis. In particular, one can adopt the {\it photon-nucleon {centre}-of-momentum frame} where ${\bf q}+{\bf P}_N=0$. The transverse components of {the} momenta are the same in all photon-hadron frames.}. The word ``photon'' in the {frame} name %
specifically refers to the one-photon-exchange approximation between the electron and the hadronic part of the process.  {In this review, we will often use the simplified notation  for the absolute value of the transverse momentum of the produced hadron: $\pT=|{\bf P}_{T{\cal H}}|$ or $
\pT^*=|{\bf P}^*_{T{\cal H}}|$.}

Even if the colliding particles are %
{unpolarised}, there could always be some dependence of the cross section on the azimuthal angle $\phi_T$ (or $\phi^*_T$) formed %
{by} the vector %
{${\bf P}_{T{\cal H}}$} (or ${\bf P}^{*}_{T{\cal H}}$) and the plane spanned by {the} initial ($\Bell$) and final ($\Bell'$)  lepton three-momenta (\cf{fig:eP_frames}), due to the exchanged-photon polarisation. If the initial %
{nucleon} and/or electron have transverse polarisation, {\it additional} angular modulations of the cross section, related \newb{to} the direction(s) of \newb{the} transverse spin vector(s) of \newb{the} colliding particles, can be generated. {The transverse polarisation vector of the initial \newb{nucleon} is denoted as ${\bf S}_T$ and the angle of this vector with respect to the lepton plane in the photon-hadron (resp. laboratory) frame is generally indicated as $\phi^*_S$ (resp. $\phi_S$).}   

If the recoil effects of {the} photons which {can} be emitted by {the} initial and final electrons during the scattering process (QED radiative corrections) %
are neglected, then the four-momentum of {the} exchanged photon is simply $q=\ell-\ell^\prime$ as stated above. In such an approximation, the variables {of \ce{eq:SIDIS:kin-var}} as well as {the} frame-dependent variables, such as ${\bf P}_{T{\cal H}}^{*}$, can be directly computed  from the measured energy and momentum of the scattered electron. However, such a {QED} Born approximation %
{might} be insufficient for precision studies. %
{Section}~\ref{sec:QED-corr} is devoted to this issue. 

The regime of the process of \ce{eq:process:e-pA}, when the quasi-real-photon approximation can be applied to the exchanged photon, \ie\ 
{when $Q$ is negligible compared to the hard scale ($m_Q$, $P_T$, $P_T^*$, ...)}, is commonly referred to as {\it photoproduction}, while the regime with \newb{$Q$ being the hard scale, or among the potential hard scales,} is called \newb{leptoproduction or} {\it (semi-inclusive) deep inelastic scattering} (SIDIS). Experimentally, photoproduction is usually defined by a fixed cut on the photon virtuality, \eg $Q < 1$ GeV. %

Beside the %
{well-known} regimes of photoproduction and {leptoproduction (or }{SIDIS)}%
, which {a priori} require %
setting some constraints on $Q^2$, it appears very valuable for quarkonium studies to consider measurements where $Q^2$ is fully integrated {over}. Such yields then contain the contributions from both quasi{-}real and off-shell photons. This proposal is described in %
section~\ref{sec:inclusive-lepto-prod-Jpsi}. 

As it was mentioned in the introduction, polarisation observables play an important role in quarkonium physics. The polarisation parameters of a spin-1 heavy quarkonium $\lambda_\theta$, $\mu_{\theta\phi}$ {and} $\nu_{\theta\phi}$ parametrise the angular distribution of decay leptons in the quarkonium rest frame:
\begin{equation}
	{{\rm d}\sigma \over {\rm d} \Omega} \propto 1 + \lambda_{\theta} \cos^2\theta + \mu_{\theta\phi} \sin 2\theta \cos \phi + {{\nu_{\theta\phi}} \over 2} \sin^2 \theta \cos 2\phi. \label{eq:pol-par-def}
\end{equation}
These parameters depend on the orientation of the axes of the coordinate system chosen in the quarkonium rest frame with popular frame choices such as the Helicity, Collins-Soper, Gottfried-Jackson and target frames (see \eg Section 2.3 of~\cite{Andronic:2015wma}). {The same definition of polarisation parameters holds for the case of exclusive production of a vector quarkonium. }

\subsubsection{On the importance of QED corrections}
\label{sec:QED-corr}

The possibility to make a distinction between {the} photoproduction and 
{electroproduction (or SIDIS)} %
regimes, together with {the} rich phenomenology provided by measurements differential in {the} variables $\xB$, $y$, $z$ as well as %
{$\pT^*$} and $\phi^{*}_{T}$, has %
{always been} considered as an advantage of lepton-hadron reactions over hadron-hadron ones. 

\begin{figure}[hbt!]
    \centering
    \includegraphics[width=0.45\textwidth]{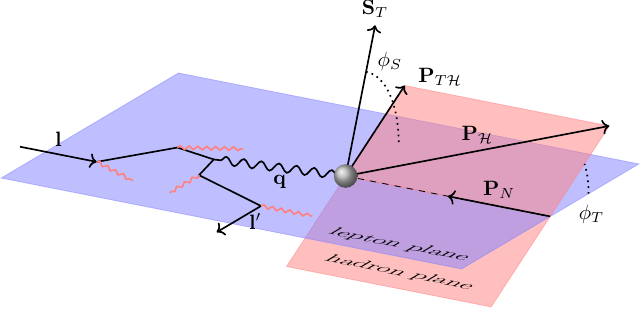}
    \caption{{Sketch of the kinematics of the process $e(\ell) + N(P_N) \to e(\ell^{\prime}) + {\cal H}(P_{\cal H}) + X$   including collision-induced photon emissions from the initial and final-state leptons. Note that the lepton and hadron planes are not (by definition) affected by such radiations, unlike the off-shell-photon momentum $q$ which is not any more in the lepton plane. Along the same lines, it is impossible to experimentally determine the photon-hadron frame (Fig 2.4 (right)), where $q$ is aligned with the $Z^*$ axis, by only measuring $\ell$ and $\ell'$.}
    }
    \label{fig:eP_bremsstrahlung}
\end{figure}

However{,} the emission of photons by initial- and final-state leptons modif{ies} the relation between the momentum $\ell^{\prime}$ of the final-state lepton {\it %
{measured} in the detector} and the four-momentum $q$ of the photon exchanged between {the} leptonic and hadronic parts of the process in \ce{eq:process:e-pA} (see the Fig.~\ref{fig:eP_bremsstrahlung}) {which in turn modifies the Lorentz-invariant variables} %
(\ce{eq:SIDIS:kin-var}) as well as ${\bf P}_{T{\cal H}}^{*}$ and $\phi^{*}_{T}$. Beyond the Born approximation of QED, this relation is no longer simply $q=\ell-\ell^\prime$ but includes {the} recoil from emitted photons. For strictly inclusive DIS measurements, as opposed to SIDIS, the application of QED radiative corrections boils down to an overall radiative correction factor to the cross section differential in $x_B$ and the inelasticity $y$ \cite{Akhundov:1994my}. In the SIDIS case, fully differential Monte-Carlo computations have to be performed, using dedicated tools such as DJANGOH~\cite{Charchula:1994kf}.

Recently, it has been shown~\cite{Liu:2021jfp} that QED radiative corrections fundamentally limit the accuracy of %
SIDIS measurements, in particular for the kinematic regime where the TMD factorisation 
is needed. {In this context,} a new approach to their treatment has been proposed. The QCD factorisation for {the} SIDIS cross section was historically discussed in the photon-hadron frame. However, as it was mentioned above, the collision-induced photon radiations change both the direction and magnitude of the exchanged virtual photon, making the photon-hadron frame and the quantities related {to} it only approximately defined.  %
The ambiguities in the definition of kinematic variables in photon-hadron frame can impact our ability to extract the TMDs and, in particular, to use the angular modulation in $\phi_T^{*}$ to separate contributions of different TMD PDFs and fragmentation functions (FFs). Since the QED radiation{s differently} affect the determination of \newb{the} angles $\phi^*_T$ and $\phi^*_S$ (\cf{fig:eP_frames}), this can affect the determination of various azimuthal (spin) asymmetries~\cite{Liu:2021jfp}. In addition to the uncertainty of the ``photon-hadron'' frame, the collision-induced photon radiation{s} also change %
the true values of $x_B$ and $Q^2$. %

 Although {the effects of the} QED radiation{s} could be calculated perturbatively, the main point of concern are those QED radiative correction effects %
 which are logarithmically enhanced due to {the }collinear and infrared sensitivity coming from the smallness of the electron mass $m_e$ %
 compared to all {the} other scales of the process.  Omitting these effects may lead to significant uncertainties in some kinematic regimes %
 where %
 {a wide} phase space is available for collision-induced radiation{s}, such as those relevant to the study of small-$x$ physics {or for} %
 two-scale observables {described by TMD factorisation}. %

In Ref.~\cite{Liu:2021jfp}, it has been argued that {a} combined QCD+QED factorisation can be performed such that the exchanged photon momentum $q$ is not fixed by the {\it measured} $\ell-\ell'$, but rather has a range of values %
{to be} integrated over.  The range is determined by the observed momentum of the scattered lepton for inclusive DIS and the momenta of both {the} scattered lepton and {the} observed final-state hadron for %
SIDIS. %
{The approach consists} %
{of} using collinear factorisation to take into account the collision-induced-QED-radiation effects {which are} enhanced by large logarithms of either $Q/m_e$, $|{\bf P}_{T{\cal H}}|/m_e$ or $|\Bell_T'|/m_e$, while either collinear or TMD factorisation can be used %
{to account for} QCD contributions depending on the hierarchy between the $|{\bf P}_{T{\cal H}}-\Bell_T'|$ and the hard scale $Q$. For the SIDIS process %
{of} \ce{eq:process:e-pA} on a proton target, the hybrid factorisation formula is given by~\cite{Liu:2021jfp}: 
\begin{eqnarray}
E_{\ell^{\prime}} E_{P_{\cal H}}
\frac{d\sigma_{\rm SIDIS}}
{d^3\Bell^{\prime} d^3{\bf P}_{\cal H}} 
&\approx & \sum_{a,b} \,
\int_{\zeta_{\rm min}}^1\frac{d\zeta}{\zeta^2}\, D_{e(\ell^{\prime})/b(k')}(\zeta,\mu_F^2)
\int_{\xi_{\rm min}}^1 d\xi\, f_{a(k)/e(\ell)}(\xi,\mu_F^2)  
\nonumber\\
&& \times\,
\left[E_{k^{\prime}}E_{p_{\cal H}} 
\frac{d \sigma^{ap}[a(k)+p(P) \to b(k^{\prime})+{\cal H}(p_{\cal H})+X]}
{d^3{\bf k}^{\prime}d^3{\bf P}_{\cal H}}
\right]_{k=\xi\ell,k^{\prime}=\ell^{\prime}/\zeta}, 
\label{eq:SIDIS-factor-QED}
\end{eqnarray}
where $a,b=e, \bar{e}, \gamma$%
, and where the active lepton/photon momenta entering or leaving the hard collision %
are defined as $k=\xi \ell$ and $k'=\ell'/\zeta$ with collinear momentum fractions $\xi$ and $\zeta$, and $\mu_F$ is the factorisation scale. The process-independent lepton distribution functions (LDFs) $f_{a/e}(\xi)$ and lepton fragmentation functions (LFFs) $D_{e/b}(\zeta)$ 
in Eq.~(\ref{eq:SIDIS-factor-QED}) resum logarithmically-{enhanced} QED contributions in the limit when the hard scale, $\max (Q,|{\bf P}_{T{\cal H}}|,|\Bell_T'|)$, {is much larger than} $m_e$. The non-logarithmically{-enhanced} part of QED radiative corrections can be included into ${d}\sigma^{ap}$ %
{order by order} in powers of $\alpha_{\rm em}$. 

The differential cross section %
${d}\sigma^{ap}$ in the second line of \ce{eq:SIDIS-factor-QED} can be further factorised by TMD or collinear factorisation in QCD depending on if the observed lepton and hadron are in the back-to-back regime or not.  {As it has been demonstrated in Refs.~\cite{Liu:2020rvc,Liu:2021jfp}, the transverse-momentum broadening from {the} collision-induced QED radiation{s} is much smaller than {the} TMD effects from QCD. Factorising out QED radiations using collinear LDFs and LFFs as done in the \ce{eq:SIDIS-factor-QED} is {therefore} a good approximation.} \ce{eq:SIDIS-factor-QED} is valid up to {Leading Power (LP), that is up to} %
{power corrections} %
{scaling as the inverse of} the hard scale. Note that the same kind of equation holds in the case of $e-A$ collisions. Note also that \ce{eq:SIDIS-factor-QED} does not account for possible hadronic/resolved contributions from the photon.

Due to the smallness of $\alpha_{\rm em}$, $\sigma^{ap}$ in \ce{eq:SIDIS-factor-QED} can be approximated  by its {QED Born} %
order, $\sigma^{ap,(0)}$ %
with $a=b=e$.  This lowest order {cross section} is %
the same as the SIDIS cross section without QED radiation {which can be parametrised in terms of the usual SIDIS structure functions~\cite{Bacchetta:2006tn}} but with different kinematics: $\ell\to k=\xi\ell$ and $\ell'\to k'=\ell'/\zeta$. Consequently, the exchanged-virtual-photon momentum between the scattered lepton and the colliding hadron is modified as $q=\ell-\ell' \to k-k'=\xi\ell - \ell'/\zeta$. %
By neglecting higher order QED %
{contributions} to $\sigma^{ap}$ in \ce{eq:SIDIS-factor-QED}, the SIDIS cross section {\it with} the %
{collision-induced} QED radiation can {thus} be obtained from the same SIDIS cross section {\it without} QED radiation plus the knowledge of the universal LDFs and LFFs. %

\subsection{On the importance of $b$ feed down}

 An important and subtle  %
 {{concept} %
 needed to understand} the quarkonium-production mechanism is the knowledge of feed downs. For instance, as shown in Ref.~\cite{Flore:2020jau}{,} in the case of $J/\psi$ photoproduction at HERA, not all the $J/\psi$ are %
 produced by the hard scattering. Indeed, a non-negligible fraction of the $J/\psi$ mesons produced at large $P_T$ come{s} from the {decay of a $b$ quark}. %
 \cf{fig:H1-data-ratio-b-FD} shows the fraction of $J/\psi$ coming from {such a} $b$ feed down {(also referred to as non-prompt yield)} as a function of $P_T^2$ in the H1 kinematics. We guide the reader to Appendix A of Ref.~\cite{Flore:2020jau} for more information about how it was estimated. 
One sees that the fraction of non-prompt $J/\psi$ steadily grows to reach over $40\%$ of the $J/\psi$ yield at the highest reachable $P_T \lesssim 10$ GeV.
Although the top energy of the EIC will be at most at $\sqrt{s_{ep}} = 140$ GeV, %
given the much higher luminosity of the EIC compared to HERA, the $W_{\gamma p}$ reach\footnote{%
$W_{\gamma p} = \sqrt{s_{\gamma p}}$ {designates} the energy in the centre-of-mass of the photon-proton system.} might be such that, at high $P_T$, similar{ly large} non-prompt fraction{s} could be observed.
{With} this respect, further dedicated studies are necessary.

\begin{figure}[htbp!]
\centering
\includegraphics[width=8cm, keepaspectratio]{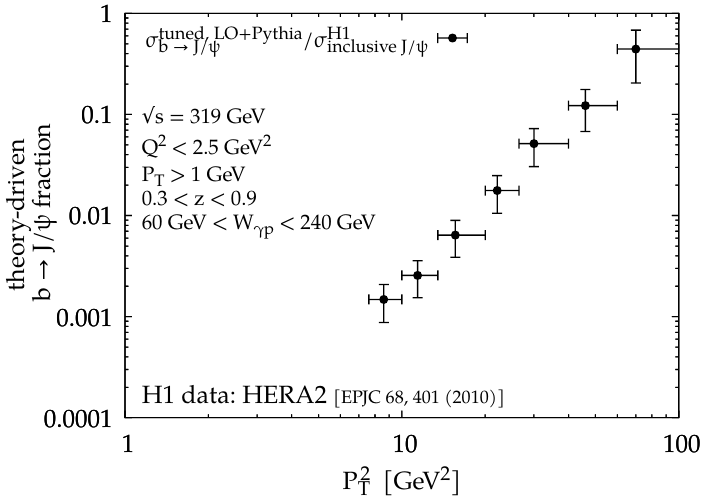}
\caption{Estimate for $b \to J/\psi$ feed-down fraction of the total cross-section for $J/\psi$ photoproduction at HERA, based on the feed-down computed in Ref.~\cite{Flore:2020jau} {as a function of the square of the transverse momentum of the $J/\psi$, $P_T^2$}.}%
\label{fig:H1-data-ratio-b-FD}
\end{figure}%
\section{EIC tools for quarkonium studies }
\label{sec:why_quarkonia}

\subsection{Quarkonium-production mechanisms}

\newcommand{\bra}[1]{\left< #1 \right |}
\newcommand{\ket}[1]{\left | #1 \right >}
\newcommand{\braket}[2]{\left< #1 | #2 \right>}
\newcommand{\sandwich}[3]{\left< #1 \right | #2 \left | #3 \right >}
\newcommand{\mean}[1]{\left< #1\right >}

As {aforementioned}, to justify the application of perturbative QCD to the studies of identified hadron production, the observable{s} %
should %
{involve} some scale $\mu\gg\Lambda_{\rm QCD}$, such that $\alpha_s(\mu)\ll 1$. %
{In such cases,} the %
cross section can %
be factorised (up to {power-suppressed corrections in} $\mu$) into a product or convolution of {a} short-distance part, which is {meant to be} computed perturbatively as a series in $\alpha_s(\mu)$ and long-distance factors. The latter %
{comprise} (TMD-) PDFs of incoming hadrons and non-perturbative quantities which describe the hadronisation of partons produced at the short-distance/perturbative stage of the process into an observed final-state hadron. 

The treatment of hadronisation differs for hadrons containing light quarks in the ``naive'' quark{-}model picture of these states as opposed to quarkonia, the primary component of which is {expected to be a} %
$Q\bar{Q}$ Fock state {with the same quantum numbers as %
quarkonium}.  In {the} case of hadrons composed of light quarks or heavy-flavoured hadrons like $D$ and $B$ mesons, {commonly denoted} ${\cal H}_Q$ { in this review,} in which relativistic (``light'') degrees of freedom play an important role, the hard-scale $\mu$ is $ \sim p_T \gg m_{{\cal H}_Q}$ and the ``final-state'' long-distance part of the cross section is usually %
{encapsulated in a} fragmentation function (FF).  Due to {the} importance of light degrees of freedom, the FFs of such hadrons can not be computed perturbatively and they are {\it parametrised} at some starting scale %
{$\mu_0$, on the order of 1 GeV, with parameters} fitted to reproduce experimental data, see e.g.\ %
{\cite{Kniehl:2000fe,deFlorian:2007ekg,Bertone:2018ecm} and~\cite{Kartvelishvili:1977pi,Peterson:1982ak,Binnewies:1998vm} for fits of respectively light and heavy-flavoured hadrons}. 
For hadrons containing two tightly-bound heavy quarks, such as ``standard'' charmonia ($\eta_c$, $J/\psi$, $\chi_{c}$, $\psi(2S)$, $\ldots$) and bottomonia ($\eta_b$, $\Upsilon(nS)$, $\chi_{b}$,$\ldots$), {denoted hereafter by $\Q$}, a deeper understanding of hadronisation is believed to be possible. %

The overall success of non-relativistic potential models in {the} description of {the} mass spectrum of these states implies %
that the contributions of QCD Fock states containing gluons or light quarks is {suppressed by the powers of the average velocity $v$ of the heavy quarks in the bound state} %
{compared to that}  of the simplest Fock state with only one heavy $Q\bar{Q}$ %
pair. %
The typical squared velocity $v^2$ is estimated in potential models to be $\sim 0.3$ for charmonia and $\sim 0.1$ for bottomonia, which turns it into a{n} {no additional n here} useful small parameter with respect to which the observable{s} can be expanded. %
{The different existing models of quarkonium production~\cite{Lansberg:2019adr} follow more or less closely the above observation which yields to somehow disparate predictions for some production observables. We review below the main features of three of the most popular ones which will follow us throughout this review.}

\subsubsection{NRQCD \& CSM}
\label{sec:NRQCD+CSM}

In the non-relativistic QCD (NRQCD) factorisation formalism~\cite{Bodwin:1994jh}{,} the cross sections and decay rates are expanded in powers of $\alpha_s(\mu)$ and $v^2$. At each order of {the} $v^2$ expansion, the short-distance part of the observable describes the production or annihilation of the ${Q\bar{Q}}$-pair {in a colour-singlet or colour-octet state with a particular value of spin, orbital and total angular momentum.}
 The hard scale{, $\mu$,} for the short-distance part %
 {can be} \newb{the heavy-quark mass $m_{Q}$,} %
 {or any other larger scale not comparable to $\Lambda_{\rm QCD}$}, justifying the perturbative calculation of this factor. The corresponding long-distance part %
 {of} the cross section is %
 a number called the  Long-Distance Matrix Element (LDME) %
 which{,} for the production case{,} can be {written up to conventional colour and spin normalisation factors, which we omit for the sake of clarity,} as:
\begin{equation}
    \langle {\cal O}^{\Q} [i] \rangle \propto \sum_{X_s}\,
\sandwich{0}{\left({\cal O}_i^{\dagger}{\cal Y}_n^\dagger \right)^{ab}(0)}{{\Q}+X_s}
\sandwich{{\Q}+X_s}{\left({\cal Y}_n {\cal O}_i\right)^{ba}(0)}{0}, \label{eq:LDME_def}
\end{equation}
where it is implied that any final state $X_s$ containing light quarks and gluons can be produced together with the quarkonium ${\Q}$. %
{The factors ${\cal Y}_n$ in %
\ce{eq:LDME_def} contain Wilson lines along the light-like direction $n$ needed for the gauge invariance of the Colour-Octet (CO) LDMEs.} {The structure of the colour indices, $ab$, connecting the amplitude and complex-conjugate amplitude in \ce{eq:LDME_def} reflects the process-dependent configuration of the Wilson lines in the factors ${\cal Y}_n$.} The local NRQCD operators ${\cal O}_i$ contain heavy-quark and antiquark fields\footnote{Denoted as $\chi$ and $\psi$ in NRQCD.} and are labelled {in the same way as the simplest Fock state $\left| Q\bar{Q}[i] + X_s \right\rangle$ which this operator} can excite from the vacuum. {The spectroscopic notation of the label $i={}^{2S+1}L_J^{[1,8]}$ is used to denote the total spin $S$, {the} orbital angular momentum $L$, {the} total angular momentum $J$ and {the} singlet (CS, ${}^{[1]}$) or octet (CO, ${}^{[8]}$) colour quantum numbers of the heavy-quark pair.} With these conventions, the complete traditional notation for the LDME becomes: $\langle {\cal O}^{\Q} \left[ {}^{2S+1}L_J^{[1,8]} \right] \rangle$.  

{NRQCD v}elocity-scaling rules~\cite{Bodwin:1994jh, Lepage:1992tx, Bodwin:2005hm} lead to the assignment of the $O(v^m)$ suppression to LDMEs, thus allowing us to truncate the velocity %
expansion at some fixed order {in $v^2$}. Usually {the contributions associated with} the LDMEs up to Next-to-Next-to-Leading Order (NNLO) 
in $v^2$ ($O(v^4)$ relative to the LDME of the ${}^3S_1^{[1]}$ state) are taken into account in phenomenological studies. %
{This} means that{,} besides the colour-singlet ${Q\bar{Q}}$ states, %
the colour-octet states ${}^1S_0^{[8]}$, ${}^3S_1^{[8]}$ and ${}^3P_J^{[8]}$ can contribute to $J/\psi$ production, for example. 

For $S$-wave quarkonia, %
the {expansion limited to} the leading order of $v^2$ corresponds to the colour-singlet ${Q\bar{Q}}$-state with the same quantum numbers as those of ${\Q}$. %
The {\it colour-singlet model} (CSM)%
{~\cite{Chang:1979nn,Berger:1980ni,Baier:1981uk}}  for {the} production of these states is nothing but the truncation of the $v^2$ expansion at this order. The CS LDMEs can be estimated from potential-model wave functions~\cite{Eichten:1995ch}, while their accurate estimation from %
{$\ell^+\ell^-$} decay rates of ${\Q}$ is rendered complicated by large NNLO QCD corrections~\cite{Beneke:1997jm} to the decay width. However, the CSM is not sufficient theoretically{~\cite{Bodwin:1994jh,Bodwin:1992ye,Cho:1995vh}} %
for the description of {the} production of the $P$-wave quarkonia, such as $\chi_{c,b}${,} beyond LO in $\alpha_s$ and can not describe inclusive hadroproduction \old{$p_T$}{$\pT$} %
spectra of charmonia and bottomonia at high %
{$\pT$}~\cite{Campbell:2007ws,Gong:2008hk}. Nevertheless, {the} NNLO corrections {in $\alpha_s$} to the short-distance part of the CSM cross section, {only partially computed so far}, may decrease %
{the existing large discrepancy between the CSM and the data from Tevatron and the LHC} ~\cite{Artoisenet:2008fc,Lansberg:2008gk,Lansberg:2010vq,Lansberg:2011hi}. {This point is still under debate~\cite{Ma:2010jj,Shao:2018adj,Lansberg:2019adr}.}

 In contrast to the hadroproduction case described above, in %
 (prompt) {\it inclusive} photo- and electroproduction of heavy quarkonia, which are relevant for the EIC experimental program, the CSM has been expected{~\cite{Kramer:1995nb,Kramer:2001hh}} and proven to be able to account %
 {for a large fraction} of the observed cross section~\cite{Butenschoen:2009zy,Flore:2020jau} {even up to the highest reachable $\pT$}. {Estimates varying from 50\%~\cite{Butenschoen:2009zy} to almost 100\%~\cite{Flore:2020jau} can be found in literature. }

For the {\it exclusive} photo- and electroproduction of single $J/\psi$ or $\Upsilon(nS)$, the CS contribution is %
{also expected} to be strongly dominating. In %
{such exclusive reactions}, no final-state radiation ($X_s$) is allowed and the NRQCD operators containing a CO $Q\bar{Q}$ pair can {only} couple %
to the higher Fock-state contributions in the expansion of the physical quarkonium eigenstate, which are velocity-suppressed, e.g. $|J/\psi\rangle= O(1)|c\bar{c}[^3S^{[1]}_1]\rangle + O(v)|c\bar{c}[^3P^{[8]}_J]+g\rangle+\ldots $ The matrix elements of {the} gauge-invariant CO operators which in principle can contribute to exclusive photoproduction, e.g. $\psi^\dagger (g_s{\bf E}\cdot {\bf D}) \chi$ where ${\bf E}$ is the chromoelectric field and ${\bf D}$ is the QCD covariant derivative, can be estimated\footnote{The scaling for ${\bf D}$ is $O(v)$ and the scaling for $g_s {\bf E}$ is $O(v^3)$ so together with the $O(v)$ suppression of the $|c\bar{c}[^3P^{[8]}_J]+g\rangle$ component of $|J/\psi \rangle$, one obtains $O(v^5)$.} to scale at least as $O(v^5)$ at the level of the amplitude using the velocity scaling rules~\cite{Lepage:1992tx}. In Ref.~\cite{Hoodbhoy:1996zg}, the same conclusion %
{has} been made about the CO contributions to the matrix element{s} of the operator $\psi^\dagger {\bf D}^2 \chi = \psi^\dagger \nabla^2 \chi + \psi^\dagger (g_s{\bf A}\cdot \nabla) \chi + \ldots$, which are {more} suppressed %
{than} the CS relativistic correction{s} $\langle J/\psi | \psi^\dagger \nabla^2 \chi |0\rangle\sim \nabla^2 \Psi(0)\sim O(v^2)$. Therefore, taking into account CS relativistic corrections to exclusive vector-quarkonium photoproduction is currently considered to be more important~\cite{Hoodbhoy:1996zg,Mantysaari:2021ryb,Mantysaari:2022kdm} than taking into account the CO corrections.    

Another success~\cite{Butenschoen:2014dra} of the CSM at NLO in $\alpha_s$ is the description of the prompt $\eta_c$ hadroproduction, measured by LHCb~\cite{LHCb:2014oii,LHCb:2019zaj}. %
{However, such a success of the CSM to describe this data set, both at moderate $\pT\sim m_{\eta_c}$ and for $\pT\gg m_{\eta_c}$ is problematic for NRQCD. Indeed, from heavy-quark{-}spin{-}symmetry (HQSS) arguments, one expects the CO contributions to $\eta_c$ cross section at $\pT\gg m_{\eta_c}$ to be on the same order of magnitude as that previously found to describe $J/\psi$ data at similar $\pT$.}

As %
{aforementioned}, at higher orders in the $v^2$ expansion, the CO LDMEs contribute, but at present they are treated as free parameters {and are} adjusted to describe experimental data. Besides order-of-magnitude constraints from $O(v^n)$ scaling and {HQSS} constraints,  the progress on their theoretical calculation has been limited so far. Recently new expressions for LDMEs in terms of potential-model %
quarkonium wave functions and certain chromoelectric-field correlators have been {proposed}  in the potential-NRQCD \newb{(pNRQCD)} formalism {in the strongly coupled regime}~\cite{Brambilla:2020ojz,Brambilla:2021abf}. These relations can be used to reduce number of free parameters in the fit {under the assumption $m_Q v^2 \ll \Lambda_{\rm QCD}$. Currently, the advantage of using pNRQCD compared to conventional NRQCD fits is still under debate as well as its applicability, since $m_Q v^2$ is naively not much smaller than $\Lambda_{\rm QCD}$ }. 

In %
Section~\ref{sec:LDMEs:coll}{,} we describe existing phenomenological fits of LDMEs within collinear factorisation, commenting on their successes and shortcomings in more details. Unfortunately at present time there is no single set of LDMEs which can %
{satisfactorily} describe the charmonium $e^+e^-$ {annihilation, $\gamma \gamma$ fusion}, hadro{-} and photoproduction data together with polarisation observables in the framework of NRQCD factorisation at NLO in $\alpha_s$, which is a serious problem for the NRQCD factorisation approach. For the case of bottomonia{,} we lack %
photoproduction, {$e^+e^-$ annihilation and $\gamma \gamma$ fusion} data, which prevents us from checking the process-independence of LDMEs for the $b\bar{b}$ family. Another important task for {the} EIC, %
{in connection with the} clarification of {the} quarkonium-production mechanism, is to perform the first measurement of $\chi_{c0,1,2}$ and $\eta_c$ {inclusive} photoproduction cross sections. %
{In this context, we discuss corresponding} phenomenological predictions {in} Section~\ref{sec:chic-etac-photoprod}. {Such} measurements will be complementary to %
{those} of $\chi_c$ and $\eta_c$ hadroproduction to check the process-independence of the corresponding LDMEs. 

 {Data at high $\pT \gg m_{\Q}$, where CS and CO contributions behave differently, are potentially very discriminant for LDME fits. This} calls for improvement of {the} perturbative accuracy of the short-distance part %
 {since, at large $\pT$,
terms proportional to} $\alpha_s^{n+k} \ln^n %
{\pT}/m_{\cal H}$ \newb {appearing} in the perturbative series for {the} short-distance part of the cross section both at LP in %
{\pT} and in  power-suppressed corrections at $\pT\gg m_{\Q}$ {need to be tackled}.
These potentially large terms can be resummed using the formalism of FFs, perturbatively evolving with the scale $\mu\sim{\pT}$. At {LP,} this formalism is analogous to the FFs for light hadrons mentioned in the beginning of this section, with %
{a sole but important} difference, {namely} that at the starting scale $\mu_0\sim m_{\Q}$ the FF %
{is assumed to be factorised into a short distance part and a} LDME. {We refer to} e.g.~\cite{Bodwin:2014gia} {as an example of the NLO study of this type as well as Refs.~\cite{Braaten:1993rw,Cacciari:1994dr,Braaten:1994xb} at LO.} At Next-to-Leading Power (NLP), new contributions with the {${Q\bar{Q}}$ pair as a whole participating in the fragmentation process} %
appear~\cite{Kang:2011mg}. These corrections %
{seem} to influence not only the cross section %
but also the evolution of leading-power FFs~\cite{Lee:2021oqr}. {However,} {the} effect of this corrections on cross sections and {the} polarisation is still under investigation {in particular for the EIC phenomenology where the $\pT$ reach{, limited to roughly 15-20~GeV,} might not be large enough for these to be relevant}.

\subsubsection{CEM \& ICEM}

Given the above mentioned phenomenological problems {along with others which we review later}, NRQCD factorisation at fixed order in $v^2$ and $\alpha_s$ is not completely satisfactory. Due to its simplicity, the {\it Colour Evaporation Model} (CEM), introduced in Refs.~{\cite{Halzen:1977rs,Fritzsch:1977ay}} {remains} an attractive alternative {mechanism to explain the formation of quarkonium}. {As the CEM is inspired from quark-hadron duality, one} postulates that any ${Q\bar{Q}}$ pair produced at short distance with invariant mass $M_{Q\bar{Q}}$ less than the invariant mass of a pair of lightest mesons (${\cal H}_Q$) with open{-}heavy flavour $Q$ (e.g. $D^0$ mesons in the case of charmonia) has to hadronise into {one of the quarkonia below this heavy-flavour-production threshold} with some {universal} probability. In {the} CEM\new{,} this probability, commonly denoted as $F_{\Q}$ for the quarkonium state ${\Q}$, is taken to be independent of spin, orbital momentum \new{and} colour quantum numbers of the pair, and \new{is fit} as \new{a} free parameter. 

In the improved CEM (ICEM)~\cite{Ma:2016exq,Cheung:2017osx,Cheung:2018tvq}{,} the kinematic effects  {arising from the} mass difference between the ${Q\bar{Q}}$-pair produced at short distance and {the} final-state quarkonium is taken into account, which roughly models the effects of soft-gluon emissions at hadronisation stage. This is done through the rescaling of the three-momentum of the pair by the mass ratio, so that the direct quarkonium-production cross section in \pp collisions in the ICEM is given by \cite{Ma:2016exq}:
\begin{eqnarray}
\label{ch6-icem-cross-section}
\sigma &=& F_{\Q} \sum_{i,j}  \int\limits^{2m({{\cal H}_Q)}}_{M_{\Q}}dM_{Q\bar{Q}} dx_i dx_j\  f_i(x_i,\mu_F)f_j(x_j,\mu_F)\cdot\hat{\sigma}_{ij\rightarrow Q\bar{Q}}(x_i,x_j,{\bf p}_{Q\bar{Q}},\mu_R,\mu_F) \Big|_{{\bf p}_{Q\bar{Q}} = \frac{M_{Q\bar{Q}}}{M_{\Q}} {\bf P}_{\Q}} \;,
\end{eqnarray}
where $i$ and $j$ are $q, \bar{q}$ and $g$ such that $ij = q\bar{q}$, $qg$, $\bar{q}g$ or $gg$, 
{$x_{i,j}$} is the momentum fraction of the parton, $f(x_{i,j},\mu_F)$ is the parton distribution function (PDF) in the proton as a function of $x_{i,j}$ at the factorisation scale $\mu_F$. Finally, $\hat{\sigma}_{ij\rightarrow Q\bar{Q}}$ are the parton-level cross sections for {the} initial states $ij$ to produce a $Q\bar{Q}$ pair of momentum ${\bf p}_{Q\bar{Q}}$ at the renormalisation scale $\mu_R$. 
In the ICEM, the invariant mass of the $Q\bar{Q}$ pair, $M_{Q\bar{Q}}$, is integrated from the physical mass of quarkonium $M_{\Q}$ to two times the mass of the lightest open heavy $Q$-flavour meson $m({\cal H}_Q)$. In the traditional CEM, see e.g.~\cite{Lansberg:2020rft}, the value of $2m_Q$ is used as the lower limit of mass-integration instead of $M_{\Q}$ {and the momentum-shift due to the mass-difference between the $Q\bar{Q}$-pair and the quarkonium is neglected}.

We emphasi{s}e  that the physical picture of {the} (I)CEM is opposite to NRQCD in {the} sense that the CS contributions play no special role at all. This assumption makes CEM incapable of describing observables where CS states are clearly dominating, e.g. the prompt hadroproduction of $J/\psi$ pairs~\cite{Lansberg:2020rft} and {the} $e^+e^-\to J/\psi + c\bar{c}$ cross section~\cite{Kang:2004zj}.
However{,} the (I)CEM still provides {a} reasonable description of single inclusive prompt quarkonium hadroproduction~\cite{Ma:2016exq,Cheung:2017osx,Cheung:2018tvq} {although the model is not capable to describe $\pT\sim m_{\Q}$ and $\pT\gg m_{\Q}$ simultaneously even at NLO~\cite{Lansberg:2020rft,Lansberg:2016rcx}}. 
 
Recent ICEM calculations~\cite{Cheung:2018tvq,Cheung:2021epq} have considered the polarisation in hadroproduction. Polarised production of quarkonium in these calculations restricts the final state quark-antiquark pair to be in the desired spin state, thus implicitly assuming that soft gluons are decoupled from heavy-quark spin. The polarisation parameters are then calculated in terms of the spin matrix elements $\sigma_{i_z,j_z}$. In these matrix elements, the quarkonium is assumed to have $J_z = i_z$ when calculating the scattering matrix element, $\mathcal{M}$. The quarkonium is assumed to take $J_z=j_z$ in calculating the conjugate, $\mathcal{M}^*$. The polar anisotropy ($\lambda_{\theta}$){, defined in the Eq.~(\ref{eq:pol-par-def}),} is given {in this model} by \cite{Faccioli:2010kd}
\begin{eqnarray}
\lambda_{\theta} &=& \frac{\sigma_{+1,+1}-\sigma_{0,0}}{\sigma_{+1,+1}+\sigma_{0,0}}\label{lambda_theta_phi_eqn}\;.
\end{eqnarray}

\begin{figure}
\centering
\includegraphics[width=8cm, keepaspectratio]{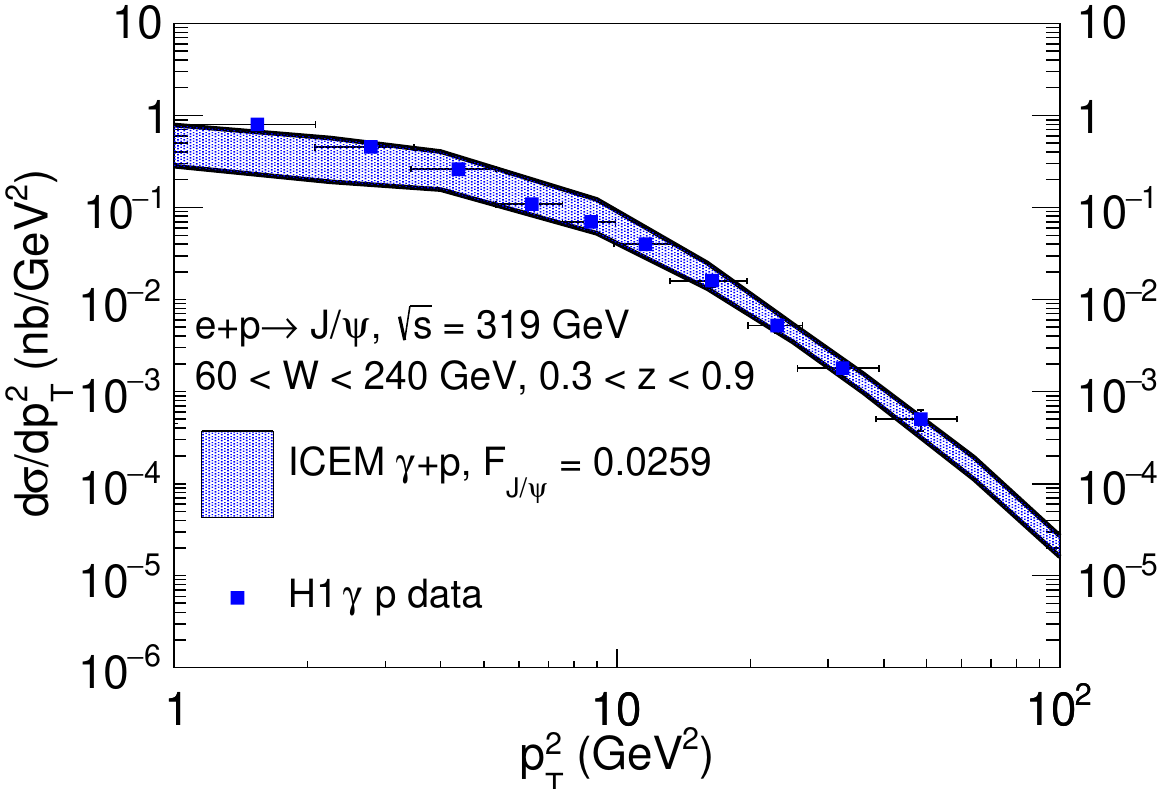}
\includegraphics[width=8cm, keepaspectratio]{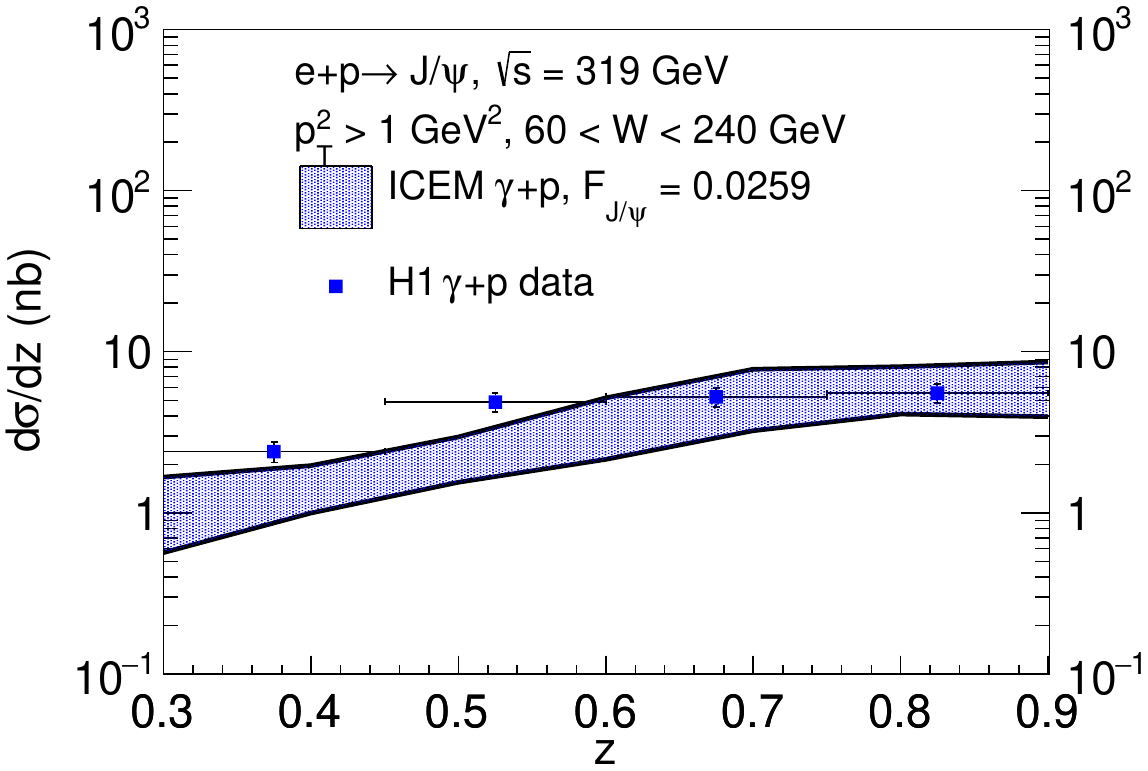}
\caption{Description of the HERA H1 data~\cite{Aaron:2010gz} on $P_T^2$ -differential (left) and z-differential (right) cross section of inclusive \jpsi photoproduction in $ep$ scattering by the ICEM calculation in collinear factorisation. The combined mass and scale uncertainties are shown in the band. Feed down contributions are not included. Taken from Ref.~\cite{Cheung:2024bnt}.} \label{fig:icem-ep-HERA}
\end{figure}

{As} the ICEM is an alternative to NRQCD in hadroproduction, developments to extend it into other collision systems are still in progress. 
{The authors of~\cite{Cheung:2018tvq,Cheung:2021epq,Cheung:2024bnt} anticipate that the value of $\lambda_\theta$ for \jpsi production in $ep$ collisions will also be very similar to the $pp$ case, which they found to be compatible with the existing Tevatron and LHC data.} 
{In addition, they also find the free parameter $F_\mathcal{Q}$ in photoproduction to be consistent with that in hadroproduction. The description of HERA H1 data~\cite{Aaron:2010gz} on \jpsi photoproduction in the ICEM~\cite{Cheung:2024bnt} is illustrated in Fig.~\ref{fig:icem-ep-HERA}.  However, the (I)CEM prediction introduces a parameter to keep the propagator at some minimum distance of $M_\psi^2$ from the pole. Thus, its} prediction of the $z$-differential spectrum in photoproduction is likely to be complicated by large radiative corrections at $1-z\ll 1$ \newb{if the parameter is removed}, which was seen already in the LO analysis of Ref.~\cite{Eboli:2002rv,Eboli:2003fr} where the agreement with data at $z\to 1$ was reached only after introduction of an {\it ad-hoc} cut $|\hat{t}|>4m_c^2$ on the partonic $\hat{t}$ variable.

{The observation that the CEM leads} to unpolarised heavy{-}quarkonium hadroproduction at high-$\pT$~\cite{Cheung:2021epq}, a result which is {non-trivial} to achieve with NRQCD fits, perhaps means that{,} in cases {where} CO LDMEs dominate, the dynamics of soft-gluon emissions should be taken into account more accurately than it is done in the fixed-order NRQCD factorisation. The recently proposed soft-gluon factorisation approach represents a progress in this direction~\cite{Ma:2017xno,Li:2019ncs}{, whose phenomenological implications, however, remain to be investigated}. 

\subsection{Legacy from HERA, the Tevatron and the LHC{,} and predictions for the EIC for cross-section and polarisation observables}

 \subsubsection{Status of  NRQCD LDME fits} \label{sec:LDMEs:coll}

 \paragraph{A side note on the positivity of the LDMEs beyond LO.} 
 Before discussing the NLO LDME fits, let us make a comment about the positivity of LDMEs. At LO in $\alpha_s$, the LDMEs have a simple interpretation as ``probabilities'' of the transition of the ${Q\bar{Q}}$-pair in a certain colour, spin and angular-momentum state into an observed quarkonium. This physical interpretation follows from the operator definition of LDMEs~(\ref{eq:LDME_def}) in terms of ``bare'' fields~\cite{Bodwin:1994jh, Nayak:2005rt} if QCD loop corrections are not taken into account and Wilson-line factors are ignored. Consequently, in LO calculations, LDMEs are typically assumed to be positive-definite. This is similar to the situation with LO (TMD)PDFs.

Already at NLO in $\alpha_s$, both {ultraviolet (UV)  and infrared (IR)}
divergences appear in the operator definitions of LDMEs, see e.g. Appendix B of Ref.~\cite{Bodwin:1994jh} as well as Section 6 of Ref.~\cite{Petrelli:1997ge} and references therein. If NRQCD factorisation holds -- which is yet to be proven beyond NNLO in $\alpha_s$~\cite{Nayak:2005rt, Bodwin:2019bpf} -- the IR divergences of the hard-scattering coefficients should cancel against the corresponding IR divergences of the LDMEs at all orders of the $v^2$ and $\alpha_s$ expansions, {while the UV divergences appearing in LDMEs are removed by the operator renormalisation.} {The {\it renormalised} LDMEs then become non-perturbative fit parameters.}  {Therefore,} these parameters do not necessarily have to be positive. Their definition involves the subtraction of the divergent part. In addition, the finite renormalised LDMEs are scheme- and scale-dependent, and mix with each %
other due to the NRQCD-scale evolution. The relation between {the} short-distance cross section and LDMEs, described above, is similar to the relation between NLO short-distance cross sections and QCD PDFs and/or fragmentation functions, which are also not necessarily positive-definite, at least if the calculation is truncated to a fixed order in $\alpha_s$. This is the reason why there are usually no positivity constraints imposed in NLO LDME fits. One of the consequences of this is that the numerical values of LDMEs obtained in fits at NLO in $\alpha_s$ have limited physical significance outside the NLO context and should only cautiously be used in LO calculations, because this could create unjustifiable cancellation between some contributions.

{In general though, it is not clear that negative NLO LDMEs would yield positive NLO cross sections for all possible measurable processes one could think of. Let us for instance mention the case of quarkonium-photon associated production for which it was shown~\cite{Gong:2012ah} that some of the NLO LDME fits which we discuss below would yield negative NLO cross sections. Such a physical constraint on LDMEs at NLO has however not been systematically investigated as it requires the complete NLO computation of the hard scatterings for all the processes one wishes to consider.}

 \paragraph{{S}urvey of existing NLO LDME fits.} 

Several groups have performed fits of CO LDMEs for charmonia~\cite{Butenschoen:2010rq,Butenschoen:2011yh,Butenschoen:2012px,Butenschoen:2012qr,Chao:2012iv,Gong:2012ug,Bodwin:2014gia,Brambilla:2022rjd} and bottomonia~\cite{Gong:2010bk,Wang:2012is,Gong:2013qka} at NLO in $\alpha_s$ for the short-distance parts. {We emphasise that the {computation at NLO in $\alpha_s$} of short-distance cross sections for the production of NRQCD states ($Q\bar{Q}[i]$) is done in exactly the same framework of collinear factorisation by most of the groups with the exception of the fit of Bodwin et al.~\cite{Bodwin:2014gia}. The latter computation includes, beside corrections at NLO in $\alpha_s$,  the resummation of logarithms of $P_T/m_{\Q}$ which \newb{become important} at $P_T\gg m_{\Q}$. Therefore the difference of \newb{the fits} boils down mostly to the choice of different experimental data to fit and approximate (up to higher-orders in $v^2$) relations between different LDMEs which \newb{are assumed or not} to hold exactly in the fitting procedure.} {For a detailed discussion, we refer to the recent review~\cite{Lansberg:2019adr}}. \ct{tab:LDME-fits_comp} briefly compares phenomenological results of %
{each fit} for the case of charmonia using benchmark observables such as the %
cross sections {and polarisation of inclusive prompt $J/\psi$ produced in $pp$ collisions as a function of $\pT$ as well as } %
photoproduction in $ep$ collisions {and the  total cross section of charmonium production in $e^+e^-$ annihilation} %
We also %
{indicate} in \ct{tab:LDME-fits_comp} whether the corresponding set of LDMEs for $J/\psi$ allows one to describe the prompt $\eta_c$ hadroproduction %
{\pT}-spectrum measured by LHCb~\cite{LHCb:2014oii,LHCb:2019zaj} using %
heavy-quark-spin-symmetry relations between  $\eta_c$ and $J/\psi$ LDMEs {which hold up to $v^2$ corrections}.

\begin{table}[ht!]
    \centering
    \begin{tabular}{|c| c|cccc|}
    \hline
    
    Acronym & Reference  & $J/\psi$ hadropr. &  $J/\psi$ photopr. & $J/\psi$ polar. & $\eta_c$ hadropr. \\ %
    &&&{and $e^+e^-$}&{in hadropr.}&{\footnotesize ($P_T>6.5$ GeV)}\\
    \hline
  BK11 & Butensch\"on et al.~\cite{Butenschoen:2010rq,Butenschoen:2011yh,Butenschoen:2012px,Butenschoen:2012qr}         & \cmark($P_T>3$ GeV) & \cmark & \xmark & \xmark  \\ %
H14    &Chao et al. + $\eta_c$~\cite{Han:2014jya}     & \cmark($P_T>6.5$ GeV) & \xmark & \cmark & \cmark \\ %
Z14   & Zhang et al.~\cite{Zhang:2014ybe}             & \cmark($P_T>6.5$ GeV) & \xmark & \cmark & \cmark \\ %
G13  & Gong et al.~\cite{Gong:2012ug}               & \cmark($P_T>7$ GeV) & \xmark & \cmark & \xmark \\ %
C12   & Chao et al.~\cite{Chao:2012iv}                & \cmark($P_T>7$ GeV) & \xmark & \cmark & \xmark \\ %
 B14  & Bodwin et al.~\cite{Bodwin:2014gia}           & \cmark($P_T>10$ GeV) & \xmark & \cmark & \xmark \\ %
   pNRQCD & Brambilla et al.~\cite{Brambilla:2022rjd,Brambilla:2022ayc}     & \cmark($P_T>15$ GeV) & \xmark & \cmark & \xmark\cmark \\ %
    \hline
    \end{tabular}
    \caption{Phenomenological comparison of {a selection of} existing %
    {$J/\psi$-$\eta_c$} LDME extractions %
    {at} NLO in $\alpha_s$. The cut on the $J/\psi$ transverse momentum, applied in each fit, is indicated in parentheses in the {third} column. {This cut is applied because all but the first fit badly fail to account for the %
    low-$P_T$ data.}}
    \label{tab:LDME-fits_comp}
\end{table}

 The %
 $\pT$ spectra of the prompt inclusive %
 quarkonia {produced} in $pp$ and $p\bar{p}$ collisions at mid and large $\pT$ at {the} Tevatron and the LHC are well described by all the fits mentioned in \ct{tab:LDME-fits_comp}; this is %
 {the} major phenomenological success of NRQCD factorisation at NLO. Note, however,  that hadroproduction data with $\pT\lesssim m_{\Q}$ (or integrated in $\pT$~\cite{Maltoni:2006yp,Feng:2015cba}) can not be simultaneously described by NLO NRQCD  {fits of large $\pT$ data.} %
 {In} fact, most of the fits have been performed with even stronger $\pT$ cuts, as indicated in \ct{tab:LDME-fits_comp}. 
 
 The only existing global NLO LDME fit~\cite{Butenschoen:2010rq,Butenschoen:2011yh,Butenschoen:2012px,Butenschoen:2012qr}, BK11, beyond hadroproduction, also provides a reasonable description of unpolarised charmonium production cross sections in $e^+e^-$, $pp$, $p\bar{p}$ and $ep$ collisions. {The description of HERA H1 data~\cite{Aaron:2010gz} on $J/\psi$ photoproduction by the BK11 fit is illustrated in the \cf{fig:LDME-fits_vs_H1-data_pT}(a) and \cf{fig:LDME-fits_vs_H1-data_z}(a).}   However{,}  this fit is not able~\cite{Butenschoen:2012px, Butenschoen:2012qr} to describe charmonium-polarisation observables, measured in hadroproduction at high-$\pT$, see e.g.\ Ref.~\cite{Andronic:2015wma} for a global survey of heavy-quarkonium-polarisation data. %
 This situation is often referred to as {the} ``heavy-quarkonium-polarisation puzzle'' in the literature. Polarisation observables {relevant for $J/\psi$ production} at the EIC will be discussed in Section~\ref{sec:testing_NRQCD_EIC}. %

\begin{figure}[hbt!]
    \centering
    \begin{tabular}{ccc}
    \includegraphics[width=0.3\textwidth]{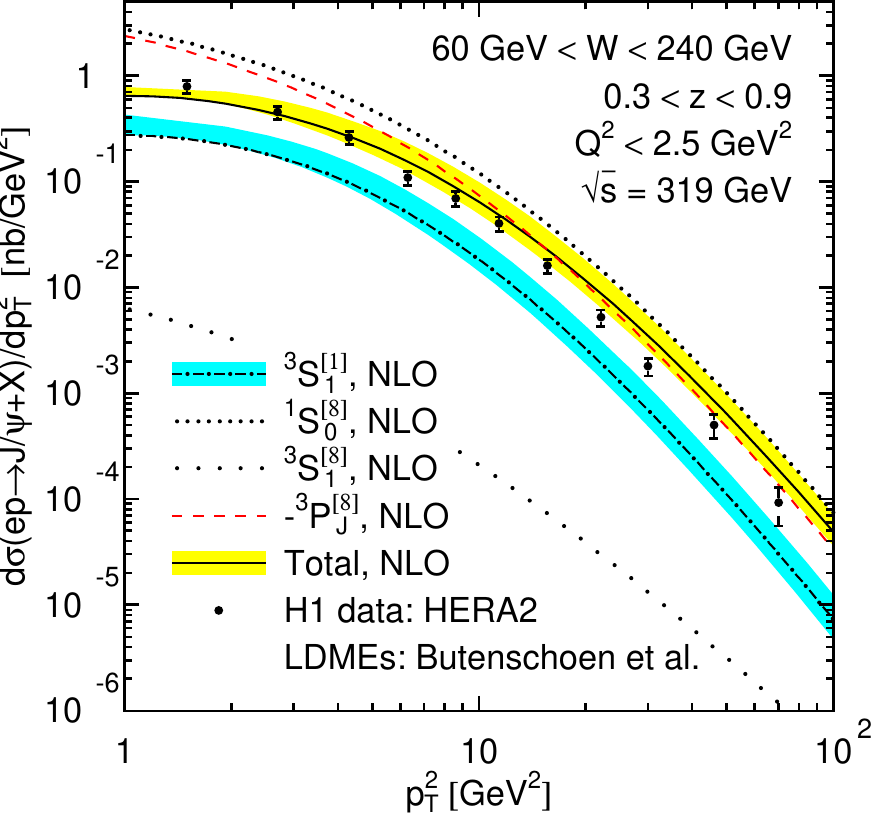}&
    \includegraphics[width=0.3\textwidth]{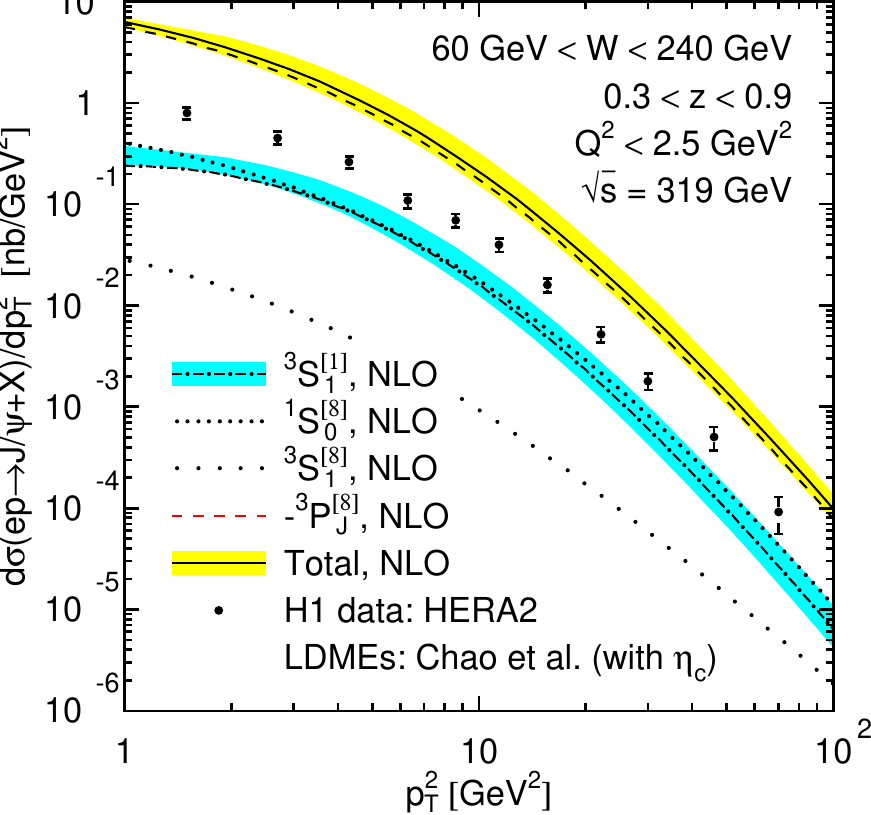}&
    \includegraphics[width=0.3\textwidth]{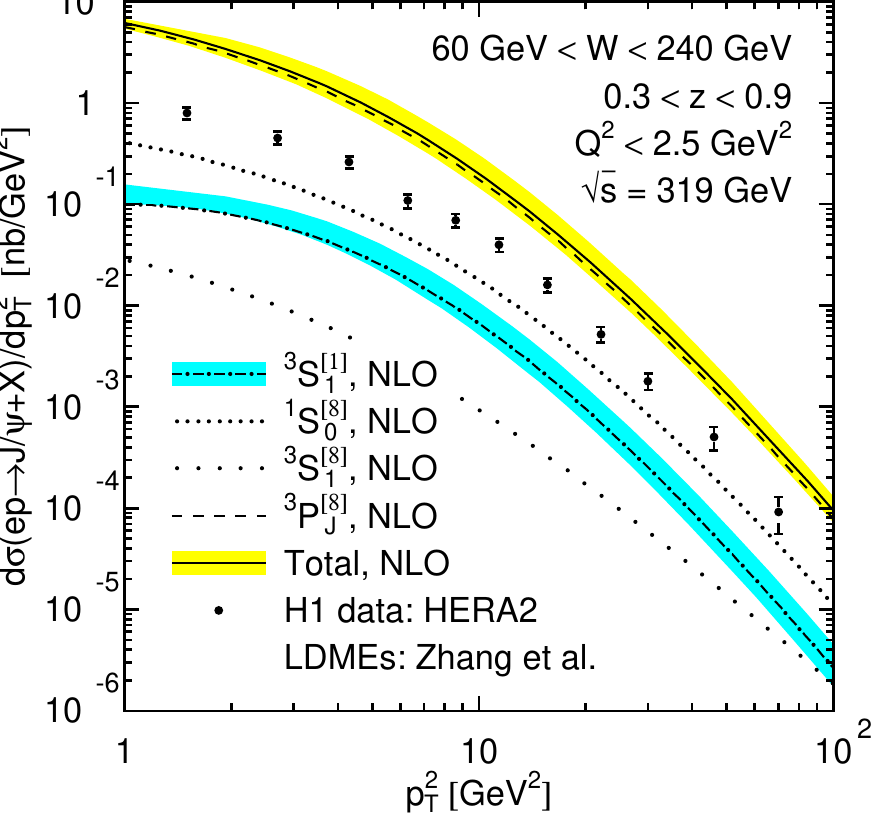} \\
    (a) BK11 & (b) H14& (c) Z14 \vspace{5mm}
    \end{tabular}
    \begin{tabular}{cc}
    \includegraphics[width=0.3\textwidth]{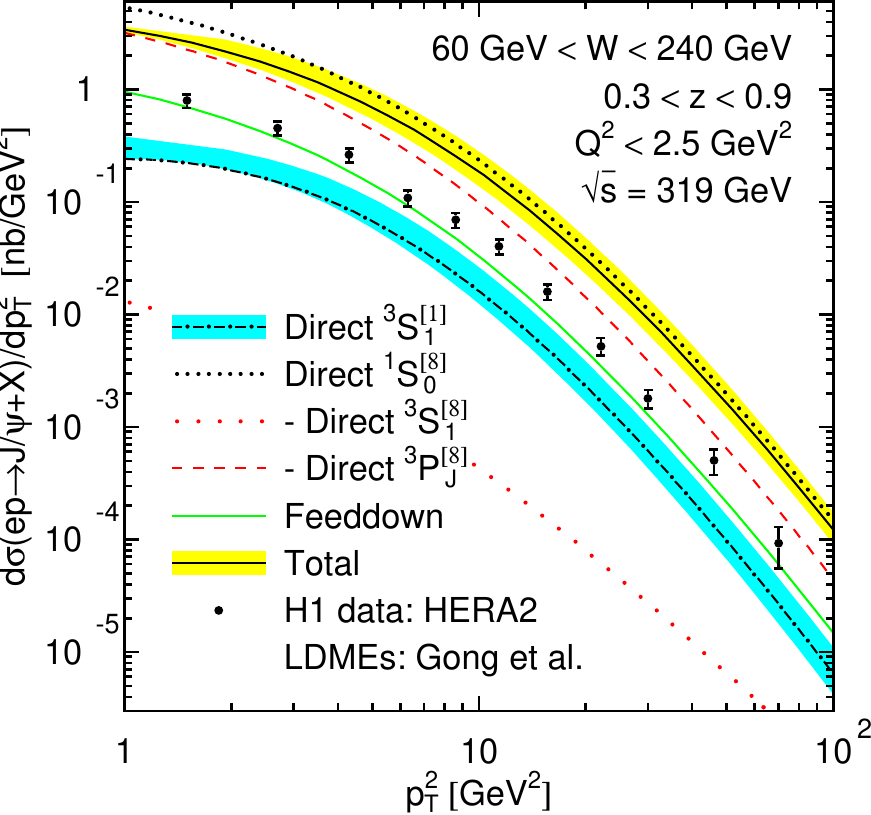}&
    \includegraphics[width=0.3\textwidth]{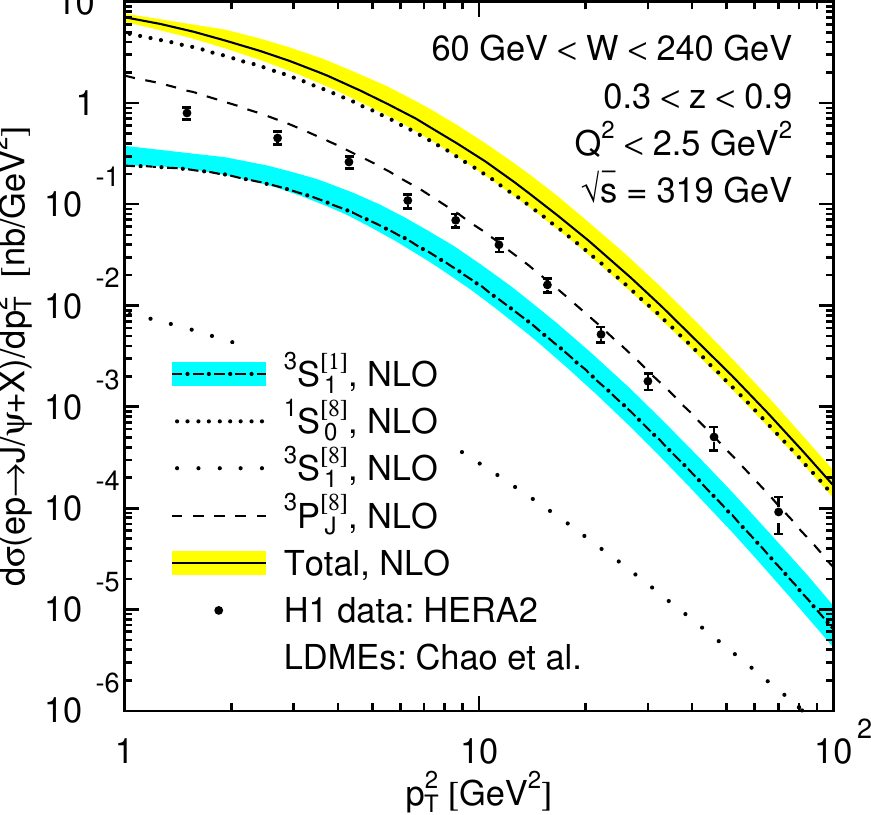}\\
    (d) G13 & (e) C12 \vspace{5mm}
    \end{tabular}
    \begin{tabular}{cc}
    \includegraphics[width=0.3\textwidth]{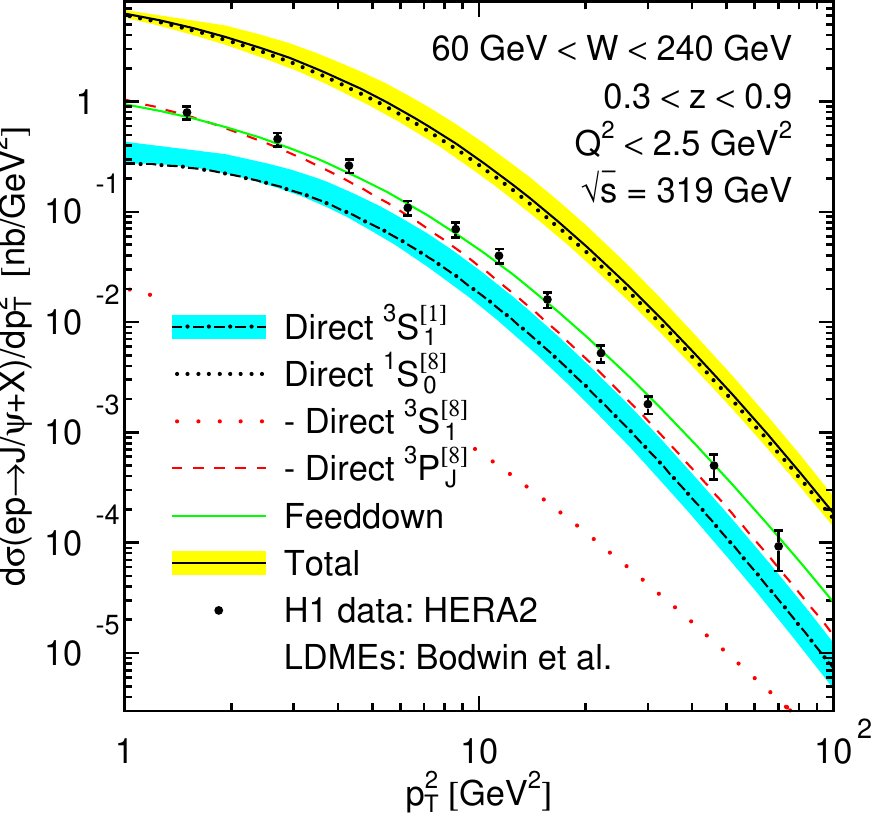}&
    \includegraphics[width=0.3\textwidth]{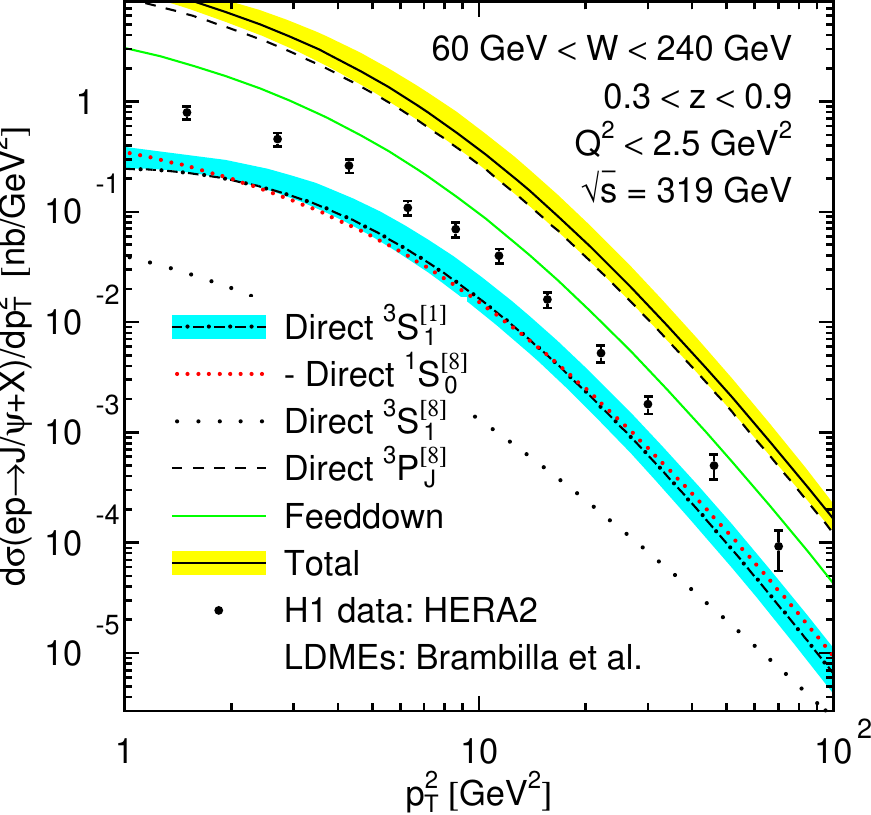}\\
    (f) B14 & (g) pNRQCD 
    \end{tabular}
    \caption{{Description of the HERA H1 data~\cite{Aaron:2010gz} for the $P_T^2$-differential  cross section of inclusive $J/\psi$ photoproduction by NLO NRQCD fits in collinear factorisation for the LDMEs listed in \ct{tab:LDME-fits_comp}. In each plot the sum of CS and CO contributions is plotted by the solid line with the yellow scale-variation band. The dash-dotted line with blue scale-variation band corresponds to the CSM contribution at NLO. Other curves in each plot correspond to the contributions to the ``total NLO'' curve from various CO states (with negative contributions being plotted in red) and to the feed down contribution, as indicated by the legend of each of the plots.  %
    }}
    \label{fig:LDME-fits_vs_H1-data_pT}
\end{figure}

Two of the fits in \ct{tab:LDME-fits_comp}, H14 and Z14, %
{turned out to be able to }simultaneously describe $J/\psi$ and $\eta_c$ hadroproduction data using heavy-quark-spin-symmetry 
relations between LDMEs. Remarkably, the $J/\psi$-polarisation observables in hadroproduction are also reasonably {well} reproduced by these fits but they significantly overestimate the HERA photoproduction cross section as can be seen in \cf{fig:LDME-fits_vs_H1-data_pT}(b,c).  {The same holds for all the other LDME fits (with the exception of BK11 discussed above), see \cf{fig:LDME-fits_vs_H1-data_pT}(d-g). The discrepancies between the NRQCD NLO predictions with these fits range from 2 at $\pT\simeq 10$~GeV up to 10 at $\pT\simeq 1-2$~GeV in the case of pNRQCD and B14. This means that the yield predictions at the EIC using these LDMEs can be overestimated by up to one order of magnitude. Since the discrepancies remain at $\pT=10$~GeV, which roughly corresponds to the maximum values which would be reached at the EIC, this should be kept in mind when considering predictions with CO contributions (except for the BK11 LDMEs) for the EIC case at any $\pT$.
 }

 {As one can seen from \cf{fig:LDME-fits_vs_H1-data_z}, all LDME fits except BK11 also strongly overestimate the $z$-differential cross section for $z>0.6$. The BK11 fit is consistent with the photoproduction data due to the cancellation between $^1S_0^{[8]}$ and $^3P_J^{[8]}$ channels. Other fits use this degree of freedom to accommodate the polarisation and/or $\eta_c$ production data and therefore lose flexibility which is needed to achieve a global fit across different collision systems.}

 \begin{figure}[hbt!]
    \centering
    \begin{tabular}{ccc}
    \includegraphics[width=0.3\textwidth]{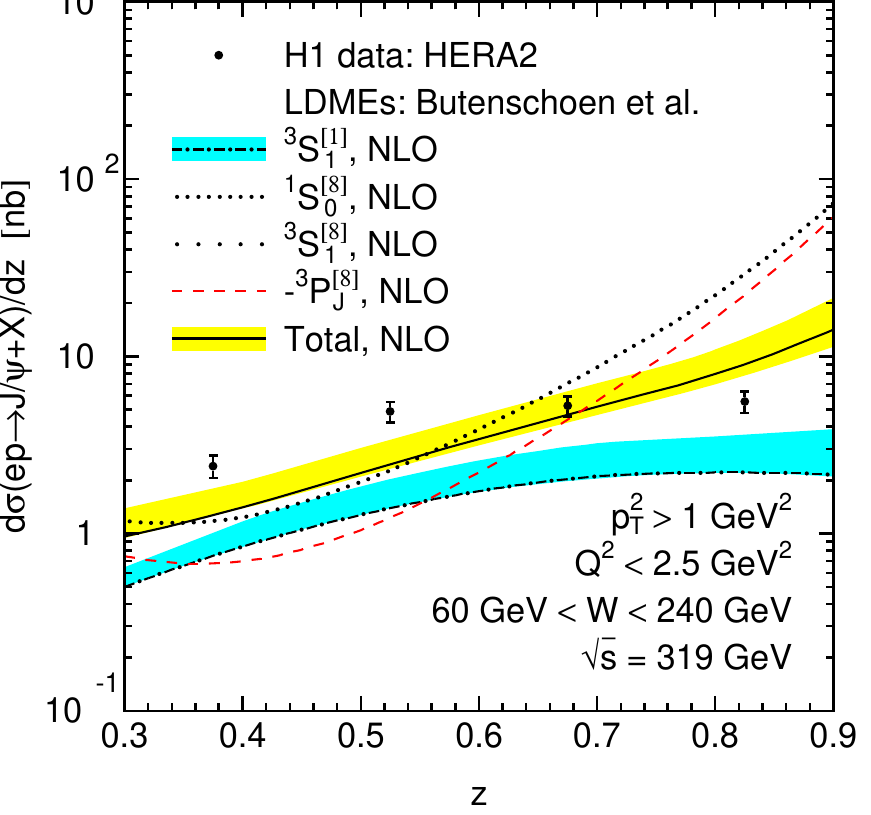}&
    \includegraphics[width=0.3\textwidth]{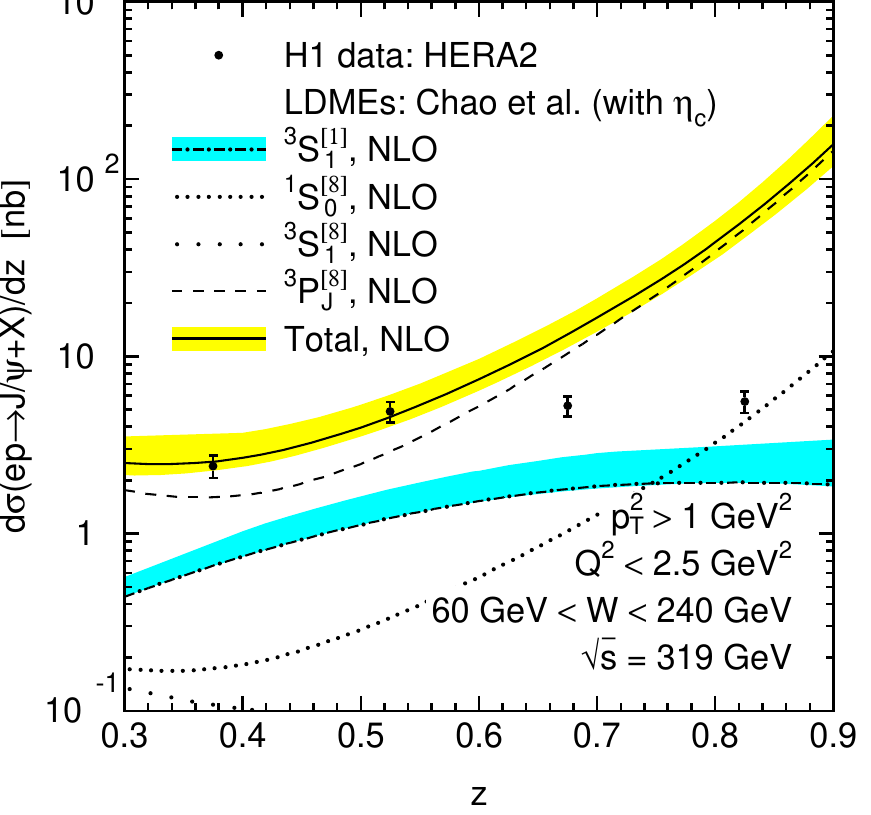}&
    \includegraphics[width=0.3\textwidth]{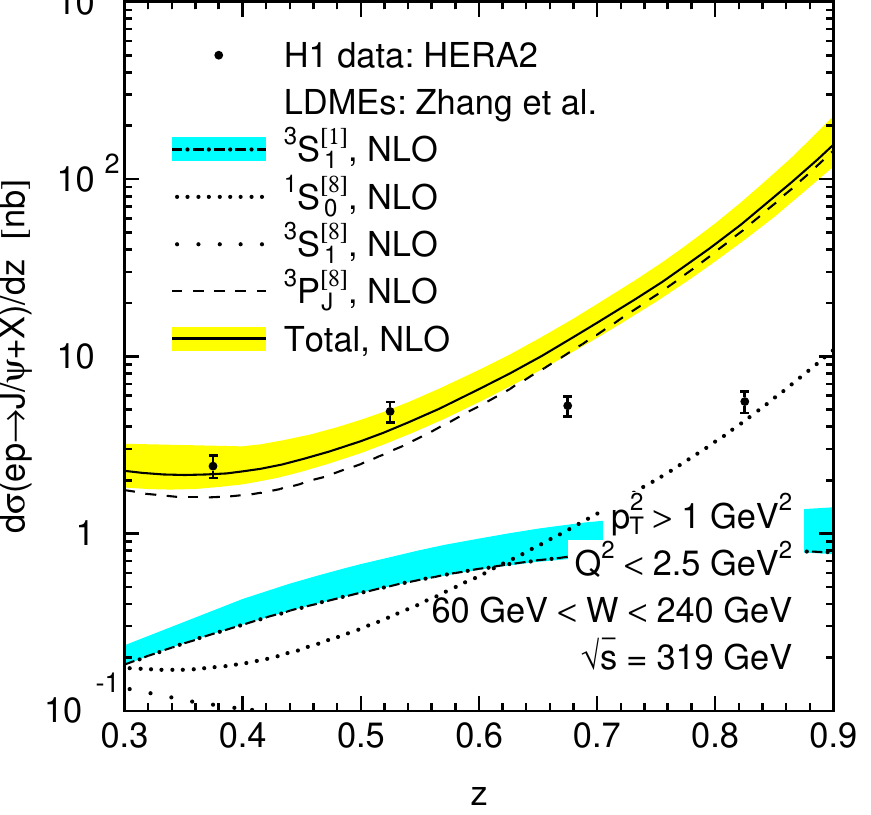} \\
    (a) BK11 & (b) H14& (c) Z14 \vspace{5mm}
    \end{tabular}
    \begin{tabular}{cc}
    \includegraphics[width=0.3\textwidth]{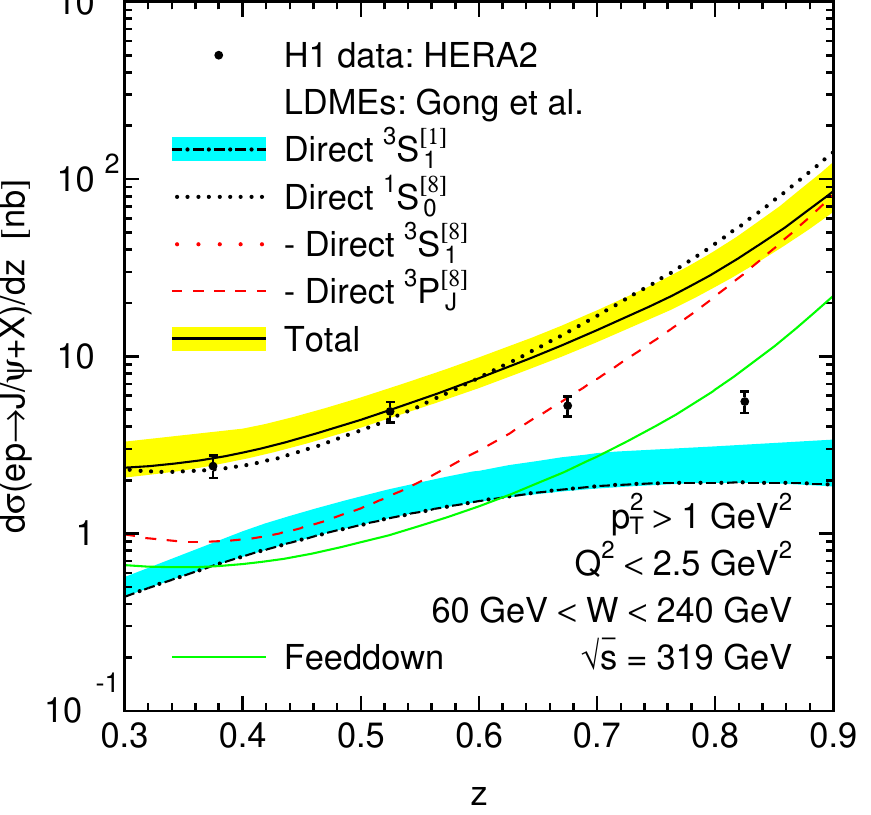}&
    \includegraphics[width=0.3\textwidth]{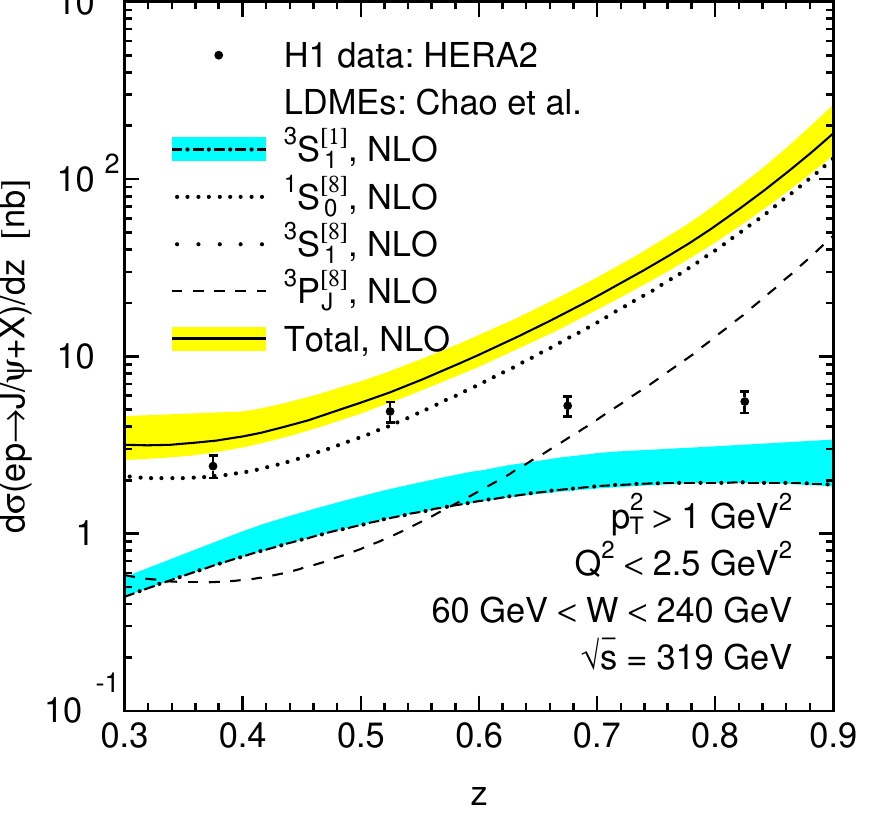}\\
    (d) G13 & (e) C12 \vspace{5mm} 
    \end{tabular}
    \begin{tabular}{cc}
    \includegraphics[width=0.3\textwidth]{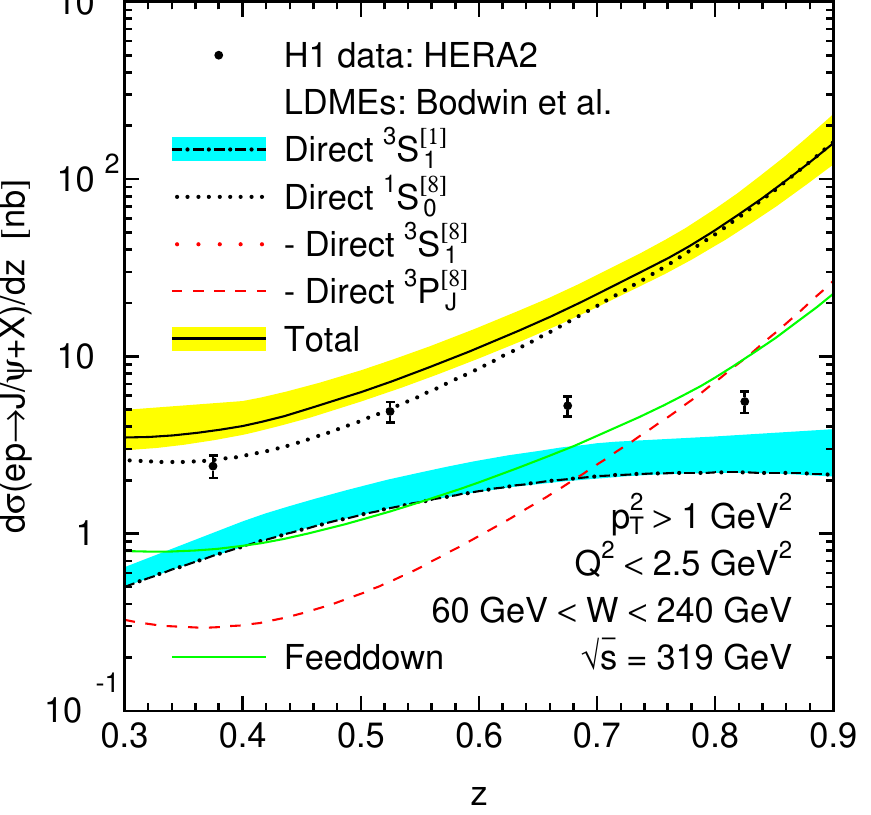}&
    \includegraphics[width=0.3\textwidth]{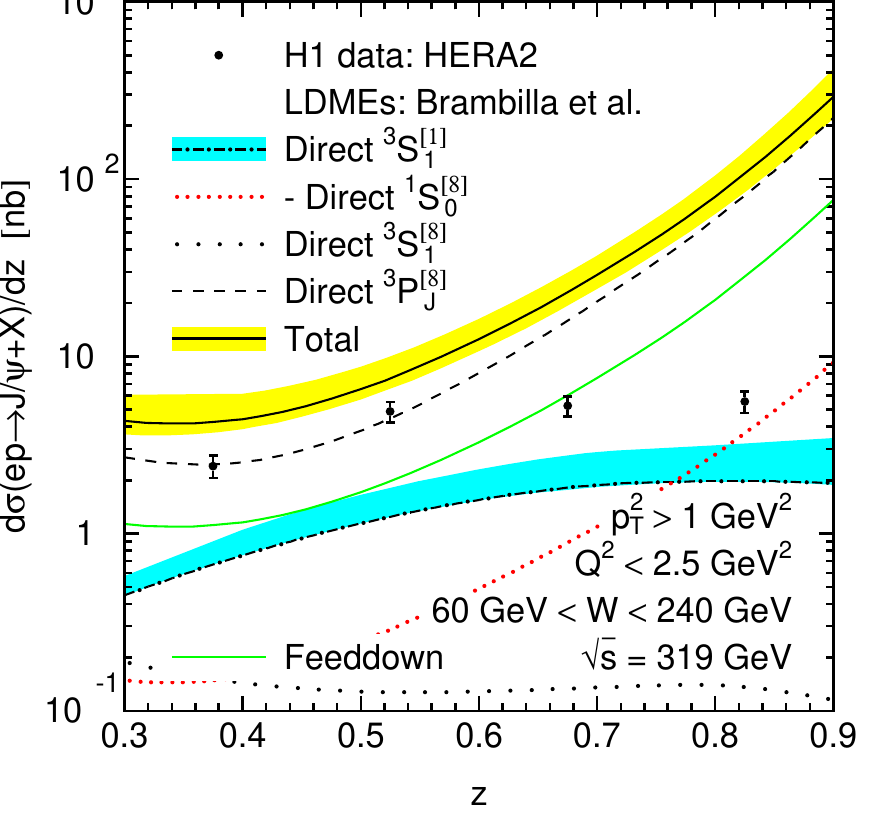}\\
    (f) B14 & (g) pNRQCD 
    \end{tabular}
    \caption{{Description of the HERA H1 data~\cite{Aaron:2010gz} for the $z$-differential  cross section of inclusive $J/\psi$ photoproduction by the NLO NRQCD fits in collinear factorisation for the LDMEs taken listed in \ct{tab:LDME-fits_comp}. The notation for the curves is the same as in the \cf{fig:LDME-fits_vs_H1-data_pT}. }}
    \label{fig:LDME-fits_vs_H1-data_z}
\end{figure}

 In a recent study~\cite{Butenschoen:2022wld}, the NRQCD cross sections of $J/\psi+Z$ and $J/\psi+W$ hadroproduction have been completely calculated at NLO. Interestingly, the only set of LDMEs found to be marginally capable {of reproducing} the $J/\psi+Z$ hadroproduction data from the LHC is the set of %
 Refs.~\cite{Brambilla:2022rjd,Brambilla:2022ayc}, {referred to in the \ct{tab:LDME-fits_comp} as ``pNRQCD''.} This fit uses potential-NRQCD relations between LDMEs to reduce {the} number of free parameters in the fit of the $J/\psi$ $\pT$ %
 spectrum in hadroproduction and which also describes polarisation observables. However{,} this set of LDMEs is not able to describe %
 $J/\psi$ photoproduction {and $e^+e^-$ annihilation} data { and is  consistent with $\eta_c$ hadroproduction data only within large uncertainties and with a $\pT$ threshold for the $\jpsi$ data large than for the $\eta_c$ data.} {As just discussed, the pNRQCD fit, like all the hadroproduction, badly fails to account for the $J/\psi$-photoproduction data {from the H1 collaboration at HERA} as shown on \cf{fig:LDME-fits_vs_H1-data_pT}(g) and \cf{fig:LDME-fits_vs_H1-data_z}(g) which cast doubts on its relevant for EIC predictions.}

Several fits of CO LDMEs for bottomonia have also been performed at  %
NLO~\cite{Gong:2010bk,Wang:2012is,Gong:2013qka}. Only the most recent one~\cite{Gong:2013qka} {considered} the $\Upsilon(1,2,3S)$ and $\chi_{bJ}(1,2P)$ LDMEs independently and systematically included the feed-down contributions from $\Upsilon(nS)$ and $\chi_{bJ}(nP)$ states with larger masses. These feed-down contributions constitute $\sim 40\%$ of the $\Upsilon(1S, 2S)$ cross section, which is significant. In the case of $\Upsilon(3S)$, the feed down from $\chi_b(3P)$ states, which were discovered by ATLAS~\cite{ATLAS:2011icy} and which lie just below the $B\bar{B}$-threshold, also turns out to be significant (see \cite{Lansberg:2019adr} for a more detailed discussion of the feed-down impact). This was{,}  however{,} not taken into account in~\cite{Gong:2013qka}. {This may explain the difficulties of the corresponding fit to account for the $\Upsilon(3S)$ polarisation.} The polarisation observables for $\Upsilon(1S,2S)$ states came out to be about consistent with data in this fit. We guide the reader to the recent Ref.~\cite{Feng:2020cvm} for a detailed discussion of the agreement with various polarisation observables. {Note that there is no bottomonium data from inelastic photoproduction nor from $e^+e^-$ annihilation.} Hence, future measurements of $\Upsilon(nS)$ inclusive electro- and photoproduction at the EIC will serve as an excellent test of the LDME process-independence in the $b$-quark case, where it has more chances to hold due to smaller $O(v^2)$ corrections.

\subsubsection{Recent developments regarding inclusive $J/\psi$ photoproduction within the CSM}
\label{sec:EIC-CSM-NLOstar}
 \paragraph{New $P_T$-enhanced contributions.}
The recent study of Ref.~\cite{Flore:2020jau}, performed within {the} CSM, is interesting regarding corrections {which} %
were not included in the NLO NRQCD analyses presented above, although they could become important at $\pT\gg M_{\jpsi}$.  The study {focused on} the leading-$\pT$ leading-$v$ next-to-leading-$\alpha_s$ corrections, within the NLO$^\star$ approximation~\cite{Artoisenet:2008fc,Lansberg:2008gk}. 
The latest HERA data from {the} H1 Collaboration~\cite{Aaron:2010gz} was first revisited, by including new contributions such as the pure QED one ($\gamma + q \to \gamma^\star + q\to \jpsi + q$  at $\mathcal{O}(\alpha^3)$ where the off-shell photon $ \gamma^\star$ fluctuates into a $\jpsi$) 
and the associated \jpsi+ charm production ($\gamma + g \to \jpsi + c +  \bar{c}$ and $\gamma +\{c,\bar{c}\} \to \jpsi+ \{c,\bar{c}\}$). The former involves quark PDFs in the initial state, while the latter is described within a LO Variable Flavour Number Scheme (LO-VFNS)~\cite{Aivazis:1993pi,Shao:2020kgj}. It was shown that the CSM at $\mathcal{O}(\alpha \alpha_s^3)$ and $\mathcal{O}(\alpha^3)$ is able to describe the latest HERA data at large $\pT$. {The NLO corrections to $\gamma + g \to \jpsi + c +  \bar{c}$ were recently computed~\cite{Feng:2024heh} and were found to increase the cross section a factor close to 2 in the HERA kinematics.}

{The corresponding p}redictions for the $\pT(\jpsi)$ spectrum in photoproduction at the EIC %
are shown in \cf{fig:EIC-CT14NLO-NLOstar} %
{with kinematical cuts on $Q^2$, the elasticity, $z$, and $W_{\gamma p} \equiv \sqrt{s_{\gamma p}}$ inspired from the latest} H1 measurements.  %
The CT14NLO proton PDF set~\cite{Dulat:2015mca} was used. The factorisation and renormalisation scales were taken to be $\mu_F = \mu_R = m_T = \sqrt{M_{\jpsi}^2 + P_T^2}$, {the transverse mass} {of the $\jpsi$}, \newb{later this is called $m_{T J/\psi}$} and the corresponding uncertainties were evaluated {by varying them} in the interval {$\mu_F, \mu_R \in [1/2, 2]\times m_T$}. The charm mass $m_c$ was set to 1.5 GeV and the corresponding mass uncertainty was evaluated by varying it by $\pm 0.1$ GeV. Moreover, the CS LDME $\langle \mathcal{O}^{\jpsi}\left[{^{3\!}S_{1}^{[1]}}\right] \rangle$ was taken to be 1.45~GeV$^3$. Finally, a $20\%$ feed-down $\psi^\prime \to \jpsi$ was taken into account.

\begin{figure}[h]
\centering
\includegraphics[width=8cm]{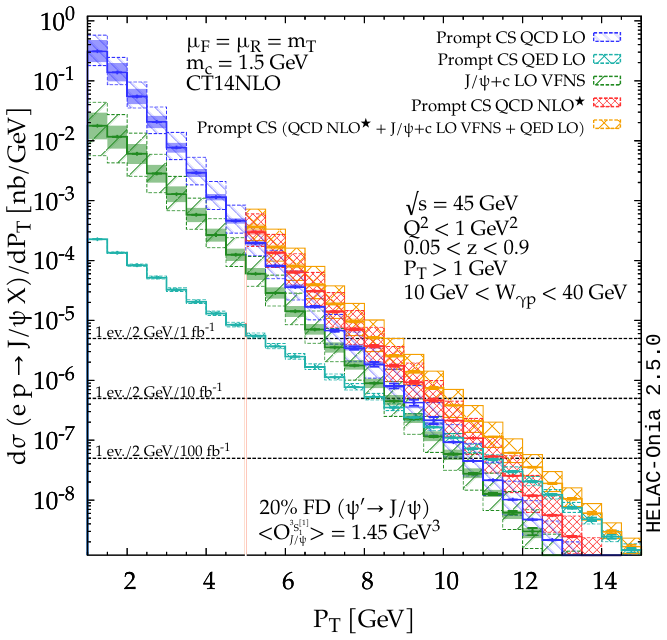}\hspace{3mm}
\includegraphics[width=8.1cm]{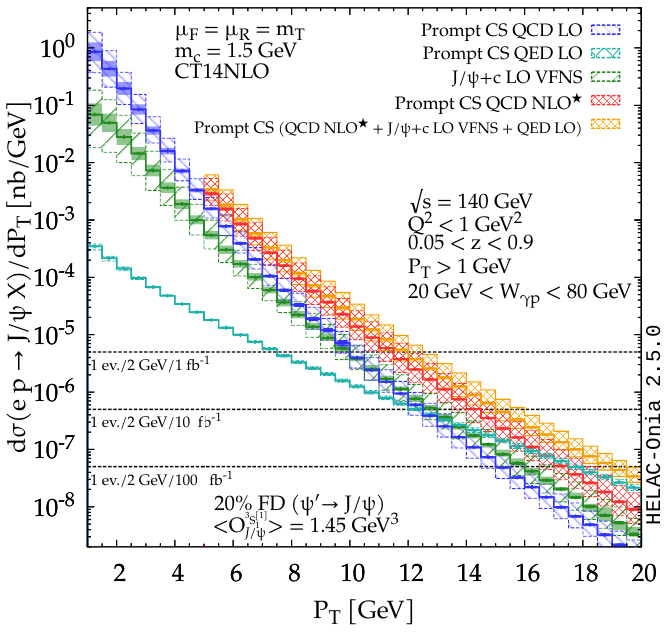}
\caption{Predictions for the future EIC at $\sqrt{s_{ep}} = 45$ GeV (left) and $\sqrt{s_{ep}} = 140$ GeV (right) {as a function of the $\jpsi$ transverse momentum, $P_T$}. The solid bands indicate the mass uncertainty while the patterns display the scale uncertainty. Figure taken from Ref.~\cite{Flore:2020jau}.}
\label{fig:EIC-CT14NLO-NLOstar}
\end{figure}

In \cf{fig:EIC-CT14NLO-NLOstar}, predictions for two {energy} configurations are presented. At $\sqrt{s_{ep}} = 45$ GeV (\cf{fig:EIC-CT14NLO-NLOstar}, left), as $\pT$ increases, one enters the valence region. This makes the QED contribution become the dominant one at the largest measurable $\pT \simeq 11$ GeV, with an integrated luminosity of $\mathcal{L} = 100$ fb$^{-1}$. Furthermore, $\gamma + q$ fusion contributes more than $30\%$ for $\pT > 8$ GeV and the $\jpsi\,+$ unidentified charm contribution is comparable to the $\gamma + g(q)$ fusion subprocesses. Hence, these so far overlooked contributions are going to be relevant at the EIC. At $\sqrt{s_{ep}} = 140$ GeV (right panel in \cf{fig:EIC-CT14NLO-NLOstar}),  the yield is measurable up to $\pT \sim 18$ GeV. The QED contribution is the leading one at the largest reachable $\pT$, while $\gamma + g$ fusion is the dominant contribution up to $\pT \sim 15$ GeV. More generally, it turns out that the production of $\jpsi+2$ hard partons (\ie $\,\jpsi+\{gg, qg, c\bar{c}\}$) is dominant for $\pT \sim 8 - 15$ GeV. This could lead to the observation of $\jpsi + 2$ jets with moderate $\pT$, with the leading jet$_1$ recoiling on the $\jpsi + {\rm jet}_2$ pair.

\paragraph{High-energy-enhanced contributions.} 
In order to study the possible effects of higher-order QCD corrections enhanced by logarithms of the partonic centre-of-mass energy ($\hat{s}$), the Leading-Twist (LT) High-Energy Factorisation~\cite{Catani:1990xk,Catani:1990eg,Collins:1991ty,Catani:1994sq} (HEF) can be used. In many phenomenological studies, it is generalised to include, not only the resummation of ${\ln} (\hat{s}/{M^2_{\Q}})$-enhanced effects in the leading-logarithmic approximation, but also the resummation of the ``Sudakov'' ${\ln} ({M_{\Q}}/P_T)$ large logarithms at $P_T\ll {M_{\Q}}$ in the next-to-leading logarithmic approximation, assuming  {CS} state production, through the use of {the} Kimber-Martin-Ryskin-Watt (KMRW formula)~\cite{Kimber:2001sc,Watt:2003mx,Watt:2003vf}. However, the systematic study of the overlap between LT HE factorisation and the TMD factorisation 
usually employed to resum {such} transverse-momentum logarithms has been initiated only very recently~\cite{Nefedov:2021vvy,Hentschinski:2021lsh}. The KMRW formula converts the set of usual collinear PDFs to the so-called unintegrated PDFs (uPDFs) of the LT HEF formalism. uPDFs depend not only on the longitudinal momentum fraction, $x$, but also on the transverse momentum of the parton. These objects can {yield} transverse momenta comparable to, or even larger than, $M_ {\Q}$ to the final state. This is indeed possible in the Regge limit $\hat{s}\gg {M_{T\Q}}$. For {a} more detailed review of the LT HEF and its connection to quarkonium physics, see %
Section 4.3 of %
Ref.~\cite{Chapon:2020heu}.

It has been shown earlier~\cite{Lipatov:2002tc,Kniehl:2006sk} that the phenomenological framework based on HEF with KMRW uPDF is capable of reproducing the $J/\psi$ photoproduction data from HERA. {This is already the case with} the HEF coefficient function {computed at LO in $\alpha_s$} and in the CS approximation of  NRQCD{,} as illustrated by the left panel of %
\cf{fig:HEF_H1-2010_EIC} {obtained with} the version of KMRW uPDF introduced in the Ref.~\cite{Nefedov:2020ugj}. We note that the transverse-momentum integral of the uPDF exactly reproduces the input gluon PDF. The %
{precise} %
fulfilment of this normalisation condition both at $x\ll 1$ and $x\sim 1$ is important to avoid contradictions between LT HEF and NLO %
{Collinear Factorisation (CF)} %
predictions for {the} $J/\psi$ prompt hadroproduction $\pT$ spectrum in $pp$ collisions at low energies, in particular at $\sqrt{s_{pp}}=24$ GeV for the planned Spin-Physics-Detector experiment at {the} NICA facility~\cite{Arbuzov:2020cqg, Karpishkov:2020wwe}.

\begin{figure}[ht]
    \centering
    \includegraphics[width=0.4\textwidth]{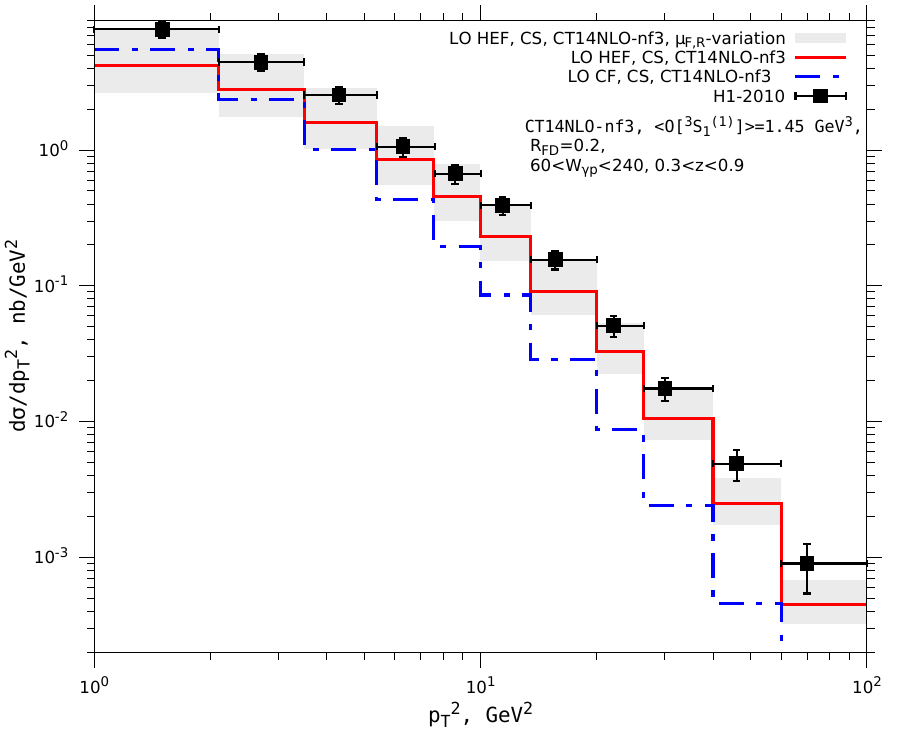}
    \includegraphics[width=0.4\textwidth]{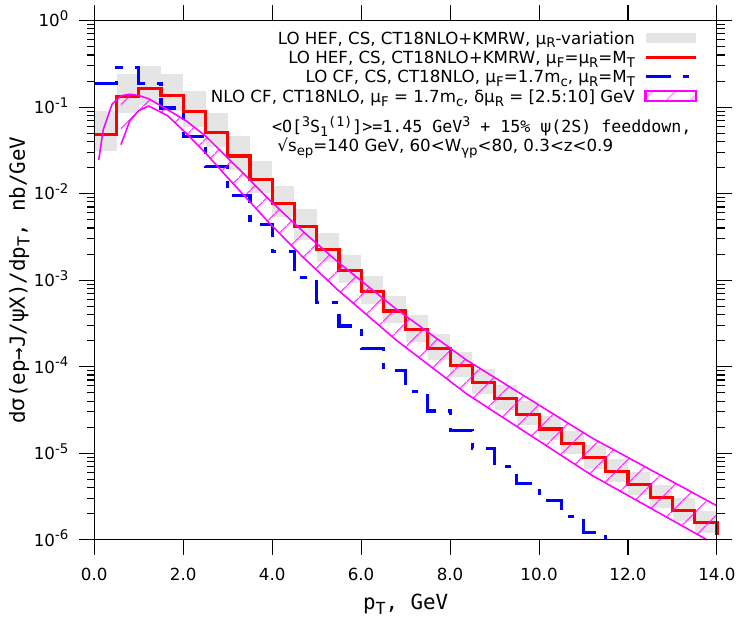}
    \caption{Left panel: LO HEF results (solid histogram) for the H1-2010~\cite{Aaron:2010gz}  {prompt} $J/\psi$ photoproduction {$\pT^2(\jpsi)$} spectrum within the CSM %
    {compared to LO CF results} (dash-dotted histogram). See the main text for details. Right panel: Comparison between the LO HEF prediction (solid histogram with $\mu_R$-variation band) for the {prompt}     ${\pT}(\jpsi)$ spectrum at the EIC with the NLO CF prediction (shaded $\mu_R$-variation band) evaluated at the optimal value of factorisation scale proposed in Ref.~\cite{ColpaniSerri:2021bla}.}
    \label{fig:HEF_H1-2010_EIC}
\end{figure}

From \cf{fig:HEF_H1-2010_EIC} (left), one can see that there is still some room for additional contributions on top of the LO CS contribution from {the fusion of a photon and a Reggeon} $\gamma(q)+R(x_1,{\bf q}_{T1})\to c\bar{c}[{}^3S_1^{[1]}]+g$%
. {These could be from} $c\bar{c}[{}^3S_1^{[1]}]+c$ {considered above} (see \cf{fig:EIC-CT14NLO-NLOstar}) and {from} CO contributions. The large scale uncertainty of the LO HEF prediction, shown in the \cf{fig:HEF_H1-2010_EIC}, comes from the variation of $\mu_R$ and $\mu_F$ around their default value of {$M_{TJ/\psi}$}. Clearly, the uncertainty %
{has} to be reduced via {the} inclusion of the NLO corrections to make such predictions {more precise}.  

The comparison between LT HEF predictions for {the} EIC energy $\sqrt{s_{ep}}=140$ GeV %
and the full NLO CF 
CSM predictions {(computed using FDC~\cite{Wang:2004du}) is shown in \cf{fig:HEF_H1-2010_EIC} (right). The latter prediction is evaluated} at %
{a} special value of {the} factorisation scale, $\mu_F=1.7 m_{c}${,} chosen~\cite{ColpaniSerri:2021bla} to minimise the NLO correction coming from the region of $\hat{s}\gg {M_{\jpsi}^2}${ (cfr. Section~\ref{sec:gluon_pdf_inclusive}). There is} %
a good agreement between these NLO predictions at the optimal scale and LO HEF predictions at the default scale $\mu_R=\mu_F={M_{T,J/\psi}}$%
{: t}his indicates that the effects of the $\ln(\hat{s}/{M_{\jpsi}^2})$ 
resummation can be reproduced by the optimal factorisation scale choice at EIC energies and that the NLO CF prediction with the optimal scale is robust. At higher photon-nucleon collision energies, a matched calculation between {LL} HEF and NLO CF predictions, similar {to that} done in Ref.~\cite{Lansberg:2021vie}, is necessary~\cite{Lansberg:2023kzf} to correctly capture the high-energy resummation effects at $\hat{s}\gg {M_{\jpsi}^2}$ while staying at NLO accuracy for $\hat{s}\sim {M_{\jpsi}^2}$.

\subsubsection{Testing NRQCD factorisation at the EIC}
\label{sec:testing_NRQCD_EIC}

\begin{figure}
     \centering
    \subfloat[]{ \includegraphics[width=0.4\textwidth]{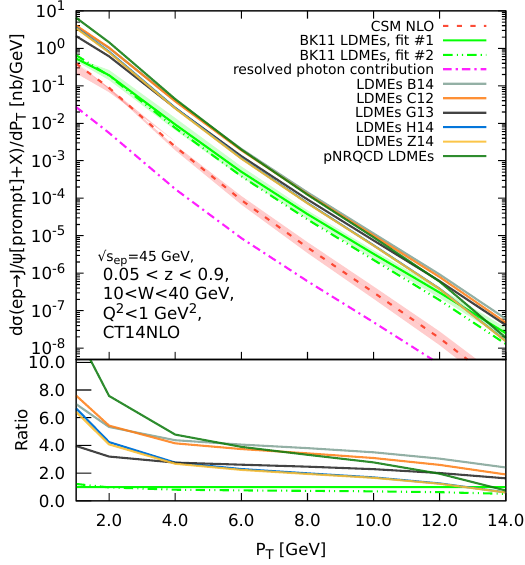}} 
    \subfloat[]{\includegraphics[width=0.4\textwidth]{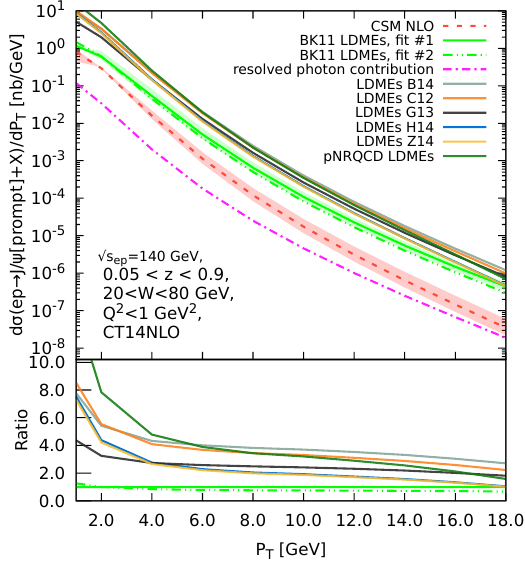}}
     \caption{
     {Predictions using NRQCD at NLO} for {the} {$\jpsi$ transverse momentum} ($\pT$) differential prompt{-}$J/\psi$ photoproduction cross section in the EIC kinematic conditions: (a) $\sqrtsep=45$~GeV and (b)$\sqrtsep=140$~GeV  {for the} various LDME sets listed in \ct{tab:LDME-fits_comp} as well as of the %
     {CSM} (dashed line, $\langle {\cal O}^{J/\psi} \left[ {}^3S_1^{[1]} \right] \rangle = 1.45$ GeV$^3$) are shown. The scale-variation uncertainty bands are {only} plotted for the prediction of {the BK} LDME set%
     ~\cite{Butenschoen:2011yh} as well as for the CSM. The resolved-photon contribution {also refers to the BK LDME set}. %
     The AGF~\cite{Aurenche:1994in} photon PDF set has been used. The calculation of the short-distance cross sections is based on~\cite{Butenschoen:2009zy,Butenschoen:2010rq}.}
     \label{fig:Jpsi-photoprod-pT_LDMEs}
 \end{figure}

\begin{figure}[hbt!]
     \centering
    \subfloat[]{\includegraphics[width=0.4\textwidth]{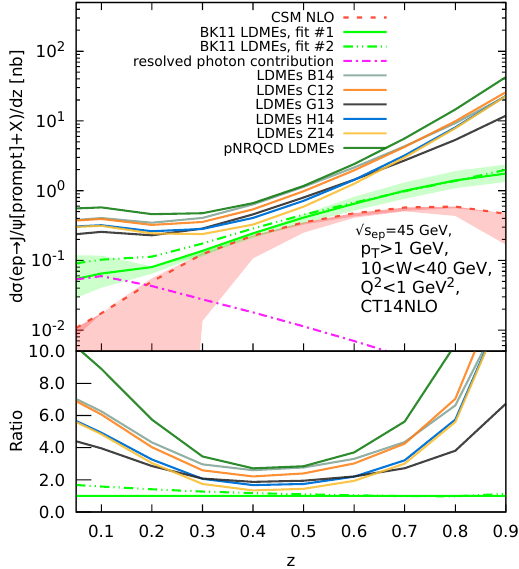}} \subfloat[]{\includegraphics[width=0.4\textwidth]{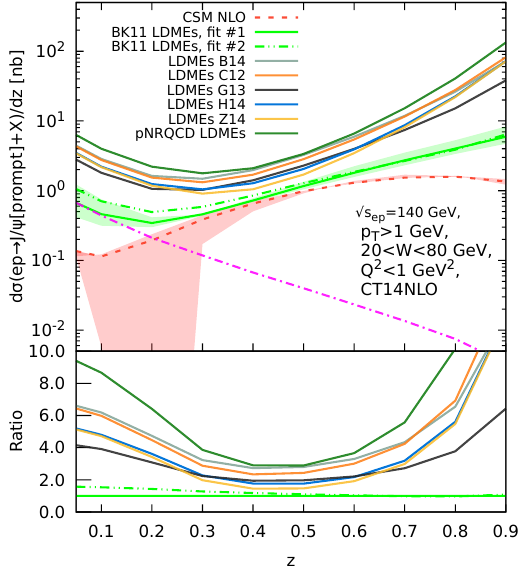}}
     \caption{
     {Same as \cf{fig:Jpsi-photoprod-pT_LDMEs} but for the $z$-differential cross section.}}
    \label{fig:Jpsi-photoprod-z_LDMEs}  
 \end{figure}

\paragraph{Prompt $J/\psi$ yields {in photoproduction}.}%
We plot in Figs.~\ref{fig:Jpsi-photoprod-pT_LDMEs} and~\ref{fig:Jpsi-photoprod-z_LDMEs} the NLO NRQCD factorisation predictions for {the} %
{$\pT$}{-} and $z$-differential photoproduction cross section of prompt $J/\psi$ mesons in the EIC kinematic conditions. These predictions have been calculated using the short-distance cross sections of Refs.~\cite{Butenschoen:2009zy,Butenschoen:2010rq} and the LDME sets listed in \ct{tab:LDME-fits_comp}. All LDME sets fitted only %
{to} %
the hadroproduction data predict {a} significantly (factor 3 to 6) higher $J/\psi$ photoproduction cross section than the LDME set of %
Table 1 of %
Ref.~\cite{Butenschoen:2011yh}, denoted as ``LDMEs Kniehl, Butensch\"on, fit \# 1'' 
{in Figs.~\ref{fig:Jpsi-photoprod-pT_LDMEs} and~\ref{fig:Jpsi-photoprod-z_LDMEs}}, which includes the photoproduction data from HERA. We also plot the predictions performed with another set of LDMEs from the same paper, denoted as ``LDMEs Kniehl, Butensch\"on, fit \# 2''. The latter set of LDMEs had been fitted %
{to} %
the prompt $J/\psi$ hadro{-} %
and photoproduction data corrected approximately for feed-down contributions from heavier charmonium states using constant feed-down fractions. For this fit, we calculate the feed-down contributions from $\chi_{c0,1,2}$ and $\psi(2S)$ decays to $J/\psi$ using the $\chi_c$ LDMEs from Ref.~\cite{Ma:2010vd} %
{and} the fit for $\psi(2S)$ LDMEs performed in %
Ref.~\cite{Butenschoen:2022qka}. Calculating the feed-down contribution in this way is consistent with the treatment of feed-down in %
Ref.~\cite{Butenschoen:2011yh}.

 {As expected, the predictions from both Kniehl-Butensch\"on LDMEs are reasonably close to each other. Yet, they differ from those obtained with the other LDME sets fit to hadroproduction data.}
This is mostly because %
{the latter sets} predict {a} more pronounced $z\to 1$ growth of the cross section (see \cf{fig:Jpsi-photoprod-z_LDMEs}) than the global fit LDME sets of Ref.~\cite{Butenschoen:2011yh} {which, when integrated over $z$, translates into larger \pT differential cross sections}. {This increase is due to both the ${}^1S_0^{[8]}$ and ${}^3P_J^{[8]}$ CO states.}

{It is important to note that} such {a} rapid increase of the spectrum towards $z\to 1$  %
{is not a feature of the} HERA data. {Including these data in LDME fits calls for a} 
compensation between contributions of $\langle{\cal O}^{J/\psi} \left[ {}^1S_0^{[8]}\right] \rangle$ and $\langle{\cal O}^{J/\psi} \left[ {}^3P_J^{[8]}\right] \rangle$ LDMEs %
{resulting in} different signs {for these as in the LDME sets of Ref.~\cite{Butenschoen:2011yh}}. Therefore the photoproduction data essentially fix the latter LDMEs and do not allow anymore to adjust them to describe the polarisation observables in hadroproduction, which leads to the polarisation puzzle discussed above. The EIC measurements will allow us to check {the} robustness of this feature of NRQCD predictions against variation of collision energy, since larger radiative corrections at $z\to 1$ could be expected at higher energies of {the} HERA collider.
 
The resolved-photon contribution manifests itself in the opposite region $z\ll 1$ (\cf{fig:Jpsi-photoprod-z_LDMEs}) and EIC data are less sensitive to it than HERA data, again due to lower collision energies. Therefore the cleaner test of process-independence of LDMEs can be performed with EIC photoproduction data {rather} than with HERA data.

\paragraph{Prompt $\jpsi$ yields in $Q^2$-integrated lepton-nucleon interactions.}
\label{sec:inclusive-lepto-prod-Jpsi}

{Another possibility to study the contributions of various LDMEs is to consider {\it single-inclusive} production of $J/\psi$ in $ep$ collisions, without detecting the final-state electron, as was pioneered recently in  Ref.~\cite{Qiu:2020xum}}: $e(\ell)+h(p)\to J/\psi(P)+X$. The rapidity ($y$) and transverse momentum (\pT) distributions of \jpsi inclusive %
production at the EIC are promising observables for both studying the production mechanism of heavy quarkonia and extracting %
{PDFs}, in particular, the gluon PDF, complementary to other observables described in Section~\ref{sec:unporlarized_N_PDFs}. %
 
When the transverse momentum of \jpsi \ {defined relatively to the lepton-hadron collision axis} %
$\pT \gg m_c$ %
the perturbative hard coefficient functions for producing the \ccbar pair receive large higher-order QCD corrections that are enhanced by powers of ${\ln}(P_T^2/m_c^2)$.  Such logarithmically-enhanced higher-order corrections can be systematically resummed and factorised into %
FFs~\cite{Berger:2001wr,Nayak:2005rt,Nayak:2006fm,Kang:2014tta,Kang:2014pya}.  
On the other hand, when $\pT\gtrsim m_c$, the perturbative hard coefficients at a fixed order in $\alpha_s$ should be sufficient.

In addition, the occurrence of {a} hard partonic collision producing the \jpsi with large transverse momentum $\pT \gg m_e$%
necessarily induces multiple photon emissions from the incoming lepton, leading to large higher-order QED corrections enhanced by powers of ${\ln}(\pT^2/m_e^2)$. {As we discussed in Section~\ref{sec:QED-corr}}, these QED corrections can also be systematically factorised and resummed into universal %
LDFs~\cite{Liu:2020rvc,Liu:2021jfp}.  %
{In order to predict} the production rate of \jpsi at the EIC, a new factorisation formalism, which takes into account both collision-induced QCD and QED radiation and provides a systematic transition from $\pT\gtrsim m_c$ to $\pT\gg m_c$, was introduced~\cite{Lee:2021oqr,QW:2024}{.} %
{T}he %
factorisation formula for the inclusive %
production cross section {is given by}:
\begin{eqnarray}
\label{eq:jpsi-lp-fac}
E_{{J/\psi}}\frac{d\sigma_{eh\to J/\psi(P_{\jpsi})X}}{d^3{\bf P}_{\jpsi}} 
&=& \sum_{a,b} \,
\int dx_a\, f_{a/e}(x_a,\mu_F^2)
\int dx_b\, f_{b/h}(x_b,\mu_F^2)
\nonumber\\
&&\times\Bigg[ 
E_{{J/\psi}}\frac{d\tilde{\sigma}^{\rm Resum}_{ab\to J/\psi(P_\jpsi)X}}{d^3{\bf P}_{\jpsi}}
+
E_{{J/\psi}}\frac{d\tilde{\sigma}^{\rm NRQCD}_{ab\to J/\psi(P_\jpsi)X}}{d^3{\bf P}_{\jpsi}}
- E_{{J/\psi}}\frac{d\tilde{\sigma}^{\rm Asym}_{ab\to J/\psi(P_\jpsi)X}}{d^3{\bf P}_{\jpsi}}
\Bigg]\, ,
\end{eqnarray}
where indices $a,b$, in principle, run{, respectively, } over all lepton and parton flavors, but in practice, as an approximation, $a$ takes into account only ($e,\gamma,\bar{e}$). The functions $f_{a/e}(x_a,\mu_F^2)$ and $f_{b/h}(x_b,\mu_F^2)$ are {the} LDFs of an electron and {the usual parton} PDFs %
respectively, depending on  partonic momentum fractions, $x_a$ and $x_b$. {The LDFs satisfy the DGLAP-like $\mu_F$-evolution  equations mixing the QED and QCD splittings~\cite{QW:2024}.}  In Eq.~(\ref{eq:jpsi-lp-fac}), the  partonic cross sections $\tilde{\sigma}_{ab\to J/\psi(P_{\jpsi})X}$ are computed with all {the} perturbative collinear singularities along the {direction of}  colliding lepton ($a$) and parton ($b$) removed. These singularities are absorbed into $f_{a/e}$ and $f_{b/h}$, respectively.  

{The cross section} d$\tilde{\sigma}^{\rm Resum}$ in \ce{eq:jpsi-lp-fac} represents {the} partonic cross section with {the} ${\ln}(P_T^2/m_c^2)$ {contributions being} resummed to describe {the} \jpsi production rate for $\pT\gg m_c${, as we have mentioned above}. In %
$\tilde{\sigma}^{\rm NRQCD}$, the production of $c\bar{c}{[^{2S+1}L^{[1,8]}_J]}$-state at the perturbative stage is computed at fixed order in $\alpha_s$ and {the} corresponding non-perturbative formation of {a} \jpsi from a produced \ccbar pair {is} taken care %
{using the} NRQCD velocity expansion and universal NRQCD LDMEs. This part of the cross section  should provide a good description of the \jpsi production rate when $\pT\sim m_c$.  Finally, $\tilde{\sigma}^{\rm Asym}$ is equal to a fixed-order expansion of $\tilde{\sigma}^{\rm Resum}$ to the same order in $\alpha_s$ as in $\tilde{\sigma}^{\rm NRQCD}$. The latter part is needed to remove the double counting between $\tilde{\sigma}^{\rm Resum}$ and $\tilde{\sigma}^{\rm NRQCD}$. %
{By including all these three terms, t}his factorisation formalism %
can be applied to both lepton-hadron and hadron-hadron collisions, as well as $e^+e^-$ collisions~\cite{Lee:2021oqr,Lee:2022anw}, providing a smooth transition when observed $\pT\sim m_c$ increases to $\pT\gg m_c$.  

The predictive power of \ce{eq:jpsi-lp-fac} relies on the factorisation of each term and our ability to calculate them. {Up to next-to-leading power corrections in $m_c/\pT$}, the $\tilde{\sigma}^{\rm Resum}$ can be factorised as~\cite{Nayak:2005rt,Nayak:2006fm,Kang:2014tta,Kang:2014pya}, 
\begin{eqnarray}
	\label{eq:jpsi-lp-resum}
	E_{{J/\psi}}\frac{d\tilde{\sigma}^{\rm Resum}_{ab\to J/\psi(P_{\jpsi})X}}{d^3{\bf P}_{\jpsi}}
	&\approx&\sum_{k} \int \frac{dz}{z^2} D_{k\to J/\psi}(z,\mu_F^2) 
	E_k\frac{d\hat{\sigma}_{ab\to k(p_k)X}}{d^3{\bf p}_k}(z,p_k=P_{\jpsi}/z,\mu_F^2)
	\\
	& & +
	\sum_{\kappa} \int \frac{dz}{z^2} {D}_{[\ccbar(\kappa)]\to J/\psi}(z,\mu_F^2) 
	E_k\frac{d\hat{\sigma}_{ab\to [\ccbar(\kappa)](p_k)X}}{d^3{\bf p}_c}(z,p_k=P_{\jpsi}/z,\mu_F^2)
	\, ,
	\nonumber
\end{eqnarray}
where $k=q,g,\bar{q}$ and $\kappa=v,a,t$ %
{for  $\ccbar$ pairs respectively} in a vector, axial-vector or tensor spin state~\cite{Kang:2014tta,Kang:2014pya}. {T}he first and second term{s} are the factorised leading power (LP) and next-to-leading power (NLP) contribution{s} to the cross section in its $1/\pT$ expansion. The corrections to \ce{eq:jpsi-lp-resum} are suppressed by $1/\pT^4$ and cannot be further factorised~\cite{Qiu:1990xy}.  The universal single-parton and double-parton (\ccbar) FFs, $D_{c\to J/\psi}(z,\mu_F^2)$ and ${D}_{[\ccbar(\kappa)]\to J/\psi}(z,\mu_F^2)$, respectively, {satisfy a closed set of evolution equations with respect to %
{changes} of the factorisation scale $\mu_F$}~\cite{Kang:2014tta,Kang:2014pya}.  Solving {these evolution equations one} resums the logarithmic contributions {scaling like $\ln(P_T^2/m_c^2)$} to these FFs. The universal FFs at an input scale $\mu_F = \mu_0{\simeq 2m_c}$ can be calculated assuming NRQCD factorisation~\cite{Bodwin:1994jh} in terms of universal NRQCD LDMEs,
\begin{eqnarray}
	D_{c\to J/\psi}(z,\mu_0^2) 
	&\approx& \sum_{\ccbar{[^{2S+1}L_J]}} \hat{d}_{c\to \ccbar{[^{2S+1}L_J]}}(z,\mu_0^2) \langle O_{\ccbar{[^{2S+1}L_J]}}^{\jpsi}(0)\rangle \, ,
	\label{eq:jpsi-ffs-c}\\
	{D}_{[\ccbar(\kappa)]\to J/\psi}(z,\mu_0^2) 
	&\approx & \int_{-1}^{1} du \int_{-1}^{1} dv\ {\cal D}_{[\ccbar(\kappa)]\to J/\psi}(z,u,v,\mu_0^2) 
	\label{eq:jpsi-ffs-cc} \\
	&\approx& \sum_{\ccbar{[^{2S+1}L_J]}} \hat{d}_{[\ccbar(\kappa)]\to \ccbar{[^{2S+1}L_J]}}(z,\mu_0^2) \langle O_{\ccbar{[^{2S+1}L_J]}}^{\jpsi}(0)\rangle
	\, {.}
	\nonumber
\end{eqnarray}
{\ce{eq:jpsi-ffs-cc} involves further approximations, neglecting possible differences between the momentum fractions carried by the \ccbar pair in the amplitude, $u$, and and its complex-conjugate, $v$, which can be taken into account through the more general FF ${\cal D}_{[\ccbar(\kappa)]\to J/\psi}$, defined in \cite{Kang:2014tta}.} The approximation in {the second line of} \ce{eq:jpsi-ffs-cc} reflects the fact that the %
{integral of this function is dominated by the vicinity of} $u=v=1/2$~\cite{Lee:2021oqr,Lee:2022anw}.

{The formalism described above has been already tested partially in the case of $pp$ collisions, where instead of LDFs in \ce{eq:jpsi-lp-fac} one substitutes  the proton PDFs.} With perturbatively calculated short-distance matching coefficients for both single-parton and \ccbar{-}pair FFs at the input scale~\cite{Ma:2013yla,Ma:2014eja} and solving the coupled evolution equations for these FFs, the factorised and resummed cross section in \ce{eq:jpsi-lp-resum} %
describe{s} %
the $\pT$ distribution of \jpsi production at the LHC and Tevatron~\cite{Lee:2021oqr,Lee:2022anw} {for  $\pT > 10$~GeV, as we note in %
\ct{tab:LDME-fits_comp}}. %
At the LHC energies, the LP contributions, {namely} the first term in \ce{eq:jpsi-lp-resum}, dominate when $\pT \gg 20$~GeV, while the NLP contributions, namely the second term in \ce{eq:jpsi-lp-resum}, are comparable at $\pT\sim 20$~GeV and become dominant when $\pT$ further decreases, which is critically important to describe the shape of the observed $P_T$ distribution.

\begin{figure}[!t]
	\begin{center}
        \includegraphics[width=0.45\textwidth]{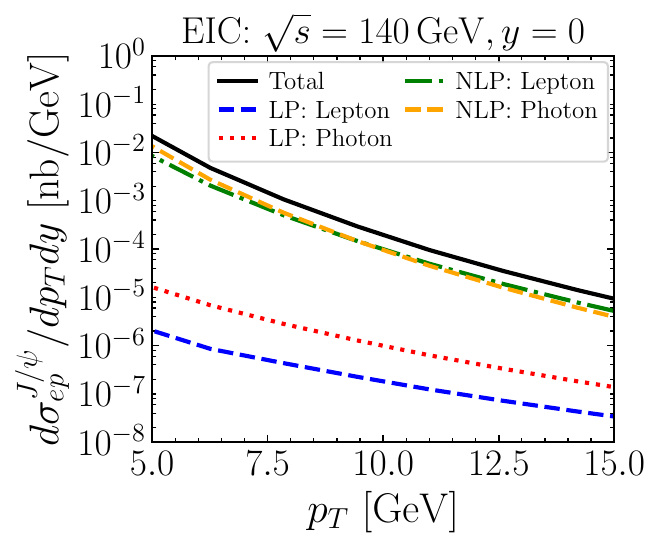} 
\caption{{The $\jpsi$ transverse momentum} ($P_T$) distribution of {the} inclusive $J/\psi$ production cross section in electron-proton collisions in the electron-hadron {centre-of-mass} frame without tagging the scattered electron, computed by using the new factorisation formalism in Eq.~(\ref{eq:jpsi-lp-fac})~\cite{QW:2024}.  The solid black line (overlap with the dashed orange line) is for {the} total contribution, which is dominated by the subprocess:  {$\gamma + g \to [c\bar{c}] + g$ (NLP Photon) and $e + g \to [c\bar{c}] + e$ (NLP Lepton)}  with the \ccbar pair fragmenting to \jpsi, while others represent contributions from other subprocesses, see the text for details.%
}
		\label{fig:jpsi-pt-eic}
	\end{center}
\end{figure}

Making predictions {of the \pT distribution of inclusive \jpsi production} at the EIC requires the knowledge of the universal LDFs.  In \cf{fig:LDFs} of \ref{sec:LDFs}, the scale dependence of the LDFs with and without the mixing of QED and QCD evolution is shown.  
Like in any factorisation approach, the perturbatively calculated short-distance partonic cross section, such as $\hat{\sigma}_{ab\to k(p_k)X}$ in \ce{eq:jpsi-lp-resum}, does not depend on the details of the hadronic state produced. It has been calculated for single hadron production at LO~\cite{Kang:2011jw}, at NLO~\cite{Hinderer:2015hra,Hinderer:2017ntk}, and at NNLO~\cite{Boughezal:2018azh,Abelof:2016pby}. The fixed-order calculation for $\tilde{\sigma}^{\rm NRQCD}$ has been carried out in NRQCD up to NLO~\cite{Qiu:2020xum}.  

In \cf{fig:jpsi-pt-eic}, we present the predictions of the \pT distribution of inclusive \jpsi production in $ep$ collisions at the EIC for {$\sqrt{s_{ep}}=140$~GeV}. {For these predictions, only the $\tilde{\sigma}^{\rm Resum}$ term in \ce{eq:jpsi-lp-fac} is used and the same LDMEs that we used for describing the \jpsi production at the LHC and Tevatron energies~\cite{Lee:2021oqr}} {are taken here}. {These LDMEs are close to those from the  Chao et al.~\cite{Chao:2012iv} fit (H14) mentioned in~\ct{tab:LDME-fits_comp}.} %
The CT18ANLO PDF central set~\cite{Moffat:2021dji} was used for the proton PDFs. Unlike \jpsi production at the LHC and {the} Tevatron, the reach in the \jpsi \pT defined with respect to the lepton-hadron axis, is much smaller due to {the} smaller collision energy. The solid line {in \cf{fig:jpsi-pt-eic}} refers to the total contribution, which is dominated by the subprocess { $\gamma + g \to [c\bar{c}] + g$ (NLP Photon) and $e + g \to [c\bar{c}] + e$ (NLP Lepton) } with the \ccbar pair fragmenting into $J/\psi$.  The lepton or photon initiated LP contribution to the production cross section, namely the first term in \ce{eq:jpsi-lp-resum}, is dominated by the lowest{-}order subprocesses, such as $e+q\to e+q$ or $\gamma+q\to g+q$, respectively, with a produced parton fragmenting into the observed $J/\psi$, and is strongly suppressed by the single-parton FFs at the EIC energy. {In summary, the LP contributions are essentially irrelevant in the EIC kinematics.} Therefore, a matching to the fixed-order calculations (described above in this section), including the second and third terms in \ce{eq:jpsi-lp-fac}, {is awaited for}.

\paragraph{Polarisation of $J/\psi$ in photoproduction.}

{Since the prediction~\cite{Cho:1994ih} in 1994 of a transversely-polarised $\jpsi$ hadroproduction yield at high \pT, much hope has been put in polarisation measurements to advance our understanding of quarkonium production, with a very limited success though~\cite{Lansberg:2019adr}. NLO CSM computations of polarisation observables in photoproduction were performed in 2009~\cite{Artoisenet:2009xh,Chang:2009uj} and subsequently completed with the COM NRQCD contributions in 2011~\cite{Butenschoen:2011ks} without clear conclusions owing to the large uncertainties in the H1~\cite{Aaron:2010gz} and ZEUS~\cite{ZEUS:2009qug} data and in the theory.} 

\begin{figure}[H]
    \centering
    \includegraphics[width= 0.6\linewidth, keepaspectratio]{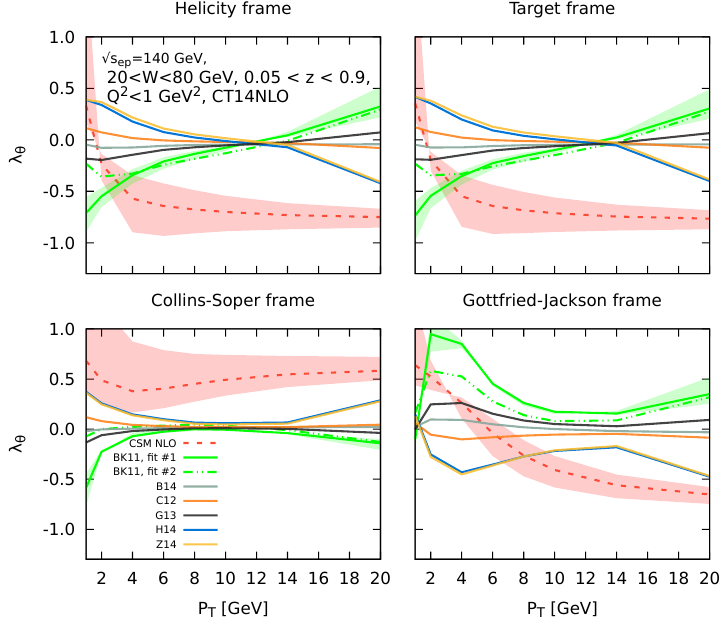}
    \caption{The NLO NRQCD factorisation predictions for {the $\jpsi$ transverse momentum} %
    {($\pT$)} dependence of the $\lambda_\theta$ polarisation parameter in prompt-$J/\psi$ photoproduction for the EIC kinematic conditions. Central predictions using the LDME sets listed in %
    {\ct{tab:LDME-fits_comp}} %
    as well as for the CSM are shown. The scale-variation uncertainty bands are plotted for the prediction of {the} LDME set of Kniehl and Butensch\"on~\cite{Butenschoen:2011yh} as well as for the CSM. The calculation of the short-distance cross sections is based on~\cite{Butenschoen:2011ks}.}%
    \label{fig:Jpsi_photoprod_lambda-pT}
\end{figure}

\begin{figure}[H]
    \centering
    \includegraphics[width= 0.6\linewidth, keepaspectratio]{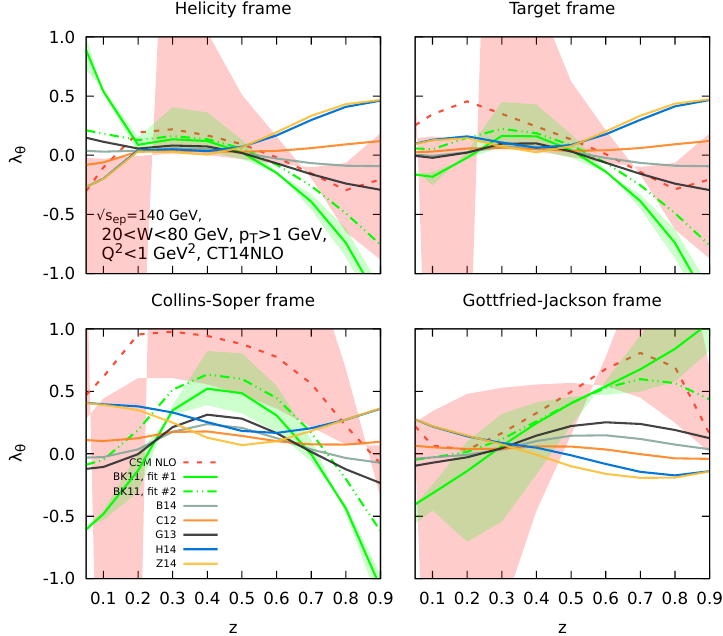}
    \caption{{Same as \cf{fig:Jpsi_photoprod_lambda-pT} but for the $z$-differential cross section.}}
    \label{fig:Jpsi_photoprod_lambda-z}
\end{figure}

In  Figs~\ref{fig:Jpsi_photoprod_lambda-pT} and  \ref{fig:Jpsi_photoprod_lambda-z}, we show the NLO NRQCD predictions for the {$\pT$} and $z$ dependence of the polarisation parameter $\lambda_\theta$ of promptly photoproduced $J/\psi$ mesons in the EIC kinematic conditions. These predictions include CS and CO contributions {using the} LDME sets discussed in  Section~\ref{sec:LDMEs:coll} as well as direct and resolved-photon interaction contributions (see the Figs.~\ref{fig:Jpsi-photoprod-pT_LDMEs} and~\ref{fig:Jpsi-photoprod-z_LDMEs} for the corresponding differential cross{-}section plots). As one can see from %
\cf{fig:Jpsi_photoprod_lambda-pT}{,} the %
{$\pT$}-dependent NRQCD predictions for all LDME sets are roughly consistent with unpolarised production ($\lambda_\theta=0$ in all frames), unlike the predictions of {the} {CSM}, which leads to significant polarisation of photo %
produced $J/\psi$ mesons. In the $z$-dependent case{,} the region of $z\to 1$ has the most %
discriminating power between different LDME sets. {We however have reasons to doubt the relevance of these predictions given that all but the BK11 LDMEs are unable to describe the corresponding HERA data.} From Figs.~\ref{fig:Jpsi_photoprod_lambda-pT} and  \ref{fig:Jpsi_photoprod_lambda-z}, one also observes that the detailed behaviour of $\lambda_\theta$ for different LDME sets is significantly different for different polarisation frames, which could be an important tool for additionally constraining the theory.

\paragraph{Polarisation of $J/\psi$ in electroproduction.}
\label{sec:polar_obs_SIDIS}
{The HERA collider experiments} provided some results on the $J/\psi$ polarisation, mostly for photoproduction~\cite{H1:2002xeb,Steder:2010zz},  
but unfortunately these data do not allow to favour or disfavour different models and/or approaches.
The reasons behind this are twofold: data were not precise enough and they were collected in regions where theoretical predictions are very close to each other~\cite{H1:2002xeb,Butenschoen:2011ks}. Furthermore in Ref.~\cite{Yuan:2000cn}, Yuan and Chao showed that the estimates for the $\lambda_\theta$ parameter in SIDIS, within both the CSM and NRQCD approaches, are overlapping for most of the values of the variable $z$.
In this respect EIC could play a crucial role: highly precise data are expected and other/extended kinematical regions could be explored. 

\begin{figure}[hbt!]
	\centering
	\includegraphics[width= 0.85\linewidth, keepaspectratio]{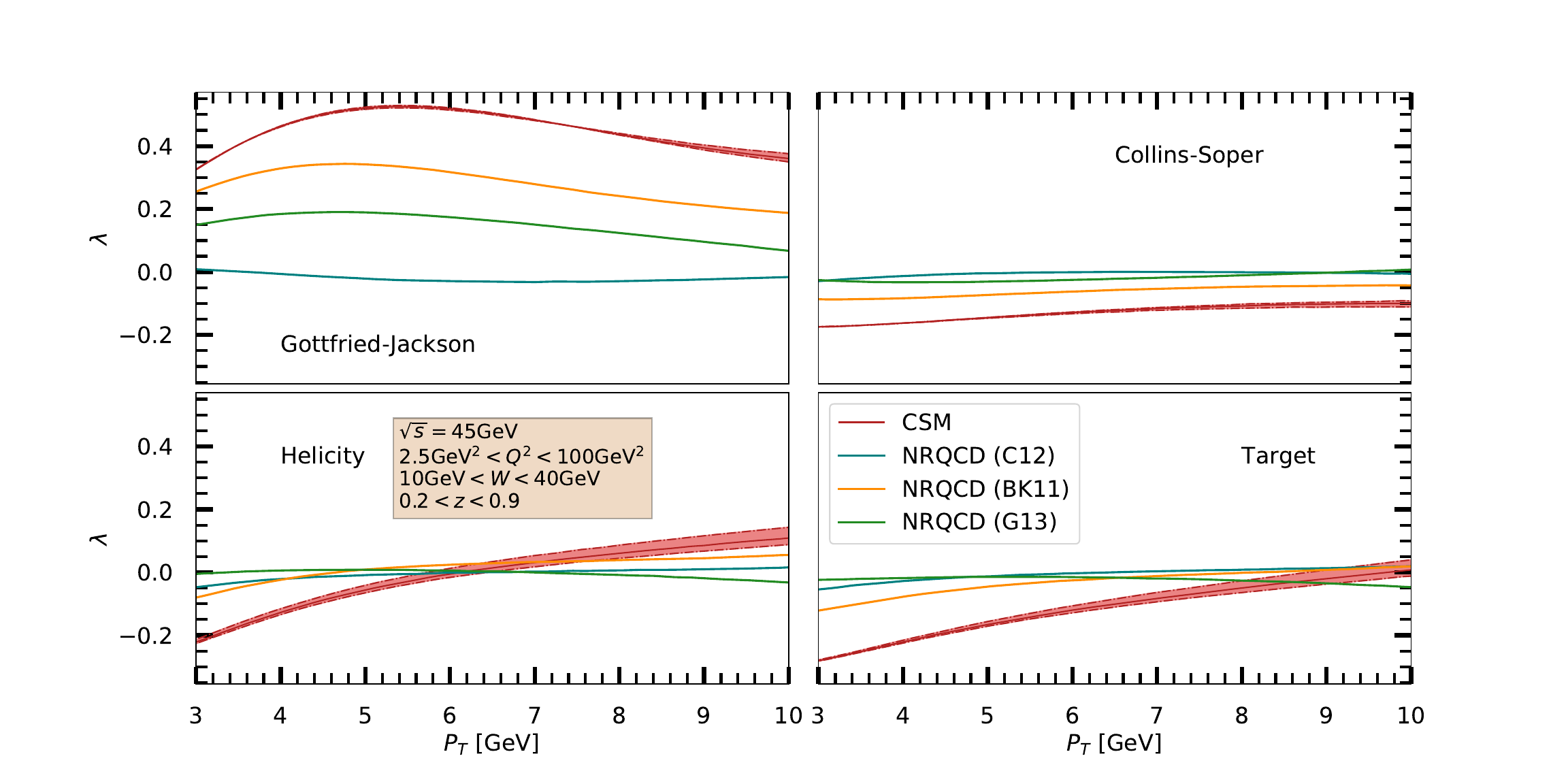}
	\caption{Predictions for the %
 {$\lambda_\theta$} parameter for  {$J/\psi$ electroproduction} (or SIDIS) at  $\sqrt{s_{ep}} = 45$~GeV as a function of {the $\jpsi$ transverse momentum,} %
 {$\pT$}, for different frames and models; bands refer to the variation of the scale %
 {$\mu_0/2 \leq \mu_F \leq 2 \mu_0$}. 
 Kinematic cuts are given in the legend. Plot based on Ref.~\cite{DAlesio:2023qyo}.}
	\label{fig: Jpsi lambda pol vs PT SIDIS}
\end{figure}

In the following, we present some predictions at {LO}, both in the CSM and NRQCD frameworks, adopting different NLO LDME sets. 
{Some comments are therefore in order: (i) as previously discussed, the combined usage of NLO hard scattering with NLO LDMEs is subject to great caution. As of now, only the CSM part of the electroproduction cross section has been computed at NLO~\cite{Sun:2017wxk}. The only full NRQCD analysis has been performed at LO~\cite{Sun:2017nly} and show mixed agreements between the different NRQCD predictions and HERA data; (ii) a number of quarkonium-production processes exhibit very large QCD corrections to polarisation observables~\cite{Gong:2008sn,Artoisenet:2008fc,Chang:2009uj,Artoisenet:2009xh,Lansberg:2010vq,Butenschoen:2012px,Feng:2020cvm}. The following LO results should therefore only be considered as a simple guidance for future measurements and certainly not as quantitative predictions to which future measurements should be confronted to. In this context, a NLO NRQCD analysis of electroproduction is eagerly awaited for}.

\cf{fig: Jpsi lambda pol vs PT SIDIS} shows some estimates for the $\lambda_\theta$ parameter at the centre-of-mass %
energy $\sqrt{s_{ep}} = 45$ GeV %
together with their uncertainty bands, visible mostly for the CSM {and} obtained by varying the factorisation scale in the range {$\mu_0/2 < \mu_F <2 \mu_0$}, with $\mu_0 = \sqrt{{M_{\jpsi}^2} + Q^2}$. The integration regions are detailed in the legend box. 
No uncertainty bands from LDMEs are included, instead predictions for different sets are presented: C12 \cite{Chao:2012iv}, BK11 \cite{Butenschoen:2011yh} and G13 \cite{Gong:2012ug}. This illustrates their impact on the results. From \cf{fig: Jpsi lambda pol vs PT SIDIS}, it is clear that the $\lambda_\theta$ value can be significantly different if we consider different frames. In particular, the \textit{Gottfried-Jackson}~frame provides the better overall separation between CSM and NRQCD curves.

\begin{figure}[hbt!]
\centering
\includegraphics[width= 0.85\linewidth, keepaspectratio]{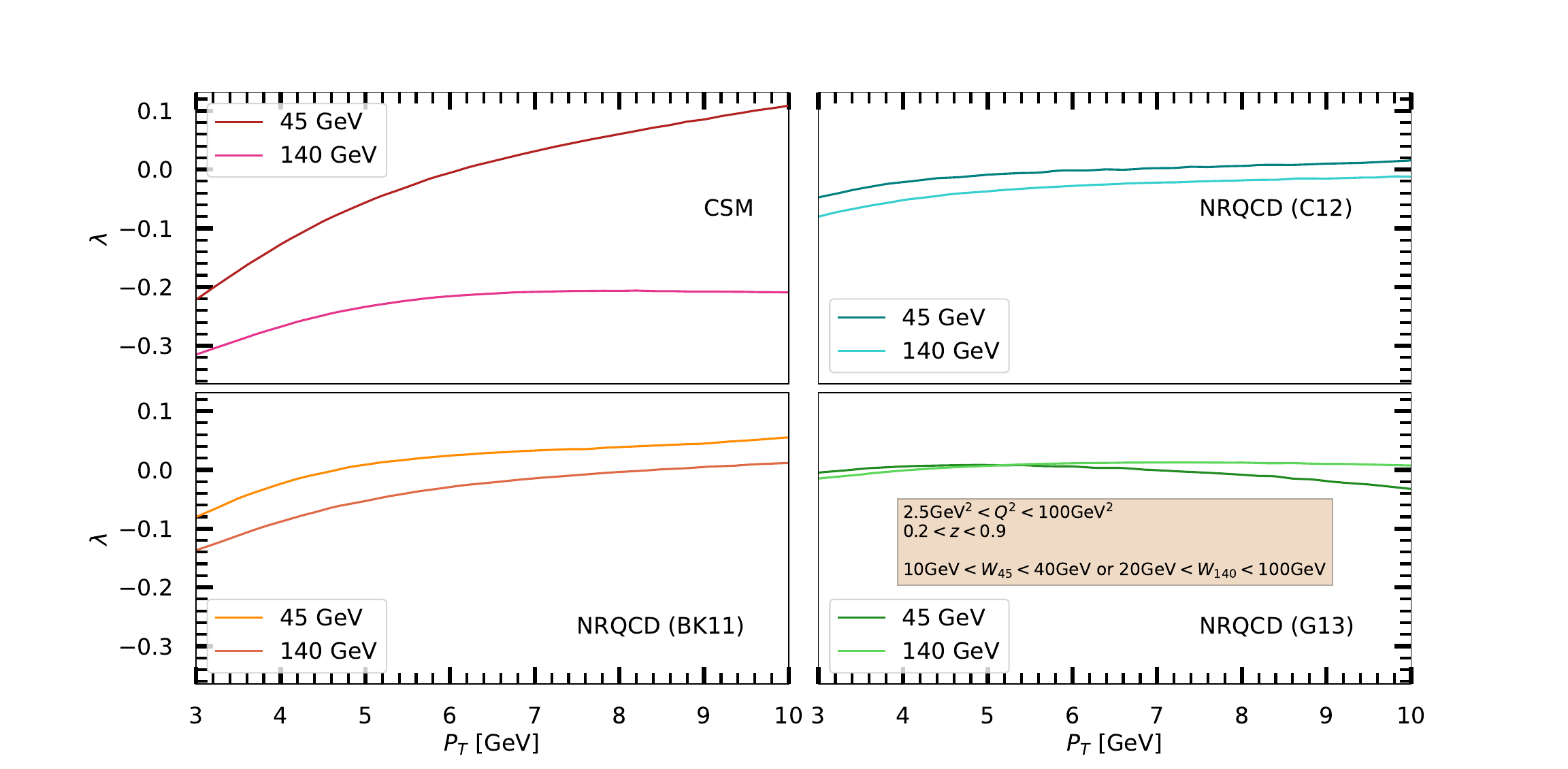}
\caption{Predictions for the {$\lambda\equiv\lambda_\theta$} parameter in the helicity frame in {$J/\psi$ electroproduction} (or SIDIS) as a function of {the $\jpsi$ transverse momentum, } %
{$\pT$} for different LDME fits at at  $\sqrt{s_{ep}} = 45$~GeV and 140~GeV .  Only central value estimates ({$\mu_F = \mu_0$}) are shown. Kinematic cuts are given in the legend.}
\label{fig: Jpsi lambda pol vs energy SIDIS}
\end{figure}

Another possibility offered by the EIC experiment is the collection of data at different energies. In \cf{fig: Jpsi lambda pol vs energy SIDIS}, the impact coming from the energy variation on CSM and NRQCD predictions is shown. In this case, only the central values are presented ({$\mu_F = \mu_0$}); for the lower energy, $\sqrt{s_{ep}} = 45$ GeV, the integration region is the same as in \cf{fig: Jpsi lambda pol vs PT SIDIS}, while for $\sqrt{s_{ep}} = 140$ GeV a wider $W$ integration is considered (see legend).
Even focusing on one specific frame, like the helicity frame in \cf{fig: Jpsi lambda pol vs energy SIDIS}, one clearly sees that the CSM is more affected by the energy shift. Note that moving to higher energies allows one to access contributions with higher virtuality, with an interesting effect: in the CSM these contributions are opposite to the lower virtuality ones (reducing the size of the estimates), while in NRQCD this phenomenon is less important. {It however remains to be shown that such discriminant effects remain at NLO.}

\paragraph{Prompt $\eta_c$ and $\chi_c$ yields.}
\label{sec:chic-etac-photoprod}
{As} mentioned in %
Section~\ref{sec:NRQCD+CSM}{,} the dominance of %
{the CS} mechanism in prompt-$\eta_c$ hadroproduction at {$\pT\gtrsim M_{\eta_c}$} was not expected by NRQCD factorisation. Therefore{,} from the point of view of studies of the heavy-quarkonium production mechanism{,} it is important to understand if this feature of $\eta_c$ production persists also in {$ep$ collisions}. If it is indeed the case, then $\eta_c$ hadro-, photo- and leptoproduction can be used as %
{a} tool for hadron{-}structure studies with %
{a reduced uncertainty stemming from the CO mechanism} compared to production of other charmonium states.

\begin{figure}[htbp!]
    \centering
    \includegraphics[width=0.4\textwidth]{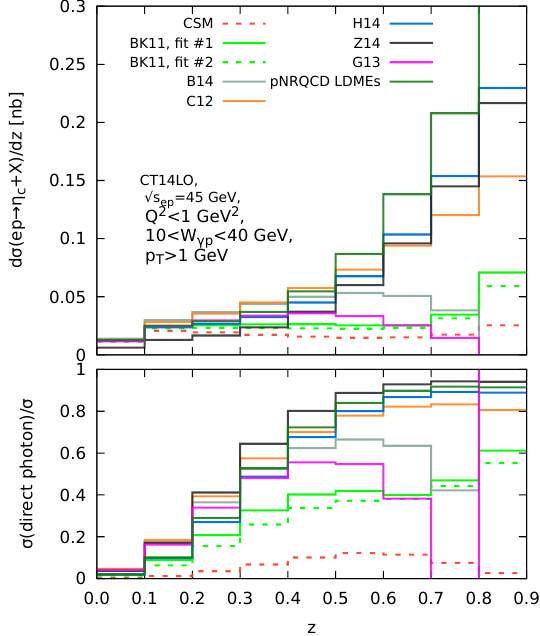}
    \includegraphics[width=0.4\textwidth]{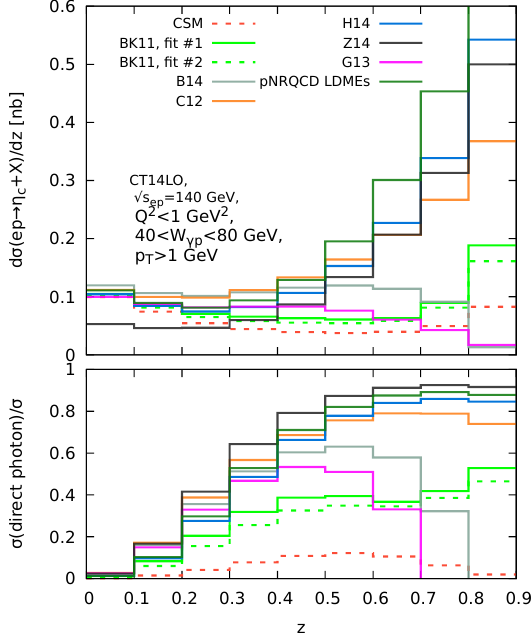}
    \caption{LO CF predictions for $\eta_c$ inclusive photoproduction distributions {as a function the elasticity $z$} in the EIC kinematics \newb{using HQSS and the} LDME sets mentioned in \ct{tab:LDME-fits_comp}. The calculation of the short-distance cross sections is based on Refs.~\cite{Zhang:2019wxo, Zhang:2021dkd}. The bottom plots show the fraction of direct-photon interaction contributions. }
    \label{fig:eta-c_photopr_z-distr}
\end{figure}

\begin{figure}[htbp!]
    \centering
    \includegraphics[width=0.9\textwidth]{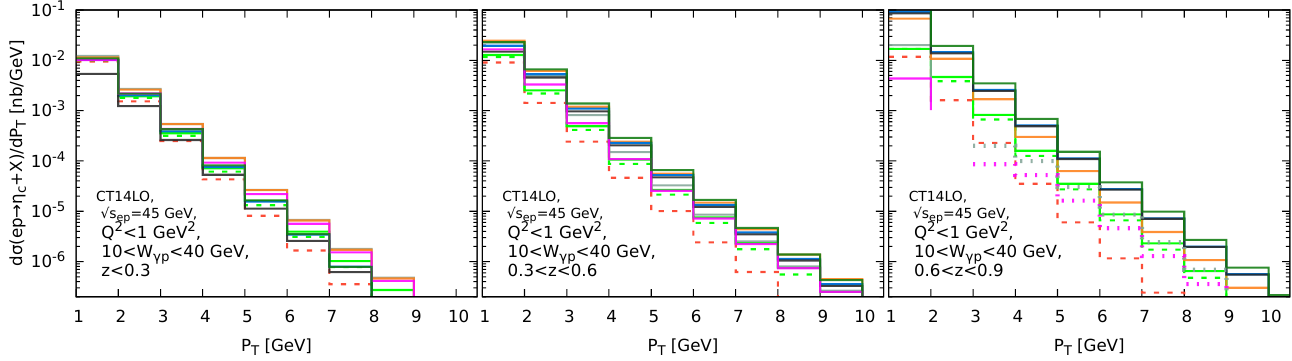}
    \includegraphics[width=0.9\textwidth]{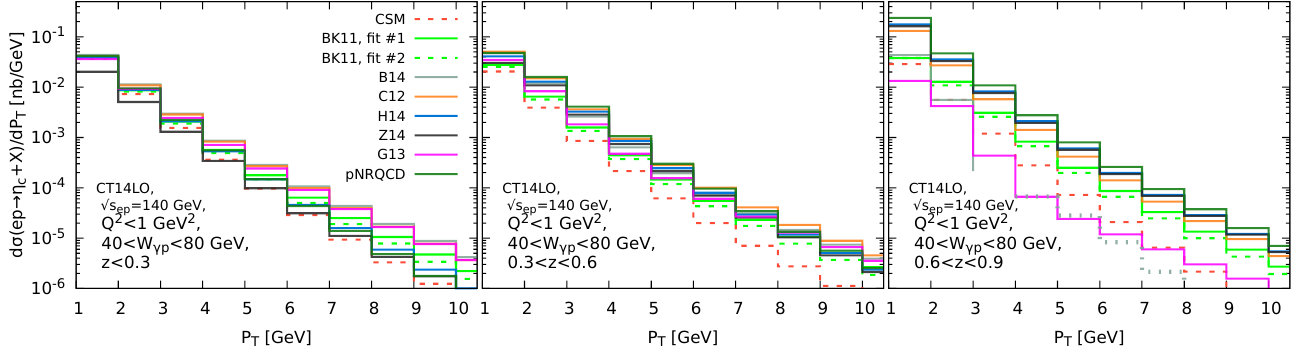}
    \caption{LO CF predictions for $\eta_c$ inclusive photoproduction distributions {in the $\eta_c$ transverse momentum {$\pT$}} in the EIC kinematics using the LDME sets mentioned in \ct{tab:LDME-fits_comp}. The calculation of short-distance cross sections is based on Refs.~\cite{Zhang:2019wxo, Zhang:2021dkd}. {The negative values of the cross sections are plotted with the dotted histograms.} }
    \label{fig:eta-c_photopr_pT-distr}
\end{figure}

In recent works~\cite{Zhang:2019wxo, Zhang:2021dkd}, $\eta_c$ photo- and electroproduction {cross sections were computed} including all the CO and CS contributions  {at} LO in $\alpha_s$. In {the} case of photoproduction~\cite{Zhang:2019wxo}{,} both direct-photon and resolved-photon interactions {were} taken into account. The CS contribution had been assumed to be negligible in earlier studies~\cite{Hao:1999kq,Hao:2000ci}, because the corresponding direct-photon interaction subprocess \newb{appears at} ${\cal O}(\alpha\alpha_s^3)$ due to the necessity of %
{two-gluon radiation in} the final state to produce {a} $c\bar{c}[{}^1S_0^{[1]}]$ pair and {because} resolved-photon contributions were assumed to be small. However, it was found~\cite{Zhang:2019wxo} that the resolved-photon subprocesses make the CS contribution to the photoproduction cross section non-negligible. These predictions, updated with the use of CT14LO PDFs, are shown in Figs.~\ref{fig:eta-c_photopr_z-distr} and \ref{fig:eta-c_photopr_pT-distr}.  {The CO contributions were computed by converting the $J/\psi$ CO LDME sets listed in \ct{tab:LDME-fits_comp} to the 
$\eta_c$ LDMEs through HQSS relations valid up $v^2$ corrections.} 
As one can see from these figures, the CO contributions are still important and {the} cross section at $z>0.5$ strongly depends on the LDME choice.%

For %
electroproduction%
~\cite{Hao:2000ci}, the CS contribution is also sizeable, but for  {a} different reason, namely an additional $Q^2$-dependent terms appearing in the short-distance cross section. Of course, the main problem of the predictions for $\eta_c$ production in %
$ep$ collisions is that they so far have been done only at %
LO in $\alpha_s$. {The NLO corrections could be particularly important for the CS ${}^1S_0^{[1]}$ state whose LO contribution is highly suppressed at $\pT\gtrsim M_{\eta_c}$ 
in photo- and leptoproduction in comparison to CO states, especially ${}^3S_1^{[8]}$}. As known from  $J/\psi$ {production}, this suppression will be lifted by large NLO corrections~\cite{Kramer:1995nb,Campbell:2007ws}. NLO calculation, at least in the CS channel, should be done before drawing conclusions about the importance of the CS mechanism in $\eta_c$ production at the EIC.

\begin{figure}[htb!]
    \centering
    \includegraphics[width=0.37\textwidth]{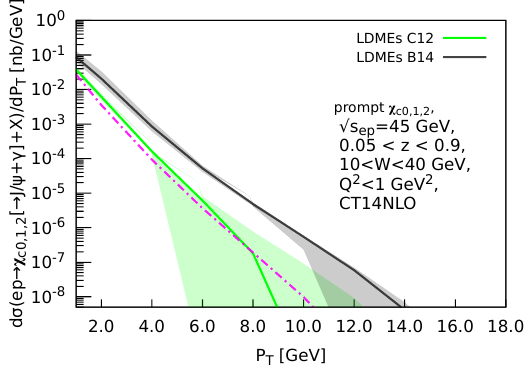}
    \includegraphics[width=0.37\textwidth]{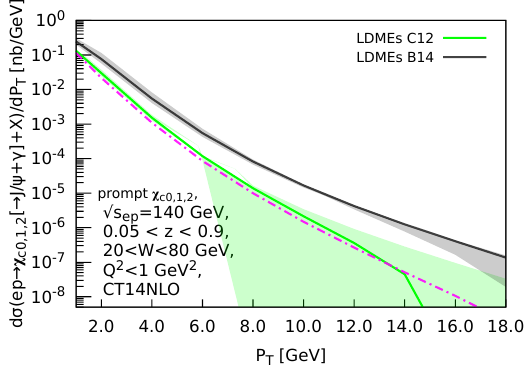} \\
    \includegraphics[width=0.37\textwidth]{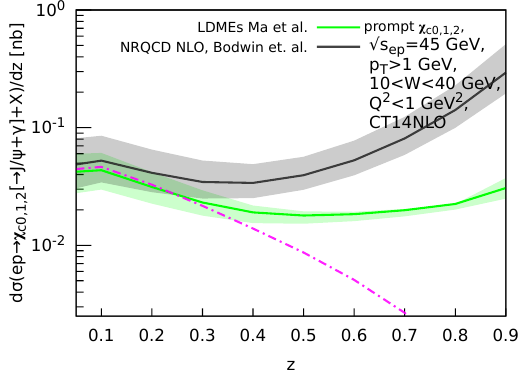}
    \includegraphics[width=0.37\textwidth]{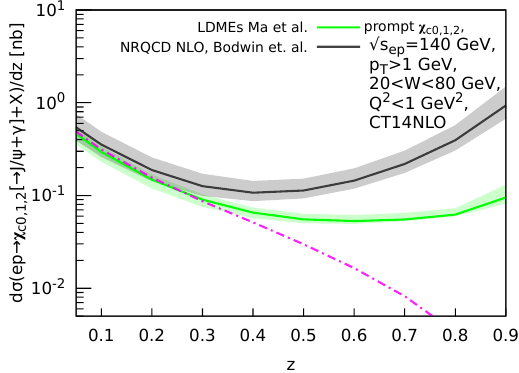}
    \caption{NLO NRQCD factorisation predictions for {cross sections differential in the $\chi_c$ transverse momentum ({$\pT$}) and the elasticity ($z$)} for the photoproduction of prompt $\chi_{c0,1,2}$ mesons at the EIC using the LDMEs obtained by Ma et al.~\cite{Ma:2010vd} (which is compatible with {the} treatment of feed down to $J/\psi$ of Ref.~\cite{Butenschoen:2011yh}) and by Bodwin et al.~\cite{Bodwin:2014gia}. Only these sets lead to positive photoproduction cross sections. The resolved-photon contribution for {the} LDME set of Ma et al. is shown by the dash-dotted line. The AGF~\cite{Aurenche:1994in} photon PDF set has been used. The calculation of the short-distance cross sections is based on~\cite{Butenschoen:2009zy,Butenschoen:2010rq}. The $\psi(2S)\to \chi_c$ feed down has been included.}
    \label{fig:chic-photoprod}
\end{figure}

{Besides $J/\psi$ and $\eta_c$ production, it is also essential to study $\chi_{c0,1,2}$ states at the EIC.}
Photo- or electroproduction of these mesons has not been observed experimentally yet.  {NLO NRQCD predictions for the photoproduction cross sections of $\chi_{c0,1,2}$ radiatively decaying to $J/\psi$ are shown in \cf{fig:chic-photoprod}. They are based on known calculations of short-distance cross sections for $\jpsi$ photoproduction~\cite{Butenschoen:2009zy,Butenschoen:2010rq} and the} $\chi_{c0}$ LDME values obtained in hadroproduction fits by Ma et al.~\cite{Ma:2010vd}, respectively Bodwin et al.~\cite{Bodwin:2014gia}. We remark that the former $\chi_{c0}$ LDME values are also those used by Gong et al. in Ref.~\cite{Gong:2013qka} and in the LDME set denoted ``Kniehl, Butenschoen, fit \#2'' in Figs.~\ref{fig:Jpsi-photoprod-pT_LDMEs} and~\ref{fig:Jpsi-photoprod-z_LDMEs}. \newb{We remind the reader that for $P-$wave production at NLO in NRQCD, one cannot make a clear distinction between CO and CS contributions as they directly depend on the NRQCD factorisation scale, $\mu_\Lambda$.}

It is an expected feature that resolved photon contributions dominate photoproduction at low $z$. Interestingly, however, the predictions of \cf{fig:chic-photoprod} are dominated by the resolved-photon contribution already for $z$ below 0.5. Moreover, it is only due to the resolved photons that the $\chi_c$ cross sections are positive at low $z$ after all. This feature of the theoretical predictions may indicate our poor understanding of $\chi_c$ %
photoproduction, but if confirmed, the photoproduction of these mesons could serve as {a} useful source of information about the poorly known gluon component of photon PDFs.

\subsection{Learning about quarkonia from TMD observables}
\subsubsection{LDME constraints from TMD observables}
\label{sec:LDMEsTMD}
One important reason to investigate quarkonium production at the EIC is the possibility to probe TMDs that have not been extracted from experiments yet. The semi-inclusive heavy vector quarkonium production process, $e \, p \to e'\, J/\psi\, (\Upsilon) \, X$ at small transverse momentum, {\pT}, is expected to offer a promising probe of gluon TMDs\footnote{Due to the presence of the large scale given by the quarkonium mass {$M_{\Q}\approx 2m_Q$}, one can consider not only electroproduction, but in principle also the photoproduction case ($Q^2\approx 0$). A large photon virtuality is expected to suppress background from diffraction and higher-twist effects \cite{Fleming:1997fq}. To our knowledge, at present there are no studies of the numerical impact of such background on the photoproduction process $\gamma \, p \to J/\psi\, (\Upsilon) \, X$ {in the TMD regime}. %
}, as will be discussed extensively in section \ref{sec:ep}. Besides gluon TMD extractions, this process may also allow for improved determinations of certain LDMEs. In this way EIC can also improve our knowledge on NRQCD. 

At small {$P_T$,}
the differential cross section is {expected to be} described in terms of %
TMDs. %
{As will be discussed in detail in the next subsection, for quarkonium production, this involves}
{TMD} shape functions{~\cite{Echevarria:2019ynx, Fleming:2019pzj}}, {rather than TMD FFs  like for light hadron production}. At the lowest order, $\alpha^2 \alpha_s$, the {process $e \, p \to e'\, J/\psi\, (\Upsilon) \, X$ at small transverse momentum} is described by photon-gluon scattering producing a heavy quark-antiquark pair in the %
CO state. The transition from the heavy-quark pair into the bound state is then described by a shape function. {If one assumes the shape function to be a delta function in transverse momentum, one can connect to the standard NRQCD expressions for this transition.} %
{To lowest order in the strong coupling,} 
but with the inclusion of the {NNLO in $v^2$} %
$^1S_0$ and $^3P_J$ $(J=0,1,2)$ CO intermediate states~\cite{Boer:2021ehu}, the resulting expression for the cross section involves %
{two of {the} CO LDMEs which were discussed above,}
$\langle{\cal O}\left[{}^{1}S^{[8]}_{0}\right]\rangle $ and $ \langle{\cal O}\left[^{3}P^{[8]}_{0}\right]\rangle ${,} {for which constraints from new types of observables are clearly welcome}. In this way{,} measurements of the transverse{-}momentum spectrum of $e \, p \to e'\, J/\psi \,(\Upsilon) \, X$ in the TMD regime can lead to improved determinations of these CO LDMEs.  However, inclusion of higher{-}order corrections{, in particular from the leading $v$ CS NRQCD contributions at $\alpha^2 \alpha_s$,} and the proper shape functions will be required for a robust extraction of these LDMEs.

\subsubsection{TMD effects from quarkonia: shape functions}
\label{sec:TMDShFs}
The NRQCD factorisation approach can only be applied {\it for transverse{-}momentum spectra} when the quarkonium state is produced with a relatively large transverse momentum compared to its mass, \ie \ %
{$\pT \gtrsim 2 m_{Q}$}. %
This is because the emissions of soft gluons from the heavy-quark pair cannot modify the large transverse momentum of the bound state. The large %
{$\pT$} is generated in the hard process through recoil off unobserved particles, while the infrared divergences are parametrised in terms of the well-known %
LDMEs, %
collinear PDFs and %
{FFs}, depending on the particular process under consideration.

On the contrary, when the quarkonium is produced with a small transverse momentum, {all} soft gluon {effects} can no longer be factorised {in terms of standard TMD PDFs}. In order to properly deal with soft{-}gluon radiation at small %
{$\pT$} in a transverse{-}momentum spectrum of quarkonium, it has recently been found that one needs to promote the LDMEs to so-called TMD shape functions (TMD ShFs)~\cite{Echevarria:2019ynx, Fleming:2019pzj}. %
{Earlier, similar} {shape functions had been introduced} in quarkonium photo-/leptoproduction in the endpoint region~\cite{Beneke:1997qw, Fleming:2003gt, Fleming:2006cd}, {which however are functions of $z$, but a more general form was discussed in \cite{Beneke:1999gq}.} 

The newly introduced non-perturbative TMD ShFs encode the two soft mechanisms present in the process at low %
{$\pT$}: the formation of the bound state and the radiation of soft gluons. As a consequence, they parametrise the transverse{-}momentum smearing of the bound state, and carry a dependence on the factorisation and rapidity scale.

Schematically, for the production of {a} single quarkonium state $\Q$ at the EIC, with mass {$M_{\Q}$}{,}  we have:
\begin{align}
\label{eq:TMD-shape}
d\sigma &\sim
F_{g/P}({b_T};\mu,\zeta)\,
\sum_{i\in \{^1S_0^{[1]},\ldots \}} 
H^{[i]}(M_{\Q}, Q;\mu)\,
\Delta^{[i]}({b_T},\mu,\zeta)
\,,
\end{align}
where $F_{g/P}$ stands for any of the eight leading-twist %
gluon TMDs~\cite{echevarria:2015uaa}, $H^{[i]}$%
are the process-dependent hard scattering coefficients and $\Delta^{[i]}$ are the %
quarkonium TMD ShFs~\cite{Echevarria:2019ynx, Fleming:2019pzj}. 
The above formula is written down in coordinate space where %
${b_T}$ is Fourier-conjugate to the quarkonium transverse momentum {$P_T^*$} {(to be specific, in the virtual photon-proton centre of mass frame)}. 
Moreover, $\mu$ and $\zeta$ are the factorisation/resummation and rapidity scales, respectively.
The summation is performed over the various colour and angular{-}momentum configurations {($i$)} of the $Q\bar{Q}$ pair.%
Similarly to LDMEs, the TMD ShFs {are of a specific order in} the relative velocity $v$ of the heavy quark-antiquark pair in the quarkonium rest frame. 
Therefore, the factorisation formula is a simultaneous expansion in $v$ and $\lambda = {\pT^*}/M_{\Q}$.
The operator definition of a bare\footnote{It is understood that the TMD ShF in the factorised cross section in \ce{eq:TMD-shape} is free from rapidity divergences, \ie it has been divided by the relevant soft factor which has also been used to properly subtract rapidity divergences in the gluon TMD $F_{g/P}$.} 
TMD ShF with {NRQCD} quantum numbers $i$ is:
\begin{align}
\Delta^{[i]}({b_T},\mu,\zeta) \propto \,\sum_{X_s}\,
\sandwich{0}{\left({\cal O}_i^{\dagger}{\cal Y}_n^\dagger\right)^{ab}({b_T})}{X_s\,\Q}
\sandwich{{\Q}\, X_s}{\left({\cal Y}_n {\cal O}_i\right)^{ba}(0)}{0}
\, ,  %
\label{eq:ShFoperator}
\end{align} %
which is just the TMD generalisation of the LDME operator definition~in \ce{eq:LDME_def}. On the r.h.s., the usual LDME operators $\mathcal{O}$ are evaluated at positions $b_T$ and $0$ and sandwiched between the vacuum $|0\rangle$ and {the state $| \Q X_s\rangle$ of the produced quarkonium together with possible soft radiation carrying away color}. 
Moreover, these operators are multiplied by %
Wilson lines $\mathcal{Y}_n$ parametrising %
{the resummation of gluons exchanged between the hard part and}  %
the %
{state $| \Q X_s\rangle$}. 

The operator definition in \ce{eq:ShFoperator} can be related to the NRQCD LDMEs by the first term in an operator product expansion (OPE) for ${b_T}  \to 0$:
\begin{equation}
\Delta^{[i]}({b_T},\mu,\zeta) %
=
\sum_n
C^{[i]}_n(b;\mu,\zeta)
\times
\langle  {\cal O}^{\Q} [n]  \rangle(\mu)
+\mathcal{O}({b_T})
\,{.}
\label{eq:ShFOPE}
\end{equation}
{In order to extend this expression to larger $b_T$, one can introduce a prescription like $b_T \to  b_T^{*} \equiv b_T/\sqrt{1+\left(b_T/b_{T,\text{max}}\right)^{2}} \leq b_{T,\text{max}}$ to ensure validity of this perturbative expression 
and include a nonperturbative overall factor $\Delta^{[i]NP}$:}
\begin{equation}
\Delta^{[i]}({b_T},\mu,\zeta) 
 {\equiv} \Delta^{[i]NP}({b_T}) \sum_n
C^{[i]}_n({b_T^*};\mu,\zeta)
\times
\langle {\cal O}^{\Q} [n]  \rangle(\mu)
\,.
\label{eq:ShFOPE-lim}
\end{equation}
{This expression involves} the usual ``collinear'' LDMEs, multiplied by perturbatively calculable Wilson coefficients $C_n^{[i]}({b_T};\mu,\zeta)$ to match the expansion on pQCD, and a non-perturbative part $\Delta^{[i]NP}$ %
{that} needs to be modelled or extracted from experimental data. Note that, in principle, at higher orders in $\alpha_s$, there might be operator mixing: \eg the $^1S_0^{[8]}$ TMD ShF could become dependent on the $^3P_0^{[8]}$ LDME, hence the sum over NRQCD states $n$ in \ce{eq:ShFOPE}.

In Ref.~\cite{Boer:2020bbd}, the OPE { of}~\ce{eq:ShFOPE} is implemented in a practical way by studying single-inclusive $J/\psi$ electroproduction. In the regime %
${\pT^{*2}}\sim Q^2\sim M_{\Q}^2$, 
with %
{$\pT^*$} {being} the transverse momentum of the quarkonium {in the virtual photon-proton centre-of-mass frame}
and $\mu$ either given by $Q$ or by the %
{quarkonium mass $M_{\Q}$}, the cross section is computed as usual in collinear factorisation. On the other hand, when %
${\pT^{*2}}\ll\mu^2$, TMD factorisation~\ce{eq:TMD-shape} applies. By comparing both cross sections in the kinematical regime $\Lambda^2_{QCD}\ll{\pT^{*2}}\ll\mu^2$, one can %
match the relevant TMD ShF onto the collinear LDMEs, 
{confirming the need for introducing shape functions}. 
{The analysis of Ref.~\cite{Boer:2020bbd} was revised in Ref.~\cite{Boer:2023zit}, modifying the obtained expression for the shape function, but not its necessity.} 

To summarise, the factorisation theorem in \ce{eq:TMD-shape} contains a convolution of two non-perturbative hadronic quantities at low transverse momenta: the gluon TMD PDFs and the TMD ShFs. It is therefore possible to perform a phenomenological extraction of gluon TMDs from quarkonium production processes.
However, to do so, one also needs to model {or} extract the involved TMD ShFs. This is analogous to SIDIS where one observes a light hadron, where one needs information on the light-hadron TMD FFs in order to extract quark TMD PDFs.

\subsubsection{Azimuthal $\cos 2 \phi_T^*$ modulation in \jpsi electroproduction}
\label{sec:cos2phiShF}
In (semi-inclusive) quarkonium electroproduction {on a{n} unpolarised proton target}, %
an azimuthal $\cos 2 \phi_T^*$ modulation (see Section~\ref{sec:kinem-SIDIS} for our kinematic definitions) of the differential cross section {will arise from} %
linearly %
polarised gluons inside %
{the} unpolarised proton. {These are} described by the TMD $h_1^{\perp \, g}$ {\cite{Mukherjee:2016qxa,Kishore:2018ugo,Boer:2021ehu,Bacchetta:2018ivt}}\footnote{Note that, in the photoproduction regime, one cannot determine the angle $\phi_T^*$ because {the} lepton plane is not defined, hence, also not the $\cos 2 \phi_T^*$ modulation. {In photoproduction, azimuthal modulations can only be seen for two-particle observables.}}. 
{In many studies the shape functions of the quarkonium are assumed  {are not considered} and then the differential cross section can be written as: %
\begin{equation}
\label{eq:cos2phi-unp-xsec-general}
\begin{aligned}
 d\sigma={} &\frac{1}{2s}\frac{d^3l'}{(2\pi)^32E_l'}\frac{d^3P_{\Q}}{(2\pi)^32E_{P_{\Q}}}
  \int dx\,d^2\mathbf{k}_{\perp}(2\pi)^4\delta(q+k-P_{\Q})\\
  &\times \frac{1}{Q^4}\mathcal{L}^{\mu\mu'}(l,q)\Phi^{\nu\nu'}
  (x,\mathbf{k}_{\perp})~\mathcal{M}_{\mu\nu}(\mathcal{M}_{\mu'\nu'})^{\ast},
 \end{aligned}
\end{equation}
where $\mathcal{M}_{\mu\nu}$ is the amplitude of  %
production {of the quarkonium $\Q$} in  the subprocess %
{${\gamma^{\ast}+g\rightarrow \Q}$}
{,} $\mathcal{L}^{\mu\mu'}$ is the leptonic tensor, and the %
{gluon correlator} is given by {\cite{Mulders:2000sh,Meissner:2007rx,Boer:2007nd}}:
\begin{eqnarray}
 \label{eq:gluon-correlator}
 \Phi^{\nu\nu'}(x,\mathbf{k}_{\perp})=-\frac{1}{2x}\bigg\{g_{\perp}^{\nu\nu'}f_1^g(x,\mathbf{k}_{\perp}^2)-\left(\frac{k_{\perp}^{\nu}k_{\perp}^{\nu'}}{M_p^2}+g_{\perp}^{\nu\nu'}\frac{\mathbf{k}_{\perp}^2}{2M_p^2}\right)h_1^{\perp g}(x,\mathbf{k}_{\perp}^2)\bigg\}.
\end{eqnarray}
Here, $g_{\perp}^{\nu\nu'}=g^{\nu\nu'}-P^{\nu}n^{\nu'}/P\cdot n-P^{\nu'}n^{\nu}/P\cdot n$, $x$ {and $\mathbf{k}_{\perp}$ are} the light-cone momentum fraction \newb{and transverse momentum} of the gluon. %
The asymmetry is defined as:
\begin{eqnarray}
\label{eq:cos2phi-asy}
\langle \cos(2\phi_T^*)\rangle&=&\frac{\int d\phi_T^* \cos(2\phi_T^*)d\sigma}{\int d\phi_T^* d\sigma},
\end{eqnarray}
where $\phi_T^*$ is the azimuthal angle of the production plane of $J/\psi$ with respect to the lepton scattering plane. 
}

{In a more complete picture of the $P_T^{*2} \ll M_{J/\psi}^2 \sim Q^2$} %
{region,} {as explained in \ref{sec:TMDShFs}, the TMD factorisation applies and LDMEs are promoted to TMD ShFs. Hence,} the differential cross section for this process can be {re}cast in the following form: 
\begin{align}
\frac{d\sigma^{UP}}{d y\, dx_B\, d^2 {\bm{P}^*_{T}}} =   {\cal N}\, \left[\sum_n A_{UP}^{[n]}\,{\cal C}\left[f_1^g \, \Delta^{[n]}\right] + \sum_n B_{UP}^{[n]}\, %
{\cal C} \left[{w} h_1^{\perp\, g} \, \Delta_h^{[n]} \right] \,\cos 2 \phi_T^* \right] , 
\end{align}
where {the subscript $UP$ on the amplitudes $A_{UP}^{[n]}$ and $B_{UP}^{[n]}$} denotes the polarisation state of the proton ($U$, since it is unpolarised) and of the quarkonium ($P=U,L,T$), respectively, {and ${\cal N}$ denotes an overall normalisation factor}. Here{,} the quarkonium polarisation is defined with respect to the direction of the quarkonium three-momentum in the virtual photon - proton %
{centre}-of-mass frame. 
Measurements of the transverse{-}momentum dependence of the above cross section at the EIC would %
allow {one} to gather information on the so-far unknown {quarkonium} shape functions. In particular, the $\cos{2\phi_T^*}$-weighted cross section would give access to a linear combination of the convolutions ${\cal C}\left[w h_1^{\perp g} \, \Delta^{[n]}_h\right]$, with $n\!=\!^1S_0^{[8]}$ or $n\!=\!^3P_0^{[8]}$. 
{Here the weight in the convolution expression 
\begin{align}
{\cal C}\left[w h_1^{\perp g} \, \Delta^{[n]}_h\right]  (\bm{q}_T) \equiv \int d^{2}{\bm{p}}_{T} \int d^{2}{\bm{k}}_{T} \, \delta^{2}({\bm{p}}_{T}+{\bm{k}}_{T}-{\bm{q}}_{T}) \, w({\bm{p}}_{T},{\bm{q}}_{T}) \, h_{1}^{\perp g}(x,{\bm{p}}_{T}^{2}) \, \Delta^{[n]}_{h}({\bm{k}}_{T}^{2})\, ,
\end{align}
is given by (in standard TMD notation, note however that ${\bm{q}}_{T}$ will correspond to $\bm{P}^*_T$ used here)
\begin{align}
w({\bm{p}}_{T},{\bm{q}}_{T}) = \frac{1}{{M_{p}^{2}}{\bm{q}}_{T}^{2}}[2({\bm{p}}_{T} \cdot {\bm{q}}_{T})^{2} - {\bm{p}}_{T}^{2} {\bm{q}}_{T}^2] \, .
\end{align}}
On the other hand, integrating over $\phi_T^*$ would single out a combination of the convolutions ${\cal C}\left[f_1^{g} \, \Delta^{[n]}\right]${$(P_T^*)$} for the same octet $S$- and $P$-waves, which could be in principle disentangled by looking at different values of {the inelasticity} $y$. Measurements of these observables should help to establish the relevance of smearing effects and, in case they turn out to be sizeable, to even perform a first extraction of the shape functions. In this way, it would be possible to compare $\Delta^{[n]}$ and $\Delta^{[n]}_h$ as well as determine some other properties, like their relations with the LDMEs and their dependence on $n$.

For unpolarised quarkonium production ($P=U$), %
{applying the above expressions} gives the following normalised asymmetry ratio:
\begin{align}
\langle \cos 2\phi_T^* \rangle \equiv \frac{\int d \phi_T^* \cos 2\phi_T^*\,\frac{d\sigma^{UU}}{d y\, d x_B\, d^2 {\bm{P}^*_{T}}}}{\int d \phi_T^* \,\frac{d\sigma^{UU}}{d y\, dx_B\, d^2 {\bm{P}^*_{T}} }} = {\frac{1}{2}} \frac{\sum_n B_{UU}^{[n]} \, {\cal C} \left[{w} h_1^{\perp\, g} \, \Delta_h^{[n]} \right]}{\sum_n A_{UU}^{[n]}\,{\cal C}\left[f_1^g \, \Delta^{[n]}\right]}. %
\label{eq:cos2phi}
\end{align}
As the matching analysis mentioned in Section~\ref{sec:TMDShFs} suggests, it is expected that the shape functions are proportional to the LDME{s} belonging to the $[n]$ state, %
{at least at} LO: $\Delta^{[n]} (\bm k_\sT^2;\mu^2) \simeq %
{ \langle  {\cal O}^{\Q} [n]  \rangle} \, \Delta(\bm k_\sT^2;\mu^2) $ and $\Delta_h^{[n]} (\bm k_\sT^2;\mu^2) \simeq %
{\langle  {\cal O}^{\Q} [n]  \rangle} \, \Delta_h(\bm k_\sT^2;\mu^2) $, for some %
$\Delta(\bm k_\sT^2;\mu^2) $ and $\Delta_h(\bm k_\sT^2;\mu^2)$. 
In this case the above asymmetry expression reduces to: 
\begin{align}
\langle \cos 2\phi_T^* \rangle = {\frac{1}{2}}
\frac{B_{UU}}{A_{UU}} \, 
\frac{{\cal C} \left[{w} h_1^{\perp\, g} \, \Delta_h \right]}{{\cal C}\left[f_1^g \, \Delta\right]}, %
\label{eq:cos2phi-red}
\end{align}
where $A_{UU} = \sum_n A_{UU}^{[n]} \, \langle  {\cal O}^{{\Q}} [n]  \rangle$ and 
$B_{UU}= \sum_n B_{UU}^{[n]} \, \langle {\cal O}^{{\Q}} [n] \rangle$. 
At LO, the {coefficients} appearing in this expression are~\cite{Bacchetta:2018ivt} :

\begin{align}
A_{UU}^{[^1S_0^{[8]}]} = 1+\bar y^2,\, 
A_{UU}^{[^3P_0^{[8]}]} = \left[2 \bar y \frac{7+3\hat{Q}^2}{1+\hat{Q}^2} + y^2 \frac{7+2\hat{Q}^2+3\hat{Q}^4}{(1+\hat{Q}^2)^2} \right] \frac{1}{m_{Q}^{2}}, \,
B_{UU}^{[^1S_0^{[8]}]} = - \bar y, \, 
B_{UU}^{[^3P_0^{[8]}]} =  \frac{3-\hat{Q}^2}{1+\hat{Q}^2}\frac{\bar y}{m_{Q}^2}. \label{eq:ABUU}
\end{align}
Here, we defined $\bar y=1-y$, {with $y$ being the inelasticity variable (see \ce{eq:SIDIS:kin-var}{)}}, and $\hat{Q}^2\equiv Q^2/(4m_{Q}^2)$ and we approximated $m_{\Q} \simeq 2 m_{Q}$, where $m_{Q}$ denotes the heavy-quark mass. 

At the EIC, one %
{could} try to determine the LDMEs together with the gluon TMDs. The $Q^2$ and $y$ dependence of the ${P_T^*}$-independent pre-factor {$B_{UU}/A_{UU}$} can then be exploited%
{{,} as it makes the observable dependent on different linear combinations of the LDMEs. This can be paralleled to the slight rapidity dependence of the LDME linear combination appearing in the polarisation of the hadroproduction yield~\cite{Lansberg:2019adr}}. 
Another option is to consider ratios in which the gluon TMDs cancel out~\cite{Bacchetta:2018ivt,Boer:2021ehu}, although that may only hold %
at LO in certain cases. An example of this will be discussed in Section~\ref{sec:polarisationTMD} where the quarkonium polarisation is used to cancel out the gluon TMDs.  

A further constraint on the LDMEs comes from the bound on the above asymmetry. {At leading order,} the bound $\bm q_T^2 | h_1^{\perp\, g}(x, \bm q_T^2) | / (2M_p^2) \leq f_1^g(x, \bm q_T^2)$ \cite{Mulders:2000sh} and the fact that $|\langle \cos 2\phi_T^* \rangle | \leq 1$ leads to the condition $|B_{UU}/A_{UU}| \leq 1$. The LDMEs that determine the ratio $B_{UU}/A_{UU}$ will have to respect this bound. In this way one can find for instance that the CO LDMEs from Ref.~\cite{Butenschoen:2010rq} do not respect this bound at LO (and $A_{UU}$ which should be positive becomes negative below the central value within the 1$\sigma$ uncertainty range), but it has to be noted that these LDMEs were obtained at NLO from hadro- and photoproduction data. 

The ratio $B_{UU}/A_{UU}$ at LO is shown in \cf{fig:ratioBUUoverAUUandratioAULoverAUU} {(left plot)} for two different CO LDME sets: {one obtained at LO}~\cite{Sharma:2012dy} (SV){,} which is very similar %
{to the NLO fit \cite{Chao:2012iv}, denoted C12 in \ct{tab:LDME-fits_comp}}{,} and {another obtained at NLO with FF}~\cite{Bodwin:2014gia} (BCKL), {denoted {B14}  in \ct{tab:LDME-fits_comp}}. The uncertainty bands are obtained assuming uncorrelated uncertainties on the $\langle{\cal O}^{{\Q}}\left[^{1}S_{0}^{[8]}\right]\rangle$ and $\langle{\cal O}_{8}^{{\Q}}\left[^{3}P^{[8]}_{0}\right]\rangle$  determinations. The ratio is shown as a function of $\hat{Q}^2$. Here $\hat{Q}^2=0.01$ is considered to be the minimum achievable value. %
{Indeed,} in order for $\phi_T^*$ to be determined, one needs {to %
be in the} electroproduction {regime where} $Q^2 \geq 1$ GeV$^2$. {In the bottomonium case, $\hat{Q}^2$ should thus be larger than} 1 GeV$^2/(4 m_b^2) \approx 0.01$. 

\begin{figure}[htb] %
   \centering
    \includegraphics[width=0.4\textwidth]{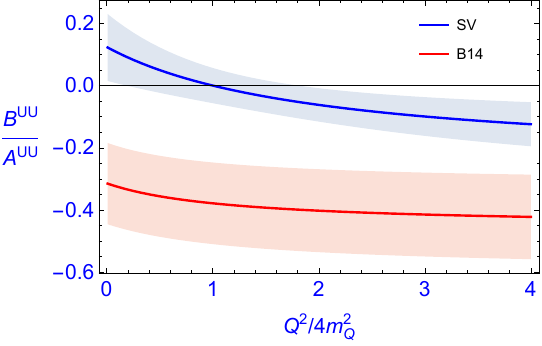}\quad  
    \includegraphics[width=0.4\textwidth]{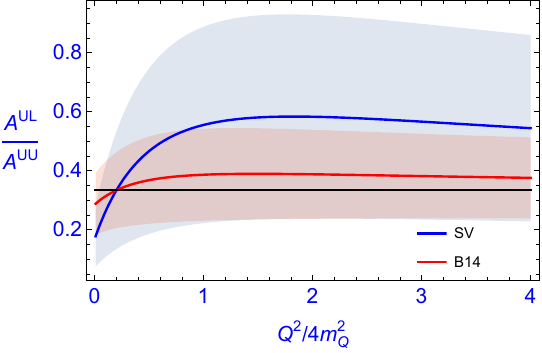} 
      \caption{(Left plot) The ratio $B_{UU}/A_{UU}$ of the asymmetry $\langle \cos 2\phi_T^* \rangle$ (\ce{eq:cos2phi}) as a function of $\hat{Q}^2=Q^2/(4m_Q^2)$ for the LDMEs at {LO by Sharma} and Vitev (SV)~\cite{Sharma:2012dy} and at NLO B14~\cite{Bodwin:2014gia} and for the {inelasticity} {value} $y=0.1$; {(Right plot) The ratio $A_{UL}/A_{UU}$ \ce{eq:cos2phibis} as a function of $Q^2/(4m_Q^2)$ for the same LDMEs and $y=0.1$. The line at 1/3 corresponds to unpolarised production. {These plots are obtained from results presented in \cite{Bacchetta:2018ivt,Boer:2021ehu}.}}}
      \label{fig:ratioBUUoverAUUandratioAULoverAUU}
\end{figure}

The figure indicates that there is much uncertainty in the LO result. It also indicates the precision needed at the EIC in order to differentiate among the various fits and to improve on them. A determination of $B_{UU}/A_{UU}$ at the 10\% level would be an improvement of the current situation. Assuming $h_1^{\perp\, g}$ is 10\% of its maximal value at EIC energies {(for a more detailed analysis, see section \ref{sec:lin})}, this translates into a percent level accuracy requirement on the measurement of $\langle \cos 2\phi_T^* \rangle$. For other $y$ values, similar conclusions hold. Needless to say, an NLO analysis of the asymmetry will be needed in order to arrive at more accurate predictions for the EIC {and for a fully coherent NLO computation with NLO LDMEs.}

For these measurements{, a} good ${P_T^*}$-resolution at small ${P_T^*}$ is an important requirement. Small ${P_T^*}$ applies to the range up to a few GeV for the EIC energies. Therefore, the transverse{-}momentum resolution in the small transverse{-}momentum region should be on the order of a few hundred MeV, such that sufficient bins can be selected to map out this region. For the determination of $\langle \cos 2\phi_T^*\rangle$, a sufficient angular resolution is needed.

\subsubsection{Quarkonium polarisation in electroproduction within TMD factorisation}
\label{sec:polarisationTMD}

If the polarisation state $P$ ($L$ or $T$) of the produced quarkonium can be determined in the semi-inclusive quarkonium production process, $e \, p \to e'\, J/\psi\, (\Upsilon) \, X$ at small transverse momentum, ${P_T^*}$, then that may offer a further possibility to improve our knowledge on LDMEs. As an illustration, here we consider the example of the ratio of the $\phi_T^*$-integrated cross sections:
\begin{align}
\frac{\int d \phi_T^* \,\frac{d\sigma^{UP}}{d y\, dx_B\, d^2 {\bm{P}^*_{T}}}}{\int d \phi_T^* \,\frac{d\sigma^{UU}}{d y\, dx_B\, d^2 {\bm{P}^*_{T}} }} = \frac{\sum_n A_{UP}^{[n]} \, {\cal C} \left[f_1^{g} \, \Delta^{[n]} \right]}{\sum_n A_{UU}^{[n]}\,{\cal C}\left[f_1^g \, \Delta^{[n]}\right]} = \frac{A_{UP}}{A_{UU}}.
\label{eq:cos2phibis}
\end{align}
{Let us stress that \ce{eq:cos2phibis} relies on the assumption that} the shape functions are equal to the corresponding LDMEs times a universal shape function {%
{that} is also polarisation independent}. %
{If so,} the ratios $A_{UL}/A_{UU}$ and $A_{UT}/A_{UU}$ are independent of the value of %
{$P^*_{T} \equiv |\bm{P}^*_{T}|$} to all orders and hence not affected by TMD evolution. The ratio will only receive contributions from higher orders through modification of the amplitudes. Thus far only the %
{LO} expressions are known~\cite{Bacchetta:2018ivt,Boer:2021ehu}: $A_{UU}$ was already given in Section \ref{sec:LDMEsTMD}, {and}
\begin{align}
A_{UL}
& = \frac{1}{3}\,[1+(1-y)^2] \, \langle{\cal O}^{{\Q}}\left[^{1}S^{[8]}_{0}\right]\rangle +  \left[2(1-y)\,\frac{1+10\hat{Q}^2+\hat{Q}^4}{(1+\hat{Q}^2)^2} +  y^2\,\frac{1+2\hat{Q}^2+\hat{Q}^4}{(1+\hat{Q}^2)^2} \right] \frac{\langle{\cal O}^{{\Q}}\left[^{3}P_{0}^{[8]}\right]\rangle}{m_{Q}^{2}}   \,, \label{eq:AUL}
\end{align}
where $A_{UT}= A_{UU} - A_{UL}$. Compared to $A_{UU}$, the {$\langle{\cal O}^{{\Q}}\left[^{3}P_{0}^{[8]}\right]\rangle$} term in $A_{UL}$ has different inelasticity $y$ and $\hat{Q}^2$ dependence{s}. This implies that there can be a significant deviation of $A_{UL}$ from $A_{UU}/3$ (and of $A_{UT}$ from $2A_{UU}/3$), signalling the production of polarised quarkonia. Likewise, one could consider the ratios $B_{UL}/B_{UU}$ or $B_{UT}/B_{UU}$ which are similar, but different linear combinations of LDMEs. 

In \cf{fig:ratioBUUoverAUUandratioAULoverAUU} {(right plot)} the ratio $A_{UL}/A_{UU}$ at LO is shown for the LDME fits \cite{Sharma:2012dy} at LO (SV) and~\cite{Bodwin:2014gia} (here denoted BCKL, B14 in~\ct{tab:LDME-fits_comp}) at NLO, including uncertainty bands, assuming again uncorrelated uncertainties on the $\langle{\cal O}^{{\Q}}\left[^{1}S_{0}^{[8]}\right]\rangle$ and $\langle{\cal O}^{{\Q}}\left[^{3}P^{[8]}_{0}\right]\rangle$ determinations. 
{In reality the uncertainties are correlated, which means that the bands are expected to be overestimations. The difference between the central values of the two different LDME sets could be viewed as another measure for the size of the involved uncertainties.}
Although both fits are compatible with unpolarised production, %
both fits also allow{, within their uncertainties,} for values considerably different from $1/3$. {It is important to recall that quarkonium-polarisation observables are very sensitive to radiative corrections~\cite{Gong:2008sn,Artoisenet:2008fc,Chang:2009uj,Artoisenet:2009xh,Lansberg:2010vq,Butenschoen:2012px,Feng:2020cvm}. Computations of the NLO corrections to the hard parts entering the ratio $A_{UL}/A_{UU}$ are therefore necessary to perform a reliable extraction of the LDMEs from these ratios.}

In Ref.~\cite{Qiu:2020xum}, the fit C12~\cite{Chao:2012iv} is used to demonstrate the dominance of the ${}^1S_0^{[8]}$ $c\bar{c}$ state in the inclusive process $e\,h \to J/\psi\, X$ (which is dominated by $Q^2 \approx 0$) described in collinear factorisation. As a result, it is concluded that the $J/\psi$ will be approximately produced in an unpolarised state. However, the above results show that due to the large uncertainties in the CO LDMEs{,} one cannot draw the same conclusion for semi-inclusive $J/\psi$ electroproduction {\it in the TMD regime} {(where $P_T^{*2}$ is much smaller than the two hard scales $ M_{J/\psi}^2$ and $Q^2$)}. %
Observation of %
{a non-zero polarisation of the} $J/\psi$ {yield} would signal the relevance of the $P$-wave LDME $\langle{\cal O}^{{\Q}}\left[^{3}P^{[8]}_{0}\right]\rangle$ or of higher{-}order contributions.

Again a 10\% level precision of the determination of the ratio $A_{UL}/A_{UU}$ at EIC would be sufficient to improve on the present situation. For this the polarisation state $L$ or $T$ of the quarkonium needs to be determined with sufficient precision of course.

\subsection{On the importance of final-state effects on quarkonium formation in electron-nucleus collisions }

Interest in quarkonium {formation} %
in reactions with nuclei goes back more than 30 years in the context of heavy-ion reactions.  {The colour interaction between heavy quarks immersed in a high temperature quark-gluon plasma (QGP), such as produced in these reactions, was predicted to be screened, preventing quarkonium states from forming as well as dissociating them~\cite{Matsui:1986dk}.}  %
Excited, weakly-bound-state solutions to the Schr\"odinger equation, such as $\Upsilon(2S)$, $\psi'$, $ \chi_c$, %
{were} expected  to melt  first in the QGP and provide a ``thermometer'' for determining the plasma temperature. Since their introduction, these ideas have evolved significantly. It was realised that dissociation and {formation suppression} of hadrons in QCD matter is not limited to quarkonia. Open heavy{-}%
{flavour} mesons have short formation times and can also be destroyed by collisional interactions in the nuclear medium~\cite{Adil:2006ra,Sharma:2009hn}, reducing the experimentally measured cross sections.  Importantly, the breakup of $J/\psi$s and $\Upsilon$s is not exclusive to the QGP and can take place in different forms of strongly-interacting matter, for example a hadron gas or a large nucleus.  
{Measurements of the modification of charmonium and bottomonium production in $d$Au, $p$Au, $p$Al and $p$Pb collisions at RHIC~\cite{PHENIX:2013pmn,PHENIX:2022nrm,PHENIX:2019brm} and at the LHC~\cite{CMS:2013jsu}, respectively, showed that production suppression increases with the multiplicity of hadrons recorded in a reaction~\cite{Andronic:2015wma,Rothkopf:2019ipj}.} 
Moreover, recent measurements of bottomonium yields in $p$Pb collisions showed~\cite{CMS:2022wfi} that excited $\Upsilon$ states are more suppressed than the ground state, and the hierarchical pattern becomes more manifest in the negative rapidity region, which is the direction of the lead nucleus~\cite{Ferreiro:2014bia}. 
Studies indicate that final-state interactions can play a significant role in reducing the rates of quarkonium production at the EIC. This quenching effect has been demonstrated for light and heavy mesons (containing a single heavy quark)~\cite{Li:2020zbk} and inclusive and heavy flavor-tagged  jets~\cite{Li:2020rqj,Li:2021gjw}. At forward rapidities and, especially at lower %
{centre}-of-mass energies, suppression in cold nuclear matter can be as large as a factor of two and serves as strong motivation to investigate these effects for quarkonium final states. 

{These observations and predictions indicate that final-state effects (interaction of quarkonium with co-moving hadrons as well as the remnant of the nucleus) require careful treatment in order to extract information about nuclear PDF and transport properties of the nuclear matter. This %
can be addressed from both experimental and theoretical perspectives.

Experimentally, one can approach this concern by studying femtoscopic correlations (two-particle correlations at low relative momentum) between quarkonium and hadron in $ep$ and $eA$ reactions. Such observables are sensitive to interactions in the final state and strong interaction parameters can be measured directly (the scattering length and effective range)~\cite{STAR:2015kha,ALICE:2020mfd,ALICE:2021cpv}.  Quarkonium-hadron elastic and inelastic scattering cross sections can be evaluated as a function of event multiplicity. Such information can be used to calculate the modification of the quarkonium yield in the hadronic environment.
}

{From the theory point of view, in order to extract nuclear PDFs} and constrain the transport properties of large nuclei using quarkonium production in {$eA$} collisions, we need to develop a theoretically well-controlled framework %
{capable of describing} final-state interactions. %
{Below, we briefly present an example of such an attempt.}

\begin{figure}[hbt!]
\centering
\subfloat[Wilson lines for dissociation.]{\includegraphics[height=1.95in]{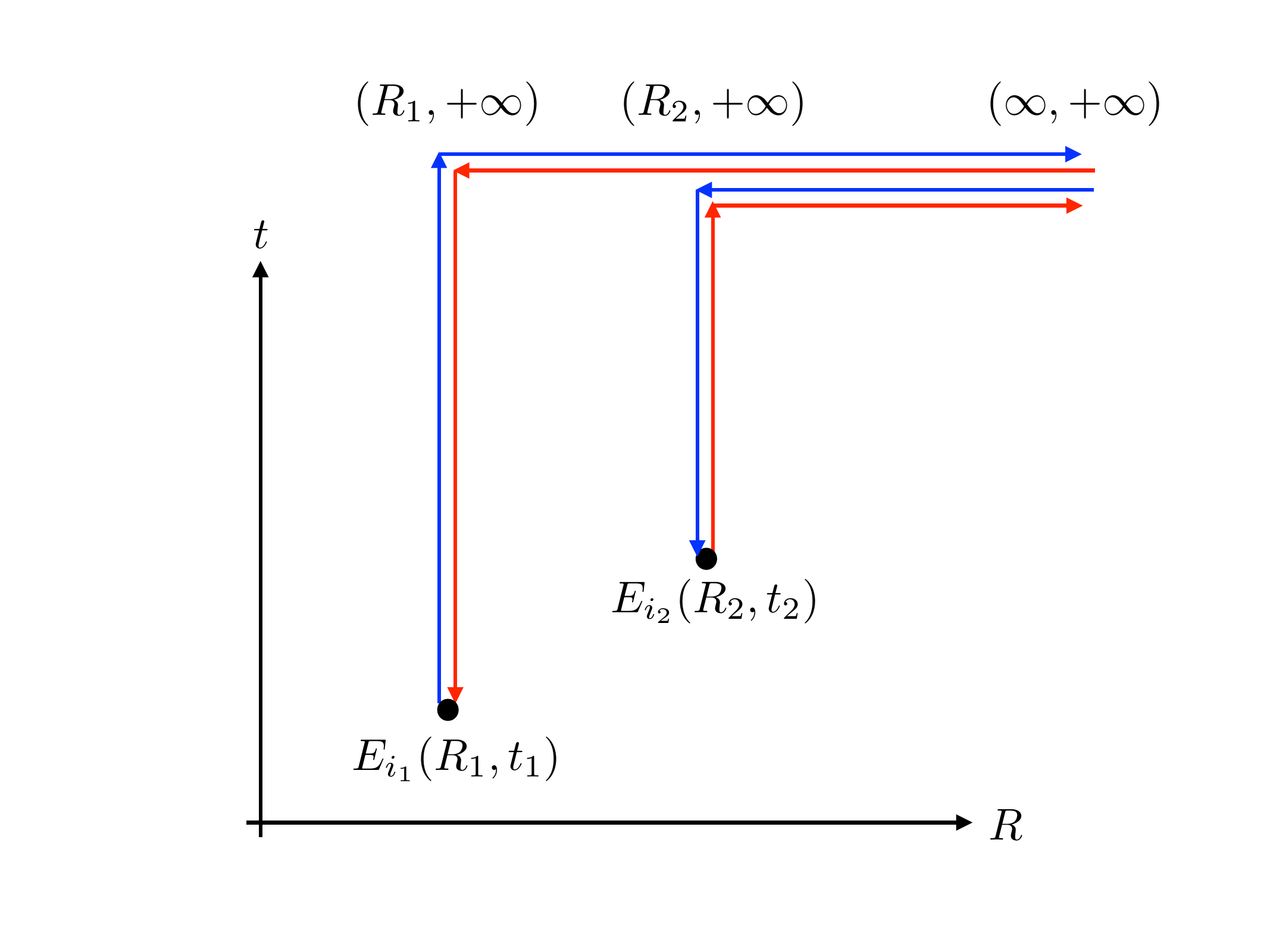}}
\subfloat[Wilson lines for recombination.]{\includegraphics[height=1.95in]{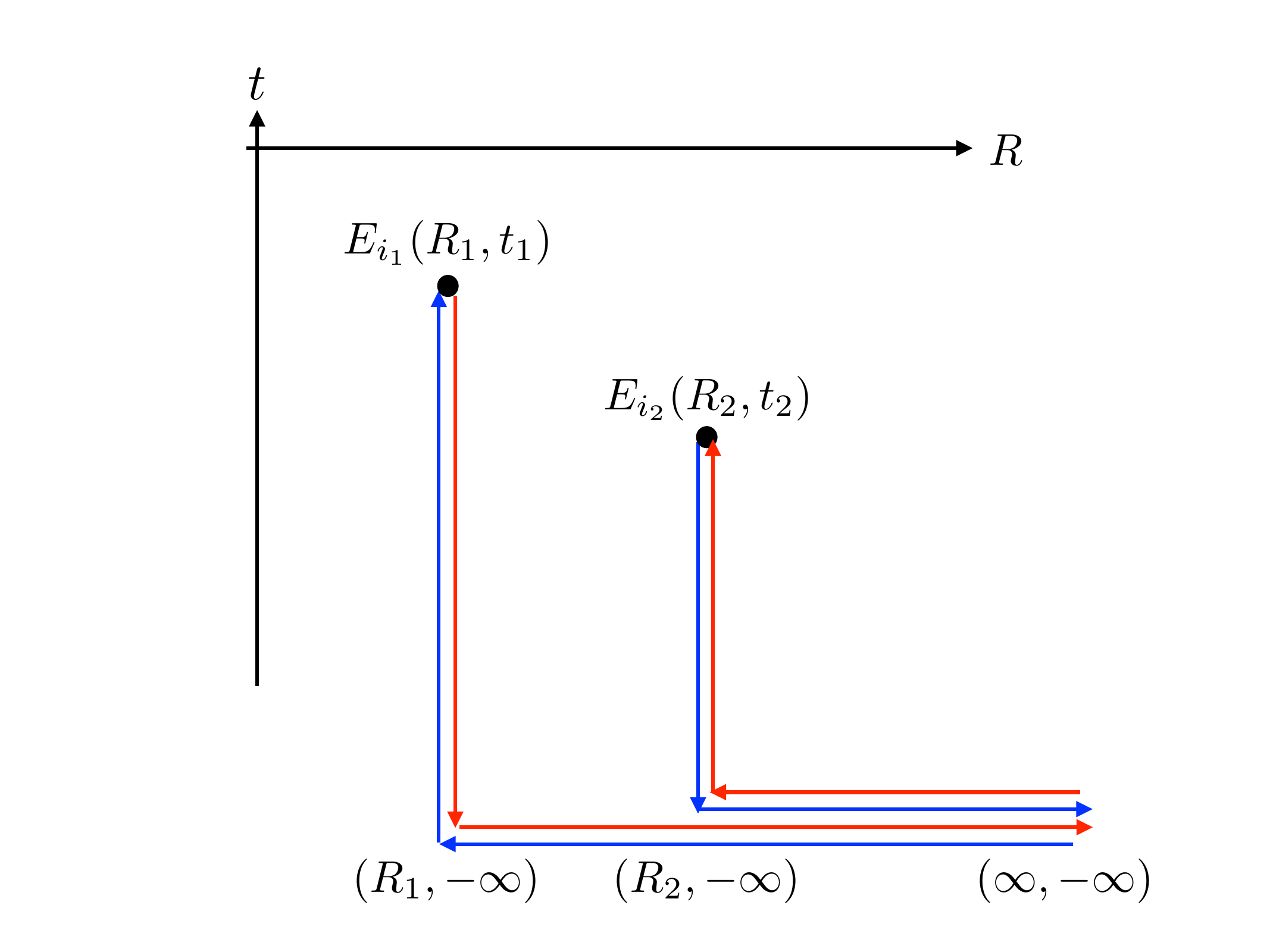}}
\caption{Staple-shape Wilson lines in the definition of the chromoelectric correlator~(\ref{eqn:eA_EEcorrelator}). Here we set $i_1=i_2$, $y=({\boldsymbol R}_1,t_1)$ and $x=({\boldsymbol R}_2,t_2)$. The plots are taken from Ref.~\cite{Yao:2020eqy}\,.}
\label{fig:eA_wilson}
\end{figure}

Since the remnant of the nucleus is a cold nuclear environment, we expect the energy transferred between the nucleus remnant and the heavy{-}quark pair traversing the nucleus to be small. {With this assumption}, one can use the open quantum system framework and the Boltzmann equation developed in Refs.~\cite{Yao:2018nmy,Yao:2018sgn,Yao:2020eqy,Yao:2021lus,Binder:2021otw,Scheihing-Hitschfeld:2023tuz,Nijs:2023dbc} to study final-state interactions. In this approach the physical quantity that encodes the essential information of the nuclear remnant relevant for a final-state interaction is the chromoelectric field correlator, which is defined in a gauge invariant way:
\begin{eqnarray}
\label{eqn:eA_EEcorrelator}
g_E^{>}(q) &=& \sum_{i=1}^3 \int {\rm d}^4 (y-x) \, e^{iq \cdot (y-x)} \, {\rm Tr}_N\Big( E_i(y) W E_i(x) \rho_N \Big) \,,
\end{eqnarray}
where $\rho_N$ is the density matrix of the remnant nucleus and $W$ denotes a staple-shape Wilson line in the adjoint representation that connects the spacetime points $y$ and $x$ such that the correlator is defined gauge invariantly. For quarkonium dissociation and formation, the two time-like Wilson lines are connected at positive and negative infinite times separately, as shown in \cf{fig:eA_wilson}. The Wilson lines involved here are similar to those involved in the definition of proton TMD{s}, with {a} %
difference in the orientation of the Wilson lines.

In a nutshell, quarkonium production in $eA$ collisions involves both initial-state and final-state effects. It will be important for the community to develop strategies for how to best separate these distinct contributions. The combination of both $eA$ and $pA$ experimental data will be useful to determine {quarkonium-hadron interaction parameters}, {nuclear PDFs}, and properties of the remnant nucleus such as the chromoelectric correlator strength.

\section{Quarkonia as tools to study the parton content of the nucleons}
\label{sec:ep}

The goal of the present section is to show that quarkonium production in lepton-hadron collisions can be an excellent observable to probe the partonic content of the nucleon. 

First, we discuss %
{how} quarkonium production measurements at the EIC can contribute to our knowledge of collinear PDFs of the nucleon. Section~\ref{sec:gluon_pdf_inclusive} is dedicated to accessing the gluon PDF from inclusive \newb{-}quarkonium photoproduction processes. In Section~\ref{sec:gluons-exclusive}, we emphasise how measurements of exclusive \jpsi and \ups electroproduction at the EIC, by extension of those from the HERA collider, can be used as an indirect probe of the gluon PDF at moderate values of the momentum fraction over a wide range of scales. 
Section~\ref{sec:light-quarks} is devoted to the sensitivity to light quark PDFs of inclusive \jpsi photoproduction, while Section~\ref{sec:charm_intr-charm} focuses on the charm PDF and the potential detection of intrinsic charm at the EIC. %
We then move to the multidimensional imaging of the partonic structure of nucleons through quarkonium-related measurements at the EIC.

Section{s}~\ref{sec:TMD-unpol} and%
~\ref{sec:TMDs-polar} are devoted to the possibility to extract information on TMD PDFs of unpolarised and polarised nucleons, respectively, from quarkonium electroproduction data at the EIC.
The systematic description of exclusive production processes is done in terms of GPDs (Section~\ref{sec:GPDs}) and GTMDs (Section~\ref{sec:GTMDs}).  We stress that the relation between GPDs and PDFs used in Section~\ref{sec:gluons-exclusive} is only an approximation, albeit a good one at the moderate and low values of the momentum fraction that we consider here. Furthermore, in Section~\ref{sec:trace_anomaly}, we touch on the %
possibility  to access the QCD trace anomaly through the measurement of exclusive \jpsi electroproduction at the threshold. 

Finally, in Section~\ref{sec:DPS-EIC}, we concentrate on double{-}parton scattering {(DPS)}, which is another interesting \newb{probe} %
of nucleon structure. First estimates for \jpsi{-}pair electroproduction at the EIC, which include DPS contributions, are presented.

\subsection{Unpolarised-nucleon PDFs}
\label{sec:unporlarized_N_PDFs}

\subsubsection{Gluon PDF from inclusive quarkonium photoproduction}
\label{sec:gluon_pdf_inclusive}
Inclusive \jpsi photoproduction, when an almost real photon hits and breaks the proton producing a \jpsi, is a useful tool to study the quarkonium-production mechanism and to %
{probe} {the} gluon PDF. %
This process has been the object of several studies at HERA~\cite{Aid:1996dn,Breitweg:1997we,Chekanov:2002at,Adloff:2002ex,Chekanov:2009ad,Aaron:2010gz,Abramowicz:2012dh}, and, in the future, it could be %
studied at the EIC. 

In Ref.~\cite{ColpaniSerri:2021bla} %
the inclusive photoproduction up to NLO {in QCD} for $J/\psi$ and $\Upsilon(1S)$ at lepton-proton colliders {was revisited,} %
{focusing} on the $P_T$- and $z$-integrated yields.
Like for other charmonium{-}production processes~\cite{Mangano:1996kg,Feng:2015cba,Lansberg:2020ejc}, %
the appearance  of negative hadronic cross sections {was observed at increasing energies, due to large negative \textit{partonic} cross sections}. There can only be two sources of negative partonic cross sections:  the interference of the loop amplitude with the Born amplitude or the subtraction of the IR poles from the initial-emission collinear singularities to the real-emission amplitude. %
Here, the latter subtraction is the source of the negative cross sections.  %
Conventionally, such divergences are removed by subtraction and included to the PDFs via {Altarelli-Parisi counterterms (AP-CT}). In principle, the negative term from the AP-CT should be compensated by the evolution of the PDFs according to the DGLAP equation. Yet, for the $\mu_F$ values on the order of the natural scale of these processes, the PDFs are not evolved much and can sometimes be so flat for some PDF parametrisations that the large $\hat s$ region still significantly contributes. This results in negative values of the hadronic cross section.%

{To solve the negative cross-section issue, }%
the $\hat \mu_F$ {prescription} proposed in \cite{Lansberg:2020ejc} {was used}, which, up to NLO, corresponds to a resummation of such collinear divergences in %
HEF~\cite{Lansberg:2021vie}. According to this prescription, one needs to choose $\mu_F$ such that{, for the partonic cross section $\hat{\sigma}_{\gamma i}$ ($i = q, \bar{q}, g$)}, $\lim\limits_{\hat{s}\rightarrow \infty}\hat{\sigma}_{\gamma i}^{\rm NLO}=0$%
. %
{It was} found that, for $z<0.9$, the optimal factorisation scale is $\hat \mu_F={0.86} M_{\Q}$~\cite{ColpaniSerri:2021bla} which falls well within the usual ranges of used values. Like for $\eta_c$ hadroproduction, such a factorisation{-}scale prescription indeed allows one to avoid negative NLO cross sections, but it of course in turn prevents one from studying the corresponding factorisation-scale uncertainties.
The NLO $\mu_R$ uncertainties {become} reduced compared to the LO ones but slightly increase around {$\sqrt{s_{\gamma p}} =$ }50~GeV%
, because of rather large (negative\footnote{{Let us stress that unless $\mu_R$ is taken very small with a large $\alpha_s(\mu_R)$, these negative contributions are not problematic, unlike the oversubtraction by the AP-CT.}}) interferences between the one-loop and Born amplitudes.
At NNLO, %
a further reduction of the $\mu_R$ uncertainties {is expected}. This is particularly relevant especially around {$\sqrt{s_{\gamma p}} =$ }$50-100$ GeV, which corresponds to the EIC region. This would likely allow us to better probe gluon PDFs using photoproduction data. Going further, differential measurements in the elasticity or the rapidity could provide a complementary leverage in $x$ to fit the gluon PDF, even in the presence of the  $v^4$ 
{CO} contributions. Indeed, these would likely exhibit a very similar dependence on $x$. 

The possibility to constrain PDFs using future $J/\psi$ and $\Upsilon(1S)$ photoproduction data~\cite{ColpaniSerri:2021bla}
{was investigated by comparing the PDF and $\mu_R$ uncertainties.}
Unsurprisingly, the PDF uncertainties get larger than the (NLO) $\mu_R$ uncertainties with the growth of the $\gamma p$ {centre-of-mass}~energy, in practice from around $300$ GeV, \ie ~for $x$ below $0.01$. Although this is above the reach of the EIC, with NNLO predictions at our disposal in the future, with yet smaller $\mu_R$ uncertainties, one could set novel constraints on PDFs with such EIC measurements.  Following the estimated counting rates for 100~fb$^{-1}$ of $ep$ collisions {given in~\cite{ColpaniSerri:2021bla}}, %
a number of differential measurements (in $\pT$, $z$ and/or $y$) will be possible to reduce the impact of highly- or even partially-correlated theoretical uncertainties, including the contamination of higher-$v^2$ corrections{,} such as the %
{CO} contributions.

\begin{table}[htbp!]\renewcommand{\arraystretch}{1.25}
    \begin{centering}\small
    \begin{tabular}{c|ccc}

        $\sqrt{s_{ep}}$
        & ${\cal L}$~(fb$^{-1})$
        & $N_{J/\psi}$ %
        & $N_{\Upsilon(1S)}$ %
        \\
   \hline
        45 & 100 & $8.5^{+0.5}_{-1.0} \cdot 10^6$%
        & $6.1^{+0.7}_{-0.8} \cdot 10^2$
\\
        140 & 100 & $2.5^{+0.1}_{-0.4} \cdot 10^7$%
        & $7.6^{+0.3}_{-0.7} \cdot 10^3$
\\

    \end{tabular}
    \caption{Expected event rates for quarkonium photoproduction at NLO  at different $\sqrt{s_{ep}}$ (in GeV) of the EIC  for $\mu_R=5$~GeV for $J/\psi$ and $\mu_R=16$~GeV for $\Upsilon(1S)$, setting $\mu_F=\hat \mu_F$ and applying the cut $z<0.9$. We asummed a detector efficiency of $\epsilon_{detect}=85\%$ for both  $\mu^+\mu^-$ and $e^+e^-$ channels. Combined with branching fractions, this yields $\epsilon^{J/\psi}_{ \ell^+ \ell^-}\approx0.1$, 
    and $\epsilon^{\Upsilon(1S)}_{ \ell^+ \ell^-}\approx0.04$. The CT18NLO PDFs~\cite{Hou:2019efy} are used. [Table from~\cite{ColpaniSerri:2021bla}]}
    \label{tab:numb_of_particles} 
    \end{centering}
\end{table}

\ct{tab:numb_of_particles} {gathers estimates} of the expected number of {$J/\psi$ and $\Upsilon(1S)$} possibly detected at the different $ep$ %
{centre-of-mass}~energies at {the} EIC. 
For $\Upsilon(1S)$, the yields should be sufficient to extract cross sections even below %
{the} nominal {EIC} luminosities. 

One can also estimate the expected number of detected $\psi'$, $\Upsilon(2S)$%
{and} $\Upsilon(3S)$ using the following relations
\begin{equation}
  \begin{aligned}
    N_{\psi'}&\simeq0.08 \times  N_{J/\psi},\\ 
    N_{\Upsilon(2S)}&\simeq0.4 \times N_{\Upsilon(1S)},\\ 
    N_{\Upsilon(3S)}&\simeq0.35 \times N_{\Upsilon(1S)},
  \end{aligned}\label{eq:yield_ratio}    
\end{equation}
derived from the values of\footnote{These contributions were estimated using $|R_{\psi'}(0)|^2 = 0.8$ GeV$^3$, $|R_{\Upsilon(2S)}(0)|^2 = 5.0$~GeV$^3$  and $|R_{\Upsilon(3S)}(0)|^2 = 3.4$~GeV$^3$ and the corresponding measured branching fractions to $J/\psi$ and $\Upsilon (1S)$~\cite{ParticleDataGroup:2020ssz}.} $|R_{\Q}(0)|^2$ {(the quarkonium radial wave function at the origin, that is related to the $^3S_1^{[1]}$ LDME)} and of the branching fractions to leptons. Using the values in \ct{tab:numb_of_particles} and~\ce{eq:yield_ratio}~\cite{ColpaniSerri:2021bla}, one can see that the yield of $\psi'$ should be measurable and the yields of $\Upsilon(2S)$ and $\Upsilon(3S)$ are close to about half of that of $\Upsilon(1S)$ and should be measurable {as well} at the EIC. %

\subsubsection{Gluon PDFs from exclusive quarkonium photo- and electroproduction}
\label{sec:gluons-exclusive}
The exclusive production of heavy vector mesons has long been a fascinating observable to study, functioning as an enticing avenue to unravelling the small-$x$ behaviour of the gluon PDF from low to moderate scales. Measured in the first instance in the fixed-target mode~\cite{Binkley:1981kv, Denby:1983az, 1993197} and in DIS events at HERA, see e.g.~\cite{H1:2000kis, ZEUS:2004yeh, ZEUS:2009asc, H1:2013okq}, and then more recently in ultra{-}peripheral collisions at the LHC~\cite{LHCb:2014acg, LHCb:2015wlx, LHCb:2018rcm}, they provide a means to explore the quarkonium production mechanism and act as sensitive probes at the frontier of small-$x$ saturation physics.

The exclusive $\jpsi$ electroproduction, $\gamma^* p \rightarrow \jpsi p$, has been measured via dilepton decays at HERA in a narrow range of photon virtualities, extending up to $\langle Q^2 \rangle = 22.4~ \text{GeV}^2$. {The corresponding photoproduction has} also been determined in ultraperipheral events at the LHC. %
There are not, as of yet however, any data from HERA and the LHC for exclusive $\ups$ electroproduction,  $\gamma^* p \rightarrow \ups p$, away from the photoproduction limit. Going forward, the EIC will extend the kinematic reach in $Q^2$,  providing a lever arm up to larger virtualities and, moreover, allow for a measurement of the $\ups$ electroproduction with off-shell photon kinematics for the first time, albeit with a projected lower $Q^2 + M_{\Q}^2$ bin coverage and event count rates due to its heavier mass~\cite{AbdulKhalek:2021gbh}. %

Recently, in Ref.~\cite{Flett:2021ghh}, the coefficient functions for exclusive heavy vector meson electroproduction were derived at NLO within the framework of collinear factorisation, with the transition from an open heavy quark-antiquark pair to a bound heavy vector meson made within LO NRQCD.

\begin{figure}[htbp]
\begin{center}
\includegraphics[width=0.6\textwidth]{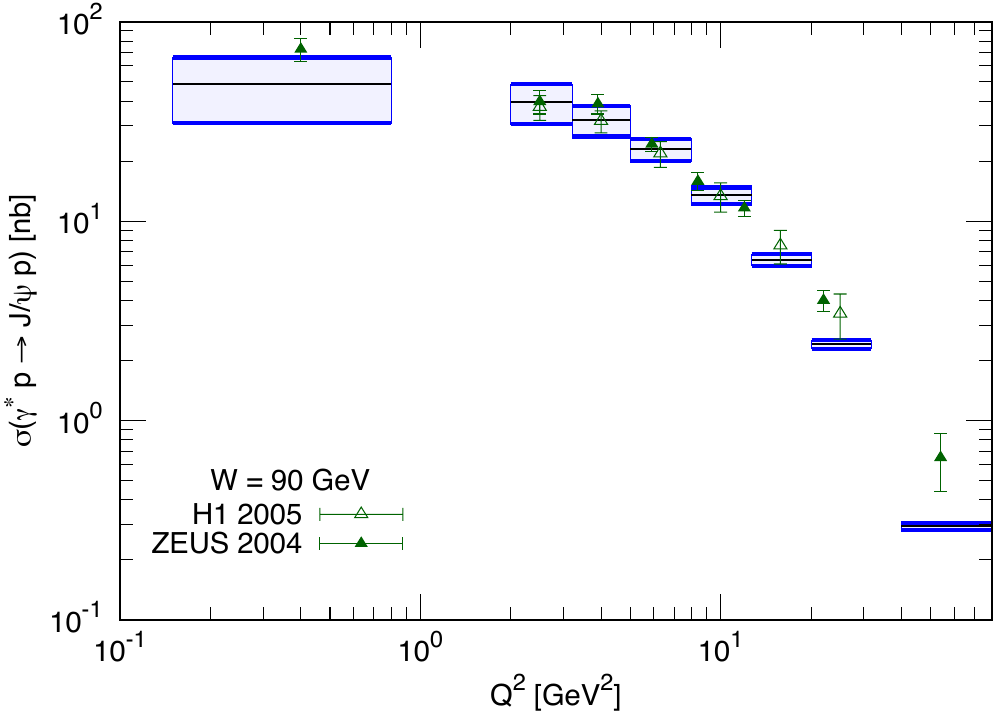}
\caption{The exclusive $\jpsi$ electroproduction cross section as a function of $Q^2$ for a fixed centre-of-mass energy $W$ = 90 GeV and compared to the data from H1~\cite{H1:2005dtp} and ZEUS~\cite{ZEUS:2004yeh}, using  results in~\cite{Flett:2021ghh} and Shuvaev-transformed MSHT20 input NLO PDFs~\cite{Bailey:2020ooq}. The black lines represent the central prediction in each bin{,} while the upper and lower blue lines are indicative of the propagation of the PDF error only. The discrepancy of the prediction from the data at the largest $Q^2$ is indicative of the need for resummation effects, see text for details.  The current data in this regime are, however, sparse and we anticipate the EIC will be able to provide more resolving power in the shape of further statistics to discern if such effects are already needed.}
\label{fig:chris_1}
\end{center}
\end{figure}

Based on the above derivation of the coefficient functions, predictions for the exclusive \jpsi electroproduction cross section have been made. They are shown in~\cf{fig:chris_1} in bins of $Q^2$ at a fixed centre-of-mass energy, $W = 90~\text{GeV}$, of the $\gamma^* p$ pair.
We use the Shuvaev transform~\cite{Shuvaev:1999fm, Shuvaev:1999ce} as a reliable means to obtain the GPD from input PDFs in the kinematic regions shown. We construct GPDs in such a way using MSHT20~\cite{Bailey:2020ooq}, NNPDF3.1~\cite{NNPDF:2017mvq} and CT18~\cite{Hou:2019efy} input NLO PDFs and the predictions based on the former are shown in the figure. The choice of input PDF has the largest effect at the lowest $Q^2$, where the choice of the initial condition of the DGLAP evolution is felt, while for larger $Q^2$, this effect washes out and the predictions based on each PDF set agree at or below the percent level.
The central values of the prediction for low to moderate $Q^2$ are in good agreement with the experimental data from H1 and ZEUS, but for larger $Q^2$ there appears to be a downward shift of the prediction from the data. The prediction in the highest $Q^2$ bin exhibits a small factorisation-scale dependency and is essentially independent of the choice of the input PDF but, as shown, the deviation from the data is sizeable. 
Interestingly, in the large $Q^2$ limit, the gluon amplitude $\propto \ln(Q^2/m_Q^2)^2$ while the quark amplitude $\propto \ln(Q^2/m_Q^2)$. This observation seems to necessitate a program of resummation for the exclusive electroproduction of heavy vector mesons for virtualities $Q^2 \gg m_Q^2$, $\ie$ those relevant for EIC kinematics, and may provide for the reconciliation of the theory prediction and experimental data at large scales.

The data statistics are currently limited for larger $Q^2$ and, in particular, there is a wide range 
where the EIC can provide a first coverage. This will help to ascertain on which front the difference between this prediction and the data at large $Q^2$ lies and if resummation effects are already needed. Other numerical effects in this framework such as the so-called `$Q_0$ subtraction'~\cite{Jones:2016ldq}, crucial for a fruitful description of the photoproduction data~\cite{Flett:2019pux, Flett:2019ept, Flett:2020duk, Flett:2021fvo, Flett:2022ues}, are not surmised to be important for electroproduction kinematics because the corresponding power correction $\mathcal O(Q_0^2/\mu_F^2)$ is no longer of $\mathcal O(1)$. See also~\cite{Eskola:2022vpi, Eskola:2022vaf} for a recent baseline study of exclusive \jpsi photoproduction in heavy-ion collisions in the collinear factorisation framework to NLO. 

\begin{figure}[htbp]
\begin{center}
\includegraphics[width=0.45\textwidth]{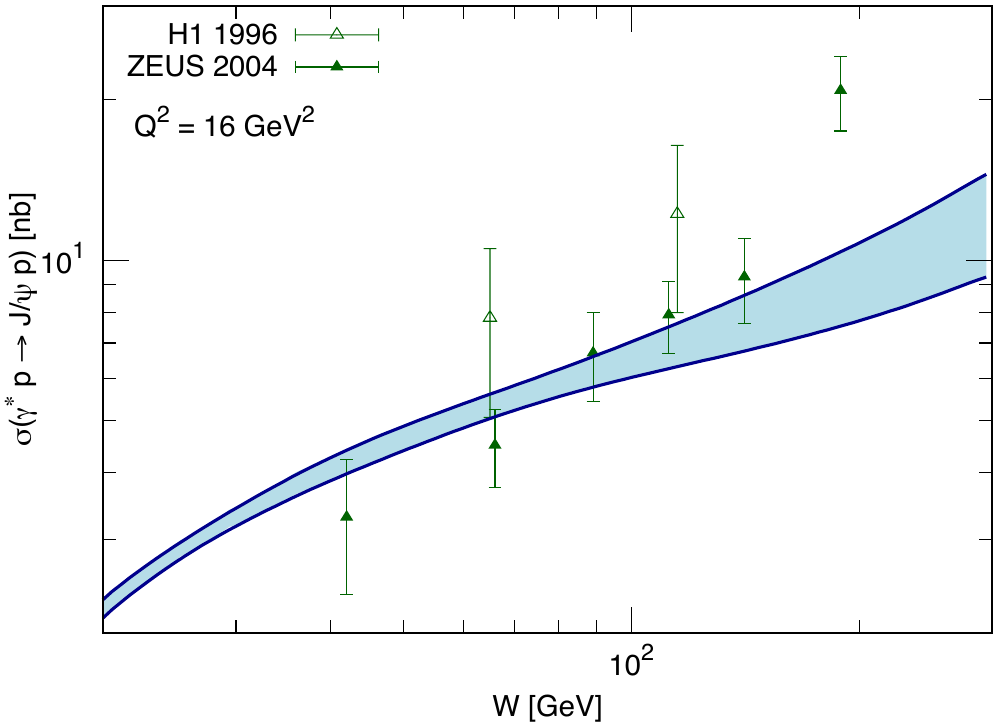}
\qquad
\includegraphics[width=0.45\textwidth]{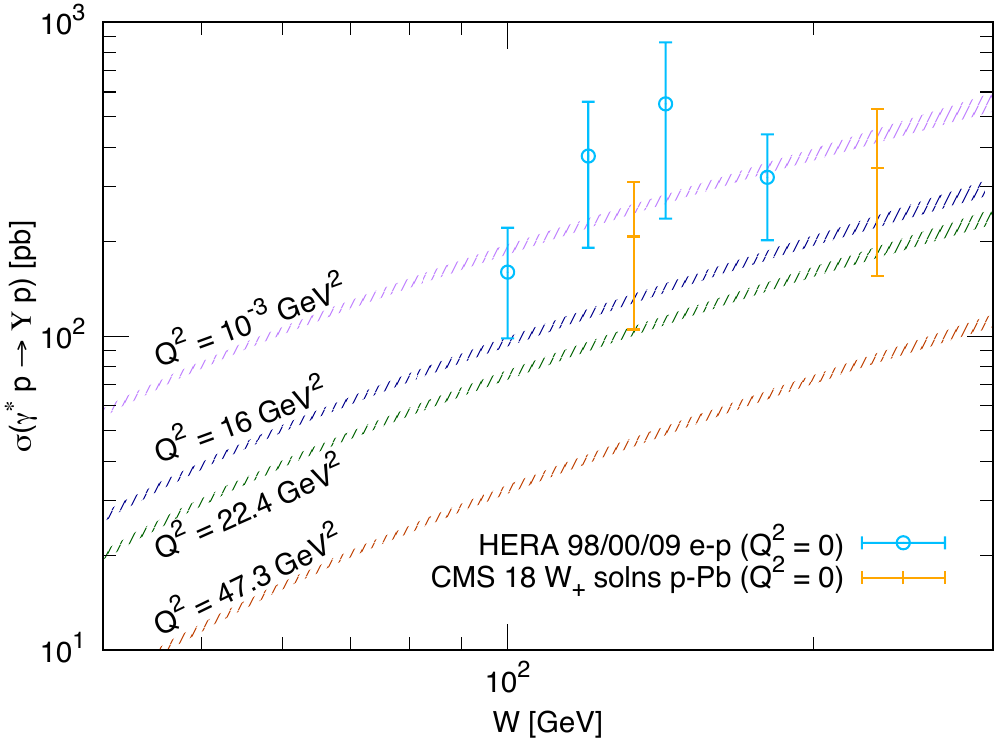}
\caption{(Left panel): The EIC will provide increased statistics for the $\jpsi$ electroproduction at current and new and unexplored virtualities. Shown is a postdiction for the exclusive $\jpsi$ electroproduction as a function of $W$ for fixed $\langle Q^2 \rangle = 16~\text{GeV}^2$. (Right panel): Prediction for the exclusive $\ups$ electroproduction as a function of $W$ for a selection of scale choices accessible by the EIC, as well as current photoproduction data from HERA and LHC in the given $W$ range. 
}
\label{fig:chris_2}
\end{center}
\end{figure}

Simulated event count projections were given for the exclusive electroproduction of the $\jpsi$ and $\ups$ in bins of $Q^2 + M_{\Q}^2$ as a function of $x$ in~\cite{AbdulKhalek:2021gbh}.  In \cf{fig:chris_2} (left panel), we show predictions for the exclusive $\jpsi$ electroproduction cross section as a function of $W$ at a fixed scale $\langle Q^2 \rangle = 16~\text{GeV}^2$ using Shuvaev-transformed MSHT20 input NLO PDFs, as well as the exclusive $\jpsi$ electroproduction HERA data that lie in this bin for comparison purposes. The prediction agrees most favourably with the more up-to-date dataset, however the EIC will be able to provide more statistics and resolve the slight tension between (and discrepancies within) the datasets. In particular, the data point at $W = 189$ GeV is around a factor of two larger than other data lying in this bin.  We also show predictions for the exclusive \ups electroproduction cross section as a function of $W$ (right panel) for $\langle Q^2 \rangle = 0.001, 16, 22.4~\text{GeV}^2$ and $47.3~\text{GeV}^2$, which may ultimately be compared with data from the EIC.\footnote{Admittedly, the expected event count rate is a lot lower than that of the corresponding $\jpsi$ production, even by three orders of magnitude in the photoproduction bin containing the most counts~\cite{AbdulKhalek:2021gbh}. Any data will therefore likely be sparse and exhibit large uncertainties, but nonetheless complement those already existing from HERA and LHC, shown in the right panel of Fig.~\ref{fig:chris_2}.} In each case the quark contribution to the total amplitude is negligible and so the forthcoming enhanced statistics and increased data coverage from the EIC will allow for refined and improved constraints on the gluon PDF at low to moderate scales. 

\subsubsection{Light quarks}
\label{sec:light-quarks}

At EIC energies, we also expect to be sensitive to quark-initiated partonic subprocesses. As shown in Ref.~\cite{Flore:2020jau}, in inclusive quarkonium photoproduction, the quark-induced subprocesses $\gamma + q \to \jpsi + q\,(+\, g)$ will be a relevant contribution to the  cross section. {Therefore, through quarkonium photoproduction, the EIC will also} %
be %
{partially} sensitive to the light-quark PDFs.

\begin{figure}[ht]
\centering
\hspace*{-3mm}\includegraphics[width=8cm]{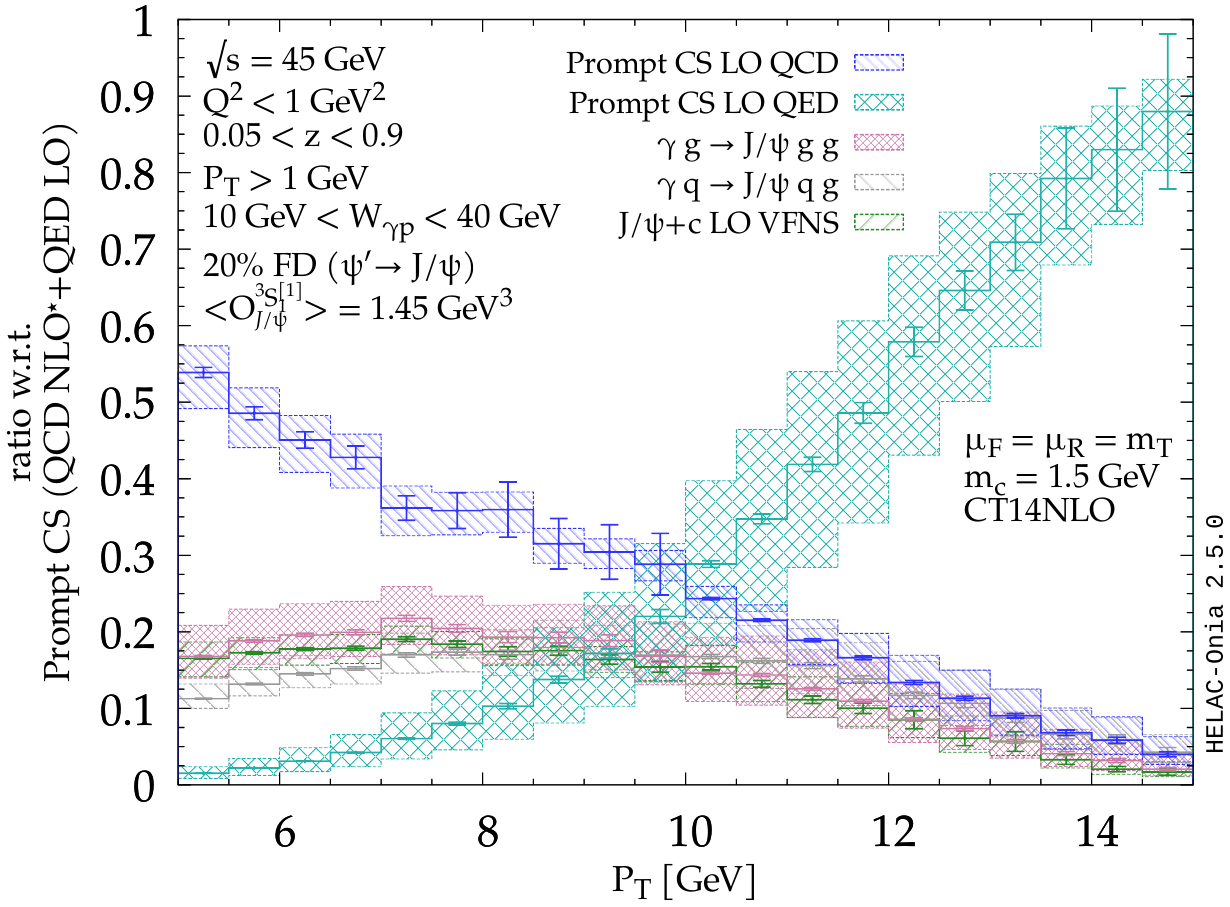}\hspace{3mm}
\includegraphics[width=8cm]{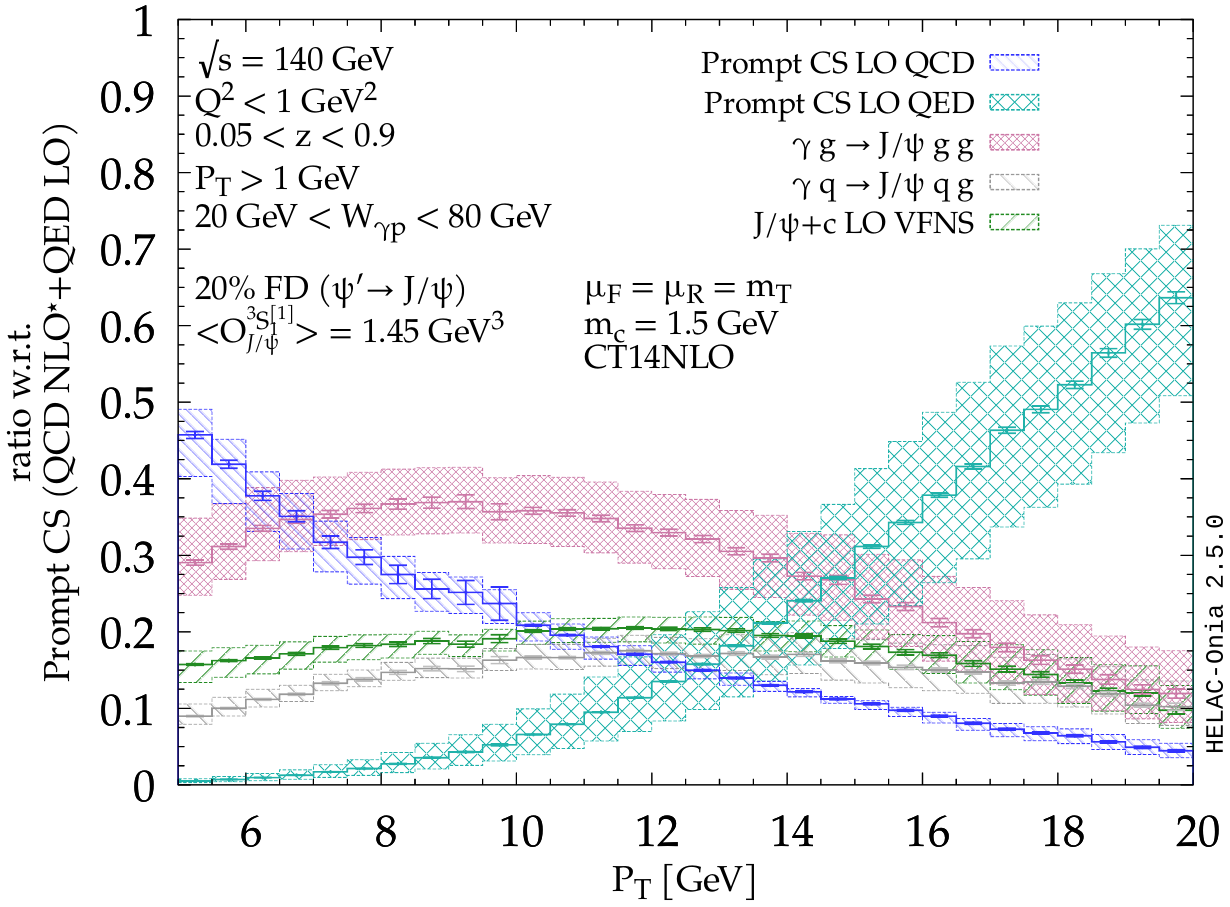}
\caption{Ratio of the different contributions to the %
cross section of \cf{fig:EIC-CT14NLO-NLOstar} at the %
EIC at $\sqrt{s_{ep}} = 45$ GeV (left) and $\sqrt{s_{ep}} = 140$ GeV (right) {as a function of the $\jpsi$ transverse momentum, $\pT$}. Figure taken from Ref.~\cite{Flore:2020jau}.}
\label{fig:EIC-CT14NLO-NLOstar-ratio}
\end{figure}

\begin{figure}[htbp]
    \centering
    \includegraphics{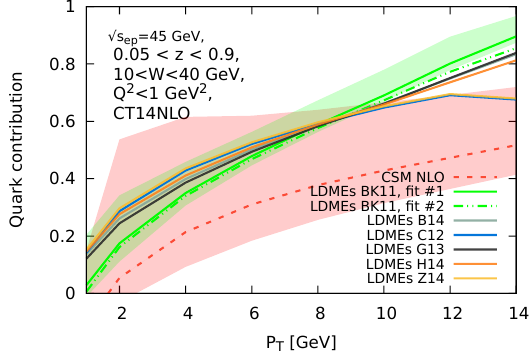}   
    \includegraphics{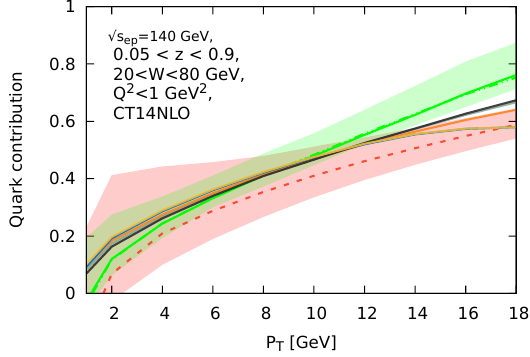}
    \caption{Plots of the fraction of light-quark-induced subprocesses in the %
    {$\pT$} spectra of $J/\psi$ photoproduction at the EIC shown in  \cf{fig:Jpsi-photoprod-pT_LDMEs} %
    {at} NLO %
    {in} NRQCD factorisation for different LDME sets.}
    \label{fig:quark_fractions}
\end{figure}

To highlight the quark-induced contribution, we show in \cf{fig:EIC-CT14NLO-NLOstar-ratio} the ratio to the %
CSM cross section for every partonic subprocess {(up to $\mathcal{O}(\alpha\alpha_s^3)$)} {depicted in} \cf{fig:EIC-CT14NLO-NLOstar}, at two different %
{centre-of-mass} energies, $\sqrt{s_{ep}} = 45$~GeV (left panel) and $\sqrt{s_{ep}} = 140$ GeV (right panel), as a function of \jpsi transverse momentum. {It is clear that} %
the pure QED quark-initiated process at $\mathcal{O}(\alpha^3)$ become dominant at high $P_T$, {accounting for} %
over half of the %
cross section at $P_T \sim 12\,(16)$ GeV at $\sqrt{s_{ep}} = 45\,(140)$ GeV. {The effect is larger at $\sqrt{s_{ep}} = 45$~GeV{,} where the valence region of the PDF is probed.} The $\mathcal{O}(\alpha\alpha_s^3)$ contribution ($\gamma + q \to \jpsi + q + g$) is {roughly} %
$5-15 \%$ and $10-15 \%$  of the %
cross section at $\sqrt{s_{ep}} = 45$ GeV and $\sqrt{s_{ep}} = 140$ GeV, respectively. We then expect that, in \jpsi photoproduction processes at the EIC, the \jpsi produced at large $P_T$ will be recoiling {off of} %
at least one quark jet.  {The significant contribution of quark-induced subprocesses at high %
{$\pT$} of the $J/\psi$ is also observed in the NLO NRQCD calculation, {as shown in} %
\cf{fig:quark_fractions}. Moreover, this conclusion depends only mildly on the %
NRQCD LDMEs {that were used}.} %

\subsubsection{Charm quark and intrinsic charm}
\label{sec:charm_intr-charm}
The existence of a nonperturbative charm{-}quark content in the proton, referred to as intrinsic charm (IC), has long been postulated~\cite{Brodsky:1980pb,Brodsky:1981se}. Intrinsic charm states are a fundamental property of hadronic bound-state wave functions~\cite{Brodsky:1980pb,Brodsky:1981se}. They differ from extrinsic charm in perturbative QCD that arises from gluon splitting and contributes to the heavy-quark PDFs { (i.e., radiatively generated)}. {The ``intrinsic" label is due to the fact that a $c\bar{c}$ pair formed by gluons from more than one quark line forces the $c\bar{c}$ parameters to be dependent upon (i.e., reflective of) the hadron that creates it.  Therefore the $c$ and $\bar{c}$ distributions are ``intrinsic'' to the identity of the proton, or the meson, or whichever hadron contains the bound quarks that emit gluons.  ``Extrinsic'' means that the sea quark pairs come from a single quark line gluon and therefore do not reflect the bound state structure they exist in, at least not in the clear way that IC of the proton does, peaking at $\sim x_B=0.4$ and imparting a difference in $c$ and $\bar{c}$ distributions, according to recent lattice calculations \cite{Sufian:2020coz}. } 

Since extrinsic {charm} contributions are due to a gluon {emitted by a single quark line which then splits into a $c\bar{c}$} pair, these charm distributions are soft, appear at low $x$ and depend logarithmically on {the mass of the heavy quark} $m_Q$. On the other hand, IC contributions dominate at higher $x$ and have a $1/m_Q^2$ dependence. They come from five-quark {(and higher) Fock-}state configurations of the proton, $|uud c \overline c \rangle$, and are kinematically dominated by the regime where the state is minimally off-shell, leading to equal{-}rapidity constituent quarks. Thus{,} the charm quarks are manifested at large $x$. When the proton in this state interacts with its collision partner, whether a hadron or a lepton, the coherence of the Fock components is broken and the fluctuations can hadronise~\cite{Brodsky:1980pb,Brodsky:1981se,Brodsky:1991dj}. In hadroproduction, the state can be broken up by a soft gluon from the target interacting with the proton. In $ep$ interactions, instead of a soft gluon, a low{-}energy photon can play the same role and bring the state on mass shell.

Several formulations of intrinsic charm in the proton wave function have been proposed. The first was proposed by Brodsky and collaborators in~\cite{Brodsky:1980pb,Brodsky:1981se}:
\begin{eqnarray}
\frac{dP_{{\rm ic}\, 5}}{dx_1 dx_2 dx_3 dx_c dx_{\overline c}} = P_{{\rm ic}\,5}^0
N_5 \int dk_{x\, 1} \cdots dk_{x \, 5}
\int dk_{y\, 1} \cdots dk_{y \, 5} 
\frac{\delta(1-\sum_{i=1}^5 x_i)\delta(\sum_{i=1}^5 k_{x \, i}) \delta(\sum_{i=1}^5 k_{y \, i})}{(m_p^2 - \sum_{i=1}^5 (\widehat{m}_i^2/x_i) )^2} \, \, ,
\label{eq:icdenom}
\end{eqnarray}
where $i = 1$, 2, 3 are the light quarks ($u$, $u$, $d$) and $i = 4$ and 5 are the $c$ and $\overline c$ quarks.
Here{,} $N_5$ normalises the $|uud c \overline c \rangle$ probability to unity and $P_{{\rm ic}\, 5}^0$ scales the unit-normalised probability to the assumed intrinsic{-}charm content of the proton. The delta functions conserve longitudinal and transverse momentum. The denominator of \ce{eq:icdenom} is minimised when the heaviest constituents carry the dominant fraction of the longitudinal momentum, $\langle x_Q \rangle > \langle x_q \rangle$. 
{In the%
first papers, the $c$ and $\overline c$ distributions were treated equally, but %
{later studies showed %
} %
an asymmetry in $c$ and $\overline c$ distributions \cite{Sufian:2020coz}.  The asymmetry is caused by QCD diagrams where, for example, two gluons from  two different valence quarks in the nucleon couple to a heavy-quark pair $gg \to Q\overline Q$ with charge conjugation value $C=+1$ \cite{Brodsky:2000zc}.
This amplitude interferes with QCD diagrams where an odd number of gluons attach to the heavy-quark pair, \eg $ g\to Q\overline Q$ and $ggg\to Q \overline Q$ with $C=-1$.  The interference of amplitudes with the same final state but different charge conjugation symmetry for the $Q \overline Q$ produces the asymmetric distribution functions.  The analogous interference term is seen in the electron and positron distributions in $e^+ e^-$ pair production \cite{Brodsky:1968rd}.}

At leading order, the charm{-}quark structure function from this state can be written as
\begin{eqnarray}
  F_{2c}^{{\rm ic}}(x_c) = \frac{8}{9} x_c\,c(x_c) = \frac{8}{9} \int dx_1 dx_2 dx_3
    dx_{\overline c} \frac{dP_{{\rm ic}\, 5}}{dx_1 dx_2 dx_3 dx_c dx_{\overline c}} \,.
\end{eqnarray}

\paragraph{Intrinsic-charm models} Intrinsic-charm distributions in the proton have also been calculated using meson-cloud models where the proton fluctuates into a $\overline D(u \overline c) \Lambda_c (udc)$ state~\cite{Paiva:1996dd,Steffens:1999hx}. A further development of this model examined all {possible} charm meson-baryon combinations  in the $|uud c \overline c \rangle$ state \cite{Hobbs:2013bia}, finding that charm mesons would predominantly {be} produced through $D^*$ mesons. In these models the charm sea contribution would be asymmetric $xc(x) \neq x \overline c(x)$.
In both the Brodsky {\it et al.} and the meson-cloud formulations, the intrinsic-charm contributions appear as an enhancement at large $x$.  On the other hand, a sea-like distribution~\cite{Pumplin:2007wg,Nadolsky:2008zw} has also been considered. In this case, the intrinsic-charm distribution is represented simply as an overall enhancement to the light-quark-mass sea. These distributions are %
symmetric, $x c(x) = x \overline c (x)$.

Intrinsic-charm distributions from these models have been included in global analyses of the parton densities~\cite{Pumplin:2007wg,Nadolsky:2008zw,Dulat:2013hea,Jimenez-Delgado:2014zga,Ball:2016neh}. Earlier analyses \cite{Hoffmann:1983ah,Harris:1995jx} focused specifically on the European Muon Collaboration (EMC) high-$x$ and high-$Q^2$ data~\cite{EuropeanMuon:1981obg}.  A range of values of $P_{{\rm ic} \, 5}^0$ were extracted, from 0.1\% to 1\%. For more details of these analyses, see~\cite{Brodsky:2015fna}.  See also the recent review in~\cite{Brodsky:2020zdq} for more applications of intrinsic-heavy-quark states. New evidence for a finite charm{-}quark asymmetry in the nucleon wave function from lattice gauge theory, consistent with intrinsic charm, was published in~\cite{Sufian:2020coz}.  {Further evidence for unequal $c$ and $\overline{c}$ distributions in the proton has recently been presented along with proposed experimental tests with the EIC using flavour-tagged structure functions \cite{NNPDF:2023tyk}.}

Note that only the 5-particle intrinsic-charm state of the proton has been discussed.  However, one can also consider higher Fock components such as $|uud c\overline c q \overline q \rangle$.  These will reduce the average momentum fraction of the charm quark and also have lower probability. See e.g.~\cite{Gutierrez:1998bc} for examples of charm hadron distributions from higher Fock states. {Finally, the possibility for an enhanced IC component in the deuteron was studied in~\cite{Brodsky:2018zdh}.}

\paragraph{Recent hints from the LHC} 
A number of experimental measurements \cite{EuropeanMuon:1981obg,NA3:1983ltt,R608:1987dyw} over the last several decades have provided tantalising hints of intrinsic charm. %
Recently LHCb announced that their measurement of $Z + {\rm charm \, jets}$ relative {to} all $Z + {\rm jets}$ is consistent with an intrinsic{-}charm component of the proton as large as 1\% at large $Z$ rapidity~\cite{LHCb:2021stx}. These results were recently confirmed by a phenomenological analysis made by the NNPDF Collaboration~\cite{Ball:2022qks}. {Measurements at lower scales than the $Z$-boson mass are therefore eagerly awaited for to advance our understanding of this higher-Fock-state phenomenon.}

\paragraph{Intrinsic charm at the EIC} The %
EIC will offer the possibility to probe the nonperturbative charm-quark content in the proton. Recent studies show that the EIC will be capable of precision studies of intrinsic{-}charm as well as gluon distribution functions in the nucleus and in the nucleon~\cite{Kelsey:2021gpk}. 

The associated production of a $\jpsi$ and a charmed particle is %
{an additional} potential probe of intrinsic{-}charm related effects. A leading order VFNS study, first made in~\cite{Shao:2020kgj} for quarkonium hadroproduction, has been extended in~\cite{Flore:2020jau} to the case of $\jpsi$ photoproduction.  Such a scheme allows a proper merging of different partonic contributions, namely $\gamma + g \to \jpsi + c + \bar{c}$ and $\gamma + \{c,\bar{c}\} \to \jpsi +  \{c,\bar{c}\}$, respectively calculated {with 3 and 4 flavours in the proton}, {using} a counter term, $d\sigma^{\rm CT}$, that avoids double counting. When the charm-tagging efficiency $\varepsilon_c$ is taken into account, the corresponding VFNS cross section is given by:

\begin{equation}\label{eq:VFNS_with_efficiency}
 d\sigma^{\rm VFNS} = d\sigma^{\rm 3FS}\left[1-\left(1-\varepsilon_c\right)^2\right] + \left(d\sigma^{\rm 4FS} - d\sigma^{\rm CT}\right)\varepsilon_c.
\end{equation}

Based on such computations, the $\jpsi+$charm yield has been calculated for two different EIC configurations: $\sqrt{s_{ep}} = 45 (140)$ GeV, taking into account a $10\%$ charm-tagging efficiency~\cite{Chudakov:2016otl}. The calculation has been done with the CT14NNLO PDF set \cite{Hou:2017khm}{,} which includes different eigensets with some IC effects: a ``sea-like'' (in green in the following){,} %
a ``valence-like'' (in red) also called ``BHPS,'' and a central eigenset with no IC effects which we refer to as ``no IC'' (in blue). 

\begin{figure}[htbp]
\centering
\includegraphics[width=8.5cm]{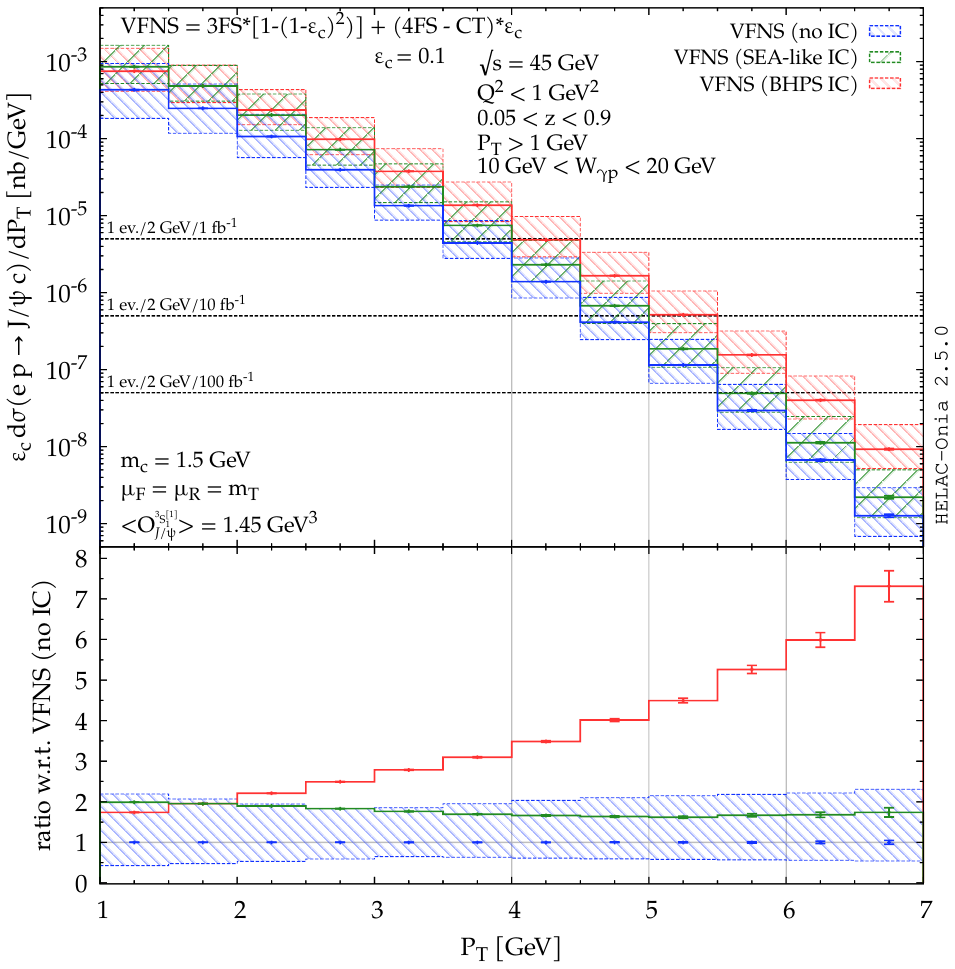}\hspace{2mm}
\includegraphics[width=8.5cm]{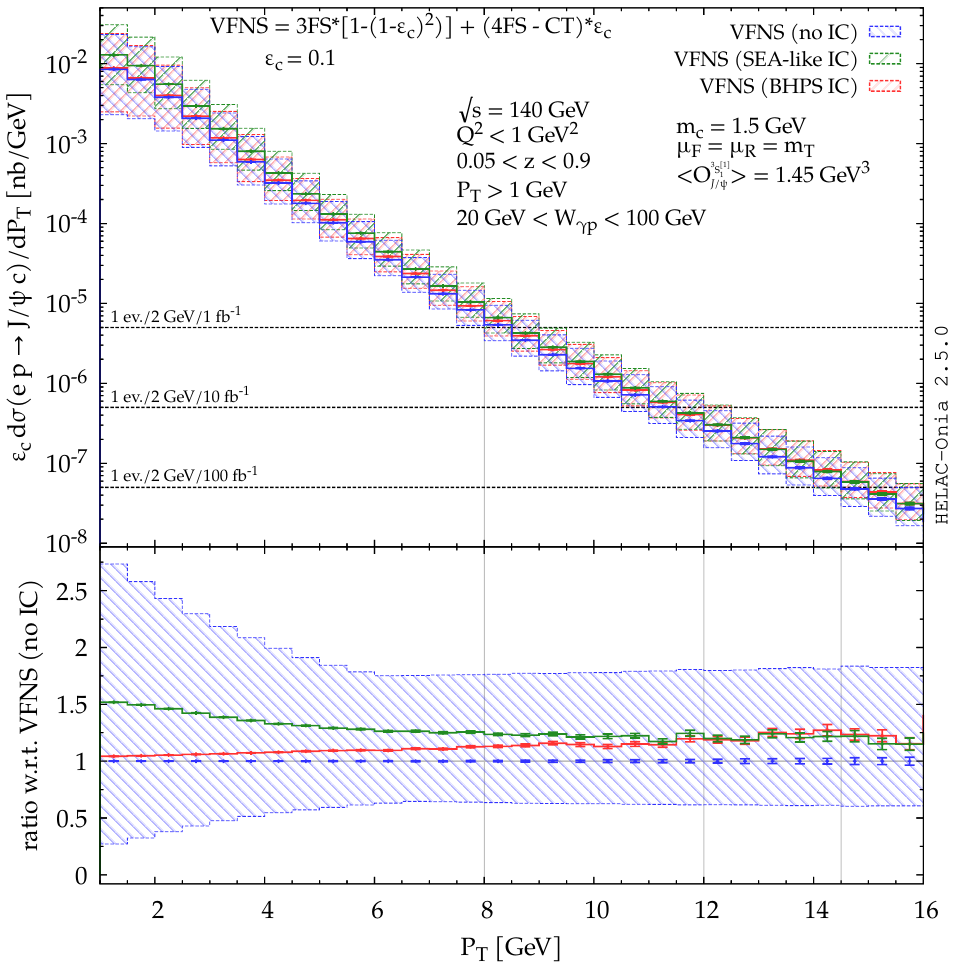}
\caption{Predictions for the $\jpsi+$charm yield at the EIC at $\sqrt{s_{ep}} = 45$ GeV (left) and $\sqrt{s_{ep}} = 140$ GeV (right) {as a function of the $\jpsi$ transverse momentum, $\pT$}. The solid bands indicate the mass uncertainty while the patterns display the scale uncertainty. Figure taken from Ref.~\cite{Flore:2020jau}.}
\label{fig:EIC-VFNS-IC}
\end{figure}

\cf{fig:EIC-VFNS-IC} shows the result for the $\jpsi$+charm yield at the %
EIC. First, we note that, at $\sqrt{s_{ep}} = 45$ GeV (left panel in \cf{fig:EIC-VFNS-IC}), the yield is limited to low $P_T$ values even with the largest {estimated} integrated luminosity. Nonetheless, it is clearly observable if $\varepsilon_c = 0.1$ with $\mathcal{O} (5000, 500, 50)$ events for $\mathcal{L} = (100,10,1)$ fb$^{-1}$. On the other hand, at $\sqrt{s_{ep}} = 140$ GeV (\cf{fig:EIC-VFNS-IC}, right panel), the $P_T$ range {extends} to $\sim 14$ GeV and we expect $\mathcal{O}(10000)$ events at $\mathcal{L} = 100$ fb$^{-1}$. Such events could be observed by measuring a charmed jet. Finally, we note that, at $\sqrt{s_{ep}} = 140$ GeV, where the valence region %
is not probed, no clear IC effect is visible, while at $\sqrt{s_{ep}} = 45$ GeV we {expect} a measurable effect, where the BHPS valence-like peak is visible with a {yield} enhancement as large as $5-6$ times the ``no IC'' yield. The EIC at $\sqrt{s_{ep}} = 45$ GeV will thus be the place to probe the nonperturbative charm content of the proton via associated $\jpsi+$charm production.

\subsection{Unpolarised-nucleon TMDs}
\label{sec:TMD-unpol}

\subsubsection{Unpolarised gluons}
\label{subsec:glue-tmds}
Quark TMDs have now been extracted from data with reasonable precision~\cite{Moos:2023yfa,Bacchetta:2022awv,Bertone:2019nxa,Bacchetta:2017gcc}. On the contrary, phenomenological studies of gluon TMDs are still very much at the beginning stage. In Ref.~\cite{Gutierrez-Reyes:2019rug}, a gluon TMD description of the Higgs{-}production transverse{-}momentum spectrum {was compared to data{,}  
which, however, suffers from very large uncertainties}. In Refs.~\cite{Lansberg:2017dzg,Scarpa:2019fol}, a gluon TMD description of the LHCb $\jpsi$-pair-production data~\cite{LHCb:2016wuo}  was obtained. Like for Higgs-boson production, the experimental errors are %
large and require the subtraction of double parton scattering contributions (see~\cite{Lansberg:2014swa,Lansberg:2019adr} and Section \ref{sec:DPSdoubleJpsi}){,} which adds an additional uncertainty. In Ref.~\cite{Dunnen:2014eta}, it was discussed that back-to-back production of a heavy quarkonium, in particular of an $\Upsilon$, and an isolated photon in proton-proton collisions at the LHC is a promising way to access the distribution of both the transverse momentum and the polarisation of gluons inside unpolarised protons. In a wide range of invariant masses of the quarkonium and photon system, gluon-gluon scattering into a photon plus a quarkonium in the CS state dominates. 

{In the aforementioned processes}, one %
however %
probes a convolution of  two gluon TMDs. At {the} EIC, one can probe gluon TMDs more directly {through the $P_T^*$ distribution}, although upon {the} inclusion of %
{ShFs} (see section \ref{sec:TMDShFs})\newb{;} {this also deals with convolutions.} %
{At the LHC, with the consideration of such %
{ShFs}, one even folds three transverse-momentum-dependent distributions}. Another possibility to study gluon TMDs at the EIC using quarkonia is to consider the transverse-momentum imbalance between the scattered lepton and the observed $J/\psi$ in the electron-hadron centre-of-mass frame $\overline{\bf p}_T = |\bm{\ell}'_T + \bm{P}_{T}| $. If large $|\bm{\ell}'_T|\simeq |\bm{P}_{T}| \gg |\overline{\bf p}_T|$ determines the hard scale of the process{,} then %
in quarkonium production at the EIC the leading subprocess is %
$e+g\to (c\bar{c})^{[8]}+e$ with the octet \ccbar pair %
{hadronising} into an observed $J/\psi$. Within the hybrid factorisation formalism for SIDIS discussed in Section \ref{sec:QED-corr}, the {$\overline{\bf p}_T$} %
should be determined by the transverse momentum $k_T$ of the colliding gluon (or its TMD distribution) and the {quarkonium} TMD %
{ShF}. %
Since gluon radiation from a heavy quark should be strongly suppressed compared to a light quark or a gluon, the observed momentum imbalance $\overline{\bf p}_T$ is expected to be dominated by the $k_T$ of the colliding gluon~\cite{Liu:2020rvc}. Therefore, the {$\overline{\bf p}_T$}%
-distribution of \jpsi production in SIDIS could be a more direct observable for {the} gluon TMD~\cite{QW:2024}.%

{It} would be very interesting to compare the {gluon TMD obtained at EIC} to that from the {$J/\psi+J/\psi$ or} $\Upsilon+\gamma$ process at LHC in the future. 
In principle, gluon TMDs are process dependent, even in the unpolarised case (see e.g.\ \cite{Dominguez:2011wm,Boer:2016fqd}).
However, provided that the CS final state dominates in %
{$J/\psi+J/\psi$ and} $\Upsilon+\gamma$ production at the LHC, %
{these} processes involve the same gluon TMD. This then would provide a nice test of TMD factorisation in combination with NRQCD and of TMD evolution, if the processes are probed at different scales. Another comparison that seems worthwhile is the extraction of gluon TMDs from open heavy{-}quark pair production at the EIC \cite{Boer:2016fqd} or from inclusive $\eta_c$ or $\eta_b$ production in proton-proton collisions~\cite{Boer:2012bt,Ma:2012hh}. Note that inclusive CS $\jpsi$ or $\Upsilon$ production from two gluons is forbidden by the Landau-Yang theorem, while inclusive CO $\jpsi$ or $\Upsilon$ production does not involve the same gluon TMD and may not even %
{factorise} to begin with.

\subsubsection{Linearly polarised gluons}
\label{sec:lin}
{As discussed in Section \ref{sec:cos2phiShF}, linearly polarised gluons lead to a $\cos 2 \phi_T^*$ asymmetry in  semi-inclusive electroproduction of $J/\psi$ in unpolarised $ep$ collisions~\cite{Mukherjee:2016qxa,Bacchetta:2018ivt,Kishore:2018ugo,Boer:2021ehu,Kishore:2021vsm,Kishore:2022ddb}.} {In this section, we present some predictions for this asymmetry %
{at low transverse momenta $\pT^*$.}} %

{Within NRQCD, contributions to the asymmetry %
comes %
through the fusion of a virtual photon and a gluon~\cite{Mukherjee:2016qxa} {already at Born order, i.e.\ $\alpha_s \alpha$}, but at NNLO in $v^2$ since via CO contributions. Such $\alpha_s \alpha$ contributions however only sit at  $z=1$.} %
{As soon as $z\neq 1$, a recoiling particle against the quarkonium is needed and Born-order contributions are at $\alpha^2_s \alpha$ both from CS and CO states. From a simple counting in $v^2$ the CS contributions~\cite{Sun:2017wxk} should be dominant at $z\neq 1$. However, the current LDME fits seem not to obey such a simple $v^2$ counting and, as a matter of fact, sometimes leads to an excess\footnote{It should be clear to the reader that such computations are as of now only carried at LO whereas the LDMEs are extracted at NLO. We refer to our introductory discussion at the beginning of Section \ref{sec:LDMEs:coll} regarding potential issues in doing so.} in describing the scarce data available from HERA~\cite{Sun:2017nly}.} %
{In principle, the asymmetry thus receives contributions from both CS and CO states}.

{The first estimate we present here is based on %
a model expression for the cross section \cite{Kishore:2022ddb}:}
\begin{equation}
\begin{aligned}
 d\sigma={} &\frac{1}{2s}\frac{d^3l'}{(2\pi)^32E_l'}\frac{d^3P_{\Q}}{(2\pi)^32E_{P_{\Q}}}\int \frac{d^3p_g}{(2\pi)^32E_g}
  \int dx\,d^2\mathbf{k}_{\perp}(2\pi)^4\delta(q+k-P_{\Q}-p_g)\\
  &\times \frac{1}{Q^4}\mathcal{L}^{\mu\mu'}(l,q)\Phi^{\nu\nu'}
  (x,\mathbf{k}_{\perp})~\mathcal{M}_{\mu\nu}(\mathcal{M}_{\mu'\nu'})^{\ast}.
 \end{aligned}
\end{equation}
{This expression is akin to the Generalised Parton Model employed to describe single-spin asymmetries in polarised proton collisions (to be discussed in section \ref{sec:TMDs-polar}). It is not of TMD-factorisation form and differs from Eq.~(\ref{eq:cos2phi-unp-xsec-general}) by considering the subprocess ${\gamma^{\ast}+g\rightarrow \Q+g}$, where the additional hard gluon in the final state generates larger transverse momenta and elasticity $z$ values below 1, %
while the dependence on the initial gluon transverse momentum is kept everywhere. In other words, no collinear expansion is performed and the obtained expression is thus not a CF expression either.}

In \cf{fig:lin}{,} we %
{show the} $\cos 2 \phi_T^*$ asymmetry as a function of $P_T$ for $\sqrtsep =140$ GeV{, for fixed values of $z$ and $Q^2$.} {Both } %
{CS and CO} contributions are included. %
We show %
the results for two different models for the TMDs{, }%
{the} Gaussian \cite{Boer:2012bt} %
and {the}  McLerran-Venugopalan  model \cite{McLerran:1993ni}{, and for two different sets of LDMEs, CMSWZ~\cite{Chao:2012iv} and BK \cite{Butenschoen:2011yh}}. The asymmetry is small and %
depends  on the {chosen} LDME set. %
Details of the calculation may be found in \cite{Kishore:2021vsm,Kishore:2022ddb}.  

\begin{figure}[htbp]
\includegraphics[width=.49\textwidth]{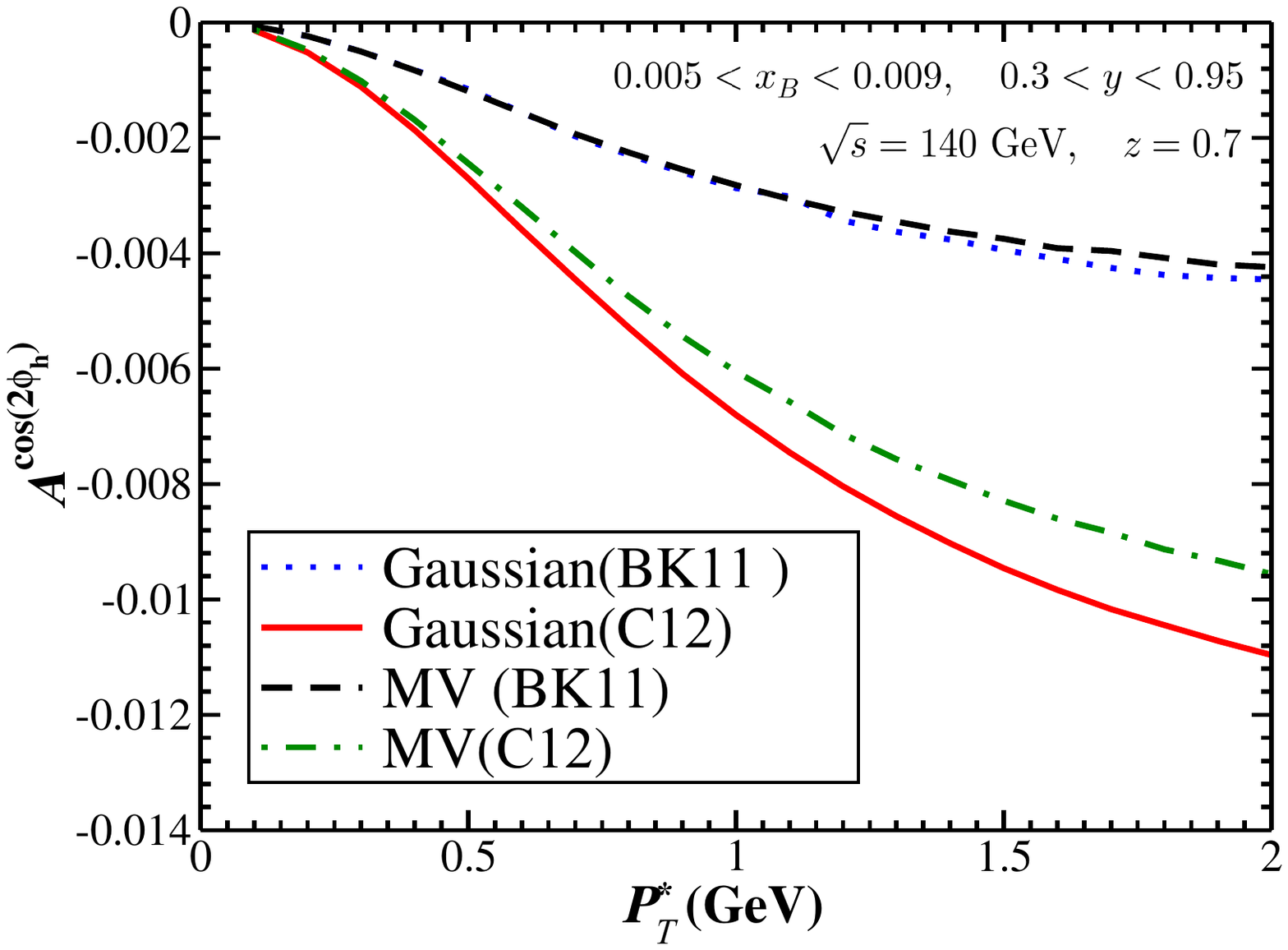}
\includegraphics[width=.49\textwidth]{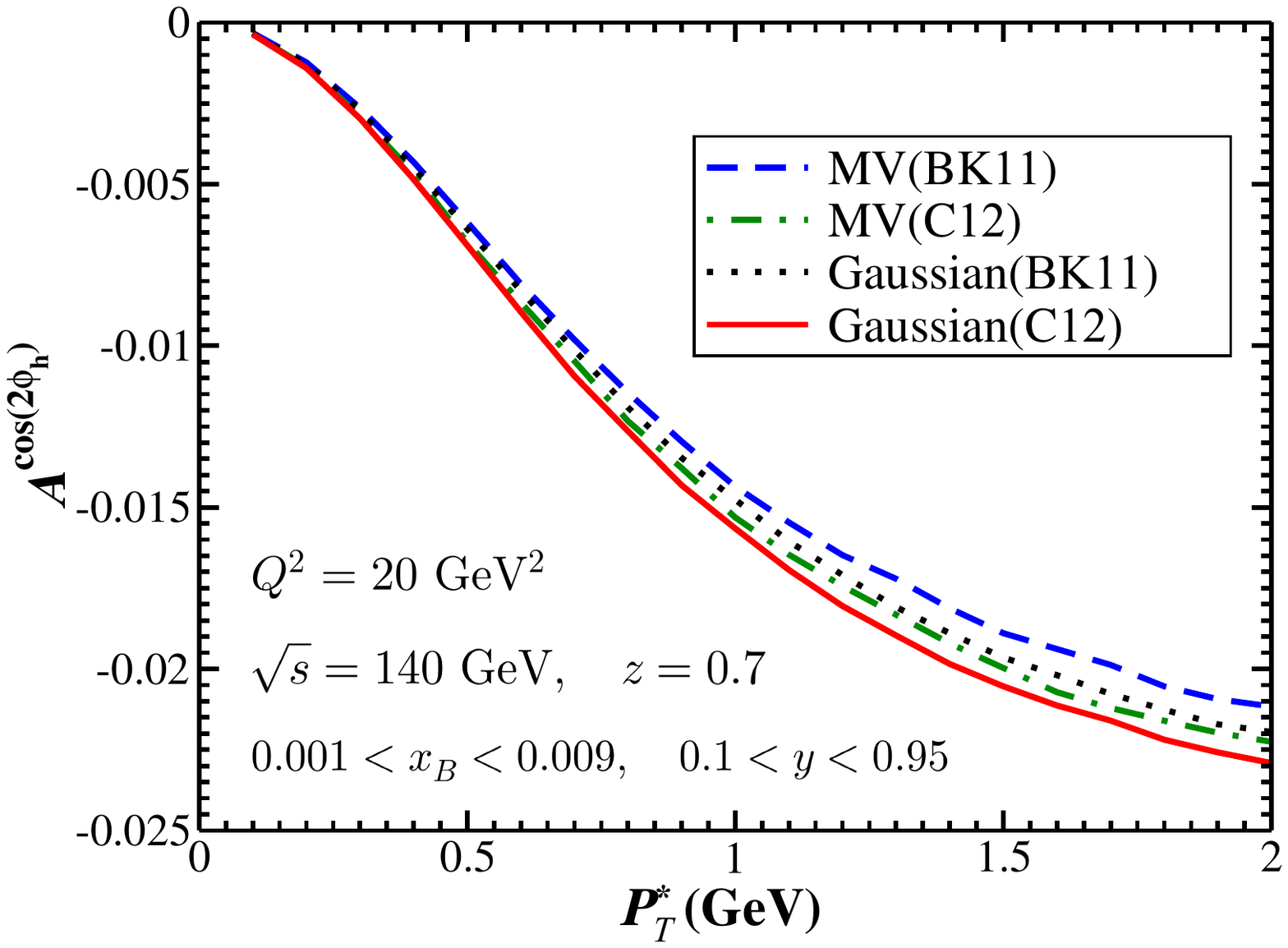}
\caption{$\cos(2\phi_h) \equiv \cos(2\phi_T^*)$ asymmetry in $e+p\rightarrow e+ J/\psi +X$ process as function of  {the \jpsi transverse momentum $\pT^*$} 
at $\sqrt{s}=140$~GeV and $z=0.7$. Left plot: asymmetry obtained by integrating over %
${x_B} \in [0.005:0.009]$ and {the} {inelasticity} $y \in [0.3:0.95]$; right plot: asymmetry obtained at fixed $Q^2 = 20~\text{GeV}^2$, integrated over %
${x_B} \in [0.001:0.009]$ with {the} corresponding $y$ range determined from $y = Q^2 / (s {x_B})$. The curves are obtained using a Gaussian parameterisation for the TMDs~\cite{Boer:2012bt} as well as McLerran-Venugopalan (MV) model~\cite{McLerran:1993ni} in small-$x$ region. Two sets of LDMEs are used: C12~\cite{Chao:2012iv} and BK11~\cite{Butenschoen:2011yh}.}
\label{fig:lin}
\end{figure} 

{A second estimate -- only relevant for $z\simeq 1$ -- is based on the %
TMD formalism involving shape functions. Although  the semi-inclusive quarkonium electroproduction is naturally described in TMD factorisation at small quarkonium transverse momentum  ($\pT^* \ll M_{\jpsi} \sim Q$), } there is large uncertainty due to the non-perturbative part of the TMD description and due to the lack of knowledge on the TMD shape functions.  However, using the leading-order shape functions in terms of LDMEs and including leading-order TMD evolution, it is nevertheless possible to obtain rough predictions for the EIC (details on the shape function can be found in Ref.~\cite{BoerBorMaxiaPisano}). Using this approach, estimates for the $\cos 2\phi_{T}^*$ asymmetry in $J/\psi$ production as a function of {$\pT^*$} can be obtained. The results are shown in \cf{fig:EICpredictioninclTMDevolution} for several LDME sets (for more predictions see Ref.~\cite{Bor:2022fga}) and for kinematics similar to that of \cf{fig:lin} (to be precise, for the same $\sqrtsep$ and $Q^2$, and comparable $x_B$, but different values of $z$). Despite the large uncertainties in these TMD results (the uncertainty bands reflect the uncertainty in the non-perturbative Sudakov factor), it is clear that within these uncertainties it allows for significantly (by more than an order of magnitude) larger asymmetries than in \cf{fig:lin}. %
{Its measurement may thus be feasible at EIC such that} further constraints on the LDMEs, and more generally on the TMD shape functions, can be obtained in this way. %

\begin{figure}[htbp]
    \centering
    \includegraphics[width=10cm]{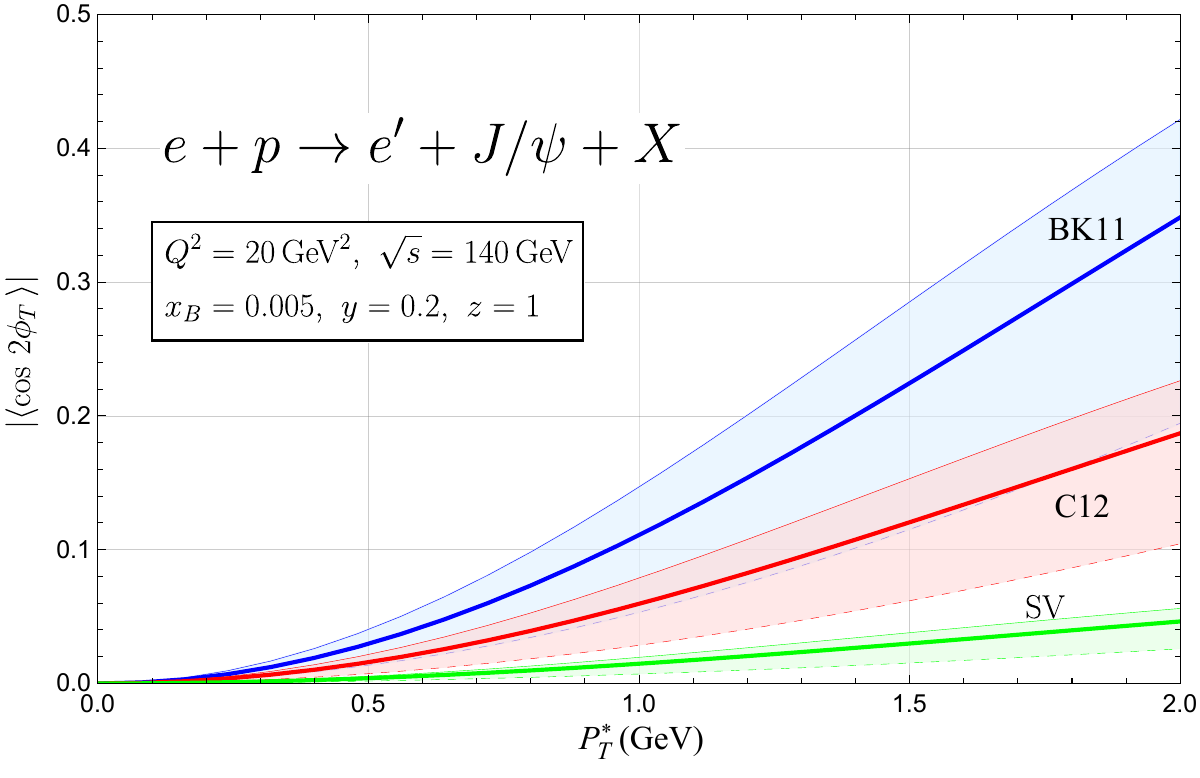}
    \caption{Estimates for the $\cos 2\phi_{T}$ {$\equiv \cos(2\phi_T^*)$} asymmetry in $J/\psi$ production as a function of %
    {the \jpsi transverse momentum $P_{h\perp}=\pT^*$} 
    for three different LDME sets (central values) and including the uncertainties from the nonperturbative Sudakov factor.}
     \label{fig:EICpredictioninclTMDevolution}
\end{figure}
Observing a nonzero asymmetry would be a signal of linear polarisation of the gluons inside an unpolarised proton, which is expected theoretically but not established experimentally thus far. The range of predictions is currently too large to draw a definite conclusion about its observability at EIC, but that makes it all the more important to obtain first data on the $\cos 2 \phi_T^*$ asymmetry. It would provide information on the distribution of linearly polarised gluons as well as on LDMEs.

\subsection{Polarised-nucleon TMDs}
\label{sec:TMDs-polar}

Among the observables that can be measured at the EIC to access polarised nucleon TMDs (\eg the Sivers function), %
{the most common are probably} the Single {Transverse} Spin Asymmetries ({STSA}), denoted $A_N$, or $A_{UT}$. %
{Two theory approaches have been pushed forward to explain STSAs observed on polarised protons~\cite{Koike:2007dg}. Both of them can in principle be extended to quarkonium production.}

{The first approach is referred to as collinear twist-3 (CT3) formalism~\cite{Efremov:1981sh,Efremov:1983eb,Qiu:1991pp,Qiu:1990xxa,Qiu:1990cu} 
and, like CF, %
{applies to} single-scale processes.  
The STSA then arises from  quark-gluon-quark or triple-gluon correlators, which are the sub-leading (in the scale) twist-3 extensions of the usual collinear PDFs (putting aside for this discussion FF contributions). %
Some CT3 analyses for $A_N$ in $ep$ collisions have been performed in the past, see e.g.~\cite{Gamberg:2014eia}}{, and only very recently this approach has been extended to STSAs in quarkonium production in polarised $ep$ collisions~\cite{Chen:2023hvu}.} %

{The second approach is TMD factorisation, thus applicable {when two very different momenta are measured, or when a small (yet perturbative) momentum is measured in a process involving a large mass} (\eg ($\Lambda_{\rm QCD}) \lesssim P_T^* \ll Q$ in SIDIS, where $P_T^*$ is the transverse momentum of the hadron in the final state and $Q^2$ is the photon virtuality). The STSA arises from the Sivers TMD PDF $f_{1T}^{\perp}$~\cite{Sivers:1989cc}, i.e. the distribution of unpolarised partons inside the transversely-polarised hadron. In the case of quarkonium production in $ep$ collisions, TMD factorisation has been assumed and used to compute the Sivers asymmetry in several cases~\cite{Kishore:2019fzb, Chakrabarti:2022rjr}.} %

In addition, a phenomenological approach, called the Generalised Parton Model (GPM)~\cite{DAlesio:2007bjf}, encapsulates the Sivers mechanism via the aforementioned TMD Sivers function, assumed to be universal, but also applied in single-scale processes. This is done by keeping track of the transverse-momentum exchanges in the partonic scattering. As such, it can be considered as a hybrid approach between strict CT3 and TMD factorisation. Its extension, called Colour Gauge Invariant GPM (CGI-GPM)~\cite{Gamberg:2010tj,DAlesio:2017rzj}, allows one to recover the modified universality of the quark Sivers function between SIDIS and Drell-Yan~\cite{Collins:2002kn,Brodsky:2002cx,Buffing:2013kca}. Moreover, for the gluon Sivers effect, similarly to the CT3 approach case, two independent gluon Sivers functions (GSFs) appear~\cite{Boer:2015vso}, dubbed as $f$- and $d$-type. %
This approach has proven to be quite successful in phenomenological analyses~\cite{DAlesio:2017nrd, DAlesio:2018rnv,Boglione:2021aha,Boglione:2024dal}. One should however be careful if one wishes to draw any conclusion about the properties of the used TMDs and the underlying phenomena. In any case, it is useful to get estimates of STSAs in single-scale processes where a CT3 analysis becomes challenging, like for quarkonium production, due to still unconstrained twist-3 functions appearing in its computation. It has been applied to the quarkonium cases in several studies~\cite{DAlesio:2017rzj, DAlesio:2017nrd,Rajesh:2018qks,DAlesio:2020eqo,DAlesio:2022qrh}. 

Below STSAs in different quarkonium-production processes are discussed, in the context of the EIC, which could perform these measurements by polarising a target. In general, it is believed that quarkonium-related STSA would be key player to underpin the Sivers mechanism for gluons.

Experimentally, one defines the so-called transverse %
{STSA} as
\begin{equation}
\label{eq:STSA}
A_{N} = \frac{1}{{\cal P}}\frac{\sigma^{\uparrow} - \sigma^{\downarrow}}{\sigma^{\uparrow} + \sigma^{\downarrow}}
\,,
\end{equation}
where $\sigma^{\uparrow\,(\downarrow)}$ is the cross section of particles produced with the target nucleon spin orientation upwards (downwards), and ${\cal P}$  is the average nucleon polarisation. In what follows, we present predictions and projections for %
{STSA} in \jpsi inclusive photoproduction and for azimuthal weighted Sivers asymmetries in \jpsi leptoproduction in SIDIS processes.

\subsubsection{EIC reach for $A^{\jpsi}_N$ for inclusive photoproduction}

{In this section, we study how to probe the GSF via the GPM approach by measuring the STSA in inclusive $\jpsi$ photoproduction ($\gamma + p^\uparrow  \rightarrow J/\psi+X$)~\cite{Rajesh:2018qks}. In such a process, only the $f$-type GSF contributes to the Sivers asymmetry.}

In photoproduction, there are contributions from direct and resolved photons. Resolved photons mainly contribute in the  {region of low elasticity $z$.} %
{At $z$ close to unity}, diffractive contributions become significant.  In inclusive photoproduction, the variable $z$ can be measured using the Jacquet-Blondel method.   The differential cross section of inclusive $J/\psi$ production in unpolarised $ep$ collisions can be written as 
\begin{equation}\label{d1}
\begin{aligned}
E_{{\Q}}\frac{d\sigma}{d^3{\bm P}_{{\Q}}}={} 
& 
\frac{1}{2(2\pi)^2}\int dx_\gamma dx_g 
d^2{\bm k}_{\perp g}
f_{\gamma/e}(x_\gamma)f_{g/p}(x_g,{\bm k}_{\perp 
g})\delta(\hat{s}+\hat{t}+\hat{u}-M{_{\Q}}^2)\\
&\times
\frac{1}{2\hat{s}}|\mathcal{M}_{\gamma+g\rightarrow \Q +g}|^2.
\end{aligned}
\end{equation}
{Here, }$x_\gamma$ and $x_g$ are the light-cone momentum fractions of {the} photon and gluon, respectively; {$\hat s$, $\hat t$, $\hat u$ are the partonic Mandelstam variables; $\mathcal{M}_{\gamma+g\rightarrow \Q +g}$ is the matrix element for the partonic subprocess $\gamma+g\rightarrow \Q +g$; $f_{g/p}(x_g,{\bm k}_{\perp 
g}$) is the unpolarised gluon TMD, while} $f_{\gamma/e}(x_\gamma)$ is the Weizs\"{a}cker-Williams distribution, giving  the density of photons inside the electron \cite{Frixione:1993yw}.  %
{For theory predictions of measurements on a transversely polarised nucleon, the {STSA}, as introduced in~\ce{eq:STSA},  is generally used}.

\begin{figure}[hbt!]
\centering
\includegraphics[width=7cm]{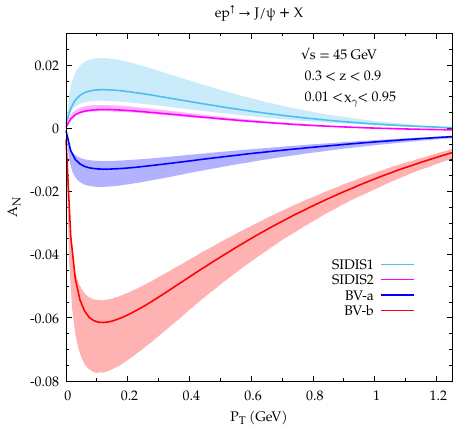}
\includegraphics[width=7.15cm]{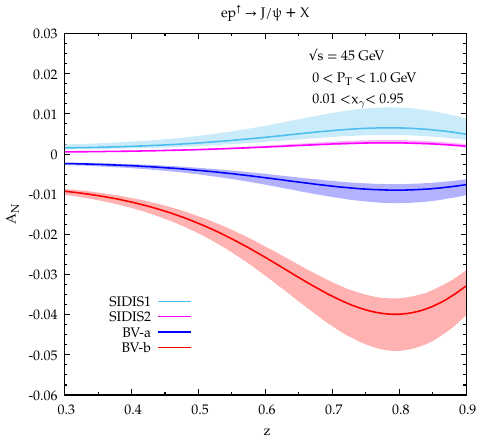}\\
\includegraphics[width=7cm]{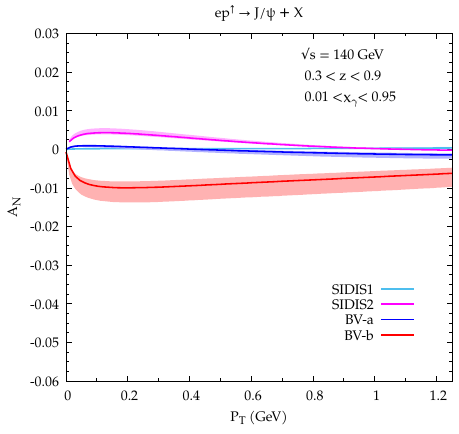}
\includegraphics[width=7.15cm]{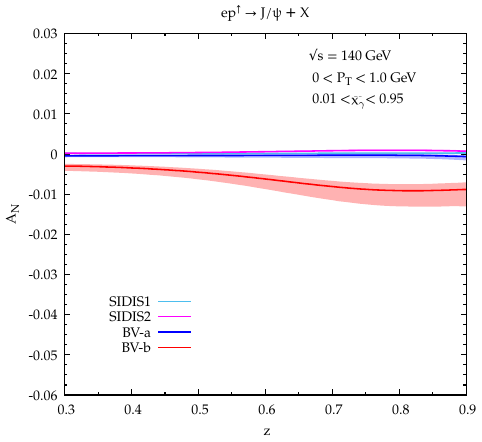}
\caption{STSA in inclusive $\jpsi$ photoproduction  ($ep^\uparrow \to J/\psi X$) as a function of { the $\jpsi$ transverse momentum, }$P_T$ (left) and $z$ (right) at $\sqrtsep=45$ GeV (top ) and $\sqrtsep = 140$~GeV (bottom). The integration ranges are $0.3 <z< 0.9$ and $0 < P_T \leq 1$~GeV, {respectively}. The uncertainty bands are obtained by varying the factorisation scale by a factor 2 {around $\mu_0=\sqrt{M_{\jpsi}^2+P_T^2}$}.}
\label{fig:Photo_Sivers}
\end{figure}

{Some GPM predictions for STSA in inclusive $\jpsi$ photoproduction at the EIC for $\sqrtsep =45 (140)~\mathrm{GeV}$ are shown in \cf{fig:Photo_Sivers}, as a function of the \jpsi transverse momentum, $\pT$, as well as a function of the elasticity $z$.} The amplitude for the $J/\psi$ production is calculated in NRQCD. Details of the calculation can be found in Ref.~\cite{Rajesh:2018qks}. 
The dominating channel of $J/\psi$ production is $\gamma g$ fusion. The contribution to the numerator of the %
{STSA} comes mainly from {the} GSF~\cite{DAlesio:2017nrd}, while the linearly polarised gluons do not contribute to the denominator for this specific process. Moreover, the numerator of the asymmetry only receives contribution{s} from CO states~\cite{Yuan:2008vn}, whereas in the denominator, both %
{CO and CS} contributions are included. 

We have used the {GSF} parametrisations %
(SIDIS1, SIDIS2) from {Ref.~}\cite{DAlesio:2015fwo}. BV-a and BV-b are parametrisations of {the} GSF in terms of up and down quark Sivers function{s} \cite{Boer:2003tx}, where parameters from {Ref.~}\cite{Anselmino:2016uie} are used. The effect of TMD evolution is not incorporated in the plot. The PDF set MSTW2008~\cite{Martin:2009iq} is used; the uncertainty bands have been obtained by varying {the factorisation scale $\mu_F \in \left[\tfrac12 \mu_0, 2\mu_0\right]$, with $\mu_0 = m_T = \sqrt{M^2_{\jpsi} + P_T^2}$ being the $\jpsi$ transverse mass. The value of $\alpha_s$ is calculated at the scale $\mu_0$ and is taken from the MSTW set}. The {used cuts are the following}: $Q^2<1$~GeV$^2$ and $0.3 < z < 0.9$. 
{Note that, in the photoproduction case, $y$ coincides with $x_\gamma$. The corresponding cut is $0.01 < x_\gamma < 0.95$.} %
{As shown in \cf{fig:Photo_Sivers}, we expect $A_N$ to be small and positive in the SIDIS1 and SIDIS2 cases, while it is larger (in size) but negative when the GSF is parametrised in terms of the up- and down-quark Sivers functions.}

{Another estimate is shown in \cf{jpsi:AN:ep}. Here, projections for statistical uncertainties for {the} \jpsi\ $A_N$ measurement as a function of transverse momentum for $ep$ collisions at \sqrtsep =45~GeV and \sqrtsep = 140~GeV for an integrated luminosity \Lint = 100 \invfb are presented.}
We consider the \jpsi reconstruction via its electron decay channel ($\jpsi \to \epem$, ${\cal B} = 5.94 \pm 0.06 \% $)%
{, and we} assume {the} single{-}electron measurement efficiency to be 80\% and constant %
{with respect to its} transverse momentum and in the pseudorapidity interval %
$|\eta|<2$. The \jpsi measurement efficiency is calculated using decay kinematic{s} simulated with PYTHIA8~\cite{Sjostrand:2014zea} (see \ref{sec:jpsi:reco:eff} for details). Based on these results, we assume the \jpsi measurement efficiency to be 64\%. Furthermore, we assume {the} signal-to-background ratio S/B = 1, and use the same method as in Ref.~\cite{Kikola:2017hnp} to estimate statistical uncertainties on $A_N$. 
For the expected cross section for {prompt} \jpsi production in $ep$ collisions at the EIC, %
{we consider the CSM predictions from Ref.~\cite{Flore:2020jau}, which were shown to approximately reproduce HERA data.} %
For illustration, the projections are compared to results from \pp\ {collisions} reported by the PHENIX experiment~\cite{PHENIX:2018qvl}. At low \pT, the statistical precision is at {the} per-cent level, exceeding the quality of {the corresponding} $pp$ data. In this range, the final uncertainty will be dominated by systematic effects. The uncertainties increase fast with increasing \pT\ of \jpsi because the \pT\ spectrum is predicted to be rather steep. Nonetheless, such a measurement would be valuable for constraining gluon TMDs at low transverse momentum. 

\begin{figure}[hbtp!]
	\centering
		\subfloat[ep \sqrtsep = 45 GeV] {\includegraphics[width=0.45\textwidth]{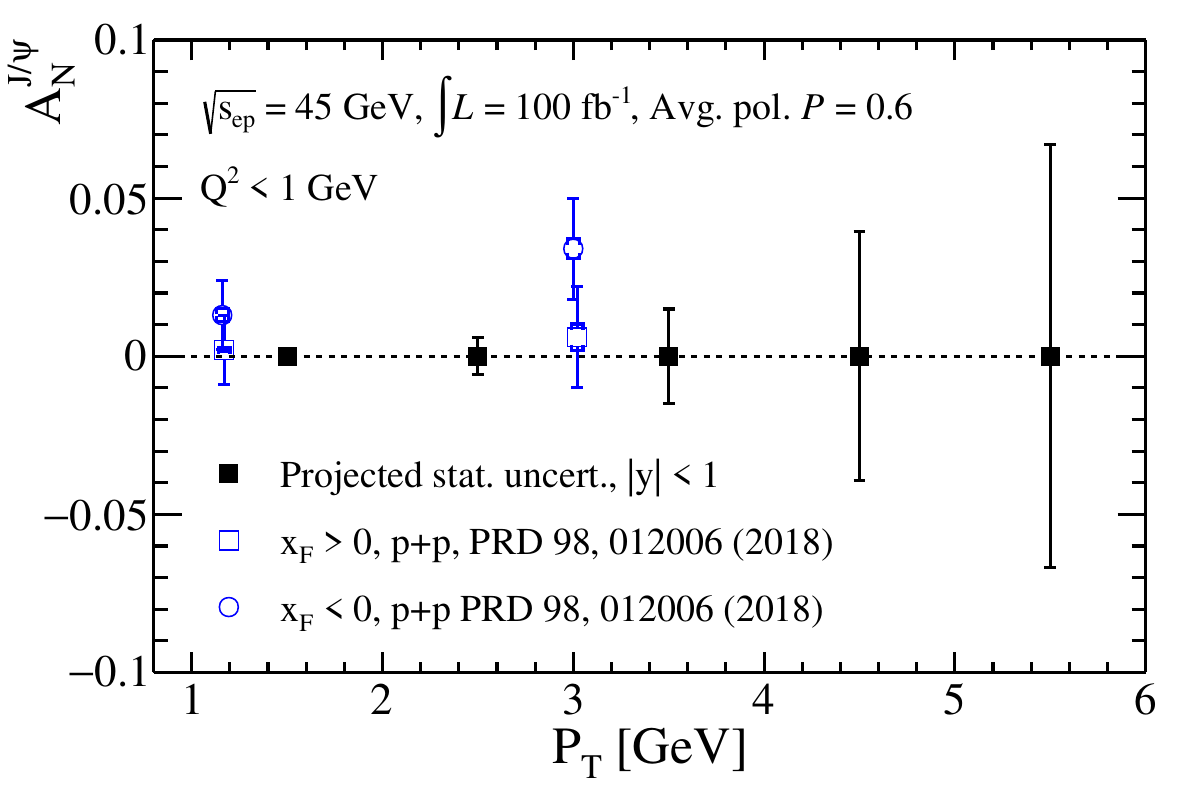}} 
		\subfloat[ep \sqrtsep =140 GeV] {\includegraphics[width=0.45\textwidth]{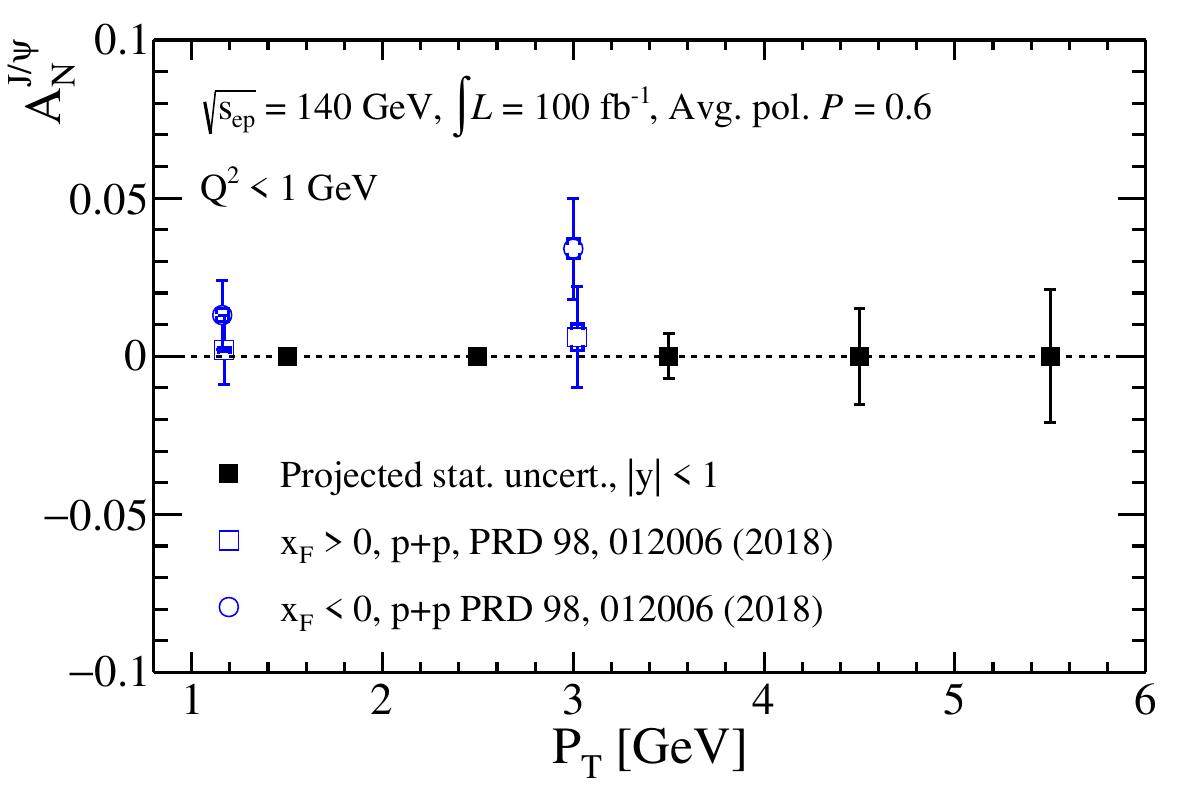}} 
	\caption{ Statistical projections for \jpsi $A_N$ as a function of {the $\jpsi$ }transverse momentum for electron+proton collisions at \sqrtsep = 45~GeV and \sqrtsep =140~GeV, compared to existing results from \pp\ %
 {interactions} reported by the PHENIX experiment~\cite{PHENIX:2018qvl}. }
	\label{jpsi:AN:ep}
\end{figure}

{Finally, we suggest that the associated} photoproduction of {\jpsi and a jet, having them} back-to-back, %
can also probe the %
{GSF}~\cite{Mukherjee:2016qxa,Kishore:2019fzb}. In this case the produced \jpsi can have large transverse momentum, and needs not be in the forward region. A wide kinematical region can be covered by varying the invariant mass of the $J/\psi$-jet pair.

\subsubsection{Azimuthal asymmetries for $\jpsi$ production in SIDIS at the EIC}

{In this section }we consider the Sivers effect in {the} {SIDIS process%
, }$e(l)+p^\uparrow(P_N)\rightarrow e(l^\prime)+ J/\psi (P_{\jpsi})+X${, that represent a promising tool to probe the GSF. The weighted Sivers asymmetry for such a process is defined as}

\begin{equation}
\label{eq:AN}
A^{\sin(\phi_T^*-\phi_S^*)}_N  \equiv  2\frac{\int \d\phi_S^*\d\phi^*_T \sin(\phi^*_T -\phi^*_S)
(\d \sigma^\uparrow - \d \sigma^\downarrow)}{\int \d\phi^*_S \d\phi^*_T (\d \sigma^\uparrow + \d \sigma^\downarrow)}
 \equiv \frac{\int \d\phi^*_S \d\phi^*_T \sin(\phi^*_T -\phi^*_S) \d\Delta\sigma(\phi^*_S, \phi^*_T)}{\int \d\phi^*_S \d\phi^*_T \d\sigma}\,,
\end{equation}
where $\d \sigma^{\uparrow(\downarrow)}=\d\sigma^{\uparrow(\downarrow)}/\d Q^2\, {\d} y\, \d ^2\bm{P}_T^{*}\,\d z$ is the differential cross section with the initial proton polarised along the transverse direction $\uparrow(\downarrow)$ with respect to the lepton plane in the $\gamma^*p$ {centre-}of{-}mass  frame (at an angle $\phi^*_S$). 

\begin{figure}[H]
\begin{center}
\includegraphics[width=9cm, keepaspectratio]{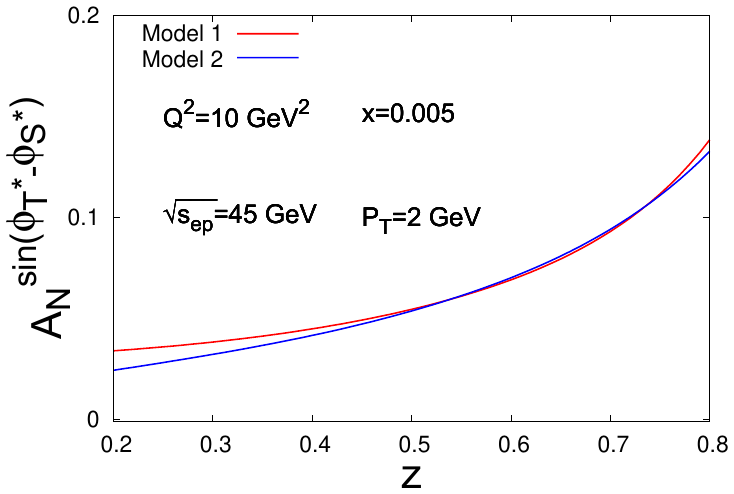}
\end{center}
\caption{CSM predictions for the gluon Sivers asymmetry in the CT3 formalism for the $e p^\uparrow\to e +J/\psi + X$ process as a function of $z$ at $\sqrtsep =45$ GeV. Predictions for two different models are given. {Figure adapted from Ref.~\cite{Chen:2023hvu}.}
}
\label{fig:EIC45z_CT3}
\end{figure}

{We start by presenting the predictions in the CT3 formalism. In Ref.~\cite{Chen:2023hvu}, the twist-3 contributions to the unpolarised and polarised cross sections (respectively denominator and numerator of Eq.~\eqref{eq:AN}) were computed in the CSM. Among the different contributions, one give access to gluon Sivers effect via the CT3 gluon Qiu-Sterman function, which at LO is realated via an integral relation to the GSF first $k_\perp$ moment. Predictions for the gluon Sivers asymmetry at the EIC at $\sqrtsep = 45$ GeV are presented in Fig.~\ref{fig:EIC45z_CT3}. They are computed at $Q^2 = 10$ GeV$^2$, $x_{(B)} = 0.005$ and $P_T = 2$ GeV, and are presented as a function of $z$ for two different models of the gluon Qiu-Sterman function. Both models are proportional to $f_{g/p}(x)$, the unpolarised collinear gluon PDF, and read}
\begin{eqnarray}
    \text{Model 1:}\quad  0.002\,x f_{g/p}(x)\,,\\ 
    \text{Model 2:}\quad  0.0005 \sqrt{x} f_{g/p}(x)\,. 
\end{eqnarray}

{Notice that, as these CSM predictions are ratios of cross sections, they do not depend on the value of the CS LDME. Both models predict a sizeable Sivers asymmetry, with a steady increase as a function of $z$, reaching up to $\sim 13-14\%$ at $z = 0.8$.}

{Another prediction for the Sivers asymmetry is performed} within %
{the GPM} at  %
$\alpha \alpha_s^2$. 
In order to study the effects of initial- and final-state interactions (ISIs and FSIs) on the Sivers asymmetry, the %
{CGI-GPM} approach~\cite{Gamberg:2010tj,DAlesio:2017rzj} is employed.
In Ref.~\cite{Mukherjee:2016qxa}, the same observable was studied at {${\mathcal O}(\alpha \alpha_s)$} within the GPM, which implies $z=1$. Here the analysis is extended to the region $z<1$.

Assuming TMD factorisation within the GPM framework, the unpolarised differential cross section, entering the denominator of \ce{eq:AN}, can be written as
\begin{eqnarray}
\label{unp:eq}
 \frac{\d\sigma}{\d Q^2\, \d y\, \d ^2{\bm P}_T^*\,\d z}=\frac{1}{(4\pi)^4 zs}
 \sum_{a}\int\! \frac{\d x_a}{x_a}\, \d^2{\bm k}_{\perp a}\,\delta\left(\hat{s}+\hat{t}+\hat{u}-M_{\jpsi}^2+Q^2\right) \sum_n\frac{1}{Q^4} f_{a/p}(x_a,k_{\perp a}) L^{\mu\nu}
H^{a,U}_{\mu\nu}[n] \langle {\cal O}^{J/\psi}[n] \rangle\,,
\end{eqnarray}
where $a=g,q, \bar q$ and $H^{a,U}_{\mu\nu}[n]$ is calculated at the perturbative order $\alpha \alpha_s^2$ using NRQCD. More precisely, it is the squared amplitude of the partonic process $\gamma^\ast+a\rightarrow c\bar{c}[n] +a$, averaged/summed over the spins and colours of the initial/final parton, with
$n={^3}{S}{_1^{[1,8]}}, {^1}{S}{_0^{[8]}}, {^3}{P}{_J^{[8]}}$, $J=0,1,2$.
$L^{\mu\nu}$ is the standard leptonic tensor { and {$\langle {\cal O}^{J/\psi}[n] \rangle$} represents the LDME of the state indicated by $n$.} %
The numerator in \ce{eq:AN} is directly sensitive to the Sivers function and within the GPM reads
\begin{eqnarray}
\label{num:GPM}
\d\Delta\sigma^\mathrm{GPM} &=&\frac{1}{(4\pi)^4 zs}\sum_{a}\int \frac{\d x_a}{x_a}\, \d^2{\bm k}_{\perp a}\,\delta\left(\hat{s}+\hat{t}+\hat{u}-M_{\jpsi}^2+Q^2\right)\,\sin(\phi^*_S-\phi^*_a)\nonumber\\
 &&{}\times\sum_n\frac{1}{Q^4} \, \left(-2 \, \frac{k_{\perp a}}{M_p}\right)\,f_{1T}^{\perp a} (x_a, k_{\perp a})\, L^{\mu\nu}
H^{a,U}_{\mu\nu}[n]\, \langle {\cal O}^{J/\psi}[n] \rangle \,,
\end{eqnarray}
where $f_{1T}^{\perp a}(x_a, {k}_{\perp a })$ is the Sivers function.

The numerator of the asymmetry in the CGI-GPM is %
given by %

\begin{eqnarray}
\label{num:CGI-GPM}
\d\Delta\sigma^{\mathrm{CGI}} &=&\frac{1}{2s}\frac{2}{(4\pi)^4 z}\int \frac{\d x_a}{x_a}\, \d^2{\bm k}_{\perp a}\,\delta\left(\hat{s}+\hat{t}+\hat{u}-M_{\jpsi}^2+Q^2\right)\sin(\phi^*_S-\phi^*_a)\,\left(-2 \, \frac{k_{\perp a}}{M_p}\right) \nonumber\\
 &&{}\times\sum_{n}\frac{1}{Q^4}\,L^{\mu\nu} \Bigg\{\sum_q f_{1T}^{\perp q} (x_a, k_{\perp a})\,H^{q,\mathrm{Inc}}_{\mu\nu}[n]+%
f_{1T}^{\perp g(f)} (x_a, k_{\perp a})\,
H^{g,\mathrm{Inc}(f)}_{\mu\nu}[n]\Bigg\} \langle {\cal O}^{J/\psi}[n] \rangle \,,\nonumber\\
\end{eqnarray}
where $H^{a,\mathrm{Inc}}_{\mu\nu}[n]$ is the perturbative square amplitude calculated by incorporating the FSIs within the CGI-GPM approach. 
{Note that, in \ce{num:CGI-GPM}, there is no contribution from the d-type GSF. In fact,} in $ep$ collisions, ISIs are absent due to the colourless nature of the virtual photon and only the $f$-type GSF is contributing to the Sivers asymmetry~\cite{DAlesio:2022qrh}. This means that quarkonium production in $ep$ collisions is a powerful tool to {directly} access the process-dependent $f$-type GSF. Moreover, the modified colour factor associated with the ${^3}{S}{_1^{[1]}}$ state is zero in the CGI-GPM approach, which leads to a vanishing Sivers asymmetry in the %
{CSM}.

By adopting a Gaussian factorised form for the unpolarised TMD distribution, a Gaussian-like Sivers distribution %
and by maximising the latter we can give estimates for the upper bounds of the Sivers asymmetry {(\ce{eq:AN})} \old{$A^{\sin(\phi^*_h-\phi^*_S)}_N$} at the EIC. %
{Results are presented in \cf{fig:EIC45ptz_BK}, and are computed using the following kinematical cuts: $2.5$~GeV$^2 < Q^2 < 100$~GeV$^2$,  $10$~GeV$ < W_{\gamma p} < 40$~GeV, $0.3 < z < 0.9$ and $\pT < 5$~GeV.} {The BK11 LDMEs set~\cite{Butenschoen:2011yh} is adopted.}%

\begin{figure}[H]
\begin{center}
\includegraphics[width=8.cm, keepaspectratio]{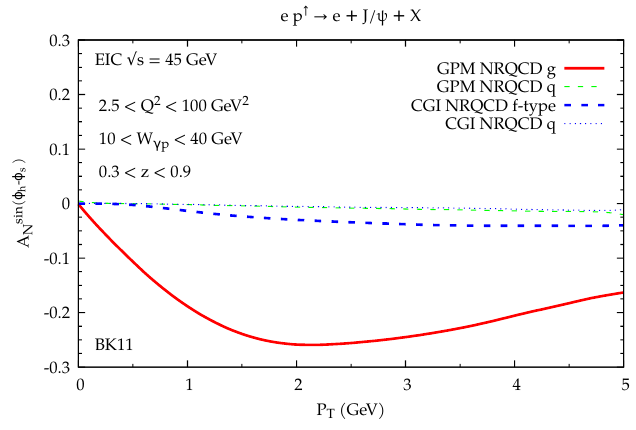}\hspace{.5cm}
\includegraphics[width=8.15cm, keepaspectratio]{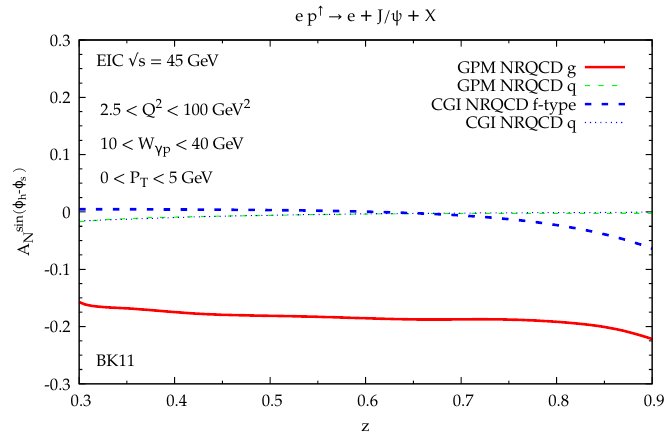}
\end{center}
\caption{Maximised contributions to the Sivers asymmetry for the $e p^\uparrow\to e +J/\psi + X$ process as a function of {the $\jpsi$ transverse momentum }$P_T$  (left) and $z$ (right) obtained with the BK11 LDME set~\cite{Butenschoen:2011yh} at $\sqrtsep =45$ GeV: GSF contribution in NRQCD for the GPM (red solid thick line) and CGI-GPM (blue{,} dashed{,} thick line); quark Sivers contribution in NRQCD for the GPM (green{,} dashed{,} thin line) and CGI-GPM (blue{,} dotted{,} thin line). {Figure adapted from Ref.~\cite{Rajesh:2021zvd}.}
}
\label{fig:EIC45ptz_BK}
\end{figure}

The asymmetry is mostly dominated by the GSF, while the quark contribution is negligible. This indicates that such an observable is a powerful tool to probe the unknown GSF. The GPM predicts negative values around $20\%$. The asymmetry is drastically reduced in size in the CGI-GPM due to colour-factor relative cancellations and the absence of the ${^3}{S}{_1^{[1]}}${-}state contribution and is essentially driven by the $f$-type GSF.

\subsection{Generalised Parton Distributions}
\label{sec:GPDs}
Information on the three-dimensional structure of the nucleon, correlating the transverse position of partons with their longitudinal momentum, 
is provided by %
{GPDs}. Processes to access GPDs include Deeply Virtual Compton Scattering (DVCS) and Deeply Virtual Meson Production (DVMP). %
A factorisation theorem has been proven for DVCS in the Bjorken limit~\cite{Collins:1998be,Ji:1998xh}. It allows one to compute the DVCS amplitude as the product of some GPDs and corresponding coefficient functions that can be calculated perturbatively. GPDs are in very solid theoretical footing: at leading-twist level, all-order QCD-factorisation theorems directly relate the GPDs to particular hard exclusive scattering processes. GPDs are thus process-independent, universal quantities.
In the case of DVMP, factorisation applies in the case of longitudinally polarised photons. The hard-scattering process includes the exchange of hard quarks and gluons, involving the strong coupling constant $\alpha_s$ and a meson distribution amplitude, which is not completely understood to date.

The GPDs do not uphold a probabilistic interpretation like PDFs %
do, but are well-defined in quantum field theory as matrix elements of bilocal quark and gluon operators at a light-like separation. In the light-cone gauge at leading twist, the quark GPD is
\begin{equation}
\begin{split}
    F^q(x,\xi,t) &= \frac{1}{2} \int \frac{\text{d} z^-}{2 \pi} e^{\text{i} x P^+ z^-} \langle p' | \bar{\psi}^q \left(-\frac{z}{2} \right) \gamma^+ \psi^q \left(\frac{z}{2} \right) | p \rangle |_{z^+ = z_{\perp} = 0} \\
    &=\frac{1}{2 P^+} \left[ H^q (x,\xi,t) \bar{u}(p') \gamma^+ u(p) + E^q(x,\xi,t) \bar{u}(p') \frac{ \mathrm{i} \sigma^{+ \mu} \Delta_{\mu}}{2m_N} u(p) \right]
    \label{quarkGPD}
\end{split}
\end{equation}
and the gluon GPD,
\begin{equation}
\begin{split}
    F^g(x,\xi,t) &= \frac{1}{P^+} \int \frac{\text{d} z^-}{2 \pi} e^{\text{i} x P^+ z^-} \langle p' | F^{+ \mu} \left(-\frac{z}{2} \right) F^{+}_{\mu} \left(\frac{z}{2} \right) | p \rangle |_{z^+ = z_{\perp} = 0} \\
    &=\frac{1}{2 P^+} \left[ H^g (x,\xi,t) \bar{u}(p') \gamma^+ u(p) + E^g(x,\xi,t) \bar{u}(p') \frac{ \mathrm{i} \sigma^{+ \mu} \Delta_{\mu}}{2m_N} u( p) \right],
    \label{gluonGPD}
\end{split}
\end{equation}
where $ z= (z^+, z_{\perp}, z^-)$ are the light-cone coordinates, {$P^{+}$ is the light-cone plus-component of the average of the incoming- and outgoing-nucleon momenta, $x$ is he fractional parton plus-component momentum of the nucleon, $\xi$ the skewness variable and $t$ the Mandelstam variable, which represents the four-momentum transfer squared to the nucleon.  The symbols $\gamma$ and $\sigma$ are the Dirac matrices,  } %
$u$ and $\bar{u}$ are nucleon spinors and $m_N$ is the mass of the nucleon. Here, $F^q$ and $F^g$ are both expressed as a Fourier transform of a matrix element of a chiral-even operator formed from either quark fields $\psi^q$ or the gluon-field strength tensor $F^{\mu \, \nu}$. The result is a decomposition into twist-2 parton-helicity conserving GPDs $H$ and $E$.

GPDs cannot be directly extracted from experimental data. Indeed, in the expression of the cross section of exclusive electroproduction processes, GPDs appear in convolution integrals known as Compton Form Factors (CFFs). These CFFs are complex quantities, the real and imaginary parts of which provide complementary constraints on GPDs. The DVCS %
{CFF} $\mathcal H$, at leading-twist and leading-order (and at fixed momentum transfer $t$ and skewness $\xi$), for example, is given by 
\begin{equation}
    \mathcal H = \int_{-1}^1 \text{d}x \frac{F^q(x,\xi,t)}{x-\xi + \mathrm{i} \epsilon} = \mathcal P \int_{-1}^1 \text{d}x \frac{F^q(x,\xi,t)}{x-\xi} - i\pi F^q(\pm \xi, \xi, t),
\end{equation}
and with 
\begin{equation}
    \mathcal \sigma(\gamma^* p \rightarrow \gamma p) \propto |\mathcal H|^2.
\end{equation}

In addition, there are also spin-dependent GPDs
and are probed in measurements in which the spin or polarisation state is fully defined. If the spin states are averaged over, as in the description of an unpolarised measurement, then there is no way to have a direct dependence on, or be sensitive to, these objects. Moreover, there are also parton-helicity{-}flip GPDs (chiral odd), in which the initial{-} and final{-}state hadrons have different polarisations.

GPDs are also connected to the distribution of pressure and shear forces inside the nucleon~\cite{Polyakov:2018zvc,Lorce:2018egm} and, furthermore, the second moment of a particular combination of GPDs is related to the 
angular momenta of quarks and gluons via Ji's relation~\cite{Ji:1996ek}. A comprehensive review on the phenomenology of GPDs in DVCS can be found in~\cite{Kumericki:2016ehc}.

\subsubsection{Gluons}

DVCS is sensitive to quarks and, at higher order and/or higher twist, also to gluons. On the other hand, the production of light mesons in DVMP probes quarks and gluons, depending on the energy scale at which the process is measured. However, \jpsi production in exclusive photoproduction (or electroproduction) reactions is a golden channel for gluon GPDs. Indeed, in this case the quark exchange plays only a minor role and due to the large scale provided by the heavy-quark mass, perturbative calculations is expected to be applicable even for photoproduction~\cite{Ivanov:2004vd}. 

At the EIC, precise measurements of exclusive cross sections will be possible in order to map out the dependence on the squared momentum transfer to the nucleon %
{$t=(P_N-P_N')^2$}  for \jpsi, $\phi$ and $K$, among others. EIC will cover the region of $0 <|t| < 1.5$ GeV$^2$, down to an impact parameter of $\sim$ 0.1 fm.

\begin{figure}[htb]
\centering
\includegraphics[width=0.5\linewidth]{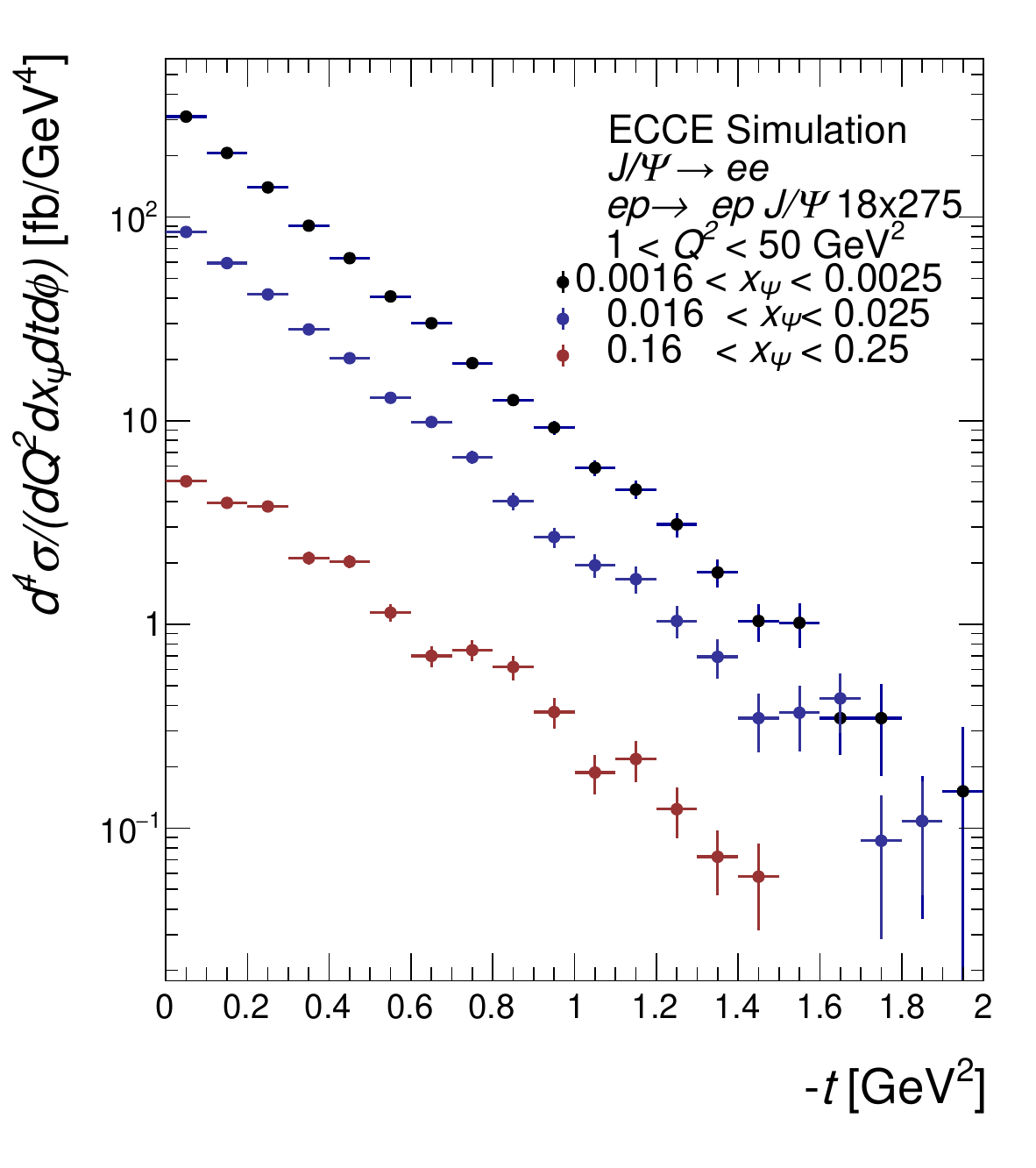}
\caption{Simulation of the \jpsi exclusive electroproduction cross section as a function of the {four-}momentum transfer {squared} $t$ for different bins in %
{$x_{\Q}$}, at {the} EIC, for lepton-proton beam energies of 18x275~GeV$^2$. %
The integrated luminosity is assumed to be  10~fb$^{-1}$.%
}
\label{fig:jpsi}
\end{figure}

Figure~\ref{fig:jpsi} shows the projected precision obtainable at {the} EIC in the exclusive \jpsi electroproduction cross section as a function of the momentum transfer $t$ to the proton, for different bins in $x_{\Q}=({Q^2+M_{\Q}^2})/({2\,p\cdot q})$, 
the $x$-Bjorken equivalent scale variable for heavy mesons. The projections are produced using the LAGER~\cite{lager} event generator and are based on the calculations presented in~\cite{Gryniuk:2016mpk}. LAGER is described as a modular accept-reject generator, capable of simulating both fixed-target and collider kinematics, and has previously been used for vector{-}meson studies at EIC kinematics, with significant recent developmental effort in support of DVMP studies.
The transverse spatial distribution of partons can be obtained by a Fourier-transform
of the cross section as a function of $t$.%

The key experimental feature of hard exclusive channels such as \jpsi electroproduction is the detection of the recoil protons in the far-forward detectors, in particular in the  B0 spectrometer and the Roman Pots. This allows %
{for accurate computation of} the momentum transfer $t$, which is the Fourier conjugate variable to the impact parameter. A wide and continuous acceptance that extends to low-$t$ is essential for a precision extraction of transverse-position distributions of partons.

{
On the other hand,  far-forward detectors can also help in detecting 
the process where the proton does not stay intact but breaks up. The dominance of this process over exclusive \jpsi production increases with increasing $t$. In \cite{deak21}, 
it has been shown that the cross{-}section \newb{measurement} of dissociative diffractive \jpsi photoproduction at large $t$ as a function of the rapidity gap between the produced \jpsi and the dissociated proton is possible at the EIC. The interest of this process lies in the presence of two comparable hard scales, the charm mass and the large $t$, and hence the possibility to probe the presence of Balitsky-Fadin-Kuraev-Lipatov (BFKL) dynamics.}

\subsubsection{Light quarks}
In~\cite{Eskola:2022vpi, Eskola:2022vaf}, it was shown that the rapidity differential cross section for exclusive \jpsi photoproduction in heavy-ion ultra-peripheral collisions at NLO decomposes into a complicated interplay of contributions from both the quark and gluon sectors as well as their interference, over the whole region of rapidities accessible at the LHC. In particular, at mid-rapidities the quark contribution was shown to be the dominant player. While such a picture remains in place under a conservative factorisation and renormalisation scale variation, and is reflected in the original work of Ivanov et al.~\cite{Ivanov:2004vd} in the context of the underlying hard scattering process, $\gamma p \rightarrow \jpsi p$, which drives the ultra-peripheral collisions, {and indeed the $eA$ collisions at the EIC}, care must be taken to interpret such results. Indeed, it was shown that such a hierarchy arises from a coincidental cancellation of LO and NLO gluon contributions together with the positive-definite quark contribution at NLO. At NNLO, when there are also interference contributions wholly within the quark sector, one may anticipate a different final picture. $\Upsilon$ photoproduction on the other hand, sitting at a higher scale, does not exhibit such a complicated interplay of contributions at NLO, see~\cite{Eskola:2023oos}, with the gluon contribution dominating over all rapidities. The \jpsi results are therefore indicative of the long-standing problem of the scale dependence and perturbative instability exhibited by low-scale processes. \newb{Indeed, after the so-called $`Q_0$ subtraction'~\cite{Jones:2016ldq} discussed in Sect.~\ref{sec:gluons-exclusive}, the quark contribution to the amplitude becomes negligible. A new study~\cite{toappear} which includes the high-energy resummation effects in the coefficient function of exclusive \jpsi photoproduction in the HEF formulism similar to one applied in the inclusive case~\cite{Lansberg:2023kzf, Lansberg:2021vie} supports this conclusion.}

\subsection{Generalised TMDs}
\label{sec:GTMDs}

The non-perturbative structure of the hadrons can be described in terms of parton correlation functions such as form factors, 1D PDFs and their 3D generalisations in terms of TMDs and GPDs. All these functions can be derived from more general objects called GTMDs~\cite{Meissner:2008ay,Meissner:2009ww,Lorce:2013pza}. Hence, GTMDs are also known as the ``mother distributions". There are several compelling reasons to study GTMDs%
{.} Firstly, GTMDs contain physics that outmatch{es} the content encoded in the TMDs and GPDs.  Secondly, via Fourier transformation, GTMDs can be related to Wigner functions, a concept that spans across other branches of physics as well. Partonic Wigner functions may allow for a hadron tomography in 5D phase-space~\cite{Belitsky:2003nz,Lorce:2011dv}. Thirdly, certain GTMDs can unravel unique correlations between parton orbital motion and spin inside hadrons~\cite{Lorce:2011kd,Lorce:2011ni,Hatta:2011ku,Lorce:2014mxa,Lorce:2015sqe}. In particular, the Wigner distribution can be used for a gauge-invariant definition of the canonical orbital angular momentum~\cite{Hatta:2011ku,Lorce:2012rr,Lorce:2012ce,Ji:2012sj,Leader:2013jra,Kanazawa:2014nha,Liu:2015xha}, which makes this quantity also accessible for calculations in lattice QCD~\cite{Engelhardt:2017miy, Engelhardt:2020qtg}.  Fourthly, there is a particular GTMD that is related to the Sivers TMD. By establishing a relation between GTMDs and the QCD odderon at small $x$, the authors in Ref.~\cite{Boussarie:2019vmk} have shown that one can access the gluon Sivers TMD through exclusive $\pi^{0}$ production in unpolarised $ep$ scattering. This finding goes against our traditional belief that the Sivers function can only be measured with a transversely polarised target.

\begin{figure}[h]
\centering
\includegraphics[width=7cm]{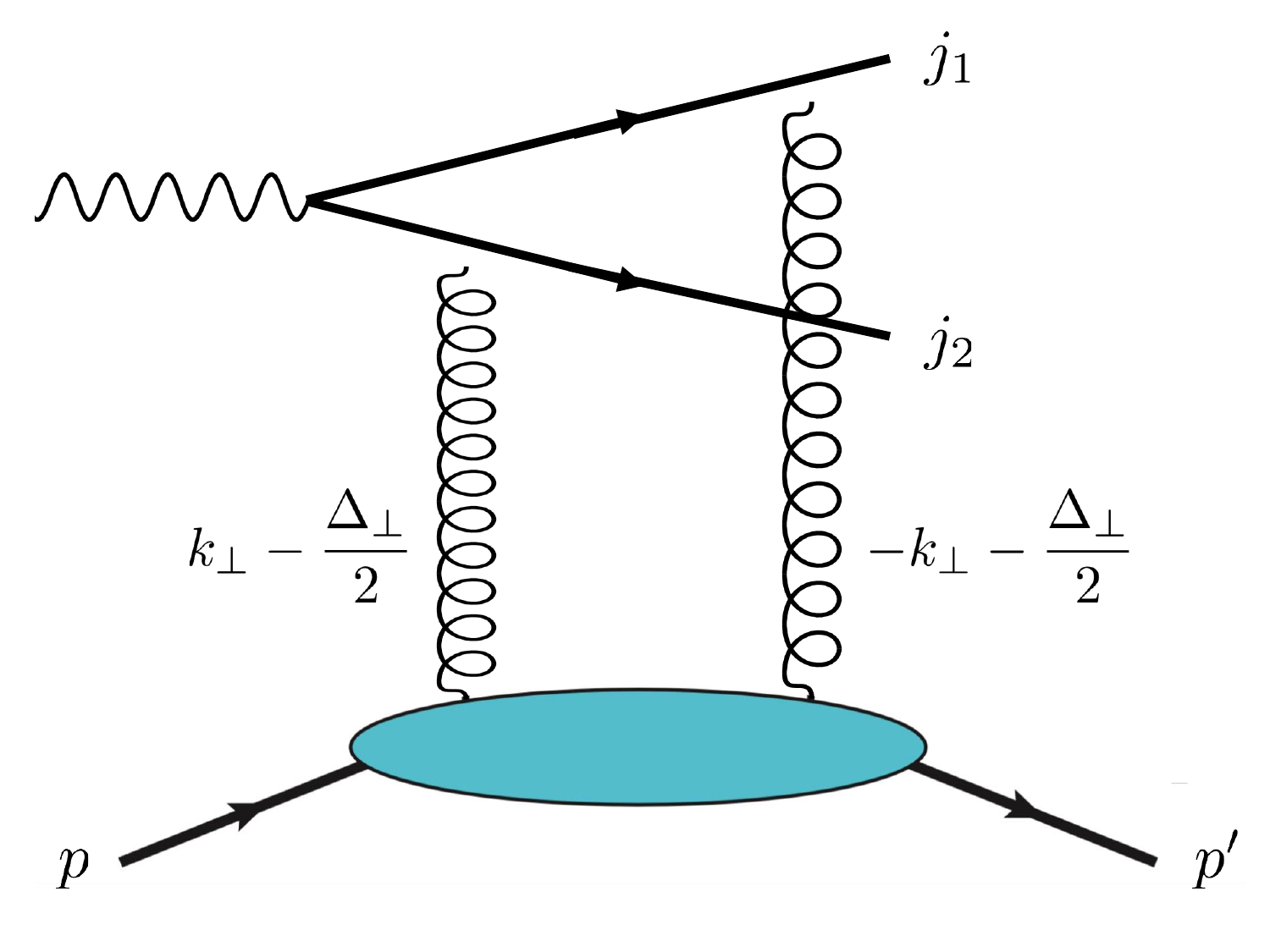}
\includegraphics[width=7cm]{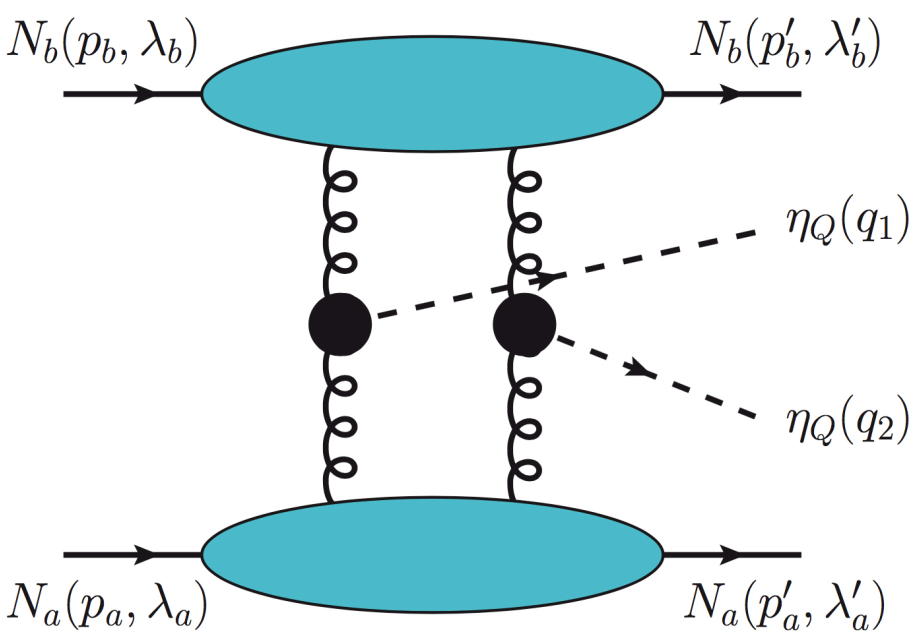}
\caption{Left panel: leading-order Feynman graph for the exclusive dijet production in lepton-nucleon/nucleus scattering. Right panel: leading-order Feynman graph for the exclusive double-quarkonium production in nucleon-nucleon collisions. The perturbative subprocess $gg \rightarrow \eta_Q$ is computed in the colour-singlet model in NRQCD.}
\label{fig:exclusive-dijet}
\end{figure}

For a long time, it was questionable whether GTMDs could be measured at all. The authors in Ref.~\cite{Hatta:2016dxp} were the first to propose addressing gluon GTMDs through exclusive diffractive dijet production in lepton-nucleon/nucleus collisions at small $x$ (see left panel of \cf{fig:exclusive-dijet}). The GTMDs depend on the average transverse parton momentum $\vec{k}_\perp$ and the transverse momentum transfer to the target $\vec{\Delta}_\perp$, and it is possible to decompose the angular correlation between these two vectors into a Fourier series. The leading angular dependent term, known as the elliptic distribution, has a characteristic cos($2\phi$) angular modulation similar to the observed elliptic flow phenomenon in relativistic heavy-ion collisions~\cite{Zhou:2016rnt,Hagiwara:2017ofm,Iancu:2017fzn}. It was shown that the cross section of this diffractive dijet process also exhibits such a cos($2\phi$) behavior where $\phi$ is now the angle between the dijet total and relative momenta. The pioneering work in Ref.~\cite{Hatta:2016dxp} gave impetus to the field of GTMDs and subsequently many other interesting ideas were put forward; see, for instance, Refs.~\cite{Hatta:2016aoc,Ji:2016jgn,Hagiwara:2017fye,Bhattacharya:2017bvs,Bhattacharya:2022vvo}.

An alternative idea~\cite{Zhou:2016rnt,Mantysaari:2020lhf} is to exclusively produce a single particle (instead of two jets) such as a $J/\psi$. The role of the second jet is now played by the scattered electron which must be detected. It has been shown that in this process the elliptic $\cos 2\phi$ correlation of the gluon GTMD manifests itself in the angular correlation between the  scattered electron and the  $J/\psi$ \cite{Mantysaari:2020lhf} (or the recoiling proton/nucleus \cite{Zhou:2016rnt}). For a proton target, a sizable $v_2$ of a few percent or larger has been predicted~\cite{Mantysaari:2020lhf}. The same effect can also be seen in DVCS, but $J/\psi$ production is more promising since there is no contamination from the Bethe-Heitler 
process. In the GPD-based approach to DVCS, the same angular correlation is known to be generated by the so-called gluon transversity GPD. The elliptic gluon GTMD is the mother distribution of the gluon transversity GPD~\cite{Hatta:2017cte}.

Quarkonium production processes are also useful to study other aspects of GTMDs. 
In Ref.~\cite{Bhattacharya:2018lgm}, it was shown that exclusive double production of pseudo-scalar quarkonia ($\eta_{c/b}$) in hadronic collisions could serve as a direct probe of GTMDs for gluons at moderate $x$ (see right panel of \cf{fig:exclusive-dijet}).  A similar idea came out in Ref.~\cite{Boussarie:2018zwg} where the authors proposed to access  the Weizs\"acker-Williams gluon GTMD at small $x$ via double  $\chi_{cJ}$ or $\eta_c$ meson production in %
{diffractive} $pp$/$pA$ collisions {where (one of) the proton(s) stays intact}.  %

At the EIC, the primary process to look for gluon GTMDs is exclusive diffractive dijet production, as mentioned above. A challenge, however, is that due to the limited %
{centre}-of-mass energy, the transverse momenta of diffractively produced particles in the forward rapidity region are often not large enough to cleanly reconstruct jets. 
As a first step to test the underlying GTMD picture of exclusive diffractive production processes, like dijet or $J/\psi$ electro{-} and photoproduction at small $x$, a GTMD model can be fitted to existing HERA data. Predictions can then be obtained for EIC in different kinematic regions. This has been considered for dijet production in \cite{Boer:2021upt}, where it was shown that a gluon GTMD model based on the impact-parameter-dependent McLerran-Venugopalan model can give a reasonably good description of diffractive dijet production data from %
H1 \cite{Boer:2021upt}. 
The same framework (slightly extended) can be applied to exclusive diffractive $J/\psi$ production to describe the H1 and ZEUS data, as shown in \cf{fig:JPsi} on the left ($\sqrtsep = 319$ GeV). With the resulting GTMD parametrisation, predictions for exclusive  diffractive $J/\psi$ production at EIC can be obtained. These are shown for $\sqrtsep = 45$ and $140$ GeV in \cf{fig:JPsi} on the right. %
 Generally, at small $x$, and in particular for nuclear targets, a GTMD-based description becomes more appropriate for exclusive and diffractive processes.
 Exclusive quarkonium production at the EIC could be used to systematically study the transition between the collinear and $k_\perp$-dependent frameworks.

\begin{figure}[!ht]
	\centering
	\includegraphics[width=0.43\textwidth]{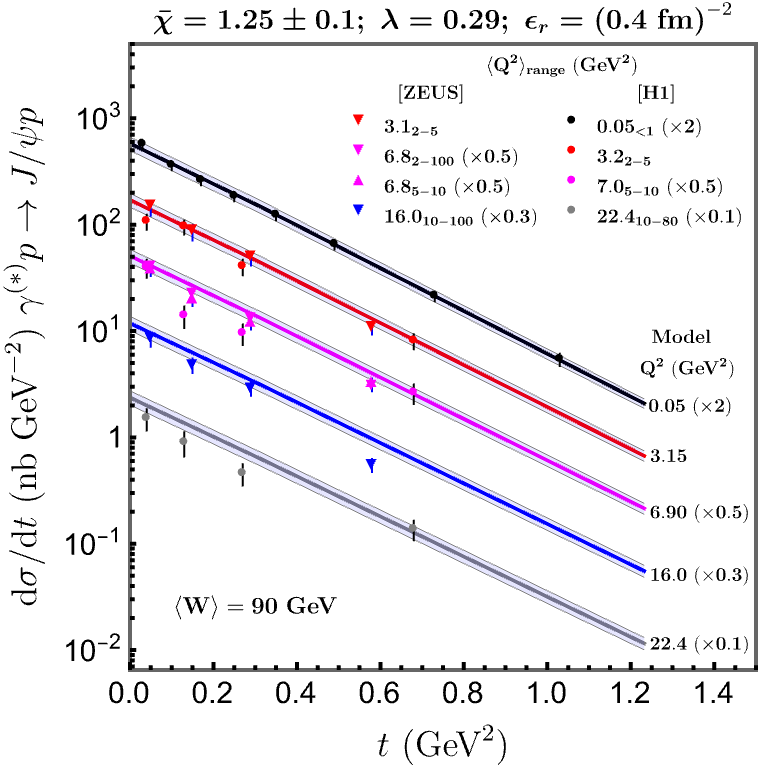} 	
	\includegraphics[width=0.43\textwidth]{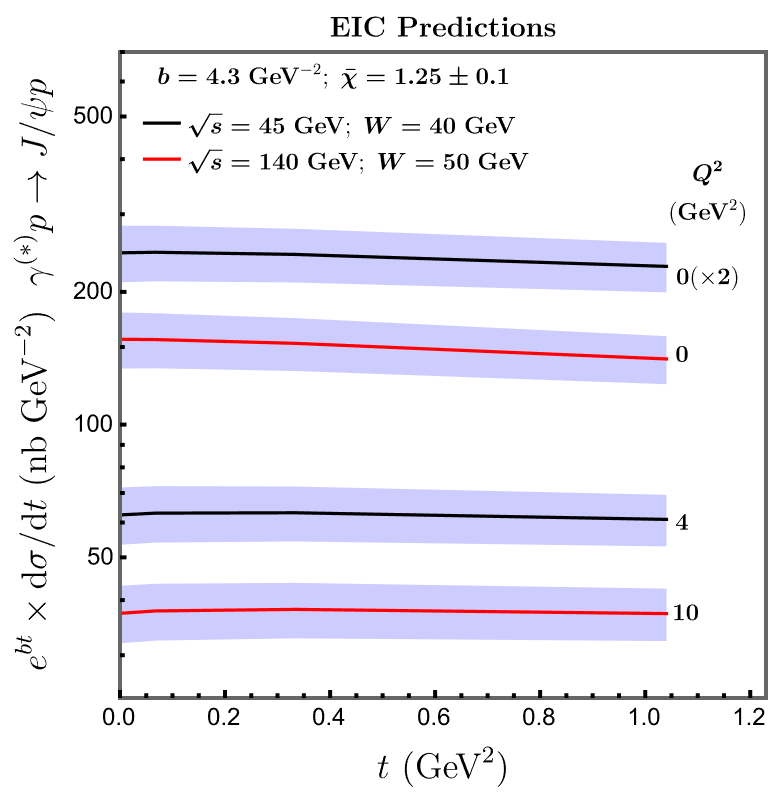} 
	\caption{Left: %
 fit of exclusive diffractive $J/\psi$ data from H1~\cite{H1:2005dtp} and ZEUS~\cite{ZEUS:2004yeh} using a gluon GTMD model%
 ~\cite{Boer:2021upt} with boosted Gaussian wave functions from~\cite{Kowalski:2006hc}. Right: EIC predictions with the same parametrisations for $W_{\gamma^* p}=40$ GeV at $\sqrtsep=45$ GeV and for $W_{\gamma^* p}=50$ GeV at $\sqrtsep=140$ GeV.}
	\label{fig:JPsi}
\end{figure}

\subsection{Exclusive quarkonium production near threshold and the trace anomaly}
\label{sec:trace_anomaly}
\label{sec:near_threshold}

It has been noticed long ago that the mass $M$ of a hadronic system can be expressed in terms of the forward matrix element of the trace of the QCD energy-momentum tensor %
as~\cite{Shifman:1978zn,Donoghue:1987av}
\begin{equation}
2M^2=\langle p|\,\frac{\beta}{2g}\,F^2+(1+\gamma_m)\,\overline\psi m\psi\,|p\rangle,
\label{tracedec}
\end{equation}
where $\beta$ and $\gamma_m$ are anomalous dimensions and the operator $\beta / (2g)\,F^2 + \gamma_m \, \overline\psi m\psi$ is the QCD trace anomaly~\cite{Collins:1976yq,Nielsen:1977sy}.  The decomposition of the r.h.s.~of Eq.~\eqref{tracedec} %
into quark and gluon contributions has been discussed in detail~\cite{Hatta:2018sqd,Tanaka:2018nae}. Other mass decompositions, based this time on the QCD Hamiltonian, have also been proposed in the literature~\cite{Ji:1994av,Ji:1995sv,Lorce:2017xzd,Rodini:2020pis,Metz:2020vxd,Ji:2021pys,Ji:2021qgo}. The latter all require the knowledge of the same four quantities, combined in different ways for the physical interpretation~\cite{Lorce:2021xku}. Two of these quantities, namely the quark momentum fraction $A_q(0)=\langle x\rangle_q$ and the gluon momentum fraction $A_g(0)=\langle x\rangle_g$, are already well known. The other two numbers $\bar C_q(0)$ and $\bar C_g(0)$ can be determined by measuring the quark and gluon contributions to Eq.~\eqref{tracedec}. While the quark condensate $\langle p|\,\overline\psi m\psi\,|p\rangle$ has already received a lot of attention over the last decades (see~\cite{Gupta:2021ahb} and references therein), little is known so far about the gluon condensate $\langle p | F^2 | p \rangle$ from the experimental side. 

Four-momentum conservation implies that $A_q(0)+A_g(0)=1$ and $\bar C_q(0)+\bar C_g(0)=0$. From a phenomenological point of view, the knowledge of $A_q(0)$ and the quark condensate  is therefore sufficient for specifying all the contributions to the various mass decompositions (see~\cite{Metz:2020vxd,Copeland:2021qni} for recent estimates). Measuring the gluon condensate is not expected to change much the current phenomenology of the nucleon mass, but it will provide a fundamental sanity check of the mass sum rules and the virial theorem~\cite{Lorce:2021xku}. Another motivation for measuring the gluon condensate is that it could shed light on the existence and nature of the recently discovered LHCb ``pentaquark'' states~\cite{Joosten:2018gyo}. 

More than two decades ago, exclusive heavy{-}quarkonium production, near the production threshold, was suggested as a promising tool for constraining the gluon condensate in the nucleon~\cite{Kharzeev:1995ij,Kharzeev:1998bz}.  
This development together with the prospect to obtain through this process further information about the gravitational structure of the nucleon, which is contained in the form factors of the {energy-momentum tensor} 
(such as the mass radius and mechanical pressure distributions~\cite{Polyakov:2018zvc,Lorce:2018egm,Freese:2021czn}), as well as the measurement of exclusive $J/\psi$ photoproduction near threshold at Jefferson Lab~\cite{GlueX:2019mkq} has stimulated a significant amount of activities in this area~\cite{Brodsky:2000zc,Gryniuk:2016mpk,Hatta:2018ina,Hatta:2019lxo,Mamo:2019mka,Wang:2019mza,Boussarie:2020vmu,Zeng:2020coc,Gryniuk:2020mlh,Pentchev:2020kao,Du:2020bqj,Kharzeev:2021qkd,Wang:2021dis,Ji:2021mtz,Hatta:2021can,Mamo:2021krl,Guo:2021ibg,Sun:2021gmi,Xie:2021seh,Mamo:2021tzd,Wang:2021ujy,Sun:2021pyw,Han:2022qet}.
Recently, it was argued that the extraction of the gravitational form factors through exclusive quarkonium photoproduction will necessarily retain model dependence~\cite{Sun:2021gmi,Sun:2021pyw}. Generally, access to the gravitational structure of the nucleon is expected to be cleaner for electroproduction~\cite{Boussarie:2020vmu,Hatta:2021can}. 
At the EIC, one would have the unique opportunity to explore photo{-} and electroproduction of both $J/\psi$ and $\Upsilon$ close to threshold ~\cite{Joosten:2018gyo}.

\subsection{Probing double parton scattering at the EIC with quarkonium pairs}
 \label{sec:DPS-EIC}

\subsubsection{A word of context}
In this section, we study the possibility to observe double-$J/\psi$ production at the EIC. In particular, we discuss both the single-parton-scattering (SPS)  and the double-parton-scattering (DPS) mechanisms{, which could lead to the observation of a pair of $\jpsi$}. In fact, the cross section for the latter case would allow one to access new information on the so-called proton %
double-parton-distribution functions {(dPDFs)}, which encode novel information on the partonic  structure {of the proton}.  

Let us recall the analysis of four-jet photoproduction  at HERA, which pointed out the relevance of  multi{-}parton interactions (MPI{s}) to account for the measured total cross section \cite{Butterworth:1996zw}.
In  Ref.~\cite{Ceccopieri:2021luf},  the DPS cross section for four-jet photoproduction was %
calculated. 
DPS are initiated by a quasi-real photon~\cite{Klasen:2002xb} splitting into a  $q \bar q$ pair. The same strategy as for $pp$ collisions \cite{Goebel:1979mi,Humpert:1983pw,Mekhfi:1983az,Mekhfi:1985dv,Humpert:1984ay,Mangano:1988sq,Paver:1982yp,Sjostrand:1986ep,Gaunt:2014ska,
Diehl:2019rdh,Diehl:2018wfy} has been used to evaluate the photoproduction cross section. {At this stage,} the only missing quantity {was} $\sigma_{eff}^{\gamma p}$, the effective size of the photon-proton interaction, which is expected to be  process independent. It was estimated %
for the first time~\cite{Ceccopieri:2021luf} and compared to that {of} the $pp$ case from Refs.~\cite{Aaboud:2016dea,CMS:2021ijt,Chapon:2020heu,Lansberg:2014swa,Lansberg:2015lva}.The four-jet DPS cross section has {then} been calculated {for the} HERA kinematics~\cite{Chekanov:2007ab} to be $\sigma_{DPS}^{4j} \geq 30$~pb, while the total one \newb{was inferred from~\cite{Chekanov:2007ab} to be}  $\sigma_{tot}^{4j} \sim 135$~pb \newb{at $x_\gamma < 0.75$}. This indicated that the DPS contribution  is  sizeable even in photon-induced reactions for the production of four jets %
{and that it could} also {be so} for other processes like quarkonium-pair production. Further analyses of the HERA data could lead to the extraction of $\sigma_{eff}^{\gamma p}$ and, in turn, provide a first access to the mean transverse distance between two partons in the proton, %
{an} unknown property {of the proton structure}. To this aim, the needed luminosity was evaluated to be $\mathcal{L} \sim 200$~pb$^{-1}$  \cite{Ceccopieri:2021luf}. Double{-}$J/\psi$ production from DPS at EIC will be %
{presented below} along the same lines.

\subsubsection{DPS at the EIC and \jpsi-pair production }
\label{sec:DPSdoubleJpsi}

{Here we discuss $J/\psi$-pair photoproduction at the EIC.}  In $ep$ collisions, the radiated quasi-real photon can interact with {the} partons %
{within} the proton in two ways, namely as a ``pointlike'' particle and via its ``resolved'' hadronic content. In the first case, the photon ``directly'' interacts with the target while, in the latter case, the  photon splits into {(colour {charged})} partons, which subsequently interact with partons in the proton. 

\begin{figure}[hbt!]
\centering
\includegraphics[width=18cm]{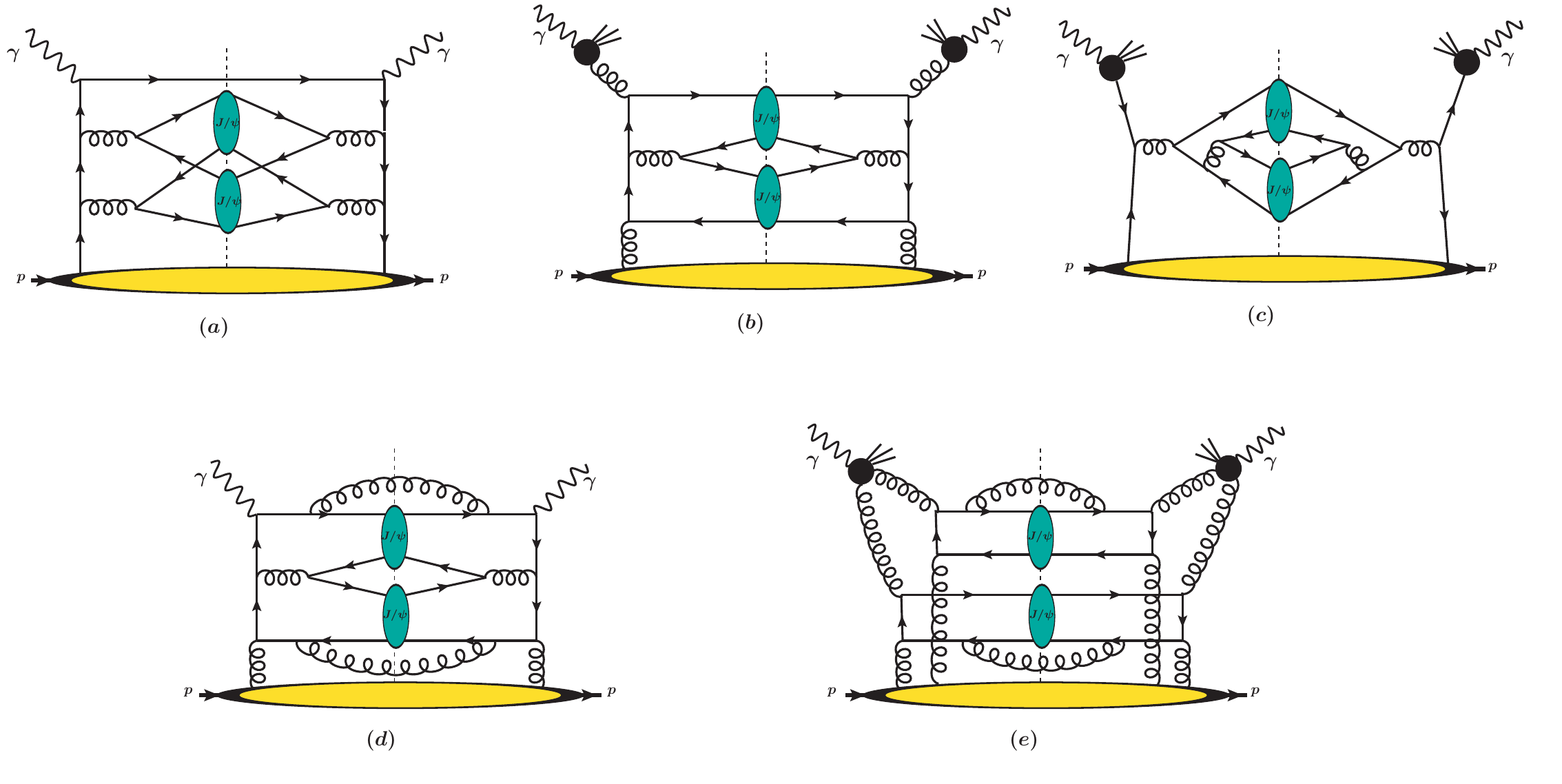}
\caption{Di-\jpsi photoproduction%
at the EIC via SPS $\mathcal{O}(\alpha_s^4)$ (a-c), $\mathcal{O}(\alpha_s^5)$ (d) and  DPS $\mathcal{O}(\alpha_s^6)$ (e).}
\label{fig:DPS_SPS}
\end{figure}

{The treatment of the interaction between a proton and such a resolved photon is %
carried out by using a PDF describing the momentum distributions of these partons inside the photon. One of these is the  GRV~\cite{Gluck:1991jc,Gluck:1994tv} set{,} which is %
{adopted} here.}
{For what regards the quarkonium-production mechanisms, the CSM (\ie\ the leading $v^2$ contribution of NRQCD) is used.} 
{\cf{fig:DPS_SPS} shows different {Feynman} graphs for SPS and DPS photoproduction.} 
In the SPS case, the contributing channels at leading order, \ie\ $\alpha \alpha_s^4$, are shown in \cf{fig:DPS_SPS}({\color{red}a-c}), namely, $\gamma q\to J/\psi+J/\psi+q$, $g g\to J/\psi+J/\psi$ and $q \bar{q}\to J/\psi+J/\psi$. However, the graph in \cf{fig:DPS_SPS}({\color{red}d}) contributes at {order} $\alpha\alpha_s^5$, i.e. {via the SPS} $\gamma g\to J/\psi+J/\psi+g+g$. The gluon{-}initiated channel in DPS for di-$J/\psi$ production {at $\alpha_s^6$} is shown in \cf{fig:DPS_SPS}({\color{red}e}), while the quark{-}initiated channel does not contribute in the CSM at \newb{order} $\alpha_s^6$. The partonic channel %
$gg\to J/\psi+g$ %
dominates for single-$J/\psi$ production.
The SPS cross section, \ie\ the squared matrix elements convoluted with single-parton PDFs, can be calculated using %
HELAC-Onia%
~\cite{Shao:2012iz,Shao:2015vga}. In order to estimate the DPS cross section, we need to {use the poorly known} proton dPDFs, which
{provide} the number densities of a parton pair with a given transverse distance $b_\perp$ and carrying  the longitudinal momentum fractions ($x_1,x_2$) of the parent hadron~\cite{Rinaldi:2018slz,Rinaldi:2016jvu,Rinaldi:2014ddl,Diehl:2011tt,Manohar:2012jr,Rinaldi:2015cya}. %
{Assuming} {that dPDFs can} be factorised in terms of %
{ordinary 1D} PDFs and a transverse part, the DPS cross section {can be expressed in terms of two SPS cross sections for the production of {each} %
of the observed particles among the pair:}
\begin{equation}
 \sigma^{(J/\psi ,J/\psi)}_{DPS}=\frac{1}{2}\frac{\sigma^{(J/\psi)}_{SPS}\,\sigma^{(J/\psi)}_{SPS}}{\sigma_{eff}^{\gamma p}}\,,
\end{equation}
which is the so-called "DPS pocket formula", {valid under the assumption of 
totally uncorrelated 
{kinematics} between both parton scatterings}. 
The $\sigma^{(J/\psi)}_{SPS}$ is the SPS contribution for single $J/\psi$ production.  
{In the present study, within the mentioned assumptions, one gets:}
\begin{align}
   \sigma_{eff}^{\gamma p}=  \left[\displaystyle \int \dfrac{{\rm d}^2\vec{k}_\perp}{(2 \pi)^2} F_2^\gamma(\vec{k}_\perp,Q^2) F_2^p(\vec{k}_\perp) \right]^{-1}
\end{align}
{where here $F^{p(\gamma)}_2(k_\perp)$ parametrises the transverse structure of the proton (photon)~\cite{Ceccopieri:2021luf}. For the photon, the only available calculation is that of Ref.~\cite{Ceccopieri:2021luf} while, for the proton, there are several models based on the data for %
DPS {in $pp$ collisions}.}
Recently, several experimental analyses on DPS  have been carried out  for the production of $J/\psi + W$~\cite{ATLAS:2014yjd}, $J/\psi+Z$~\cite{ATLAS:2014ofp}, $J/\psi+\mathrm{charm}$~\cite{LHCb:2012aiv}  in $pp$ and $J/\psi+J/\psi$~\cite{D0:2014vql} in $p\bar{p}$ %
processes. A comprehensive comparison between theory and experiments for di-$J/\psi$ production at {the} Tevatron and the LHC has been presented in~\cite{Lansberg:2014swa,Lansberg:2019adr}, and it %
{was} observed that DPS dominates the yield at large $J/\psi$-rapidity difference. %
DPS has been {also} studied for $J/\psi${-}pair production for the \newb{LHC fixed-target (also referred to as AFTER$@$LHC) kinematics} \newb{in~\cite{Lansberg:2015lva}}.

{At the EIC, LO computations using HELAC-Onia show that measurements are possible at $\sqrtsep = 140$ GeV with SPS contributions generally dominant over the DPS ones, but there are certain regions (low $z$ and large $\Delta y$) in the phase space where DPS cannot be disregarded. If $\sigma_{eff}^{\gamma p}$ is not too small, DPS events could be measured. In these regions, there is thus a compelling opportunity to distinguish between the resolved and unresolved contributions in the cross section and thereby to gain valuable insight into the internal structure of photons and protons.}

\section{Quarkonia as tools to study the parton content of nuclei}
\label{sec:eA}

\subsection{Nuclear PDFs}
Decades of experimental and theoretical studies showed that the distributions of partons in a nucleus are considerably modified compared to the nucleon ones. While significant progress has been made since the initial observation of the modification of PDFs {in bound nucleons} by the EMC Collaboration~\cite{EuropeanMuon:1983wih}, our understanding of nuclear PDFs (nPDFs) is still not satisfactory, {most} notably in the case of gluons. Measurement{s} of quarkonium production in %
{$eA$} reaction{s} can bridge this knowledge gap.

One of the main EIC goals is a high-precision survey of {the} %
{partonic structure of the nucleus} %
to significantly advance %
{our} quantitative understanding of %
{nPDFs}. {The} EIC %
{will offer} the possibility %
{to} study nPDF{s} over a broad range of momentum transfer{s}
~\cite{AbdulKhalek:2021gbh}. An improved knowledge of nPDFs will enable more precise theoretical calculations for %
nuclear effects and increase the scientific benefit of already successful heavy-ion programmes at RHIC and LHC.     

A widely accepted approach {to quantify} nuclear effects in %
PDFs is to start with proton PDFs and use a function $R(x,Q^2)$ that captures the modification of a given PDF in a nucleus. Experimentally, such a modification could be studied by a ratio of structure functions $F_2$ %
or by the so-called nuclear modification factor {as done by} RHIC and LHC experiments. In the case of $eA$ collisions, {$R(x,Q^2)$ is} defined as  
\begin{equation}
\label{eq:R_eA}
    R_{eA} = \frac{1}{A} \frac{(d)\sigma_{eA}}{(d)\sigma_{ep}},
\end{equation}
where $(d)\sigma_{eA}$ and $(d)\sigma_{ep}$ are the cross sections for %
{the process under consideration, respectively,} in $eA$ and $ep$ reactions{,} while mass number $A$ serves as a normalisation factor. {Note} that these cross sections can be differential in different kinematical variables.  With the definition of \ce{eq:R_eA}, $\ReA = 1$ in the absence of nuclear effects.  {In the following, we review and quantify prospects for nuclear-PDF determination at EIC via  $\ReA$ measurements.}

\subsubsection{Gluons}
In order to give {an} %
estimate of the potential impact of the EIC on {nPDF determination}, %
the nuclear modification factor %
$R_{e\rm{Au}}$, 
which can be measured in inclusive $J/\psi$ photoproduction in %
{$e\rm{Au}$} reactions, %
{is compared} with projected statistical uncertainties. Such a prediction %
{is} shown in \cf{fig:R_eAu-45GeV} and \cf{fig:R_eAu-90GeV} at two different values of the {centre}{-}of{-}mass energy%
, $\sqrt{s_{eN}}$,  %
45{~GeV} and 90~GeV, as a function of the $J/\psi$ rapidity in the $N\gamma$ {centre}{-}of{-}mass frame%
\footnote{Note that we adopt the same kinematical configuration as the EIC Yellow Report, with the proton(ion) moving along $+\hat{z}$ and the electron along $-\hat{z}$ (see also Fig.~\ref{fig:eic:generic:det}).} and as a function of $W_{\gamma N}$. Kinematical cuts are applied on the %
elasticity ($0.2 < z < 0.9$) and on the pseudorapidity of the electron pair coming from {the} $J/\psi \to e^+e^-$ decay ($\lvert\eta_{ee}\rvert < 3.5$). Different cuts on $W_{\gamma N}$ are applied for the rapidity spectra at the two different $\sqrtseN$ energies. 

\begin{figure}[htb!]
\centering\vspace*{-4cm}
\includegraphics[width=8cm,keepaspectratio]{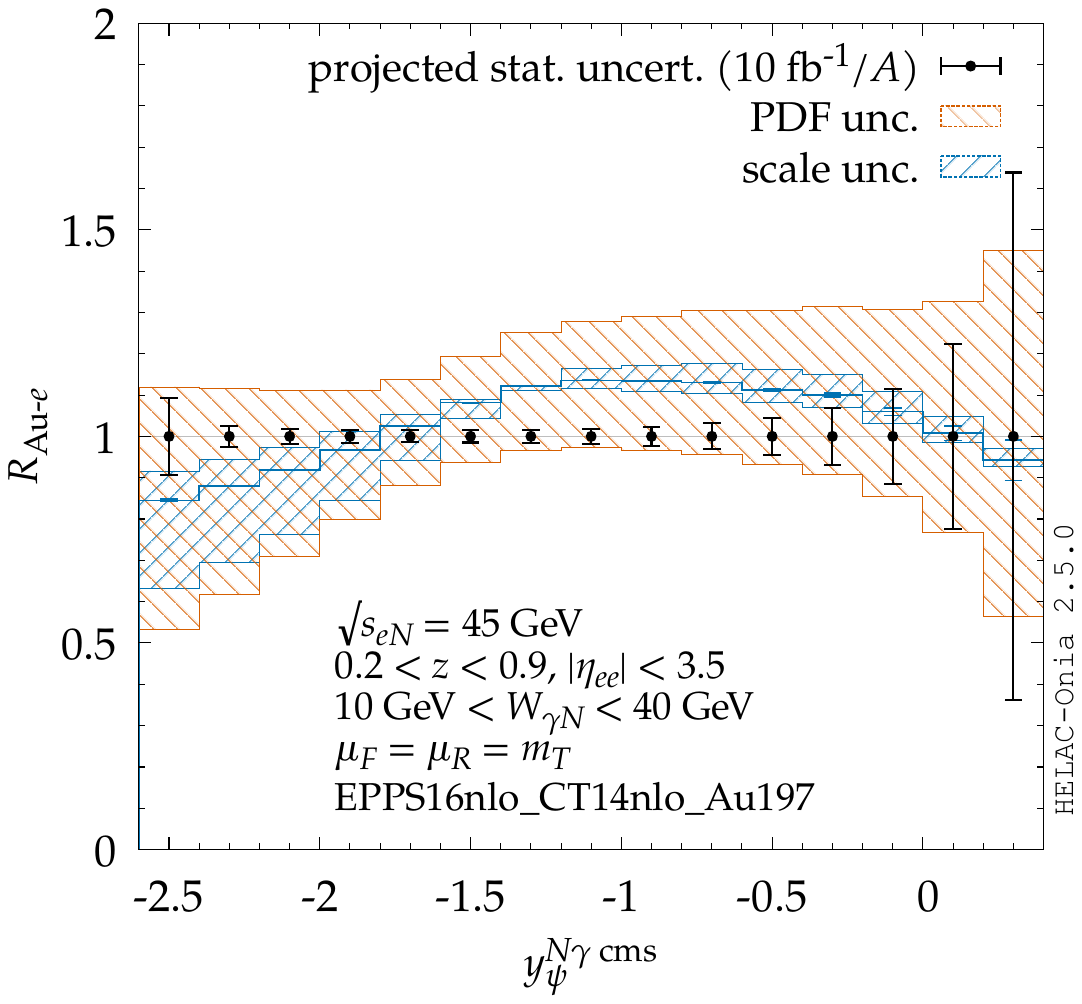}\hspace{-1cm}
\includegraphics[width=8cm,keepaspectratio]{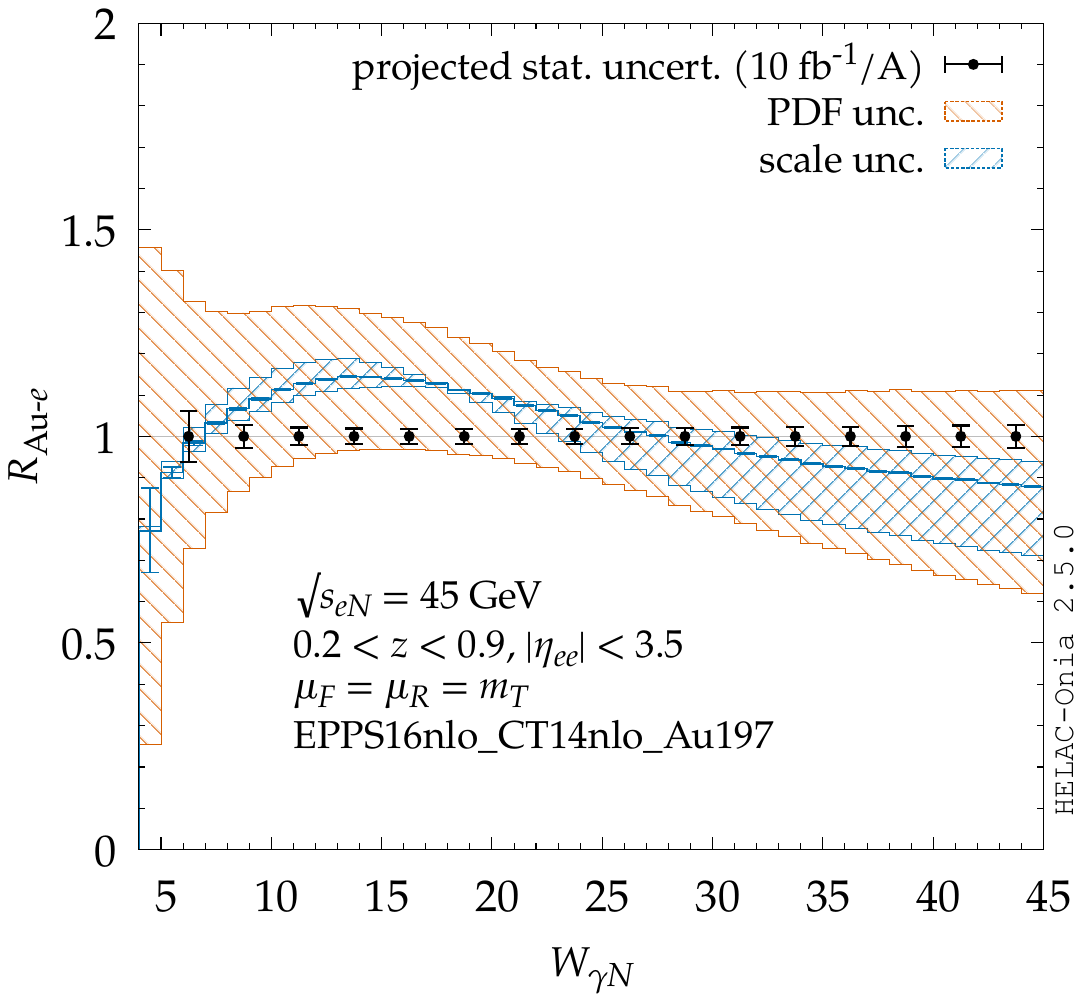}\\[-5cm]
\includegraphics[width=8cm,keepaspectratio]{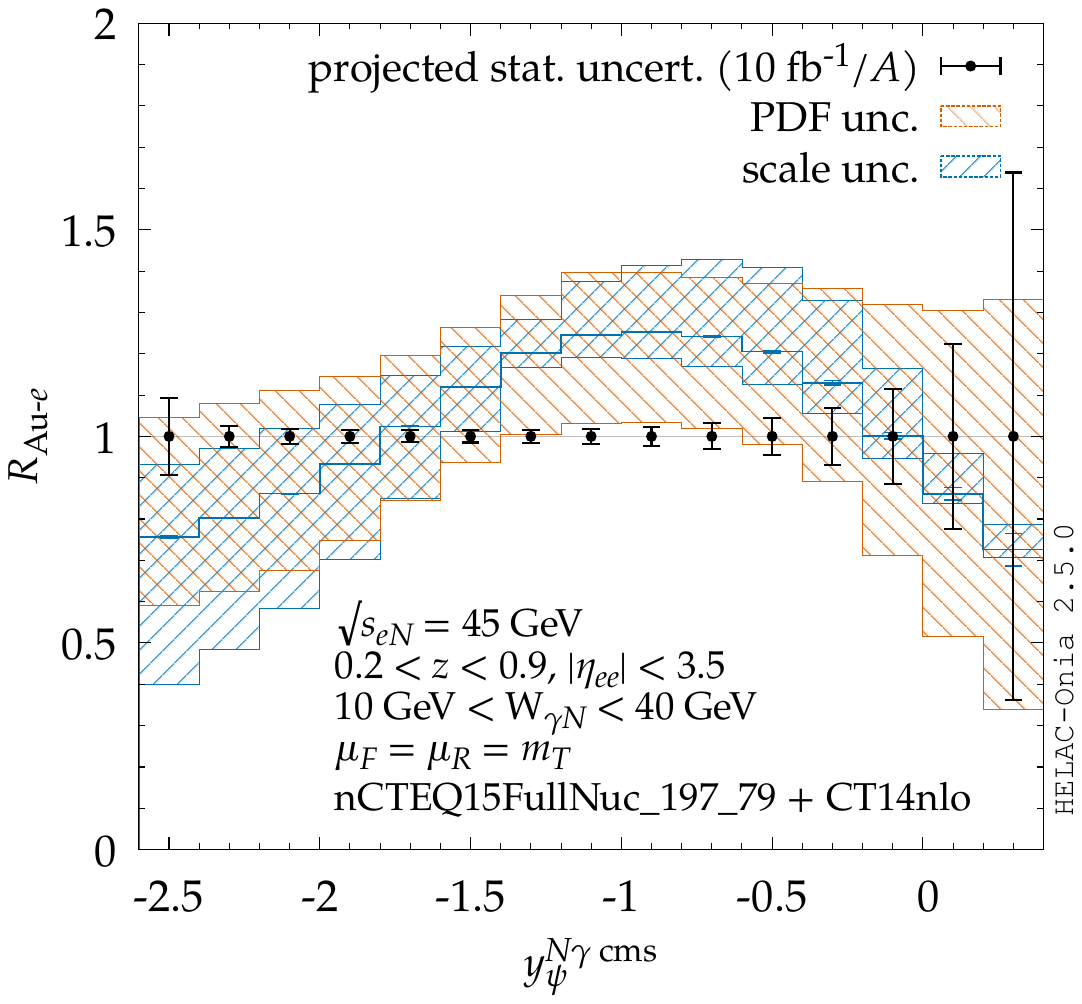}\hspace{-1cm}
\includegraphics[width=8cm,keepaspectratio]{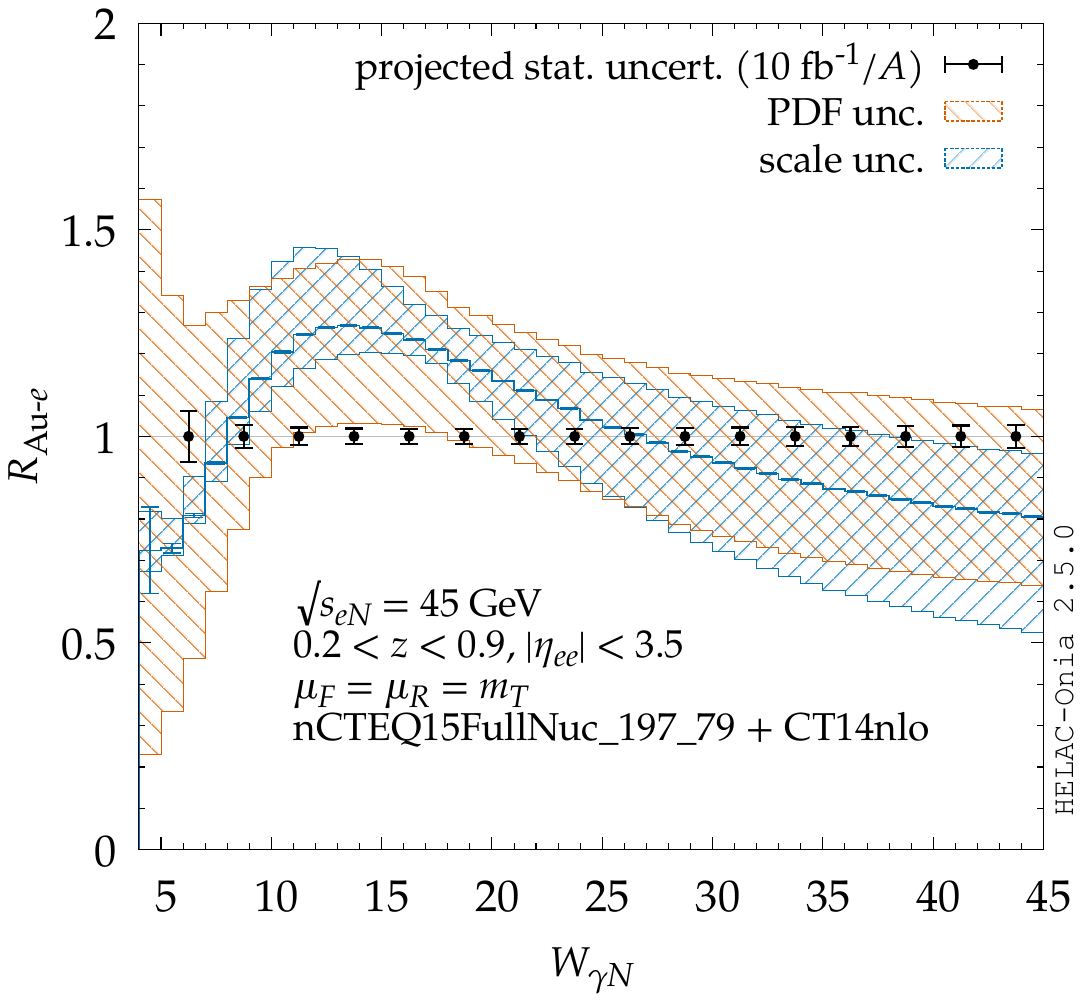}\vspace*{-1cm}
\caption{ %
{$\ReAu$}  LO CSM prediction at %
{$\sqrtseN=45$}~GeV as a function of the $J/\psi$ rapidity in the %
{$\gamma N$} %
{centre-}of{-}mass frame (left panels) and as a function of $W_{\gamma N}$ (right panels). Calculations are based on HELAC-Onia~\cite{Shao:2012iz,Shao:2015vga} with  the cuts $0.2 < z < 0.9$, $|\eta_{ee}| < 3.5$ and the nPDFs EPPS16NLO+CT14nlo (top plots) and nCTEQ15FullNuc+CT14nlo (lower plots). Projections are calculated assuming %
{$\ReAu=1$} and for an integrated luminosity of {10 fb$^{-1}/A$}.}
\label{fig:R_eAu-45GeV}
\end{figure}

The nuclear-modification-factor predictions are calculated using HELAC-Onia~\cite{Shao:2012iz,Shao:2015vga}, adopting the CT14nlo set~\cite{Dulat:2015mca} as a proton PDF baseline and using two different nuclear PDF sets for the gold nucleus, namely EPPS16nlo~\cite{Eskola:2016oht} and nCTEQ15FullNuc~\cite{Kovarik:2015cma}. Factorisation and renormalisation scale{s} are taken to be the $J/\psi$ transverse mass, $\mu_F = \mu_R = m_T=\sqrt{M^2_{J/\psi}+\pT^2}$. Note also that, since these predictions are calculated at LO in the CSM, where the {only} partonic subprocess is $\gamma + g \to J/\psi + g$, they can be directly interpreted as $R_g$, the nuclear modification factor for the gluon nPDF. The statistical projections are calculated assuming %
{$R_{e\rm{Au}}=1$} (using the central value of CT14nlo) and assuming an integrated luminosity of $10\,\text{fb}^{-1}/A$. %
The branching ratio for {the} $J/\psi \to e^+e^-$ decay was taken to be 5.94\% and a \jpsi\ reconstruction efficiency of 64\% was assumed (considering an average identification efficiency of the electrons from the \jpsi\ decay to be approximately $80\%$).

\begin{figure}[htb!]
\centering\vspace*{-4cm}
\includegraphics[width=8cm,keepaspectratio]{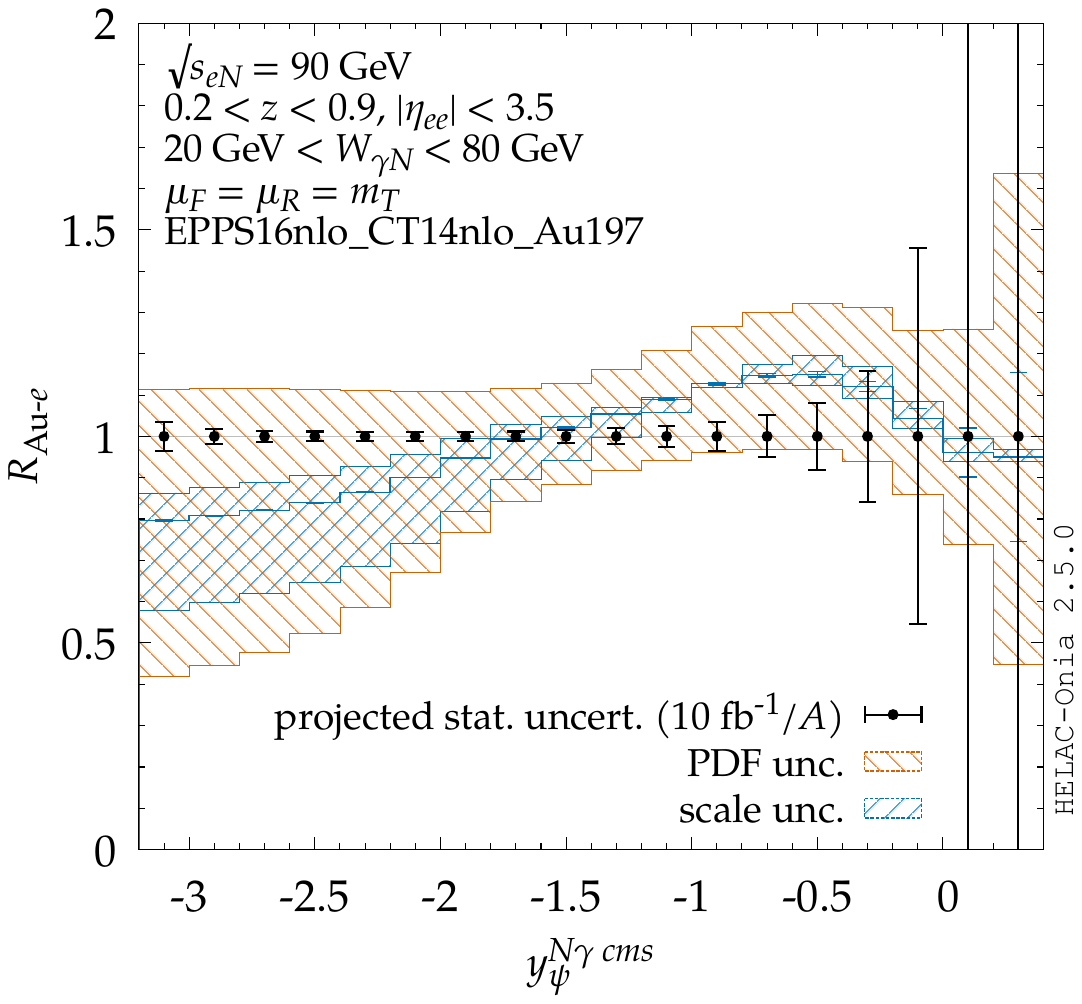}\hspace{-1cm}
\includegraphics[width=8cm,keepaspectratio]{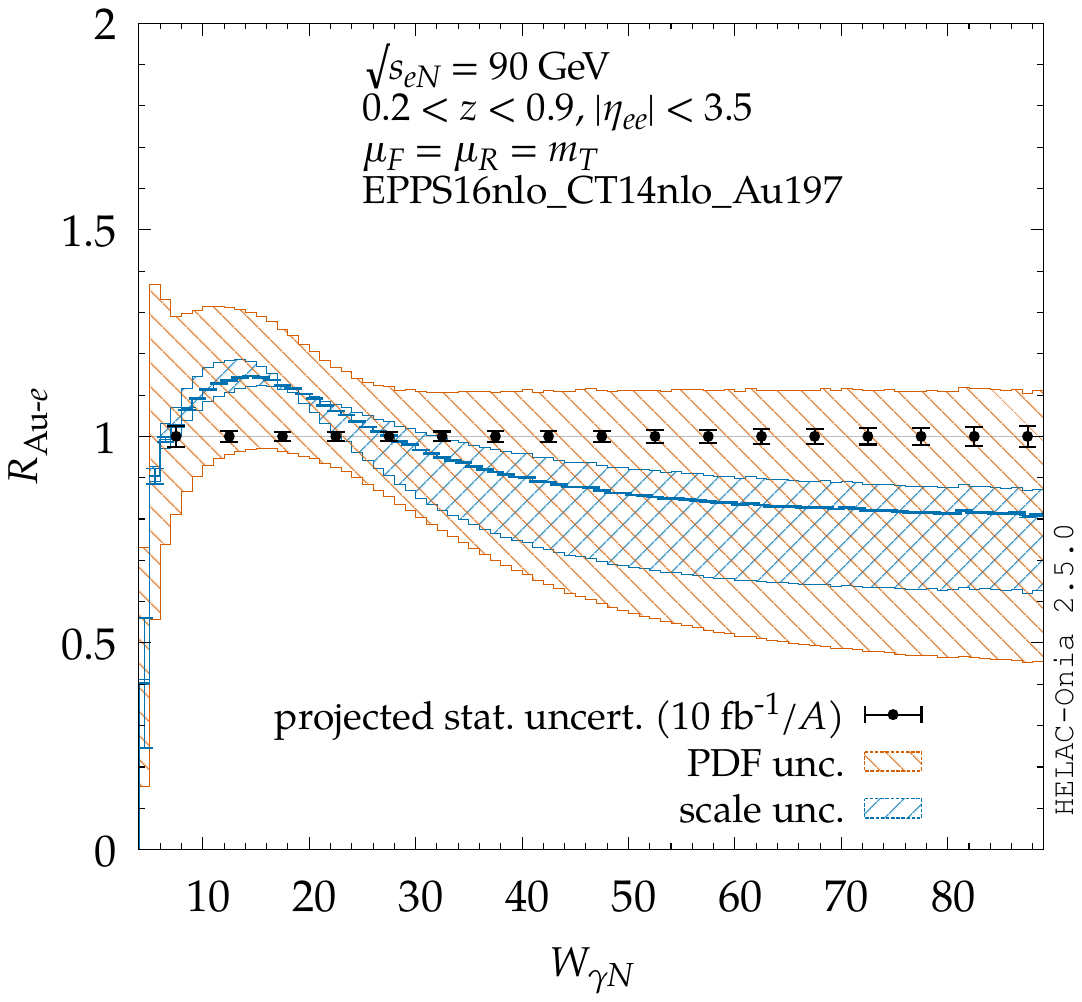}\\[-5cm]
\includegraphics[width=8cm,keepaspectratio]{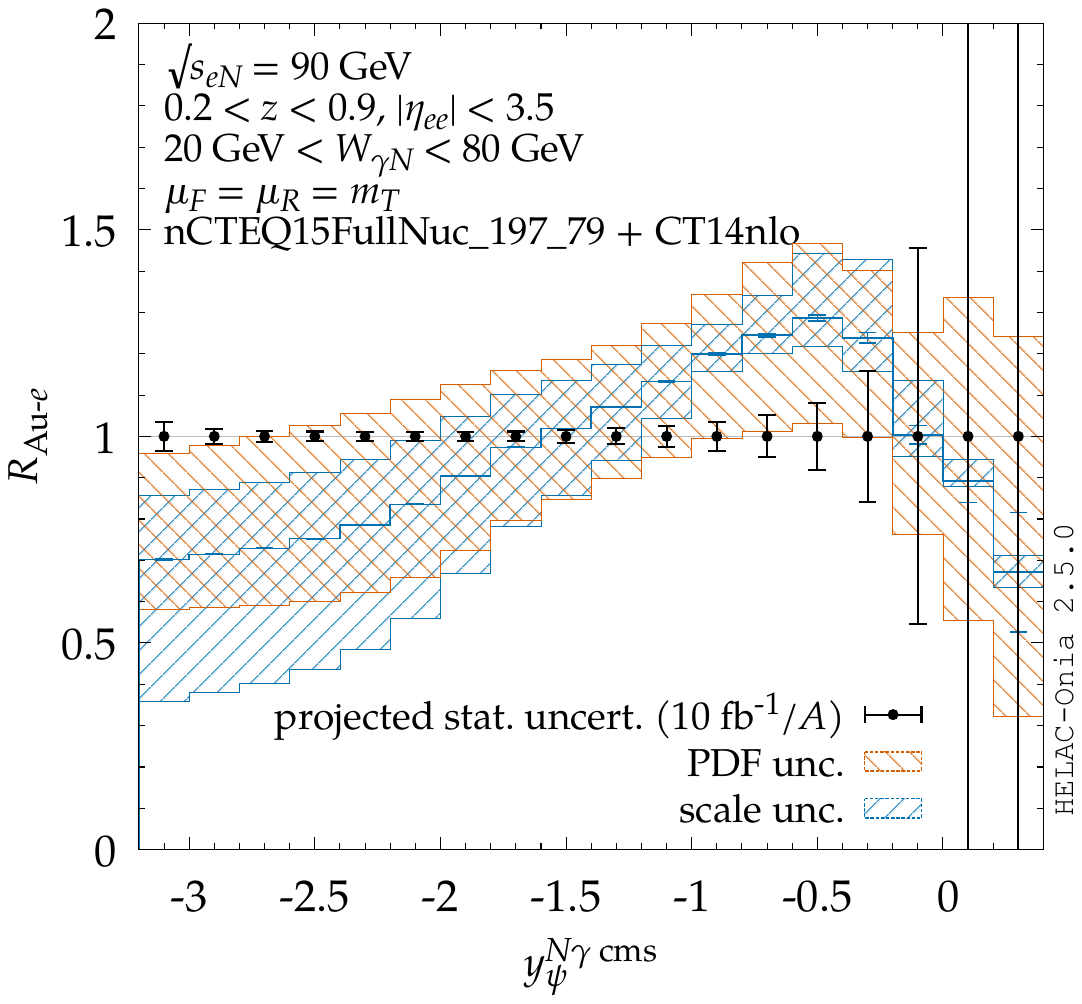}\hspace{-1cm}\vspace*{-1cm}
\includegraphics[width=8cm,keepaspectratio]{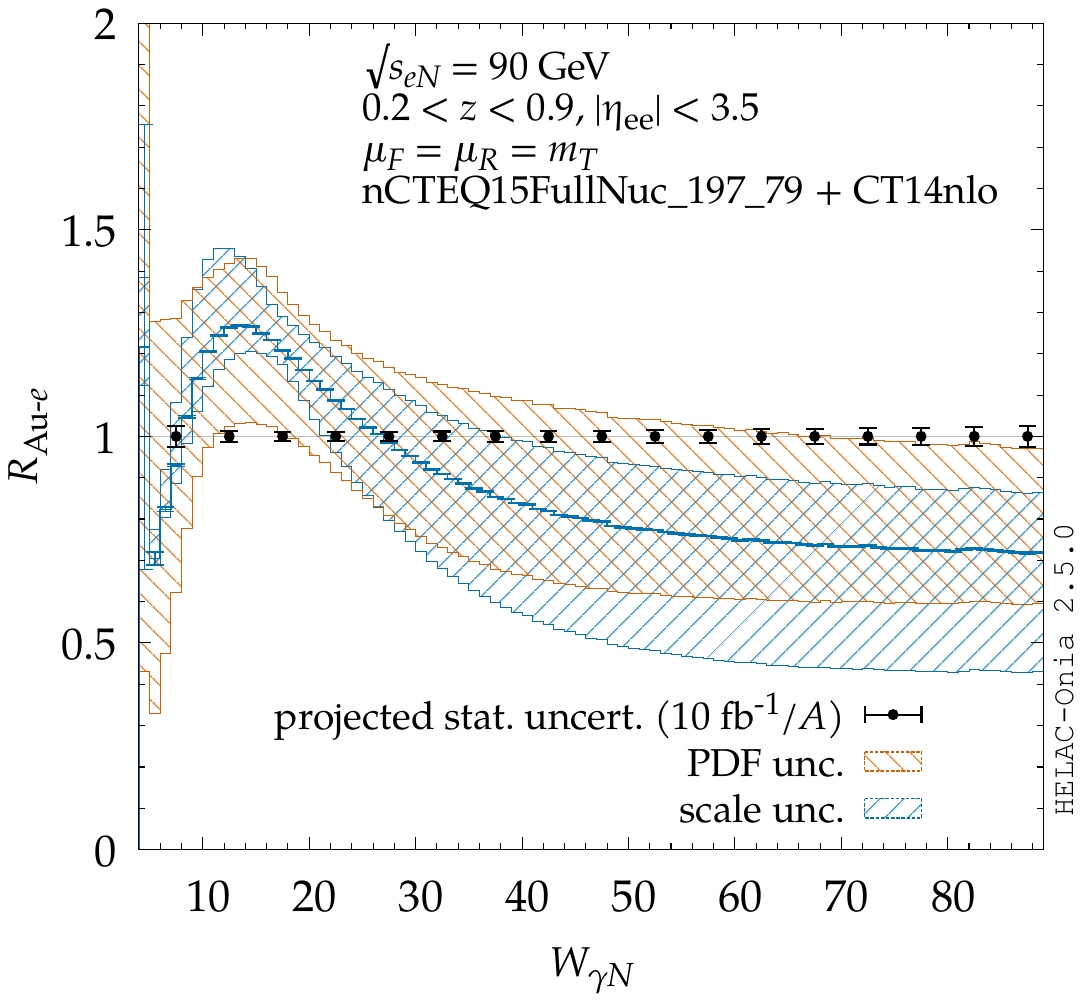}
\caption{ %
{$\ReAu$} LO CSM prediction at %
{$\sqrtseN=90$}~GeV as a function of the $J/\psi$ rapidity in the %
{$\gamma N$} %
{centre-}of{-}mass frame (left panels) and as a function of $W_{\gamma N}$ (right panels). Calculations are based on HELAC-Onia~\cite{Shao:2012iz,Shao:2015vga} with the cuts: $0.2 < z < 0.9$, $|\eta_{ee}| < 3.5$ and the nPDFs EPPS16NLO+CT14nlo (top plots) and nCTEQ15FullNuc+CT14nlo (lower plots). Projections are calculated assuming %
{$\ReAu=1$} and for an integrated luminosity of {10 fb$^{-1}/A$}.
}
\label{fig:R_eAu-90GeV}
\end{figure}

Some comments are in order. First, {as  can be seen in \cf{fig:R_eAu-45GeV} and \cf{fig:R_eAu-90GeV},}  %
$J/\psi$ is expected to be mostly produced in the backward region in the %
{$\gamma N$ centre-of-mass}~frame %
{as the yield essentially vanishes at positive rapidities (see the increase of the statistical uncertainties of our projections)}. This happens for both energy configurations.  {Second,} %
the regions where shadowing {(relative parton depletion at $x$ smaller than 0.01)}, antishadowing  {(relative parton excess at $x$ around 0.11)} and the EMC effect {(relative parton depletion for $ 0.3 < x < 0.7$)} take place %
{can be} probed at the %
EIC {via $J/\psi$ photoproduction}. The antishadowing peak is expected to be observed at moderate backward rapidity in the %
{$\gamma N$ centre-of-mass}~frame, while the shadowing region would be probed at larger negative rapidities. Such regions are also %
{those} where the projections point to a smaller {statistical} uncertainty compared to the PDF and scale {uncertainties}, %
 \textit{i.e.}~{the gluon nPDFs would be the most constrained}. %
 The $W_{\gamma N}$ dependence of the nuclear modification factor would also be a very interesting tool to probe gluon nPDFs. A large shadowing tail is expected to be probed for larger values of $W_{\gamma N}$, while clear antishadowing peaks are expected in the region $W_{\gamma N} \in [10:20]$ GeV, in both energy configurations. The projected uncertainties are also %
 small, and seem to have an interesting constraining power for the gluon nPDFs. More detailed dedicated studies are surely required and would help in motivating new measurements to probe gluon nPDFs at the %
 EIC.

\cf{fig:jpsi-pt-eAu-eic} presents predictions for %
{the \pT dependence of \ReAu at $\sqrtseN=100$~GeV} %
by using the same factori{s}ation formalism in \ce{eq:jpsi-lp-fac}, with proton PDFs replaced by nuclear PDFs for {the} $eA$ collision. {The total, LP and NLP contributions are shown.}  %
The EPPS21nlo central set~\cite{Eskola:2021nhw} is used as nPDF. %
Since the production rate is dominated by the $\gamma+g\to [c\bar{c}] + g$ subprocess, this ratio is directly sensitive to the nuclear dependence of the gluon PDF. {At EIC energies,} %
the \pT distribution of \jpsi production is sensitive to the gluon at a relatively large momentum fraction due to the soft{-}photon distribution in %
{the} incoming electron. The enhancement of {the} \jpsi production rate in %
{$e\rm{Au}$} over $ep$ collisions in \cf{fig:jpsi-pt-eAu-eic} is a direct consequence of the ``antishadowing" behavior of {the} nuclear gluon distribution from the EPPS21nlo nuclear PDF set.  Since the quark-initiated subprocesses dominate the LP contribution, the ratio of {the} LP contribution (blue dashed and red dotted lines) shows the well-known EMC-type effect from nuclear quark PDFs. However, this feature of the LP contribution does not have a real impact on the observed nuclear dependence of the \pT dependence of \jpsi production at the EIC energies (the solid line), since the LP contribution is strongly suppressed; that is, the \pT distribution of \jpsi production at the EIC should {also} be an excellent observable for probing the nuclear gluon PDF.

\begin{figure}[htbp]
	\begin{center}
		\includegraphics[width=0.5\textwidth]{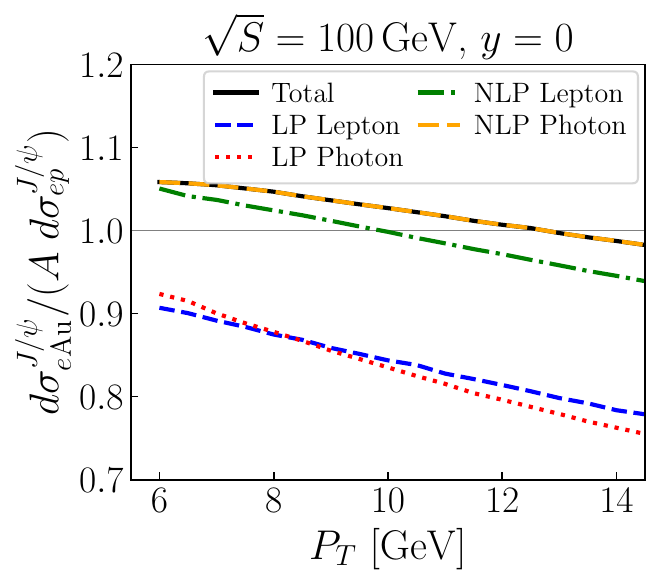}
\caption{$\ReAu$ as a function of the \jpsi transverse momentum, $P_T$, for inclusive production in electron-gold collisions %
without tagging the scattered electron, computed by using the new factorisation formalism in \ce{eq:jpsi-lp-fac}~\cite{QW:2024}.
The solid black line (overlap with the dashed orange line) %
{represents the} total contribution, which is dominated by the subprocess $\gamma+g\to [c\bar{c}] + g$ (NLP Photon) with the \ccbar pair hadonising to \jpsi, while others represent contributions from other subprocesses, see the text for details.
}
		\label{fig:jpsi-pt-eAu-eic}
	\end{center}
\end{figure}

\subsection{Nuclear GPDs}
\label{sec:eA_gpd}
\begin{figure}[h]
\centering
\includegraphics[width=0.45\textwidth]{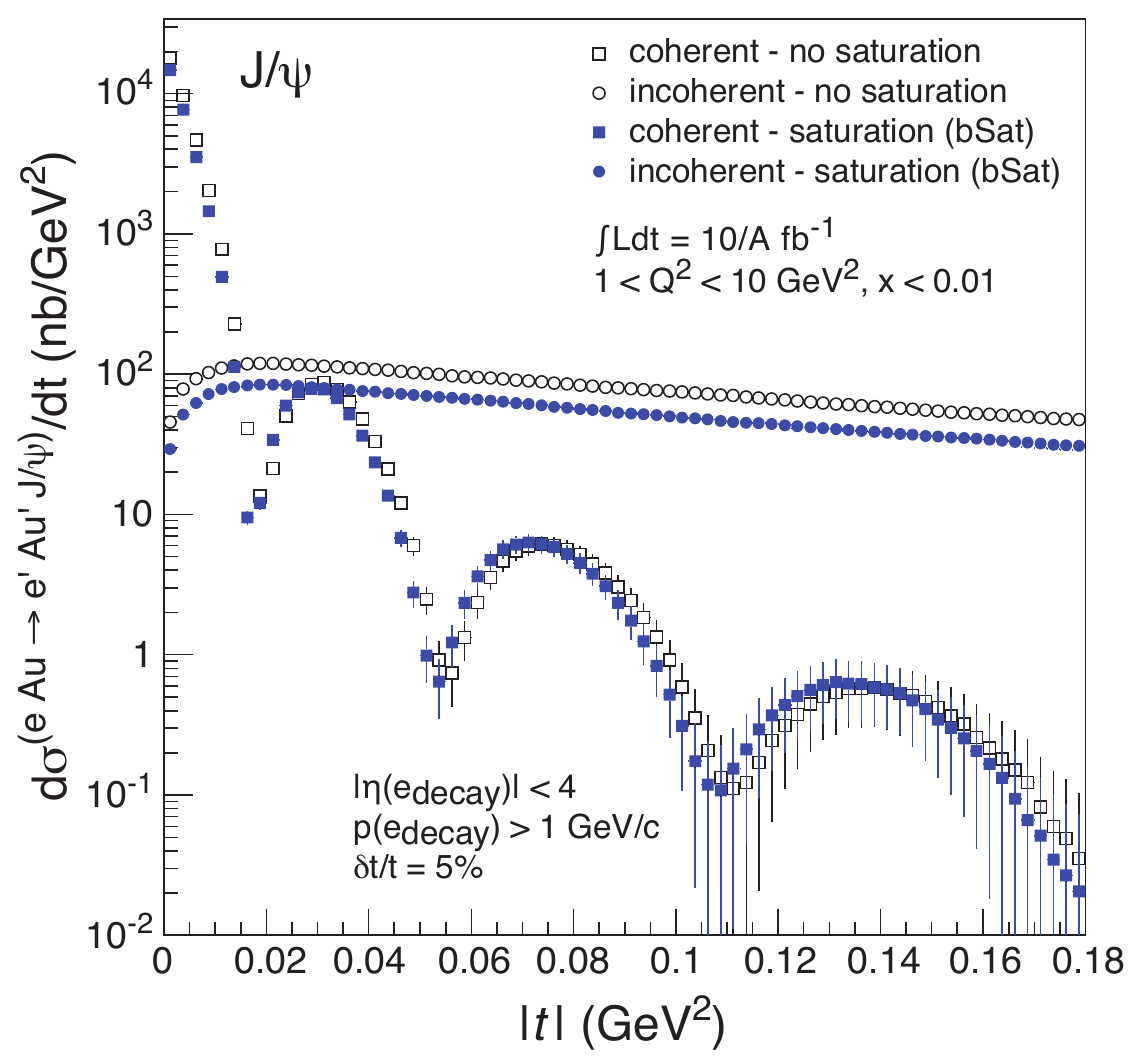}
\caption{ Simulation of the differential cross section of coherent (squares) and incoherent (circles) $J/\psi$ production in %
{$e\rm{Au}$} collisions at the EIC~\cite{Toll:2012mb}, where a 5\% resolution effect from experimental conditions is included. Predictions without saturation (open symbols) and with saturation (closed symbols) are shown.}
\label{fig:jpsi_diff}
\end{figure}

\noindent In coherent diffractive production of vector mesons off a nucleus, the light (photon) generated by the electron interacts, similarly to optical experiments of diffraction, with the nucleus as a whole, resulting in the production of a vector meson in the final state. 
This process has been proposed as a tool to investigate gluon saturation dynamics~\cite{accardi2012electron}. Here, the production of lighter vector mesons, such as the $\phi$ meson, is expected to be sensitive to saturation effects. On the other hand, the production of quarkonia would because of the heavier quarkonium mass (and thus smaller size of the dipole formed by the quark--anti-quark pair that evolves into the vector meson) not be optimal to study gluon saturation and rather serve as a baseline free from saturation effects. Diffractive production also gives access to the spatial distribution of partons inside the nucleus. While coherent diffractive production provides information on the average spatial distribution of partons, incoherent production, where the nucleus does not stay intact, probes local fluctuations of this spatial distribution~\cite{Mantysaari:2020axf}.
For the study of the spatial distribution of gluons in heavy nuclei, in particular, the diffractive production of a quarkonium, such as a $J/\psi$, is most adequate. For the coherent process, the momentum transfer distribution %
$\sqrt{|t|}$ from the photon to the target nucleus is expected to exhibit a diffractive pattern, where the details of the shape of this pattern encode information on the gluon GPD~\cite{accardi2012electron, Toll:2012mb, Kowalski:2006hc, Mantysaari:2018nng}. An example of such a diffractive pattern is shown in Fig.~\ref{fig:jpsi_diff}, as represented by the square symbols. The data points have been simulated using the Sartre Monte-Carlo event generator~\cite{Toll:2013gda}. Results including (filled symbols)  and excluding (open symbols) saturation effects are shown. In addition to the diffractive coherent production, the expected incoherent contribution (circles) is shown.
As can be seen, apart from the very low $|t|$ region, the incoherent contribution dominates
the coherent one.

Elastic and inelastic diffractive quarkonium production off the proton has been studied at the HERA lepton-proton collider experiments H1~\cite{Alexa:2013xxa,H1:2005dtp,H1:2003ksk} and  ZEUS~\cite{Alexa:2013xxa,ZEUS:2012qog,Chekanov:2009ab,ZEUS:2004yeh}, while a first measurement of exclusive $J/\psi$ photoproduction at threshold has been performed in the fixed-target experiment GlueX at Jefferson Lab~\cite{GlueX:2019mkq}.
At hadron-collider experiments, diffractive quarkonium production has been investigated in \ppbar collisions~\cite{CDF:2009xey} at the Tevatron, in \pp~\cite{Aaij:2013jxj,LHCb:2014acg,LHCb:2018rcm,LHCb:2015wlx}, \pPb~\cite{ALICE:2014eof} and \PbPb~\cite{Abelev:2012ba, ALICE:2021gpt,Khachatryan:2016qhq,LHCb:2021hoq} collisions at the LHC and in %
{$d\rm{Au}$}~\cite{STAR:2021wwq} and %
{$\rm{AuAu}$}~\cite{PHENIX:2009xtn} collisions at RHIC.
The existing measurements off nuclei are at present restricted in statistical precision, while only offering a rough determination of the momentum transfer  $\sqrt{|t|}$ and in general a limited separation of coherent and incoherent production. Hence, the knowledge on the gluonic structure of nuclei is at present poor, with many fundamental questions unanswered.

\begin{figure}
\centering
\includegraphics[width=0.47\textwidth]{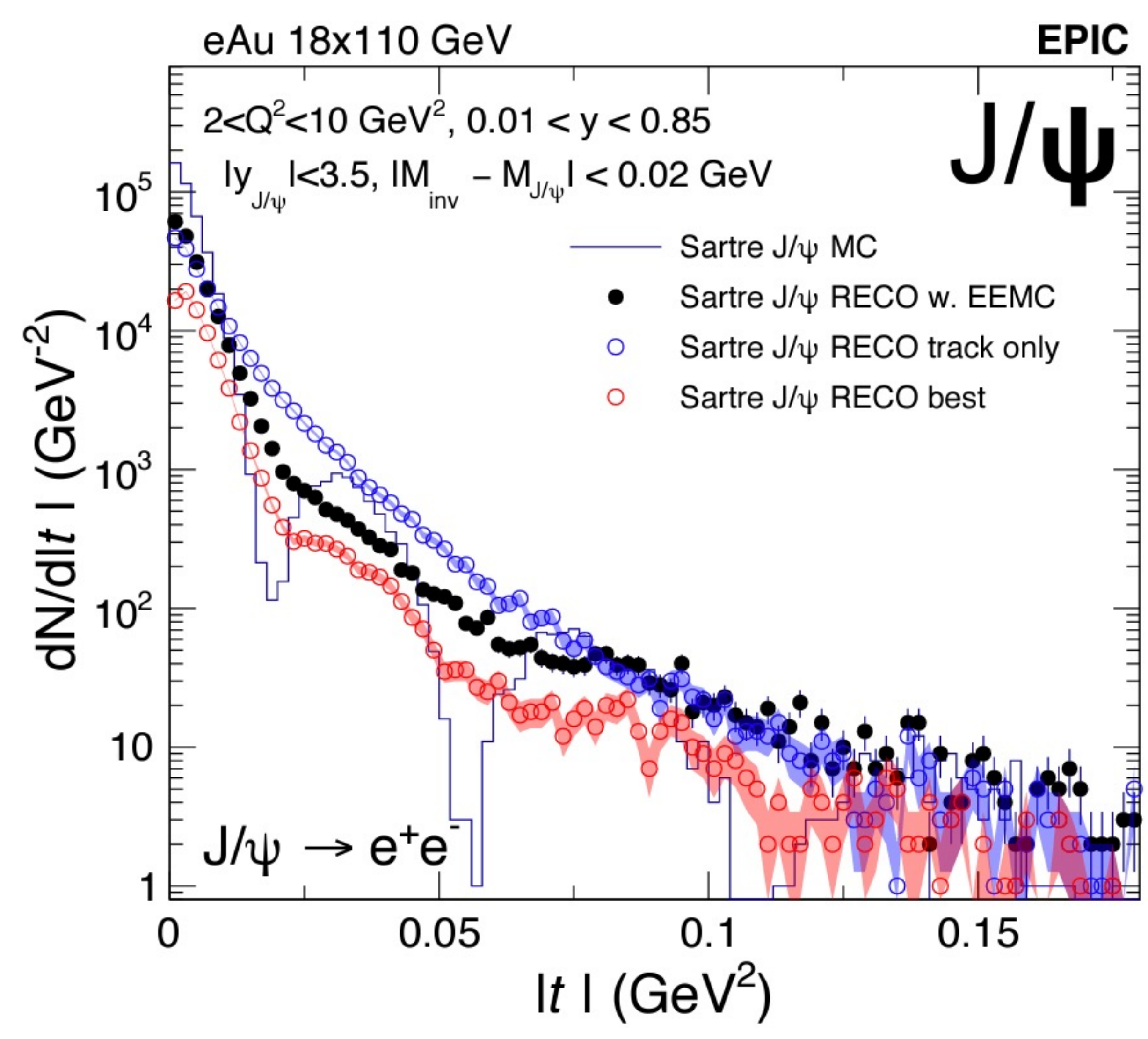}
\hspace*{0.5cm}
\includegraphics[width=0.45\textwidth]{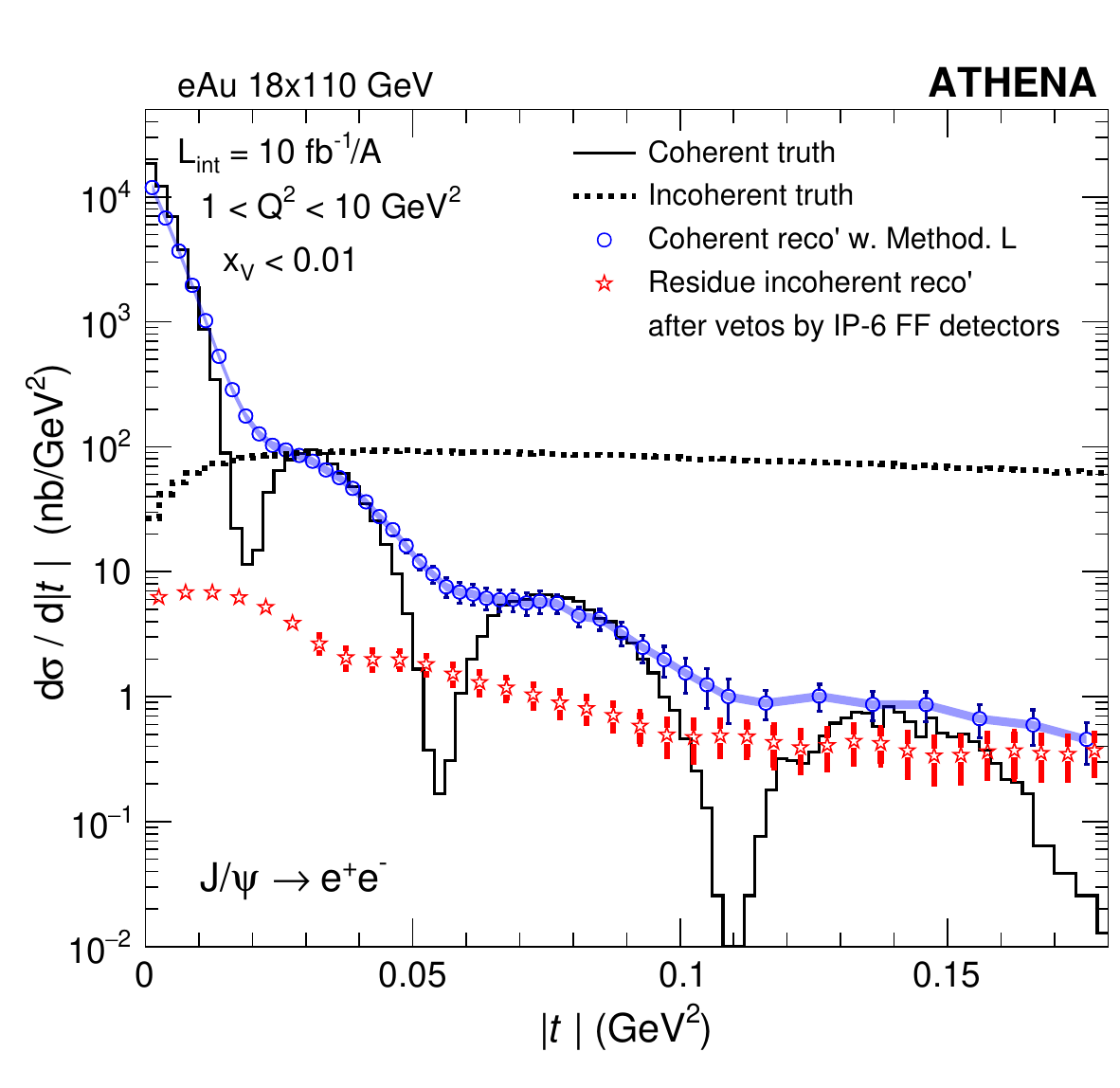}
\caption{The distribution in generated and reconstructed $-t$, with the reconstructed $-t$ being the squared sum of the transverse momenta of the scattered beam lepton and of the lepton pair originating from $J/\psi$ decay, in diffractive production off gold nuclei. The panel on the left-hand side illustrates the influence of the quality of the scattered-lepton reconstruction on the determination of $-t$, as studied by ePIC. The panel on the right-hand side shows the level of suppression of incoherent production (see text), as studied by ATHENA. Figs. taken from Ref.~\cite{dis:Kong} and from the supplementary material provided in the evaluation process of~\cite{athena}, respectively.}
\label{fig:t_epic}
\end{figure}

The EIC is expected to perform measurements of diffractive vector-meson production off light and nuclear ions with unprecedented precision. The two experimental challenges consist in determining $t$ with high precision and in
distinguishing coherent from incoherent events~\cite{Chang:2021jnu}. Recently, the capability of proposed EIC detectors in reconstructing $t$ and
their ability to suppress incoherent production have been examined~\cite{Chang:2021jnu},~\cite{athena},~\cite{ecce}.
The variable $t$ needs to be reconstructed from the scattered lepton and reconstructed vector meson, since in coherent production the trajectory of the ion after the interaction is nearly unmodified and thus the ion cannot be detected, while in the case of incoherent production not all fragments from the nuclear break up can be detected. The distribution in $|t|$ for coherent diffractive $J/\psi$ production off gold ions is shown in  Fig.~\ref{fig:t_epic}, left.
Here, $|t|$ is reconstructed  as the squared sum of the transverse momenta of the scattered lepton and of the lepton pair originating from the $J/\psi$ decay.
It forms a good approximation for the true $-t$.
The data have been simulated again with Sartre and subsequently passed through a full simulation of the ePIC detector.
The histogram represented by the continuous line is the generated distribution, while the other curves represent the reconstructed distribution, with beam effects. 
The latter include an angular divergence originating from the focussing and defocussing quadrupoles in the interaction region and
a small angular kick from the crab cavities. The crossing angle from the beams in principle also influences the $t$ distribution, but contrary
to the other effects it can be corrected for.
For the curve indicated by the open, blue circles only information from tracking detectors is used for the reconstruction of the scattered lepton, while for the curve indicated by the black, closed circles only information from the backward electromagnetic calorimeter is used for the reconstruction of the scattered lepton. The curve indicated by the red, open circles selects the best of the two methods. As can be seen, the quality of the reconstruction in $t$ is strongly dependent on the quality of the reconstruction of the scattered beam lepton. In the diffractive process the beam lepton generally is scattered under a small angle and covers a region where the tracking performance is degraded. Using in addition the electromagnetic calorimeter in the backward region for the reconstruction of the scattered lepton improves the reconstruction in $t$ vastly.

The spatial distribution of partons in impact-parameter space is
related to a Fourier transformation, with $t$ going from 0 to infinity~\cite{Burkardt:2002hr}. Experimentally, one is limited by a maximal momentum transfer, which preferably extends as far as possible. In practice,
studies have shown that it is necessary to resolve the minima up to the third one for the evaluation of the spatial distribution~\cite{AbdulKhalek:2021gbh}. This dictates the needed level of suppression
of the incoherent contribution. The suppression of incoherent events includes the requirement of exactly three reconstructed lepton tracks with the correct charge in absence of any other signal in the main detector and various criteria corresponding to the absence of signal in a series of far-forward detectors, which can tag protons (Roman Pots for protons with energy close to the beam energy and the B0 spectrometer and off-momentum detectors for nuclear-breakup protons), neutrons (Zero-Degree Calorimeters) and photons (B0 and Zero-Degree Calorimeters). The capability to suppress incoherent production is illustrated in Fig.~\ref{fig:t_epic}, right, which shows the $-t$ distribution for coherent and incoherent production off gold nuclei.
The former is again simulated using Sartre, while for the latter the BeAGLE generator~\cite{beagle} is used. 
The generated coherent (incoherent) contribution is represented by the continuous (dotted) line. The generated data are passed through 
a full simulation of the ATHENA detector. The effect of data selection requirements  on the event activity in the main detector and on the absence of activity in the far-forward detectors,
based on the studies in Ref.~\cite{Chang:2021jnu}, is represented by the
blue, open circles. As can been seen, the obtained distribution lies close to the distribution from coherent events simulated by Sartre. The remaining contribution from incoherent events is given by the red, star symbols. The largest suppression of the incoherent process comes from the requirement on the absence of any neutron signal in the Zero-Degree Calorimeter, while the requirement on the absence of photon signals in this Zero-Degree Calorimeter 
also has an impact. 
Ways to further improve the reconstruction of $t$ and the suppression of incoherent production are at present under investigation.

The study of light nuclei can offer additional insights into the %
{internal} structure of the nuclear medium.
{In contrast to} %
measurements with heavy nuclei, the total final state in incoherent diffractive production off light nuclei can be
unambiguously identified through tagging of the spectator nucleons.
Such measurements are of interest when studying the short-range correlation (SRC) of a nucleon pair, which is the temporal fluctuation of two nucleons into a strongly interacting pair in close proximity {and large measured relative momentum} \cite{Frankfurt:1993sp,CLAS:2018qpc}.
SRC pairs are suggested as a possible explanation for the nuclear modification of the {momentum} distribution of high-$x$ partons, {known as} the EMC effect,
with a strong correlation between the two phenomena suggested by measurements by the CLAS experiment at Jefferson Lab~\cite{CLAS:2019vsb} {and a quark-level QCD basis for SRC has been proposed for the lightest nuclei \cite{West:2020tyo} and $A\geq4$ nuclei \cite{West:2020rlk}.}

The simplest nuclear system consists of deuteron and the first measurement of incoherent diffractive production with spectator tagging was performed in the measurement of incoherent diffractive $J/\psi$ production in ultra-peripheral %
{$d\rm{Au}$} collisions by the STAR experiment at RHIC~\cite{STAR:2021wwq}, with tagging of the spectator neutron in the Zero-Degree Calorimeter.
At the EIC, similar measurements can be performed with enhanced precision, and studies of incoherent diffractive $J/\psi$ production off the deuteron at the EIC have been proposed to study the nuclear modification of the
gluon distribution and its possible link with the SRC~\cite{Mantysaari:2019jhh,Tu:2020ymk}.
For the proposed measurement, the scattered lepton and $J/\psi$ decay leptons are reconstructed in the main detector, while both the leading
and spectator nucleon (neutron and proton) can be detected in the far-forward detectors. The detection of both nucleons instead of {only} one %
offers certain advantages in the reconstruction of the event and some
kinematic variables~\cite{Tu:2020ymk}.

\begin{figure}[h]
\centering
\includegraphics[width=7cm]{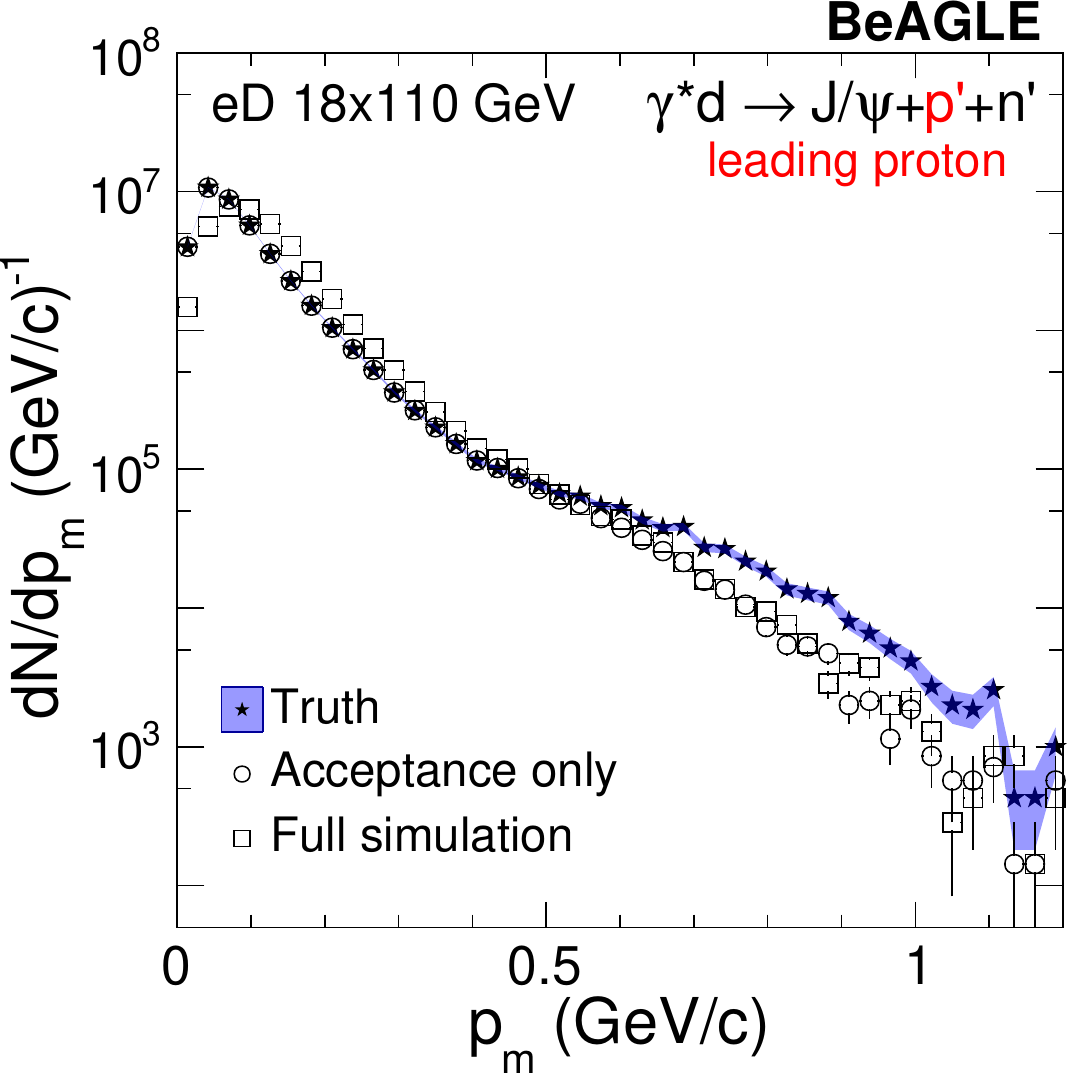}
\includegraphics[width=7cm]{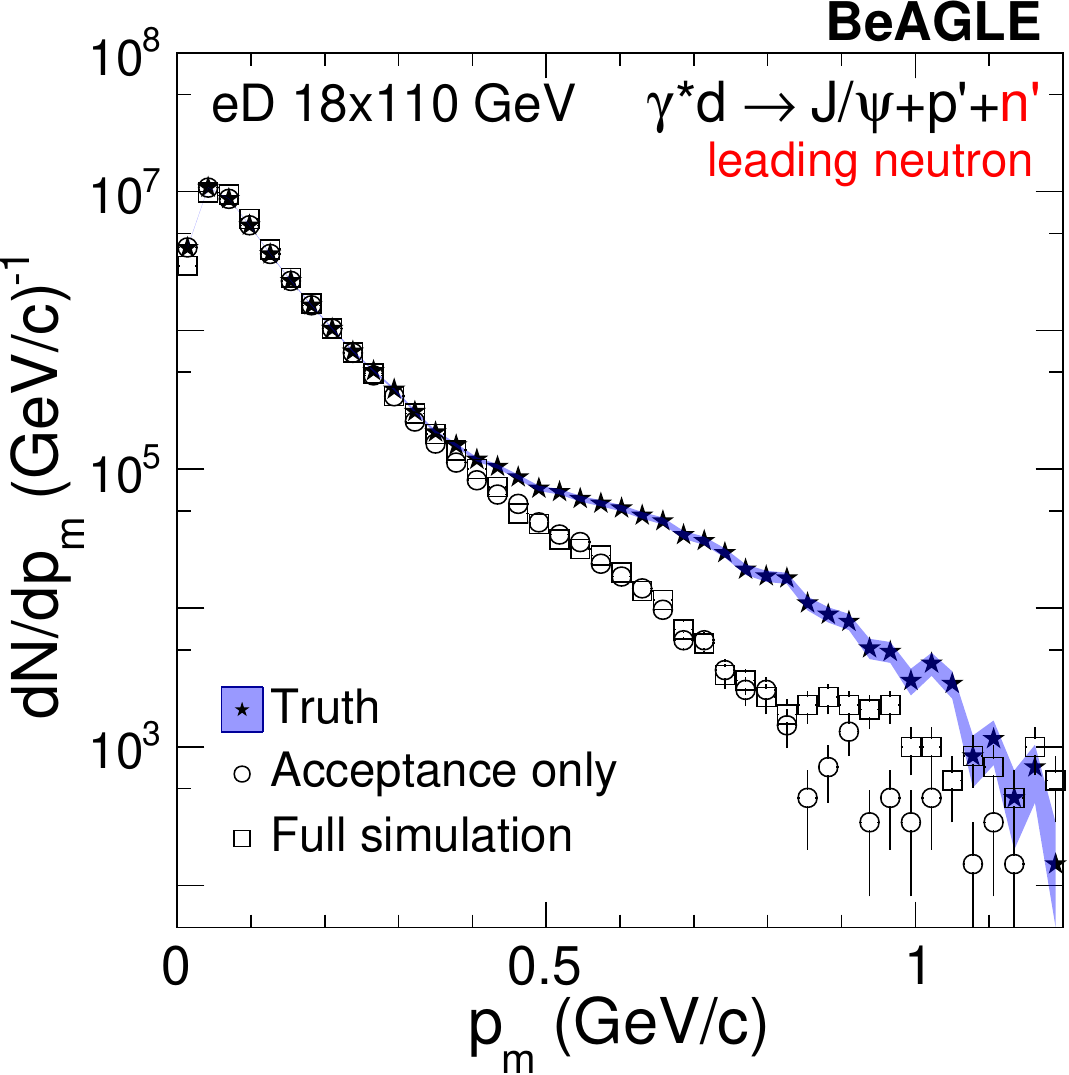}
\caption{\label{fig:eD_beagle_figure_02} The three-momentum distribution in the deuteron rest frame of the spectator neutron (left) and
  spectator proton (right) for the incoherent diffractive production of $J/\psi$ in lepton-deuteron collisions at the EIC.
  The distribution is generated with BeAGLE. The star symbols represent the generated distribution, the open circles represent the distribution
including acceptance effects of the main and far-forward detectors, and the open squares take in addition the finite
detector resolution and beam effects into account. Figures are taken from Ref.~\cite{Tu:2020ymk}. }
\label{Figures}
\end{figure}

In Fig.~\ref{fig:eD_beagle_figure_02}, the three-momentum distribution of the tagged neutron (left) and tagged proton (right) in the deuterium rest frame is illustrated for incoherent diffractive production of $J/\psi$ in the scattering of 18 GeV electrons off 110 GeV deuterons at the EIC, as
simulated with BeAGLE~\cite{Tu:2020ymk}.
The star symbols represent the generated distribution, the open circles represent the distribution including acceptance effects of the main and far-forward detectors, and the open squares {also} take %
the finite
detector resolution and beam effects into account. The momentum distribution of the tagged nucleon reflects the initial-state momentum
of the nucleons inside the deuteron.
The region above 300 MeV corresponds to the region of the SRC, and as visible in the figures, the EIC will be able to provide a good reconstruction of the tagged-nucleon momentum. A similar statement holds for the reconstruction of
other variables of interest~\cite{Tu:2020ymk}.

\subsection{Study of transport properties of nuclear matter}
\label{sec:eA_transport}

The vital element of portraying nuclear matter is {to get} information on how the medium responds to a parton traversing the matter. It is characterised by transport coefficients, %
{ \eg~}a diffusion coefficient %
or %
$\hat{q}$, which is the mean squared momentum transfer between the propagating particle and the medium per unit length. Transport coefficients are an essential ingredient in the modelling of nuclear reactions %
{, and d}etermining these parameters is one of the main goals of high-energy nuclear physics experimental and phenomenological efforts.

Measurements of hadron production in \pA\ collisions have shown a broadening of the transverse momentum distribution at %
intermediate %
{hadron transverse momentum} compared to \pp\ reactions. This phenomenon is visible over a wide range of hadronic collision energies, starting from collisions at $\sqrtsnn \approx 20$~GeV~\cite{Cronin:1974zm,Antreasyan:1978cw} up to 200~GeV at RHIC~\cite{PHENIX:2021dod}. The Cronin effect is also anticipated for quarkonium production in \pA\ collisions~\cite{Andronic:2015wma}. A similar effect was observed also in semi-inclusive deep-inelastic scattering off nuclei by the HERMES experiment~\cite{HERMES:2007plz}. One possible source of this effect is the multiple scattering of the struck parton while traversing the nucleus, which broadens the parton momentum \kT. Under this assumption, the modification of \kT can be related to the transport properties of matter, expressed by the transport coefficient $\hat{q}$.  %
{O}ther effects, like nuclear absorption and parton energy loss, are also expected to contribute {when studying particle production in nuclear matter}. 

Additional measurements of the \pT\ spectrum in %
{$ep$} and %
{$eA$} at the EIC can help to discriminate between models and constrain their parameters, including the relative role of multiple scattering and nuclear absorption. Such a program{me} %
{will} greatly extend the studies pioneered by the HERMES collaboration.

We present {here} an example of the calculation of the expected modification of the quarkonium energy spectrum in %
{$eA$} collisions due to multiple scattering of the parton in the medium. The study is based on an earlier work~\cite{Aronson:2017ymv}, where a microscopic approach was adopted for the calculation of the decay of $J/\psi$ and $\Upsilon$ in the QGP. Here the QGP medium is replaced with cold nuclear matter, specifically with a large gold %
nucleus, and its properties are constrained taking into account various nuclear effects: nuclear shadowing~\cite{Qiu:2003vd,Qiu:2004qk}, coherent QCD multiple scattering~\cite{Qiu:2004da}, initial- and final-state parton energy loss~\cite{Neufeld:2010dz, Kang:2015mta}, and initial and final scattering effects (including multiple scattering)~\cite{Kang:2014hha,Kang:2011bp}.  

To study the nuclear modification, the ratio of cross sections for quarkonium production in reactions that involve a nucleus %
and a proton baseline is used:
\begin{equation}
R_{AA} = \frac{1}{\langle N_{\rm bin} \rangle} \frac{d\sigma_{AA}}{ d\sigma_{pp}} \, ,  \qquad  
R_{eA} =  \frac{1}{A} \frac{d\sigma_{eA}}{ d\sigma_{ep}} \, . 
\label{eq:Rdefinition}
\end{equation}
Here, $A$ and the average number of nucleon-nucleon collisions $\langle N_{\rm bin} \rangle$ provide the relevant normalisation factors such that in the absence of  nuclear modification the ratios are unity. The $R_{AA}$ presents suppression {from QGP}, including thermal dissociation in the QGP, while $R_{eA}$ offers the cold nuclear-matter counterpart.  

A preliminary study demonstrates that most %
quarkonium states show a larger $R_{eA}$ compared to $R_{AA}$, and thus a decreased suppression{, with the exception of the \jpsi state, which sees an increase in suppression by roughly 20\%, and of the $\chi_b(1P)$ state, which sees a relatively low increase in suppression of roughly 10{\%}}. %
The $\chi_c$ state experiences a significant decrease of {about} 50{\%} %
in the suppression factor. The %
{$\Upsilon$} states follow %
{an analogous trend}, with { decreased suppression of around 25{\%} for $\Upsilon(1S)$ and $\Upsilon(2S)$ and 90{\%} for $\Upsilon(3S)$.} {Finally, $\chi_b(2P)$ and  $\chi_b(3P)$ show decreases in their suppression factors of roughly 55\% and >95\%, respectively.} %
The overall trend seems to indicate that highly suppressed states see the largest decrease in suppression, while the least suppressed states show either a small decrease or a slight increase in their suppression factors. All states retain a similar amount of $E$ dependence, which is not surprising given that it is assumed that the time for the onset of the interaction is $\tau_{\rm form.}=1$~fm. We direct an interested reader to \ref{sec:eA_transport:plots} for more details. 

These preliminary results show that one can expect a significant modification due to cold nuclear{-}matter effects, which should allow for experimental investigation of these effects at the EIC. Thus, %
quarkonium studies in %
{$eA$} collisions at the EIC will help to understand the impact of different transport coefficients on quarkonium production in reactions that involve heavy nuclei and, in turn, help to calibrate quarkonium as a probe of the properties of matter created in high-energy %
{$pA$} and {$AA$} %
collisions.

\section{Summary}\label{sec:summary}

Quarkonium is an extremely useful tool to probe the internal structure of matter, namely one of the main goals of the Electron Ion Collider. In this review, we argue that studies of quarkonium production and correlations in (polarised) electron-proton and electron-nucleus collisions can produce unprecedented insights into the 3D structure of the nucleon and into the partonic content of the nuclei as well as help to settle the long-lasting debate on how quarkonia form.

Section \ref{sec:eh_general} briefly introduced the EIC project, its key parameters, and requirements for an EIC detector. We also defined conventions and basic kinematical quantities useful for describing lepton-hadron reactions. Finally, we made a case for a muon detector for quarkonium studies at the EIC.

Studies of collinear PDFs, form factors, TMD PDFs, GPDs, GTMDs and even double-parton distribution functions can be done at EIC using quarkonium production on a nucleon. In Sections \ref{sec:why_quarkonia}, \ref{sec:ep} and \ref{sec:eA}, we reviewed the physics case for quarkonium measurements at the EIC. Quarkonium production at large transverse momenta in proton-proton and electron-proton collisions has been studied extensively within the frameworks of NRQCD and collinear factorisation. As discussed in Sections 3.1 and 3.2, it remains a challenge to obtain a simultaneous description of all HERA, LHC and Tevatron data for $J/\psi$ photo- and hadroproduction, $\eta_c$ hadroproduction, $J/\psi+Z$ hadroproduction, $J/\psi$ polarisation as well as inclusive production in $e^+e^-$ annihilation at $B$ factories. 

Further data from the EIC can help but its $p_T$ reach is limited to 10-15 GeV for charmonia and much less for bottomonia. The focus would then be on low-$p_T$ data. The latter needs to be described within the framework of transverse momentum dependent parton distributions (TMDs) and requires the inclusion of so-called shape functions, which are the subjects of Section 3.3. In this way the EIC will provide new data to further unravel the quarkonium production mechanism, while at the same time offer new ways to employ quarkonium production as a tool to study TMDs and other parton distributions (the subjects of Section 4). This applies especially to gluon TMDs about which currently very little is known. 
Analogous studies can be performed in electron-nucleus collisions (including, among others, insights into transport properties of nuclear matter), which is the subject of section 5. $J/\psi$ polarisation studies can be done, as well as various spin asymmetry measurements, where the electron, proton and light nuclei can be polarised.
All these observables can contribute to our understanding of hadron structure and hadron formation, in particular those involving heavy quarks.

Overall, the physics case for quarkonium physics at the EIC is very extensive and promising.

\section*{Acknowledgements}
We thank M. Chithirasreemadam,  M.A. Ozcelik and H.F. Zhang for useful comments and inputs.
This project has received funding from the European Union’s Horizon 2020 research and innovation programme under the grant agreement No.824093 (STRONG-2020). 
This project has also received funding from the French ANR under the grant ANR-20-CE31-0015 (``PrecisOnium'').
This work was also partly supported by the French CNRS via the IN2P3 project GLUE@NLO, via the Franco-Chinese LIA FCPPL (Quarkonium4AFTER), via the IEA No.205210 (``GlueGraph") and ``Excitonium'', by the Paris-Saclay U. via the P2I Department and by the GLUODYNAMICS project funded by the "P2IO LabEx (ANR-10-LABX-0038)" in the framework "Investissements d’Avenir" (ANR-11-IDEX-0003-01) managed by the Agence Nationale de la Recherche (ANR), France.
C.V.H.\ has received funding from the European Union's Horizon 2020 research and innovation programme under the Marie Sklodowska--Curie grant agreement No 792684 and from the programme Atracción de Talento, Comunidad de Madrid (Spain), 
under the grant agreement No 2020-T1/TIC-20295.
{M.N. has been supported by the Marie Sk{\l}odowska-Curie action ``RadCor4HEF'' under grant agreement No.~101065263.}
The work of U.D. and C.P. is supported by Fondazione di Sardegna under the projects “Quarkonium at LHC energies”, No. F71I17000160002 (University of Cagliari) and  ''Proton tomography at the LHC”, No. F72F20000220007 (University of Cagliari).  The work of C.F.~and C.P. is supported by the European Union ''Next Generation EU" program through the Italian PRIN 2022 grant n. 20225ZHA7W.
D.K. was supported by the National Science Centre, Poland, under the research grant no. 2018/30/E/ST2/00089.
P.T. is supported by a postdoctoral fellowship fundamental research of the Research Foundation Flanders (FWO) no.~1233422N.
The work of X.Y. was supported by the U.S. Department of Energy, Office of Science, Office of Nuclear Physics grant DE-SC0011090 and currently by the U.S. Department of Energy, Office of Science, Office of Nuclear Physics, InQubator for Quantum Simulation (IQuS) (https://iqus.uw.edu) under Award Number DOE (NP) Award DE-SC0020970 via the program on Quantum Horizons: QIS Research and Innovation for Nuclear Science.
The work of V.C. and R.V. was supported by the Office of Nuclear Physics in the U.S. Department of Energy at Lawrence Livermore National Laboratory under Contract DE-AC52-07NA27344 and the LLNL-LDRD Program under Project No. 21-LW-034 and No. 23-LW-036.
The work of I.V. was supported by the Laboratory Directed Research and Development program at Los Alamos National Laboratory. The work of C.S. was supported by the Indonesia Endowment Fund for Education (LPDP). 
S.~B. and Y.~H. are supported by the U.S. Department of Energy under Contract No. DE-SC0012704, and also by  Laboratory Directed Research and Development (LDRD) funds from Brookhaven Science Associates.
The work of A.~M. was supported by the National Science Foundation under grant number PHY-2110472, and also by the U.S. Department of Energy, Office of Science, Office of Nuclear Physics, within the framework of the TMD Topical Collaboration. 
J.R.W. was supported by the EIC Center at Jefferson Lab, the LDRD programs of LBNL and by the U.S. Department of Energy, Office of Science, Office of Nuclear Physics, under contract number DE-AC02-05CH11231.
L.C. and S.Y. were supported by the Guangdong Major Project of Basic and Applied Basic Research under the project No. 2020A1515010794.
The work of C.E.H. was supported by the U.S. Department of Energy, Office of Science, Office of Nuclear Physics grant DE-FG02-96ER40960.
J.W.Q. is supported in part by the U.S. Department of Energy (DOE) Contract No.~DE-AC05-06OR23177, under which Jefferson Science Associates, LLC operates Jefferson Lab.
The work of M.B. was supported by the German Research Foundation DFG through Grant No. BU 3455/1-1.

\clearpage

\appendix
\section{Estimation of \jpsi measurement efficiency}
\label{sec:jpsi:reco:eff}

The \jpsi measurement efficiency is calculated using {the} decay kinematic simulated with PYTHIA8~\cite{Sjostrand:2014zea} and two cases for \newb{the} minimum transverse momentum of \newb{the} electron measurable in the experiment: $\pT^{ele} > 0.2$~\newb{GeV} for a detector with \newb{a} magnetic field $B=1.5$~T, and $\pT > 0.4$~\newb{GeV} for $B = 3$~T~\cite{AbdulKhalek:2021gbh}. The \newb{single} electron tracking efficiency is assumed to be 80\%. \cf{jpsi:efficiency} shows the efficiency as a function of \jpsi\ rapidity and \pT: it approximately constant, and for the $B = 3$~T case, there is a mild decrease of efficiency with increasing \pT due to decay kinematic. For high-$\pT$ \jpsi, one of the electrons tends to carry the majority of the momentum; thus the \pT of the other falls below the reconstruction threshold. Based on these results, we assume \jpsi measurement efficiency to be 64\%. 

\begin{figure}[hbt!]
	\centering
		\subfloat[magnetic field B = 1.5~T, $\pT^{ele} > 0.2$~{GeV} ] {\includegraphics[width=0.45\textwidth]{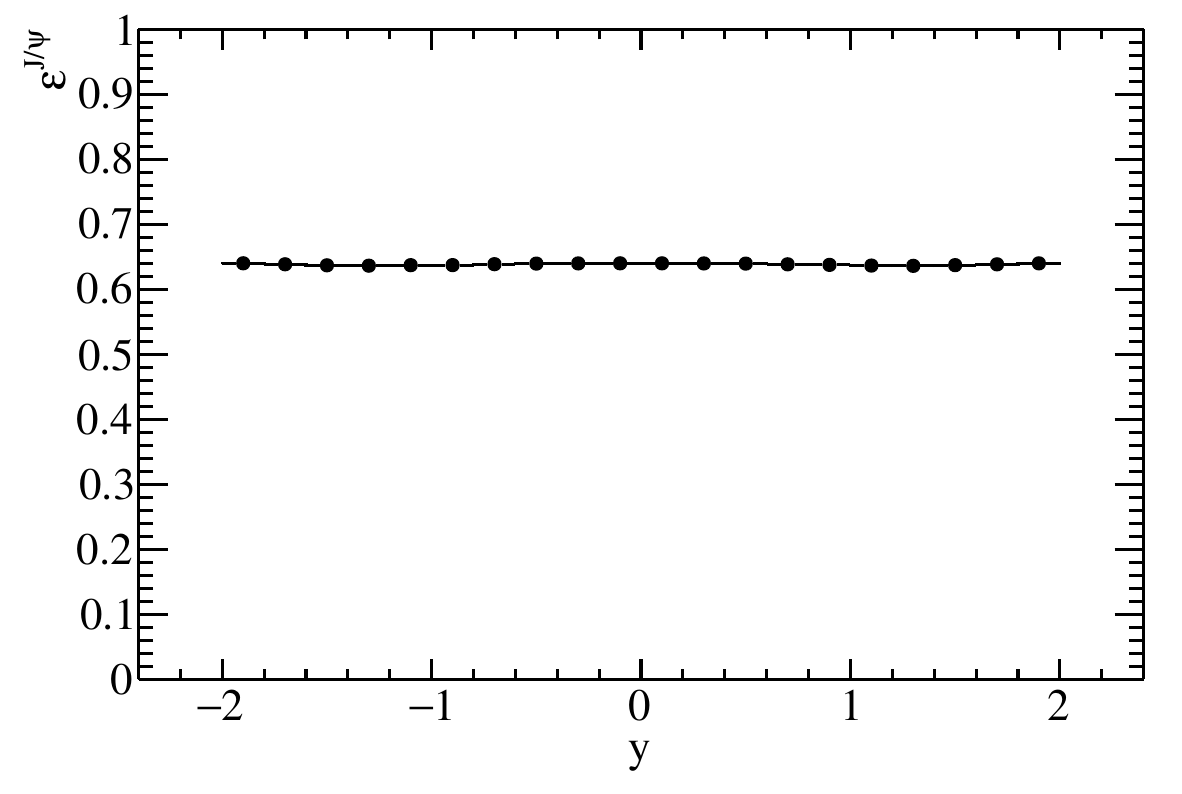}} 
		\subfloat[magnetic field B = 1.5~T, $\pT^{ele} > 0.2$~{GeV}] {\includegraphics[width=0.45\textwidth]{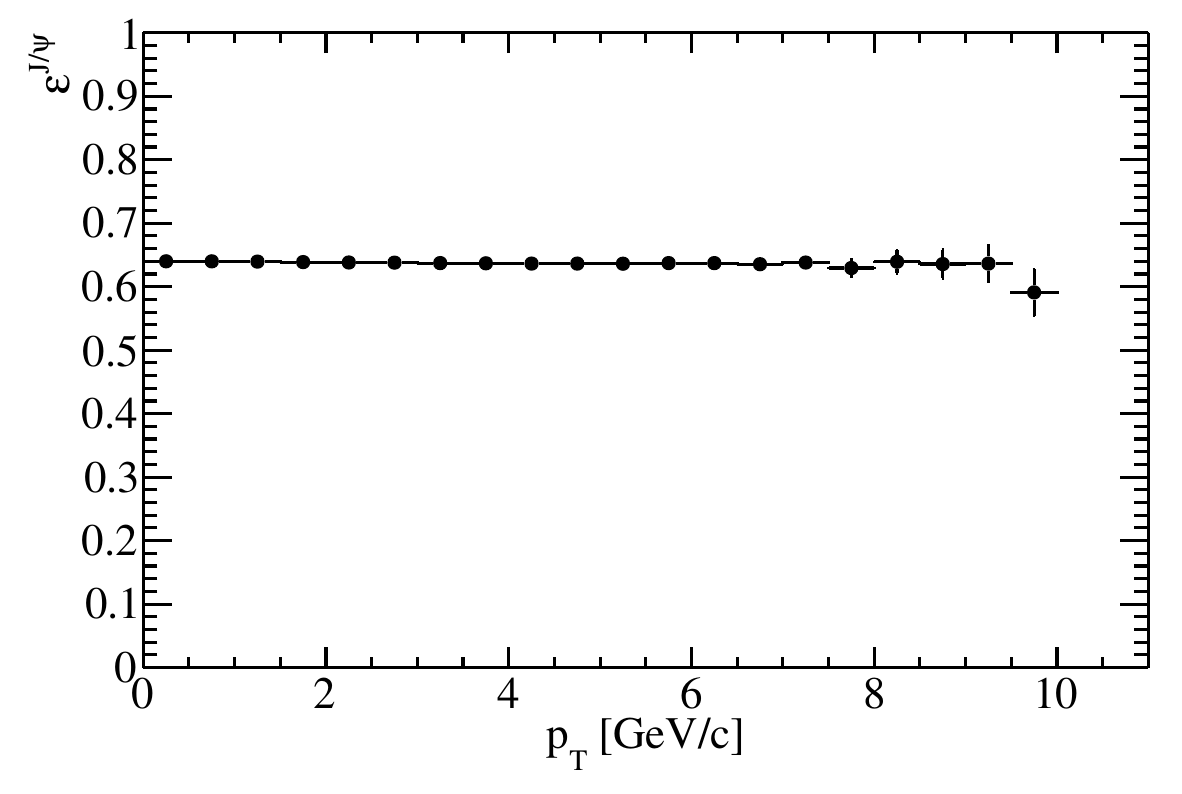}} \\
		\subfloat[magnetic field B = 3~T, $\pT^{ele} > 0.4$~{GeV}] {\includegraphics[width=0.45\textwidth]{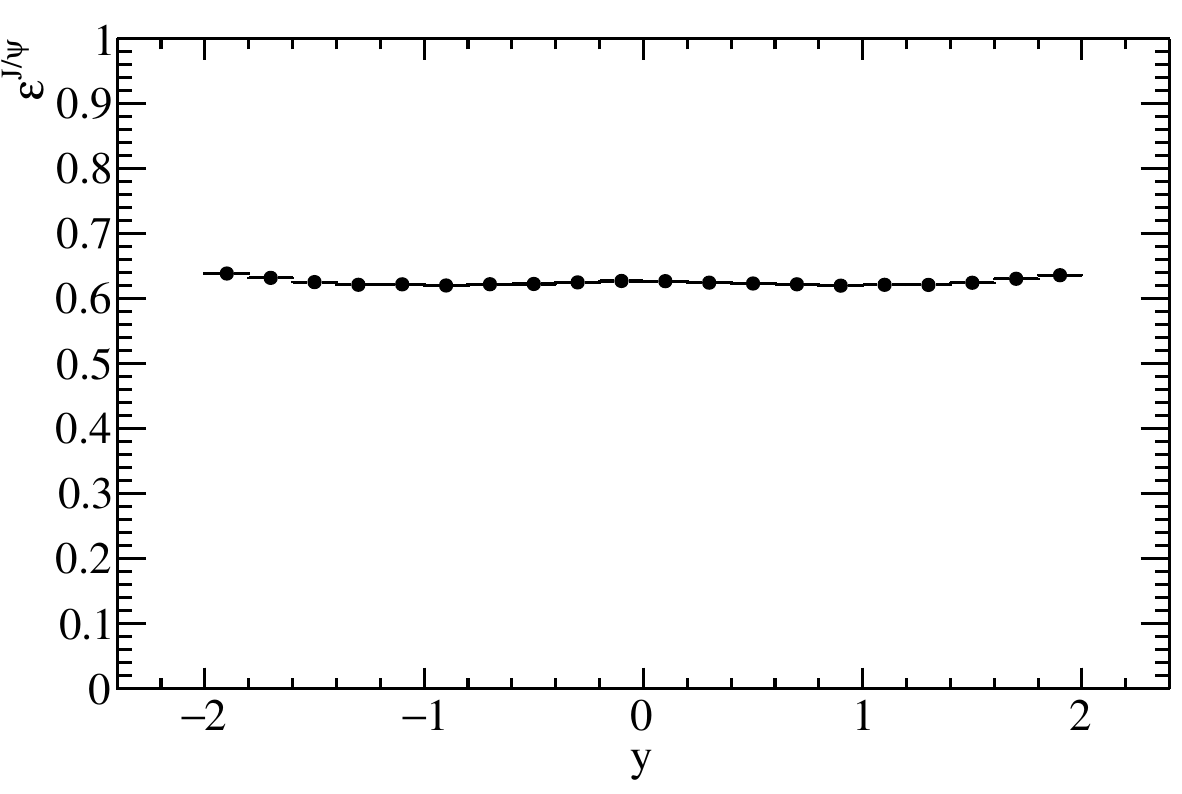}} 
		\subfloat[magnetic field B = 3~T, $\pT^{ele} > 0.4$~{GeV}] {\includegraphics[width=0.45\textwidth]{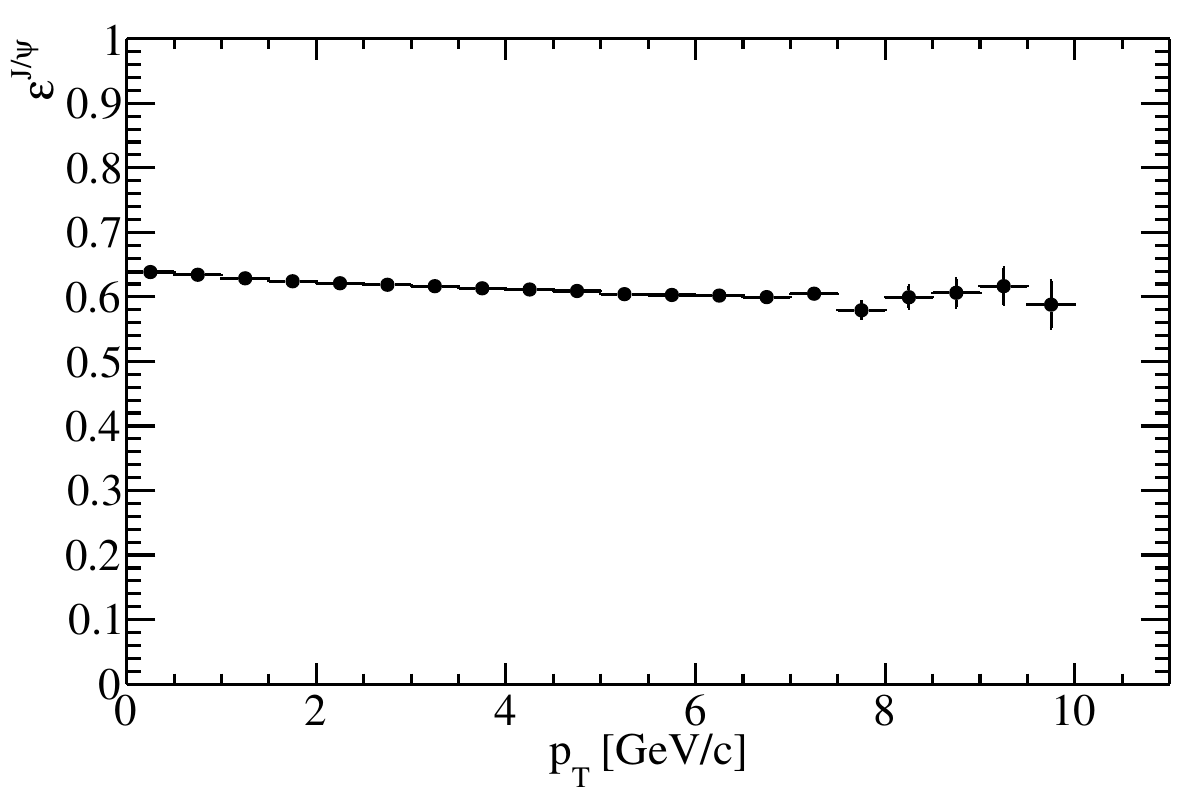}} 
	\caption{\jpsi measurement efficiency as a function of \jpsi rapidity and transverse momentum for a generic EIC detector using magetic field B = 1.5 T or B = 3 T.}
	\label{jpsi:efficiency}
\end{figure}

\clearpage
\section{Numerical results for nuclear modification  $R_{AA}$ and $R_{eA}$ for quarkonium production within the microscopic model presented in Sec.~\ref{sec:eA_transport}}
\label{sec:eA_transport:plots}

\begin{figure}[htb]
\begin{center}
\includegraphics[width=0.45\textwidth]{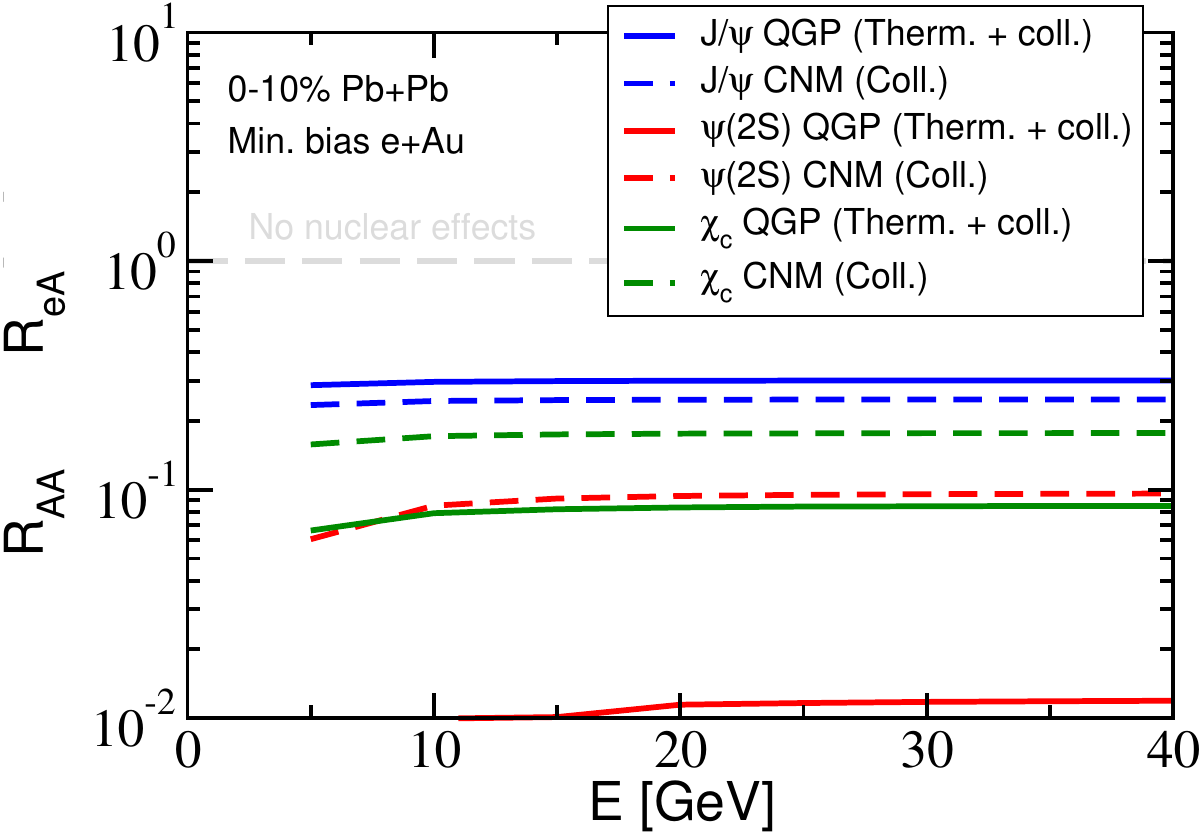}
\end{center}
\caption{Nuclear modification of the $\jpsi$, $\psi(2S)$ and $\chi_c$ states as a function of {their respective} energy $E$ in the hadron centre-of-mass frame. {The} %
{s}olid lines indicate that the calculation was done using thermal wave-function effects while traversing  the  QGP {and correspond to $R_{AA}$ in the centrality class 0-10\% in PbPb LHC collisions}. The dashed lines indicate that the calculation was done without thermal effects (only Cold Nuclear Matter (CNM) effects) \newb{and correspond to $R_{eA}$ for minimum bias $e$Au collisions}. $J/\psi$ curves are shown in blue, $\psi(2S)$ states are shown in red, and $\chi_c$ states are shown in green. All calculations are done using \newb{direct} production and ignoring feed-down effects.}
\label{fig:CharmQGPCNM}
\end{figure}

\begin{figure}[htb]
\begin{center}
\includegraphics[width=0.45\textwidth]{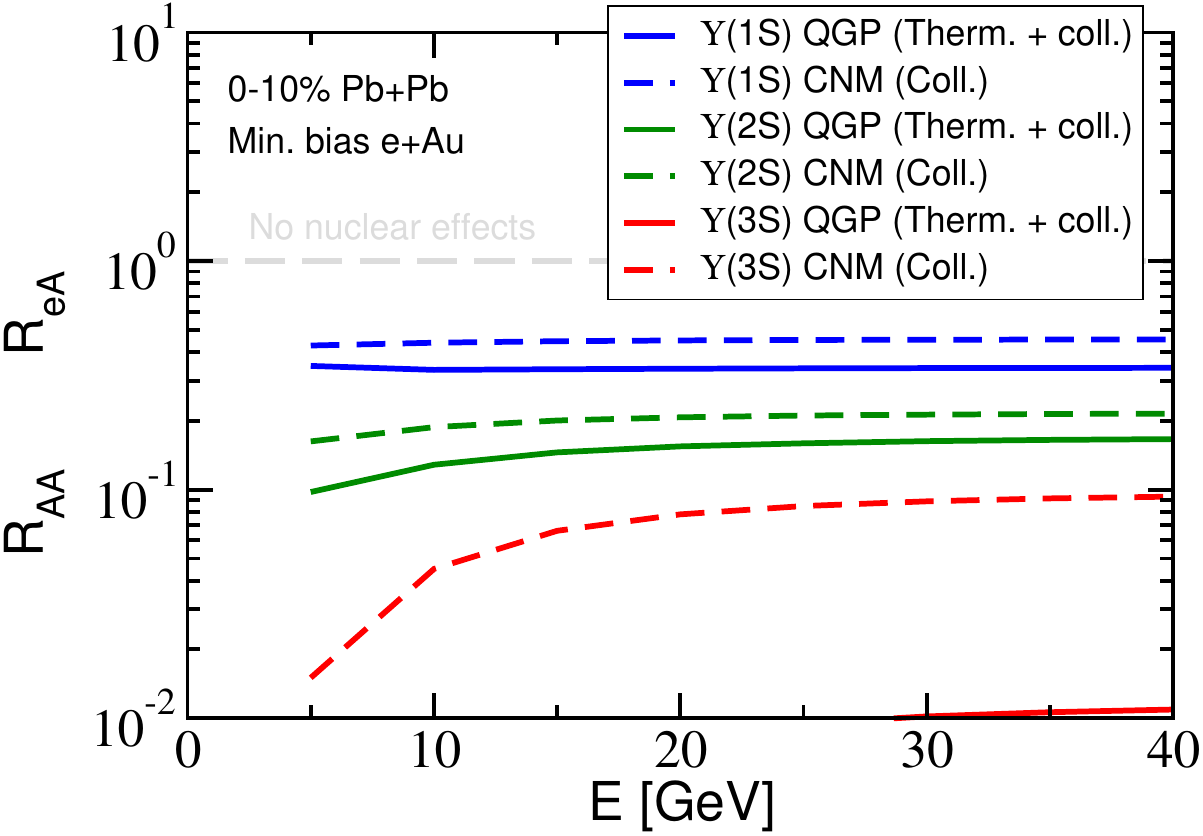}
\includegraphics[width=0.45\textwidth]{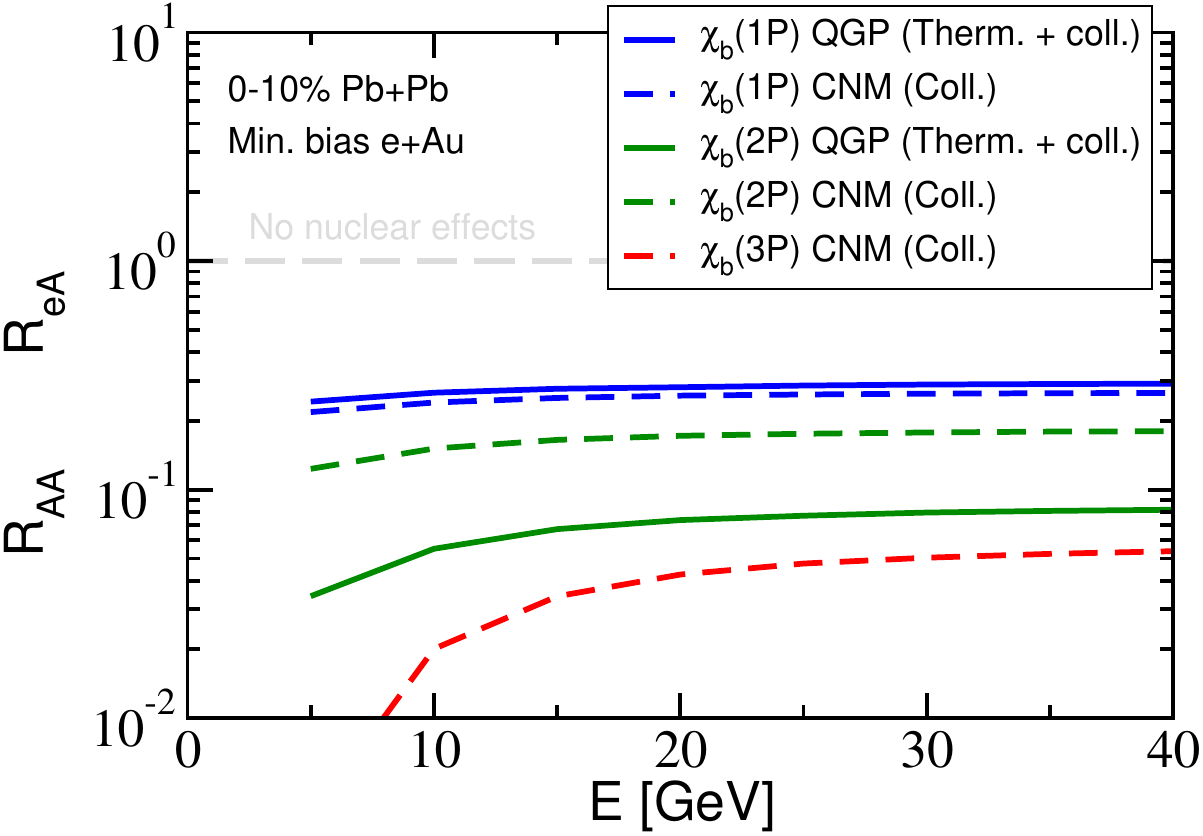}
\end{center}
\caption{Left: nuclear modification of the $\Upsilon$ states as a function of {the  $\Upsilon$} energy $E$ in the hadron centre-of-mass frame. The solid lines indicate that the calculation was done using thermal wave-function effects while traversing  the  QGP {and correspond to $R_{AA}$ in the centrality class 0-10\% in PbPb LHC collisions}. {The} dotted lines indicate that the calculation was done without thermal effects only Cold Nuclear Matter (CNM) effects) \newb{and correspond to $R_{eA}$ for minimum bias $e$Au collisions}. \newb{Results for} $1S$  {states} are shown in blue, $2S$ states are shown in green, and $3S$ states are shown in red. Right: the same ratios but for  $\chi_b$ states. \newb{The c}olour coding is similar but for $1P$, $2P$ and $3P$ states.  All calculations were done using \newb{direct} and ignoring feed-down effects. The initial suppression of $\chi_b(3P)$ is not shown because it has a very low $R_{AA}$ value, far lower than any other state pictured.}
\label{fig:BottomQGPCNM}
\end{figure}

\clearpage

\section{The lepton, photon and parton distribution in an unpolarised electron}
\label{sec:LDFs}
\begin{figure}[htbp]
	\begin{center}
		\includegraphics[width=\textwidth]{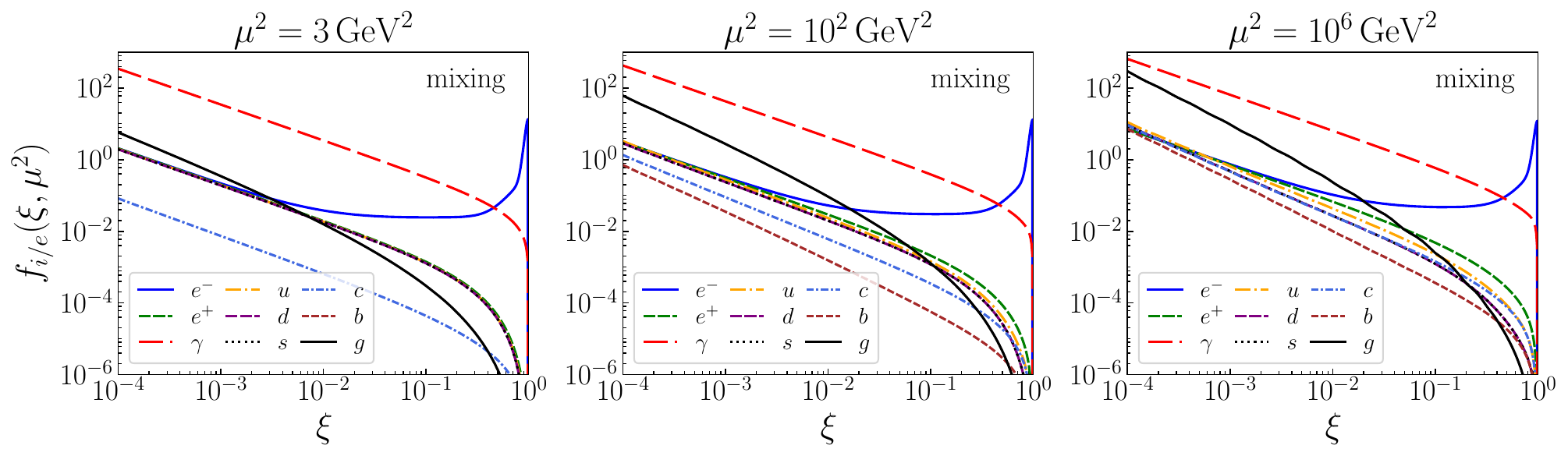}
		\includegraphics[width=\textwidth]{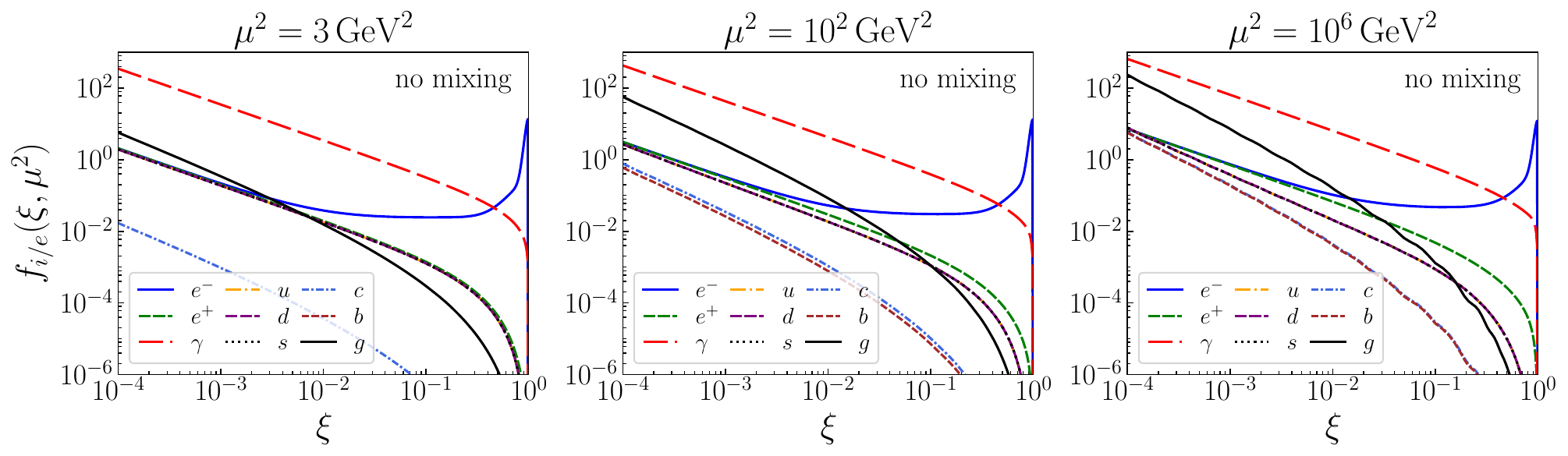}
\caption{
The lepton, photon and parton distribution in an unpolarised electron at $\mu^2=3~{\rm GeV}^2$, $10^2~{\rm GeV}^2$, and $10^6~{\rm GeV}^2$ are presented as a function of the longitudinal momentum fraction $\xi$~\cite{QW:2024}.  The upper (lower) figures represent LDFs with (without) the mixing of QED and QCD evolution. 
}
	\label{fig:LDFs}
	\end{center}
 \end{figure}
\clearpage

\bibliography{references}

\end{document}